\pdfoutput=1
\documentclass[
    aps,
    twocolumn,groupedaddress,superscriptaddress,
    footinbib,
    floatfix,
    showpacs,showkeys
    ]{revtex4-2} 
\usepackage[utf8]{inputenc}
\usepackage{amssymb,amsmath,graphicx}
\usepackage{bm}
\usepackage{makecell}
\usepackage{morefloats}

\usepackage[bookmarksnumbered,
            colorlinks = true,
            linkcolor = black,
            urlcolor  = blue,
            citecolor = blue,
            anchorcolor = blue]{hyperref}

\usepackage{xstring}
\usepackage{bbm}
\usepackage[capitalise]{cleveref}
\crefname{figure}{Fig.}{Figs.}
\usepackage{mathtools}

\graphicspath{{figures}}

\usepackage{orcidlink}

\DeclarePairedDelimiterX\abs[1]{\lvert}{\rvert}{#1}
\DeclarePairedDelimiterX\ket[1]{\lvert}{\rangle}{#1}
\DeclarePairedDelimiterX\bra[1]{\langle}{\rvert}{#1}
\DeclarePairedDelimiterX\braket[2]{\langle}{\rangle}{#1\,\vert\,#2}
\DeclarePairedDelimiterX\ketbra[2]{\lvert}{\rvert}{#1\rangle\langle#2}
\DeclarePairedDelimiterX\expval[2]{\langle}{\rangle}{#1\,\vert\,#2\,\vert\,#1}
\DeclarePairedDelimiterX\matel[3]{\langle}{\rangle}{#1\,\vert\,#2\,\vert\,#3}
\DeclarePairedDelimiterX\ep[1]{\langle}{\rangle}{#1}
\DeclarePairedDelimiterX\projector[1]{\lvert}{\rvert}{#1\rangle\langle#1}

\newcommand{\ccite}[1]{%
\IfSubStr{#1}{,}{refs.}{ref.~}\cite{#1}%
}
\newcommand{\Ccite}[1]{%
\IfSubStr{#1}{,}{Refs.~}{Ref.~}\cite{#1}%
}

\begin{document}
\pagestyle{plain}

\title{Few-body bound states in the anyon-Hubbard model}

\newcommand{\ITPTUB}{Institut f\"{u}r Physik und Astronomie, Technische Universit\"{a}t Berlin,
Hardenbergstr.~36, D-10623 Berlin, Germany}
\newcommand{\Harvard}{Department of Physics, Harvard University, Cambridge, MA 02138, USA}
\newcommand{\Hannover}{Institut für Theoretische Physik, Leibniz Universit\"{a}t Hannover, Hannover, Germany}
\newcommand{\LKB}{Laboratoire Kastler Brossel, Coll\`ege de France, CNRS, ENS-Universit\'e PSL, Sorbonne Universit\'e, 75005 Paris, France.}
\newcommand{\JILA}{JILA, NIST, and Department of Physics, University of Colorado, Boulder, CO, USA}

\author{Isaac Tesfaye\,\orcidlink{0009-0001-4194-3916}}
\email{i.tesfaye@tu-berlin.de}
\affiliation{\ITPTUB}

\author{Christina Mascherbauer}
\affiliation{\ITPTUB}

\author{Joyce Kwan}
\affiliation{\JILA}

\author{Perrin Segura}
\affiliation{\Harvard}

\author{Yanfei~Li}
\affiliation{\Harvard}

\author{Markus Greiner}
\affiliation{\Harvard}

\author{Luis Santos\,\orcidlink{0000-0001-7652-9574}}
\email{santos@itp.uni-hannover.de}
\affiliation{\Hannover}

\author{Andr\'{e} Eckardt\,\orcidlink{0000-0002-5542-3516}}
\email{eckardt@tu-berlin.de}
\affiliation{\ITPTUB}

\author{Brice Bakkali-Hassani\,\orcidlink{0009-0001-1615-1474}}
\email{brice.bakkali-hassani@lkb.ens.fr}
\affiliation{\LKB}

\begin{abstract}
    Quantum statistics in low-dimensional systems predicts anyonic particles with fractional exchange statistics which are neither that of bosons nor fermions. 
    While anyons are typically found in two dimensions as excitations of topologically-ordered states of matter, anyon-like exchange statistics has also been discussed in one dimension, for instance, in the context of the anyon-Hubbard model (AHM), the physics of which has recently been observed in experiment~\cite{Kwan2024,Dhar2025,Bakkali-Hassani2026}. 
    The AHM can be formulated in terms of bosons featuring density-dependent Peierls phases, described by a statistical phase angle $\theta$, which controls asymmetric transport and the formation of dynamically bound pairs at finite momentum. 
    Here, we show theoretically that the AHM also hosts exact two-body bound states in the continuum (BICs) for arbitrary $\theta\neq 0$, and genuine three- and four-body bound states. 
    Unlike conventional bound states stabilized by attractive (or repulsive) interactions, which are energetically localized with a large effective mass, these clusters here are bound by a purely kinematic mechanism 
    endowing them with fast chiral transport properties.
    We provide a simple variational approximation to the three-body bound states and explain their binding mechanism. 
    Moreover, we show that the signatures of three-body bound states in the AHM can be directly probed experimentally from the expansion dynamics starting from three localized particles. 
\end{abstract}

\date{\today}
\maketitle

\makeatletter
\let\origaddcontentsline\addcontentsline
\renewcommand{\addcontentsline}[3]{}
\makeatother

\hypersetup{linkcolor=blue}
\section{Introduction}

Unlike in three spatial dimensions, where quantum particles are either bosons or fermions, non-standard exchange statistics can emerge in lower dimensions.
A prime example is given by Abelian anyons~\cite{Wilczek1982} in two dimensions (2D), which acquire a phase $\theta$ upon exchange, giving rise to fractional statistics~\cite{Leinaas1977,Wilczek1982,Wu1984,Wu1984a,Greiter2024} and thus generalizing the cases of bosons ($\theta=0$) and fermions ($\theta=\pi$)~\cite{Frohlich1988,Frohlich1976}. These 2D braid anyons arise as the configuration space formed by the relative position of two particles becomes not simply connected at the point where the two particles coincide, such that the exchange paths with different windings around this defect of the space become topologically distinct. Such anyons can emerge as quasiparticle excitations in topologically ordered phases of matter such as fractional quantum Hall states~\cite{Tsui1982,Laughlin1983,Arovas1984,Bartolomei2020,Nakamura2020,Halperin1984} and spin liquids~\cite{Coldea2001,Kitaev2006a,Semeghini2021,Fradkin2013}. 
Recently, interest in 2D anyons has gained momentum as their non-Abelian versions hold promise for intrinsically fault-tolerant topological quantum computation~\cite{Nayak2008,Kitaev2003,Bravyi2006,Clarke2013,Andersen2023}. 

Although fractional exchange statistics seem to be absent in one dimension (1D), as particles need to pass each other in violation of the defect introduced by two-body hard-core constraints giving rise to the braid group, many 1D anyon models (as well as concrete realizations) have been proposed both in the continuum~\cite{Harshman2020,Bonkhoff2021,Kundu1999,Rabello1995} and on the lattice~\cite{Zhu1996,Keilmann2011,Greschner2015a, Nagies2024}. 
Starting with spinon excitations in spin chains satisfying generalized (Pauli) exclusion statistics~\cite{Haldane1991,Shastry1988,Haldane1988}, many 1D models featuring anyonic excitations, including the Kuramoto-Yokoyama model~\cite{Arikawa2001,Kato1998,Kuramoto1995,Kuramoto1991}, the Calogero-Sutherland model~\cite{Ha1994,Ha1995,Murthy1994}, and the Kundu-Lieb-Liniger model~\cite{Kundu1999,Rabello1995,Batchelor2006,Patu2007,Posske2017,Pattu2019,Hao2012,Zinner2015,Valiente2021,Bonkhoff2021}, have been proposed and investigated.
These developments have led to various experimental proposals for 1D anyon models, which are, for instance, based on the engineering of density-dependent gauge fields~\cite{Greschner2014,Greschner2015a,Cardarelli2016,Strater2016,Greschner2018a,Yannouleas2019,Gorg2019,Schweizer2019,Clark2018,Lienhard2020,Yao2022,Barbiero2019}. 

One particular model, which aims to realize Abelian anyons, is the paradigmatic 1D anyon-Hubbard model (AHM)~\cite{Zhu1996,Keilmann2011}.
It has received significant attention in recent years due to
its interesting properties~\cite{Keilmann2011,Keilmann2011,Greschner2015a,Arcila-Forero2016,Lange2017,Zhang2017b,Bonkhoff2025,Theel2025,Yang2026,Bonkhoff2026,Wei2026,Wang2026d,Zhou2026,Lyngfelt2026}, 
such as the continuous buildup of Friedel oscillations~\cite{Strater2016,Bonkhoff2021,Lange2017a}, asymmetric expansion dynamics~\cite{Liu2018b,Greschner2018a}, anyonic quasi-condensation at finite momenta~\cite{Tang2015,Keilmann2011,Greschner2015a} and the stabilization of two-body bound states \cite{Greschner2015a,Cardarelli2016,Greschner2018a,Zhang2017b}.
The interest in the AHM also lies in its connection to topological gauge theories~\cite{Kundu1999,Aglietti1996,Chisholm2022,Bonkhoff2021}, such as the chiral BF theory~\cite{Chisholm2022,Bonkhoff2021}, as recently realized in weakly interacting Raman-coupled Bose-Einstein condensates~\cite{Chisholm2022,Frolian2022}.

The AHM itself has recently been implemented with ultracold atoms in a driven optical lattice~\cite{Kwan2024,Bakkali-Hassani2026}, where the density-dependent Peierls phases were realized by using a three-tone Floquet protocol~\cite{Cardarelli2016}.
These experiments have studied quantum-walk-type expansion dynamics of two particles, revealing a crossover of bunching to anti-bunching in second-order correlations, asymmetric transport phenomena, as well as the formation of chiral two-particle bound states.
More recently, in an entirely different setting, the asymmetric quasimomentum distribution of 1D anyons has been observed by injecting a mobile impurity, whose momentum encodes the statistical angle $\theta$, into a strongly-interacting 1D Bose gas~\cite{Dhar2025,Wang2025}.

The AHM is known to give rise to exact two-particle bound states
\cite{Kwan2024,Bakkali-Hassani2026,Greschner2015a,Cardarelli2016,Greschner2018a,Zhang2017b}. 
Different from the bosonic case ($\theta=0$), where bound states can solely arise 
for non-zero on-site interactions \cite{Piil2007,Valiente2008,Valiente2010a,Winkler2006,Fukuhara2013,Kranzl2023,Krutitsky2016}, here exact two-particle bound-state solutions are also present in the absence of on-site interactions ($U=0$) for arbitrary statistical angles $\theta \neq 0$~\cite{Cardarelli2016,Zhang2017b,Greschner2018a} since the density-dependent Peierls phases mediate interactions between the particles.
In both recent optical lattice realizations of the AHM~\cite{Kwan2024,Bakkali-Hassani2026}, expansion dynamics starting from either a two-particle Fock state or an initial two-particle ground state confined to three lattice sites have revealed strong signatures of these chiral two-particle bound states. 
There, the narrowing of the density distribution close to the center of the lattice and the asymmetric transport in the expansion dynamics are both directly related to the high population of the two-body bound states, whose dynamics can be intuitively understood as a two-step tunneling process, corresponding to a specific path in two-particle configuration space~\cite{Bakkali-Hassani2026}. 
The formation of such chiral bound pairs is particularly striking since they are kinematically bound and remain fast and mobile, unlike their energetically bound bosonic counterparts, which become heavy localized objects once the on-site interaction is strong enough to bind them tightly~\cite{Winkler2006,Valiente2008}.
Based on these recent results for the two-particle problem in the AHM showing signatures of the anyonic two-body bound states~\cite{Bakkali-Hassani2026,Kwan2024}, several open questions remain.
First, one may ask whether genuine three-(or more) body bound states are also present in the AHM (in the absence of on-site interactions) and whether they can equally be understood intuitively via selected paths in the configuration space.
Second, if this is the case, can such three-body bound states be probed via the expansion dynamics of suitably prepared wave packets? 
Third, do these effects survive the impact of deviations from the exact AHM, 
as they result from the three-tone-drive implementation~\cite{Cardarelli2016} used in the recent Harvard experiments~\cite{Kwan2024,Bakkali-Hassani2026}?

In the following, we will answer these questions affirmatively and show that the AHM exhibits genuine three-body bound states, whose signatures can be directly probed in simple expansion dynamics of three-particle wave packets.
Interestingly, we also find two-body bound states in the continuum~\footnote{While preparing this manuscript, a complementary study of these two-body bound states in the continuum appeared~\cite{Bonkhoff2026}.} 
and that the three-body bound states become long-lived resonances when entering the continuum of scattering states. 
Lastly, we reveal that the AHM also hosts four-body bound states and end with brief remarks on the generalization to $N$-body bound states in the AHM. 

Moreover, we find that the clusters in the AHM remain light and travel at speeds comparable to those of single particles, captured by matrix elements of order $J$. 
This distinguishes them from few-body clusters of $N$ particles bound by strong attractive or repulsive on-site interactions, $\abs{U}\gg J$, which are tightly bound but move very slowly, with effective tunneling matrix elements of order $J^N/\abs{U}^{N-1}$ resulting from $N$th-order processes in perturbation theory. 
The essential difference is that for on-site interactions mobility is tied to the binding energy, both being set by $J/\abs{U}$, whereas the kinematic binding mechanism of the anyon clusters allows for a dispersion of order $J$ independent of the size of their binding gap.

\section{Three-particle problem}
\label{sec:3body-problem}
\begin{figure*}[t]
    \centering
    \includegraphics[width=2\columnwidth]{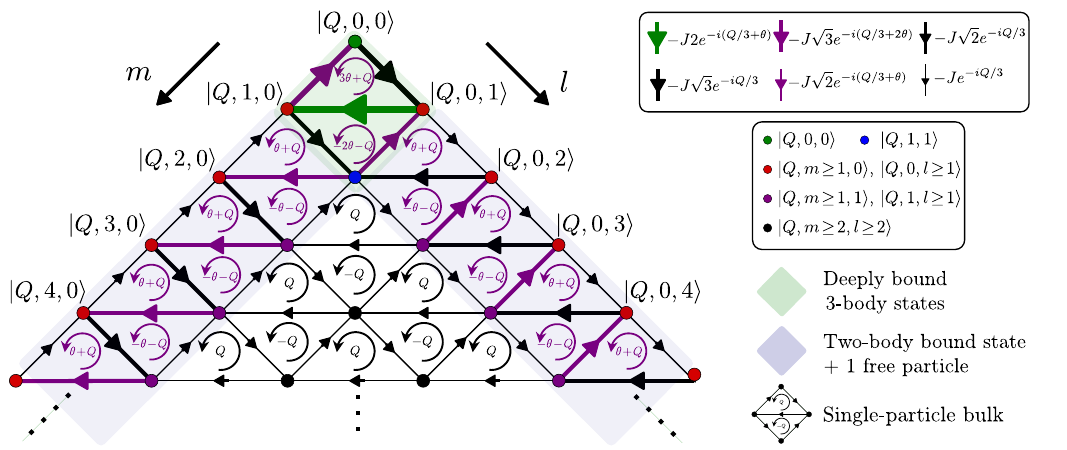}
    \caption{
    Three-particle relative coordinate space.
    Illustration of the semi-infinite 2D lattice associated with the relative coordinate space corresponding to the Hamiltonian $\hat{H}_Q$~\eqref{eq:3body-AHM-repr-hybrid}, which for each COM quasimomentum $Q$ is spanned by the basis states $\{\ket{Q,m,l}\}$.
    The matrix elements of $\hat{H}_Q$ define the tunneling processes between the different sites of this lattice, which are depicted by the arrows, whose color and thickness indicate the Peierls phase and the amplitude of the tunneling process, respectively.
    The values inside encircling arrows indicate the flux through the plaquettes of the lattice, corresponding to the phase that is acquired when tunneling around the plaquette along the direction of the arrow. 
    The green-shaded area indicates the region with deeply bound 3-body states with small relative coordinates $m,l$, the purple-shaded areas indicate the region where two-body bound states are present with the remaining particle being free, characterized by either of the relative coordinates $m$ or $l$ being small. 
   }
    \label{fig:Relative-coordinate-space}
\end{figure*}
In the bosonic representation, the AHM~\cite{Zhu1996,Keilmann2011,Greschner2015a,Cardarelli2016,Strater2016} is described by the Hamiltonian
    \begin{align}
        \hat{H}=-J\sum_{j}\left(\hat{b}_{j+1}^{\dagger} e^{-i \theta \hat{n}_{j+1}} \hat{b}_j + \text{h.c.}\right)
        + \frac{U}{2} \sum_j \hat{n}_j (\hat{n}_j-1),
        \label{eq:AHM}
    \end{align}
where $\hat{b}_j$ ($\hat{b}^{\dagger}_j$) denotes the annihilation (creation) operator for a boson at site $j$, $U$ the on-site interactions, $J$ the tunneling amplitude, and $e^{-i \theta \hat{n}_{j+1}}$ the number-dependent Peierls phases with the anyonic (statistical) exchange angle $\theta$. 
Here, we study the three-particle problem of the AHM in the bosonic representation~\eqref{eq:AHM}. 
Any state in the bosonic Fock space of three particles can be expressed in the configuration space basis $\ket{n,m,l}$. 
Here, $n$, $n+m$, and $n+m+l$ denote the position of the left, middle, and right particle, respectively, with $m\geq 0$ ($l\geq0$) denoting the relative distance between the two leftmost (rightmost) particles~[cf.~\cref{fig:three-body-spectrum-effective-AHM-U0-Binding-mechanism}(a)].
We can represent the AHM~\eqref{eq:AHM} for three particles as
    \begin{align}
        \hat{H}=\sum_{\substack{n,m,l\\ n',m',l'}} H_{n,m,l,n',m',l'} \ket{n,m,l}\bra{n',m',l'},
        \label{eq:3body-AHM-repr}
    \end{align}
where the matrix elements are specified in~\cref{eq:3body-AHM-repr-full} of Appendix~\ref{app:3body-AHM-repr}.
This allows us to study the three-particle problem as a single-particle problem in a three-dimensional configuration space given by a 3D simplex.

Due to the discrete translational invariance of the 1D lattice, we introduce the basis of states with fixed center-of-mass (COM) quasimomentum $Q$:
    \begin{equation}
    |Q,m,l\rangle = \sum_n e^{iQ\left( n+\frac{2m}{3} + \frac{l}{3}\right)} |n,m,l\rangle.
            \label{eq:COM-MOM-hybrid-basis}
    \end{equation}
This block diagonalizes the Hamiltonian $\hat H = \sum_Q \hat H_Q $, with 
    \begin{equation}
    \hat H_Q = \sum_{m,l;m',l'} H_{Q,m,l,m',l'} |Q,m,l\rangle \langle Q,m',l'|, 
    \label{eq:3body-AHM-repr-hybrid}
    \end{equation}
where an explicit expression for the matrix elements $H_{Q,m,l,m',l'}$ 
is visualized in~\cref{fig:Relative-coordinate-space} [see also~\cref{eq:3body-AHM-repr-hybrid-full} of App.~\ref{app:3body-AHM-repr}]. 
We can see that for each value of $Q$ the problem is mapped to a single-particle problem on a triangular lattice with staggered magnetic flux $\pm Q$ in the bulk and with open boundary conditions along two edges. 
These edges correspond to coincidences of the leftmost or rightmost particles ($m=0$ or $l=0$). 
The plaquettes in their immediate vicinity carry additional magnetic fluxes proportional to $\pm\theta$, and they meet at the upper corner ($m=l=0$) corresponding to three-body coincidences. 
In the following, we will refer to the full (quasi 3D) configuration space spanned by the basis states $\{\ket{n,m,l}\}$ simply as ``configuration space'' and to the two-dimensional subspace spanned by the basis states $\{\ket{Q,m,l}\}$ for each fixed COM quasimomentum $Q$ as ``relative coordinate space''.
\section{Three-body bound states}
\label{sec:3body-bound-states-AHM}
\begin{figure*}[t]
    \centering
    \includegraphics{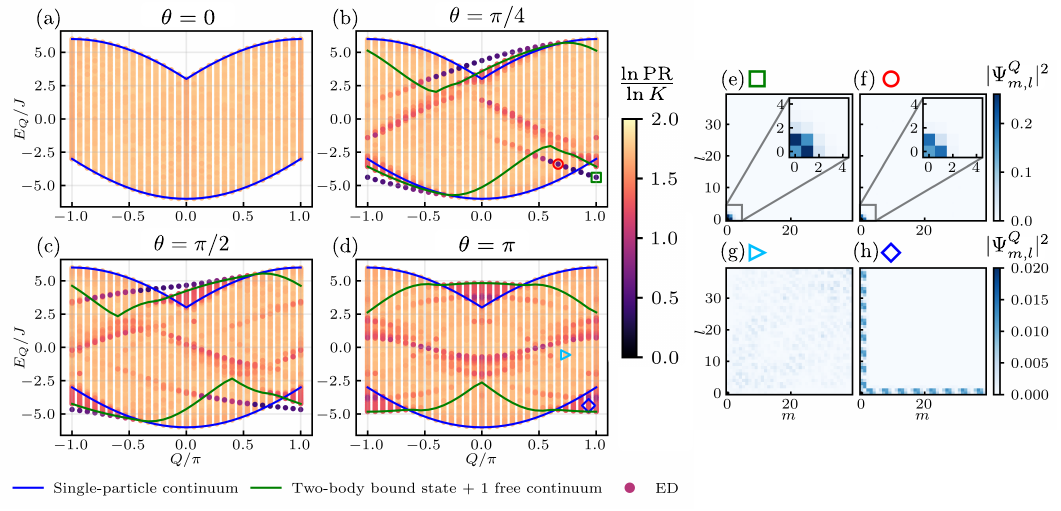}
        \caption{
        Three-particle energy spectrum of the AHM for $U=0$.
        Three-particle energy spectrum $E_Q$ as a function of the COM quasimomentum $Q$ for $\theta=0,\pi/4,\pi/2,\pi$ is shown in (a), (b), (c), and (d), respectively.
        The blue solid lines indicate the edges of the continuum of three independent particles, whereas the green solid lines provide the boundaries of the continuum spanned by two-particle bound states and one free particle.
        The exact diagonalization (ED) results for each $Q$ value are shown via the dots, whose color indicates the scaled participation ratio $\ln (\mathrm{PR})/\ln(K)$~\eqref{eq:participation-ratio}, whose value roughly indicates the dimension of the subspace over which the corresponding eigenstate $\ket{\Psi}$ is delocalized. Four characteristic states are indicated by the green square and red circle in (b) and by the cyan triangle and blue diamond in (d), whose density distribution $|\Psi^{Q}_{m,l}|^2$ is shown in the panels (e), (f), (g), and (h) respectively.
        They correspond to: (e) ``regular'' bound state outside the continuum, (f) bound state inside the continuum (BIC), (g) delocalized scattering state with three free particles (type-I), and (h) scattering states with a two-particle bound state and one free particle (type-II).
        Other parameters are $K=40$, and $31$ grid points for the COM quasimomentum $Q$ centered around $Q=0$.
        Due to visually overlapping data points from ED in (a)-(d), the appearance of data points is sorted according to their participation ratio $\mathrm{PR}$, i.e., states with smaller $\mathrm{PR}$ are plotted on top of states with larger $\mathrm{PR}$.
       }
    \label{fig:three-body-spectrum-full-AHM}
\end{figure*}
We diagonalize $\hat H_Q$~\eqref{eq:3body-AHM-repr-hybrid} and analyze the corresponding three-particle eigenspectrum as a function of the COM quasimomentum $Q$. When performing the exact diagonalization~(ED) of $\hat H_Q$, we truncate the possible values of the relative coordinates to inter-particle distances of up to $K$, i.e., $m, l \in \{0, \ldots ,K -1 \}$, 
so that $\hat H_Q$ becomes a $K^2 \times K^2$ matrix. 
This, roughly, corresponds to systems of $2K$ lattice sites. 
In \cref{fig:three-body-spectrum-full-AHM}, we depict the energy spectrum $E_Q$ for different statistical angles $\theta \in \{0,\pi/4,\pi/2,\pi\}$ and $U=0$.

The eigenstates of the three-particle AHM Hamiltonian $\hat H_Q$ can be scattering- or bound-state solutions. The bound-state solutions are localized around the top corner $(m,l)=(0,0)$ of the relative state space [cf.~Fig.~\ref{fig:Relative-coordinate-space}], i.e., 
their wave functions $\Psi^{Q}_{m,l}$ decay with increasing relative coordinates $m$ and $l$. 
The scattering states belong to two types of continua (see also Appendix~\ref{app:3body-scattering-state-continuum}). 

Type-I scattering states describe three free particles and are fully delocalized in the bulk of relative coordinate space~($m,l\gg 1$ in~\cref{fig:Relative-coordinate-space}) such that the continuum energy spectrum of these states is bounded by
 $\min_{\sum_i k_i=Q}\big[\sum_{i=1}^3\epsilon(k_i)\big] 
    \leq E_Q \leq 
    \max_{\sum_i k_i=Q} \big[\sum_{i=1}^3\epsilon(k_i)\big]$, 
with $\epsilon(k_i)=-2J\cos(k_i)$ being the single-particle dispersion relation and $k_i$ ($i\in\{1,2,3\}$) being the individual quasimomenta of the three particles, which themselves are constrained by the total COM quasimomentum $Q=k_1+k_2+k_3$.
The minimum and maximum energies of these bounds are given by $E_{\text{min}}(Q)=-6J\cos(Q/3)$, $E_{\text{max}}(Q)=6J\cos\big(\frac{\pi-\abs{Q}}{3}\big)$ for $Q\in[-\pi,\pi)$.

Type-II states describe a two-particle bound pair plus one free particle, and are localized near the edges defined by $m=0$ or $l=0$ in~\cref{fig:Relative-coordinate-space}. As the energy spectrum $E_{\pm}(q)$ of the two-particle bound state with two-particle COM quasimomentum $q$ is analytically known for arbitrary statistical angles $\theta$ and on-site interactions $U$~\cite{Kwan2024}, we can express the total energy as the sum of the two-particle bound-state spectrum and the kinetic energy of a single free particle, such that the bounds of this continuum spectrum are given by
$\min_{q+k_3=Q}[\epsilon(k_3)+E_{\pm}(q)] \leq E \leq \max_{q+k_3=Q}[\epsilon(k_3)+E_{\pm}(q)]$, where
$Q=q+k_3$ is the total three-particle COM quasimomentum, and $\epsilon(k_3)$ is the kinetic energy of the third free particle. Different from the type-I continuum described above, this continuum explicitly depends on $\theta$ and $U$.

Type-I and type-II continua are shown in~\cref{fig:three-body-spectrum-full-AHM} for different $\theta$ and $U=0$. 
States lying outside both continua are identified as three-body bound states. 
In \cref{fig:three-body-spectrum-full-AHM}(b) and (c) for $\theta=\pi/4$ and $\theta=\pi/2$, respectively, we indeed observe that such three-body bound states outside the continua exist.

To gain a more quantitative measure of how localized the solutions are in relative coordinate space, we compute for each eigenstate $|\Psi\rangle = \sum_{m,l} \Psi_{m,l}^Q |Q,m,l\rangle$, the participation ratio (PR), 
    \begin{equation}
        \mathrm{PR} \equiv \bigg[\sum_{m,l} |\Psi_{m,l}^Q|^4\bigg]^{-1}.
        \label{eq:participation-ratio}
    \end{equation}
We then introduce the scaled participation ratio $\mathrm{SPR}\equiv \ln(\mathrm{PR})/\ln(K)$ (also called the fractal dimension in the context of localization transitions~\cite{Evers2008}). 
We may expect three characteristic values for the SPR, which reflects the dimension of the subspace in the relative coordination space, over which a state is spread: 
(i) for scattering states of type-I, which are delocalized across the entire 2D relative coordinate space,  $\mathrm{SPR}\sim 2$; (ii) for scattering states of type-II, delocalized along only one direction in the relative coordinate space, $\mathrm{SPR}\sim 1$; and (iii) for three-body bound states, fully localized in the relative coordinate space, $\mathrm{SPR} \sim 0$. 
This is confirmed in~\cref{fig:three-body-spectrum-full-AHM}. 

For $\theta=0$~(a), the AHM at $U=0$ describes non-interacting bosons so that no two- or three-particle bound states exist. 
All states belong to the type-I continuum~(bounded by the blue solid line) and~$\mathrm{SPR}\sim 2$. 
For $\theta=\pi$~(d), we find no solutions outside both the type-I continuum and the type-II continuum~(green solid line). 
Solutions inside both continua have $\mathrm{SPR}\sim 2$, indicating type-I states. 
This is visible in the density distribution shown in (g) for the example state marked by a light blue triangle in (d). 
In contrast, solutions lying in between the two continua correspond to type-II states, with $\mathrm{SPR}\sim 1$. 
The density distribution in relative coordinate space plotted in subfigure (h) for the example state marked by a blue diamond in (d) shows the expected localization along the edges defined by $m = 0$ and $l = 0$. 
For $\theta=\pi/4$~(b), and $\theta=\pi/2$~(c), we find solutions lying outside both continua, which present $\mathrm{SPR}\sim 0$, as expected for three-body bound states. We plot in (e) the density distribution of one of these states, marked by a green square in (b). It clearly reveals localization at small $m$ and $l$.

Interestingly, the three-body bound states appear to survive when entering the continuum, hinting at the possible presence of bound states inside the continuum (BIC)s~\cite{Hsu2016,vonNeuman1929,Friedrich1985,Plotnik2011,Zhang2012,Zhang2013,Longhi2007,Cerjan2019,Cerjan2020,Huang2024,Qin2024,Qian2024}. 
This seems to be confirmed also by the density distribution of~\cref{fig:three-body-spectrum-full-AHM}(f).  However, we will argue below that these states cannot be genuine BICs and that in the limit of infinite system-sizes they will correspond to slowly decaying resonances. 
It is, nevertheless, remarkable to see that for $K$ as large as $50$ we still find localized states with $\mathrm{SPR}\sim 0$. 
This indicates very long lifetimes $\tau$ such that the decay rate $\tau^{-1}\sim \Gamma$ of these states is much smaller than the finite-size level spacing, so the resonances are well resolved and do not hybridize with the scattering continuum. 
For the typical experimental system sizes of about 10 to 100 lattice sites, these states will, therefore, not be distinguishable from true BICs. We, thus, call them quasi-BICs (qBICs) in the following.
Note that for fractional values of $\theta\neq 0,\pi$, the existence of qBICs cannot be explained by symmetry-decoupled subspaces~\cite{Hsu2016}, highlighting the non-trivial nature of these qBICs in the AHM~\footnote{Note the parity symmetry is broken in this case~\protect~\cite{Liu2018b} and neither the anti-unitary time reversal symmetry~\protect~\cite{Lange2017a} nor the chiral symmetry (for $U=0$)~\protect~\cite{Theel2025} of the AHM, block-diagonalize the Hamiltonian to allow for symmetry-decoupled subspaces. The qBICs are also present when the chiral symmetry is broken by any finite on-site interaction $U$ (also see Fig.~\ref{fig:three-body-spectrum-U4-AHM}).}.

\begin{figure}[t]
    \centering
    \hspace*{-.2cm}
    \includegraphics[width=\columnwidth]{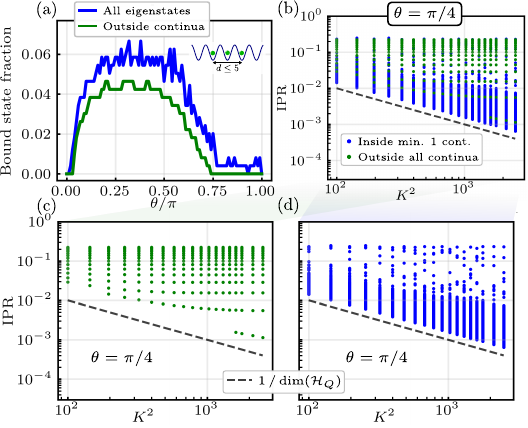}
    \caption{
    Three-body bound state fraction and finite-size scaling of the three-particle spectrum.
    (a) Bound state fraction as a function of the statistical angle $\theta$ for the entire three-particle spectrum~\eqref{eq:3body-AHM-repr-hybrid} defined as the fraction of states with $\mathrm{PR}$~\eqref{eq:participation-ratio} below that of states where the left and right particle are at most separated by five sites for all eigenstates in blue and for eigenstates outside the continua in green. 
    (b) Inverse participation ratio $\mathrm{IPR}$~\eqref{eq:participation-ratio} for a fixed statistical angle $\theta=\pi/4$ as a function of the size of the relative coordinate space $K^2$ for the entire three-particle spectrum~\eqref{eq:3body-AHM-repr-hybrid} where $K^2$ ranges from $10^2$ to $50^2$ sites.
    The green (blue) markers correspond to eigenstates lying outside both continua (inside at least one of the two continua).
    (c) Same as (b) but only for states outside both continua. 
    (d) Same as (b) but only for states inside at least one of the two continua. 
    All other parameters are the same as in Fig.~\ref{fig:three-body-spectrum-full-AHM}, where in (a)~$K=40$ sites along each of the relative coordinate directions $m$ and $l$ are used.
    }
    \label{fig:Part-Ratio-Bound-state-fraction-U0}
\end{figure}
Let us have a closer look at these qBICs. In a finite system, we cannot sharply distinguish between the bound and scattering states. Instead, we compute the maximum possible PR that one can obtain for the left and right particles to be at most separated by five sites, i.e., $d\equiv m + l \leq 5$, and define all states with a PR below this threshold as three-body bound states. 
Although this threshold is somewhat arbitrary, qualitatively similar results are obtained for other threshold values. 
The fraction of states with PR below the threshold is plotted in~\cref{fig:Part-Ratio-Bound-state-fraction-U0}(a). 
Most of the three-body (quasi) bound states are found for $\theta<3\pi/4$, where the difference between bound state fraction for all states (blue) and for the states outside both continua (green) indicates the qBIC fraction. 

To estimate finite-size effects, we compute, for an exemplary fixed statistical angle of $\theta=\pi/4$, the inverse participation ratio $\mathrm{IPR}=1/\mathrm{PR}$~\eqref{eq:participation-ratio} 
for the entire spectrum as a function of the system size $K^2$ [\cref{fig:Part-Ratio-Bound-state-fraction-U0}(b) and (c)-(d)]. 
We indicate states lying outside both continua by green dots, whereas states lying inside at least one of the two continua are marked by blue dots. 
We find that most scattering states in the bulk of the relative coordinate space are delocalized across the entire relative coordinate space, as their $\mathrm{IPR}$ decreases with the inverse of the dimension of the relative coordinate space, i.e., $\mathrm{IPR} \sim 1/\dim{\mathcal{H_Q}}=1/K^2$, 
while the three-body qBICs remain localized up to system sizes as large as $K^2=(50)^2$, making them indistinguishable from true BICs on experimentally relevant system sizes. 

Generalizing the experimental realization of the AHM for two particles~\cite{Kwan2024,Bakkali-Hassani2026} using a three-tone drive~\cite{Cardarelli2016} to the case of three and more particles, we find a model that slightly deviates from the AHM whenever three or more particles come close to each other. 
Namely, three-particle coincidences are excluded and the density-dependent Peierls phase of the AHM for processes $\ket{\dots21\dots} \leftrightarrow \ket{\dots12\dots}$ is missing. 
We refer to this model as the \emph{experimentally constrained} AHM in the following. 
We show in Appendix~\ref{app:experimentally-constrained-3body-AHM} that the three-body bound states and the qBICs survive this deformation of the model so that they can readily be investigated using existing experimental techniques.

\section{Explanation of three-body bound-state solutions}
\begin{figure}[!t]
    \centering
    \includegraphics[width=\columnwidth]{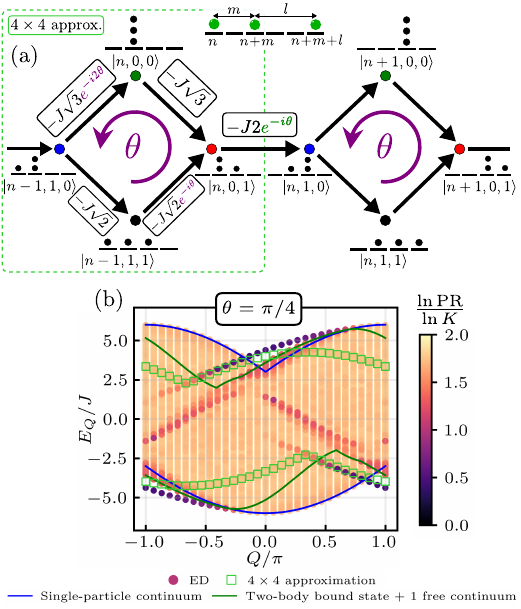}
    \caption{
    Effective three-particle energy spectrum for (potential) three-body bound states in the AHM for $U=0$.
    (a) Ansatz for three-body bound states, including only configurations with particles sitting on neighboring sites and the tunneling matrix elements connecting them, which is obtained from truncating the three-particle configuration space to basis states $\ket{n,m,l}$ with $0\leq m,l\leq1$. 
    In the experimentally constrained version of the AHM~\cite{Cardarelli2016,Kwan2024,Bakkali-Hassani2026} three-particle coincidences $\ket{n,0,0}$ (green-colored sites) and the Peierls phases for tunneling from an originally doubly to a singly occupied site (red- to blue-colored sites) are excluded and set to zero, respectively.
    (b) An effective energy spectrum for potential three-particle bound states is obtained by identifying a four-site unit cell in the configuration space (enclosed green rectangle in (a)), which in the relative coordinate space corresponds to the green-shaded area of deeply bound 3-body states region in Fig.~\ref{fig:Relative-coordinate-space}. 
    The effective and complete three-particle energy spectra for the statistical angles $\theta=\pi/4$ with light green squares indicate the effective spectrum $E_{Q,n}^{\text{eff}}$ (where only the lowest ($n=1$) and highest effective band ($n=4$) are shown). 
    As before,
    the color scale indicates the scaled participation ratio $\ln(\mathrm{PR})/\ln(K)$~\eqref{eq:participation-ratio} of the exact eigenstates.
    Other parameters for (b) and (c) are the same as in Fig.~\ref{fig:three-body-spectrum-full-AHM}.
    }
    \label{fig:three-body-spectrum-effective-AHM-U0-Binding-mechanism}
\end{figure}
\subsection{Binding mechanism for three-particle bound states outside the continuum.}
\label{sec:Binding-mechanism-outside-AHM}
The mechanism for three-particle binding in the AHM with $U=0$ below the scattering continuum can be intuitively understood by truncating the three-particle configuration space, such that, modulo COM translations, only four Fock states are kept, in which the particles remain close together. 
This truncation defines an ansatz for potential three-body bound states, which is depicted in~\cref{fig:three-body-spectrum-effective-AHM-U0-Binding-mechanism}(a). 
It defines a 1D subspace along the direction of the COM position. 
According to the Rayleigh-Ritz variational principle, it is sufficient to show that the ansatz provides energies below those of the two scattering continua (which are both analytically known), to prove the existence of three-body bound states outside the continua. 
For such bound states, the kinetic energy can only be lowered by delocalizing along one direction in configuration space, rather than along three directions as for type-I scattering states (or two as for type-II scattering states). This must be compensated by other effects to stabilize these bound states at energies below the scattering continua. 
The first effect is that, since the three particles stay close together, tunneling processes are bosonically enhanced by factors of either $\sqrt{2}$, $\sqrt{3}$, or $2$~[\cref{fig:three-body-spectrum-effective-AHM-U0-Binding-mechanism}(a)]. 
The second effect is that those tunneling processes are additionally associated with a Peierls phase of $\pm \theta$ or $\pm 2\theta$.
As a result, these bound states minimize their kinetic energy at a finite $Q$ determined by $\theta$, which differs from the values of $Q$ for which scattering states minimize their energy. 
Hence, for finite $\theta$ the kinetic energy of the three-body bound states can lie for some $Q$ values below the type-I and type-II continua. 
This is confirmed in Fig.~\ref{fig:three-body-spectrum-full-AHM}(b) and (c).  

The above reasoning can also be viewed on the level of the relative coordinate space depicted in~\cref{fig:Relative-coordinate-space}. 
Here, the bosonic enhancement gives rise to tunneling matrix elements that are stronger along the edges (where two particles are close together) and strongest at the upper corner (where all three particles meet). 
Moreover, for COM quasimomenta $Q$ close to $\pi$ the effective plaquette fluxes of $\pm Q$ in the bulk induce kinetic frustration, preventing delocalized type-I states from fully minimizing their kinetic energy.
This frustration can be reduced by the additional $\theta$-dependence of the plaquette fluxes at the edges as well as at the upper corner. 
For appropriate $\theta$, the latter effect can favor type-II states or three-body bound states compared to type-I states as well as with respect to each other.

\subsection{Effective three-particle spectrum for three-body bound states in the AHM}
\label{sec:approximate-spectrum-AHM}
We will now quantify these arguments using the truncated three-particle configuration space depicted in~\cref{fig:three-body-spectrum-effective-AHM-U0-Binding-mechanism}(a) to capture the physics of three-body bound states.
Moving to the relative coordinate space representation~\eqref{eq:COM-MOM-hybrid-basis}, we obtain for each COM quasimomentum $Q$ an effective $4\times 4$ Hamiltonian $H_Q^{\mathrm{eff}}$. 
It acts on the four sites inside the green-shaded area depicted in Fig.~\ref{fig:Relative-coordinate-space} at the origin of the relative coordinate space. 
Diagonalizing this problem, we obtain an effective three-body spectrum, $E_{Q,n}^{\mathrm{eff}}$ with $n=4$~energy bands (see App.~\ref{app:3body-effective-model-4x4} for details). 
We depict the lowest and highest energy bands of this spectrum in~\cref{fig:three-body-spectrum-effective-AHM-U0-Binding-mechanism}(b) for $\theta = \pi/4$ together with the entire exact spectrum of~\eqref{eq:3body-AHM-repr-hybrid}. 
The upper and lower branches of the effective spectra mirror the three-body bound states outside the continuum for those $Q$ values where they exist. 
However, due to the variational nature of the approach, the ansatz only captures qualitatively the bound states outside the continuum for $Q$ close to $0$ and $\pi$. 
Away from these points, the effective bands show considerable deviations from the exact bound-state energies. 
Interestingly, the uppermost and lowermost states of the truncated model also provide fairly good approximations to the exact energies of some of the qBIC branches. 
Finally, we note that the truncated model described in this section also provides a good approximation for three-body bound states for the case of $U \neq 0$ (see App.~\ref{app:3body-spectrum-finite-U}). 

\subsection{Two-body bound states in the continuum}\label{sec:2BodyBIC}
\begin{figure}[t]
    \centering
    \includegraphics[width=\columnwidth]{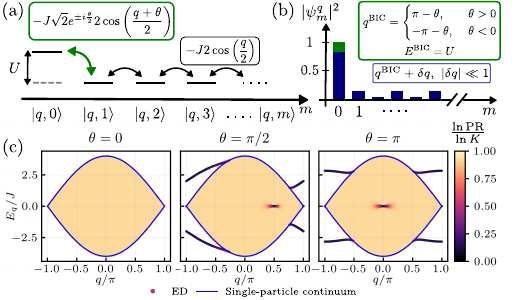}
    \caption{
    Two-body bound states in the continuum.
    (a) Illustration of the Hamiltonian $\hat{H}_q$~\eqref{eq:2body-AHM-repr-hybrid} acting in the relative coordinate space of the two-particle AHM with COM quasimomentum $q$, whose states $\{\ket{q,m}\}$ form a semi-infinite chain. 
    (b) Sketch of the weights $|\psi^{q}_{m}|^2$ for two types of bound states of the wave function in the relative coordinate space, for the special bound states: (i) perfect BIC (green) where $q^{\text{BIC}}=n\pi-\theta$ for $\theta\neq 0$ and $n\in\{-1,1\}$ and (ii) quasi-BICs in the proximity of the perfect BIC condition $q=q_{\mathrm{BIC}} + \delta q$ (blue), which are highly localized for $\abs{\delta q}\ll 1$, and
    leak into the continuum as a traveling wave $\psi^{q}_{m\geq 1} \sim  \psi^{q}_{0}e^{-i\kappa m}v^*(q)/t(q)$ 
    with the amplitude $|v(q)/t(q)|$ controlling the degree of localization~(see App.~\ref{app:2body-BIC} for details). 
    (c) Energy spectrum $E_{q,\kappa}$ of the two-particle AHM as a function of the COM quasimomentum $q$ (evaluated at $201$ grid points). The (scaled) participation ratio $\mathrm{PR} = \left[\sum_{m} |\psi^{q}_{m}|^4\right]^{-1}$ of the eigenstates is indicated by color. The chain is truncated after $K=100$ sites and $U=0$.}
    \label{fig:2-body-BICs}
\end{figure}
Before considering the case of qBICs, it is helpful to briefly discuss the two-body problem. 
We may write $\hat H = \sum_q \hat H_q$, with
\begin{align}
    \hat{H}_q & =  U\ketbra{q,0}{q,0}+\big[v(q)\ketbra{q,0}{q,1} +\text{h.c.}\big] \nonumber \\
    &+ \sum_{m\geq2}t(q)\big(\ketbra{q,m}{q,m\!-\!1}  +\text{h.c.}\big),
    \label{eq:2body-AHM-repr-hybrid}
\end{align}
where $\ket{q,m}$ denotes the two-body state with two-particle COM quasimomentum $q$, and relative distance $m\geq 0$. 
The Hamiltonian $\hat{H}_q$, which is visualized in~\cref{fig:2-body-BICs}(a), describes an effective site, given by $\ket{q,0}$, with on-site energy $U$ that is coupled via~$v(q)=-2\sqrt{2}J\cos([q+\theta]/2)e^{-i\theta/2}$ to a semi-infinite 1D chain of sites $\ket{q,m}$ with $m\geq1$ and nearest-neighbor hopping $t(q)=-2J\cos(q/2)$.  Diagonalizing $\hat H_q$, we find both scattering and bound states. 
Two-body bound states are localized in the vicinity of the edge of the chain, $m=0$. 
The regular bound states are located outside the two-particle scattering state continuum, which has energies $|E_{q}|\leq |2t(q)| $~\cite{Greschner2018,Cardarelli2016,Zhang2017b,Kwan2024}.
The AHM may, however, also host exact two-body BICs resulting from a perfect destructive interference effect. 
In particular, if $\cos((q + \theta)/2) = 0$, then $v(q)=0$, and $\ket{q,0}$ decouples, resulting in an on-site bound state, with $\mathrm{PR}=1$ ($\mathrm{SPR}=0$) and energy $U$, which is a BIC if $\abs{U}<4J|\sin(\theta/2)|$, see~\cref{fig:2-body-BICs}(b)-(c)~\footnote{We note that two-particle anyonic BICs in a 1D lattice were previously discussed in~\cite{Zhang2023a}. However, as pointed out in~\cite{Zheng2024}, the analysis of Ref.~\cite{Zhang2023a} relies on an overcomplete, non-orthogonal basis for the two-anyon Hilbert space. This leads to non-physical redundant eigenstates whose wave function amplitudes violate the anyonic commutation relations~\cite{Zheng2024}, complicating the identification of genuine BIC there. 
Our analysis avoids this issue entirely by working directly in the \emph{bosonic} relative coordinate space representation~\eqref{eq:2body-AHM-repr-hybrid}, which is formulated in terms of the orthogonal and complete basis $\ket{q,m}$ so that only physical eigenstates are obtained ab initio.
This representation also makes the BIC condition analytically transparent:
a perfect two-body BIC arises exactly when $\cos\!\left([q+\theta]/2\right)=0$, as derived below.}. 
In the proximity of the perfect BICs, $q=q_{\mathrm{BIC}} + \delta q$, we find quasi-BICs with $\mathrm{SPR}\sim  0$, for which the coupling to the rest of the chain is small but non-zero~[\cref{fig:2-body-BICs}(b)-(c)].
These two-body quasi-bound states have a decay rate $\Gamma \propto \delta q^2$. 
Thus, in any realistic finite-size system, the quasi-BICs will be highly localized and well-resolved as resonances in the spectrum, as long as the finite-size level spacing $\delta E \sim J/L>\Gamma$~(see App.~\ref{app:2body-BIC} for details). 
We briefly note that our findings regarding the exact two-body BIC
solutions are fully consistent and complementary to the recent analysis of~\Ccite{Bonkhoff2026}, where a similar low-entangled ``Doublon State'' was found for open boundary conditions. 

\subsection{Three-body quasi-bound states in the continuum}{\label{sec:3BodyBIC}}
\begin{figure}[t]
    \centering
    \includegraphics[width=\columnwidth]{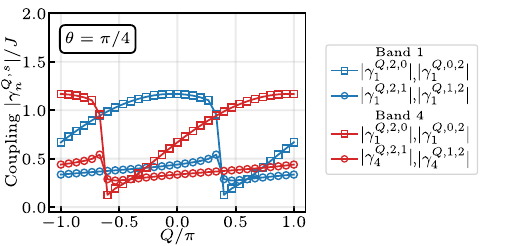}
    \caption{Three-body quasi-bound states in the continuum.
    Individual coupling amplitudes $|\gamma^{Q,s}_{n}|$ of the energy eigenstates of the deeply bound region~[cf.~Fig.~\ref{fig:Relative-coordinate-space}], corresponding to the lowest (Band $n=1$, blue) and highest (Band $n=4$, red) energy, to the neighboring four bulk sites $s\equiv m,l \in \{(2,0),(0,2),(1,2),(2,1)\}$ as a function of $Q$ for $\theta=\pi/4$. 
    }
    \label{fig:3-body-BICs}
\end{figure}
In the following, we investigate the nature of the (quasi-) three-body bound states in the continuum, showing that for very large but finite systems they will have a large but finite lifetime. 
Three-body bound states, i.e., localized states at the top upper corner of the relative coordinate space~[\cref{fig:Relative-coordinate-space}], with energies above or below the scattering continua, can have a finite exponentially decaying weight, when moving away from the top corner, either along the edges or into the bulk. 
In turn, three-body bound states in the scattering continuum would require perfect localization at the upper corner, such that tunneling away from this subspace to another neighboring state is completely suppressed via destructive interference. 
However, such an interference would require that any neighboring state couples to at least two states of the localized subspace. 
It is however easy to see from~\cref{fig:Relative-coordinate-space}, that regardless of which subset of states one picks as the localized manifold, there are always neighboring states which are coupled to the subspace by a single tunneling process. As a consequence, there cannot be true three-body BICs in the AHM. 

However, as shown above, in a large but finite system, we may find inside the scattering spectrum three-particle bound states essentially localized on the top four corner states in~\cref{fig:Relative-coordinate-space}. This may be understood by considering the energy eigenstates $|Q,n\rangle$ on the truncated subspace spanned by these four corner states (with $n$ labeling the four eigenstates). For the neighboring states $|Q,s\rangle$~($|Q,2,0\rangle$, $|Q,0,2\rangle$, $|Q,2,1\rangle$, and $|Q,1,2\rangle$), we compute the couplings $\gamma^{Q,s}_n=\langle Q,s|H_Q|Q,n\rangle$. 
If for some $n$, all $\gamma^{Q,s}_n$, although non-zero, remain small compared to the finite-size level splitting of the scattering states, such a state will remain localized (provided its energy is not too close to that of a bulk scattering state) and cannot be distinguished from a true BIC, even though it will slowly decay in an infinite system~(see App.~\ref{app:3body-BIC-mechanism}). 

In Fig.~\ref{fig:3-body-BICs} we depict $\gamma^{Q,s}_n$ for the lowest~($n=1$) and highest~($n=4$) eigenstates, corresponding to the lowest  and highest effective bands in Fig.~\ref{fig:three-body-spectrum-effective-AHM-U0-Binding-mechanism}(b). 
The coupling is indeed strongly suppressed at those $Q$ values where qBICs were found in the upper and lower branches of the full model, see~\cref{fig:three-body-spectrum-full-AHM} and~\cref{fig:three-body-spectrum-effective-AHM-U0-Binding-mechanism}(b). 
Note that the existence of highly-localized qBICs is particularly remarkable, since it demands the near-cancellation of all four complex-valued couplings $\gamma^{Q,s}_n$, rather than just one as in the two-body problem, with only one free parameter~$Q$. 

\section{Four-body bound states}
\begin{figure*}[t]
    \centering
    \includegraphics{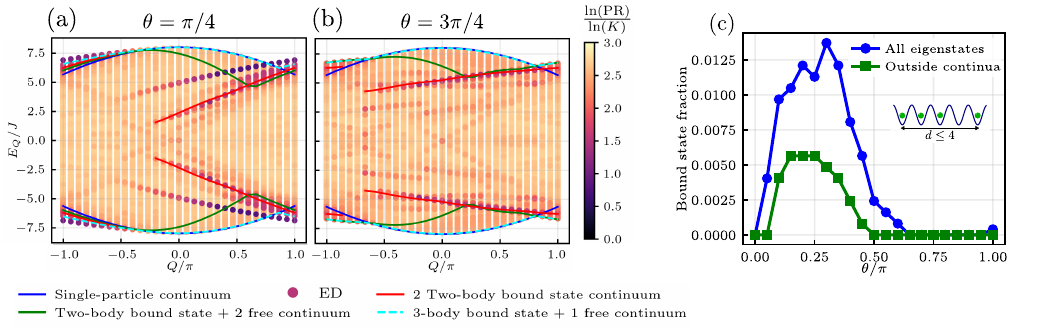}
        \caption{
        Four-particle energy spectrum of the AHM for $U=0$ and four-body bound state fraction.
        Four-particle energy spectrum $E_Q$ as a function of the COM quasimomentum $Q$ for $\theta=\pi/4,3\pi/4$ is shown in (a) and (b), respectively. 
        The blue solid lines indicate the edges of the continuum of four independent particles, the green solid lines provide the boundaries of the continuum spanned by two-particle bound states and two free particles, the red solid lines indicate the boundaries of the continuum of two independent two-particle bound states, and the cyan dashed lines indicate the boundaries of the continuum of a three-particle bound state with one free particle.
        The ED results for each $Q$ value are shown via the dots whose color indicates the scaled participation ratio $\ln(\mathrm{PR})/\ln(K)$ of the corresponding eigenstate of the Hamiltonian $\hat{H}_Q$. 
        (c) Bound state fraction as a function of the statistical angle $\theta$ for the entire four-particle spectrum of the AHM, which is defined as the fraction of states with $\mathrm{PR}$ below that of states where the distance between the left- and rightmost particles is at most four sites, for all eigenstates in blue and for eigenstates outside the continua in green.
        Other parameters are $K=20$ (maximum value of the relative coordinates $m,l,k$) and $Q$ evaluated at $31$ grid points.
        }
    \label{fig:four-body-spectrum-U0-AHM}
    \end{figure*}
A similar analysis may be performed for clusters with larger number of particles, as illustrated in this section for the case of four particles. 
The four-particle model becomes a single-particle problem in the four-dimensional configuration space spanned by the basis states $\ket{n,m ,l,k}$, where the four particles are placed at sites $n$, $n+m$, $n+m+l$ and $n+m+l+k$. 
Here the relative distances are again non-negative as a consequence of indistinguishability, $m, l, k \ge 0$. 
The single-particle Hamiltonian acting on the four-particle configuration space can be derived from the original AHM Hamiltonian $\hat{H}$, and is given by~\cref{eq:4body-AHM-repr-full} in App.~\ref{app:4body-AHM}. 
Again $\hat H=\sum_Q \hat H_Q$, with $Q$ denoting the COM quasimomentum. 
The Hamiltonian $\hat H_Q$, which may be found in~\cref{eq:4body-AHM-repr-hybrid-full} in App.~\ref{app:4body-AHM}, acts on a 3D relative coordinate space spanned by the basis states $\ket{Q,m,l,k} = \sum_n e^{iQ(n+3m/4+l/2+k/4)} \ket{n,m,l,k}$. 
The four-particle spectrum is obtained by exact diagonalization of $\hat H_Q$. 
Figure~\ref{fig:four-body-spectrum-U0-AHM} shows the spectra for $\theta=\pi/4$ (a) and $\theta = 3\pi/4$ (b) with $U=0$, where we have truncated the relative distances to maximum values of $K=20$.
Eigenstates with $\mathrm{SPR} \sim 0,1,2,3$ correspond to states localized on zero-, one-, two-, and three-dimensional subspaces of the relative coordinate space, respectively. 
The blue, green, and red solid, and cyan dashed lines indicate the edges of the four different scattering-state continua which are available in the case of the four-particle problem, see App.~\ref{app:4body-scattering-state-continuum} for details. 

Interestingly, true four-body bound states are found outside the continuum, as can be clearly seen from the highly localized solutions with low PR in the case of $\theta=\pi/4$~[\cref{fig:four-body-spectrum-U0-AHM}(a)] for instance, while for~$\theta=3\pi/4$~[\cref{fig:four-body-spectrum-U0-AHM}(b)] no bound-state solutions outside the continuum are found.
As for the three-particle case, we also find strong indications of some highly localized quasi-BICs, prominently visible in the case of $\theta=\pi/4$ (a).

To gain a more quantitative measure of the presence of four-body bound states, we define highly localized states as those with $m+l+k\leq 4$, and plot in~\cref{fig:four-body-spectrum-U0-AHM}(c) the bound-state fraction for the finite system with $K=20$ as a function of $\theta$. 
Most four-body bound states roughly occur around $\theta \approx \pi/4$, where, interestingly, the most localized solutions are those lying inside the continuum. 
\begin{figure*}[t]
    \centering
    \includegraphics[width=2.08\columnwidth]{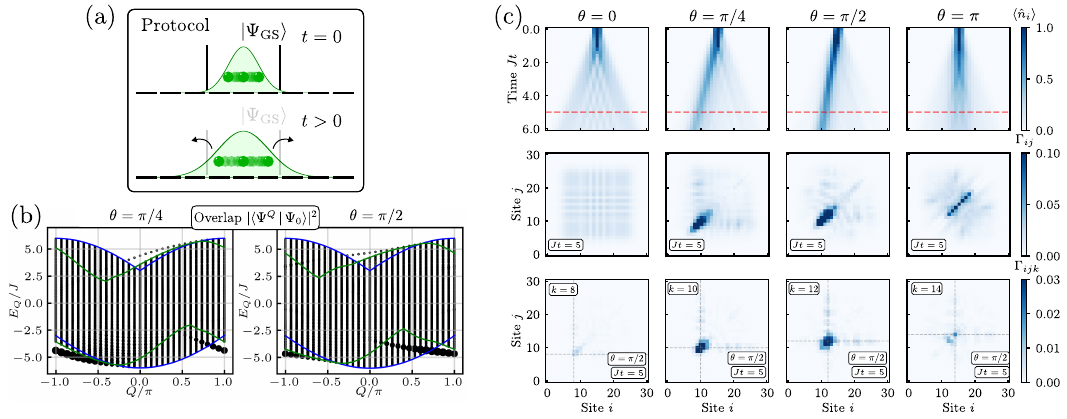} 
    \caption{
    Asymmetric expansion dynamics and density correlators of three-particle wave packets.
    (a) Illustration of the expansion protocol starting from the ground state of three particles on three sites. 
    (b) The three-particle spectrum $E_Q$ (cf.~\cref{fig:three-body-spectrum-full-AHM}(b)-(c)) with the overlap $\mathcal{O}=\abs{\braket{\Psi^{Q}}{\Psi_0}}^2$ of the initial state $\ket{\Psi_0}$ with the entire three-particle spectrum $\{\ket{\Psi^{Q}}\}$~\eqref{eq:3body-AHM-repr-hybrid} for $\theta=\pi/4$ (left) and $\theta=\pi/2$ (right). 
    The overlap is illustrated by the size of the area of the filled circles, which scale quadratically with the magnitude of the overlap. 
    The first row in (c) shows the spatio-temporal density distribution 
    $\ep{\hat{n}_{j}}(t)=\expval{\Psi(t)}{\hat{n}_{j}}$ for the statistical angles $\theta=0$ (bosons), $\theta=\pi/4$, $\theta=\pi/2$, and $\theta=\pi$.
    The second row shows the second-order density correlator $\Gamma_{ij}=\ep{\hat{b}^{\dagger}_{j}\hat{b}_{i}^{\dagger}\hat{b}_{i}\hat{b}_{j}}$ at the fixed time $Jt=5$ (red dashed line in first row) for the same statistical angles $\theta$.
    The third and fourth row show the third-order density correlator $\Gamma_{ijk}=\ep{\hat{b}^{\dagger}_{k}\hat{b}^{\dagger}_{j}\hat{b}_{i}^{\dagger}\hat{b}_{i}\hat{b}_{j}\hat{b}_{k}}$ at the fixed time $Jt=5$ (red dashed line in the first row) for the fixed statistical angle $\theta=\pi/2$, where the index $k=8,10,12,14$ increases from left to right. 
    The gray dashed lines are a guide to the eye, with the vertical and horizontal lines corresponding to $i=k$ and $j=k$, respectively, and their crossing point indicating the three-particle coincidence point $i=j=k$.
    The initial state $\ket{\Psi_0}$ is the ground state of the AHM of $N_b=3$ bosons on $L_{\mathrm{small}}=3$ sites for $U=0$. 
    For the quench dynamics, the system is then evolved under the AHM~\eqref{eq:AHM} with $L=31$ sites and $U=0$. 
    }
    \label{fig:5-6-density-vs-time-ground-state-expansion}
\end{figure*}
This further hints at the presence of long-lived four-body qBICs.

Comparing the results of the three- and the four-particle cases, we find that the gap between the bound states outside the continuum and the scattering-state continua becomes smaller as the particle number $N_b$ is increased. 
This can be understood from an energetic perspective (similar to the arguments discussed in the $N_b=3$ case~[\cref{sec:Binding-mechanism-outside-AHM}]). The number of directions in the relative coordinate space along which the different scattering states can delocalize increases with the particle number, while the bound states always remain localized at the origin of the relative coordinate space, such that the energy of the bound states is expected to be less separated from the continua as $N_b$ grows. 
Hence, we expect that there will be a finite particle number beyond which no bound states outside the continuum (for zero on-site interaction $U=0$) are found (see App.~\ref{subsec:Nbody-bound-states-outside-continua} for more details). Whether we can still find qBICs in that regime remains an interesting question for future work.

\section{Expansion dynamics of three-particle wave packets}
\label{sec:expansion-dynamics}
We will now show that the presence of three-body bound states can be readily probed by means of a quantum-walk expansion protocol similar to the one used in the experiment of Ref.~\cite{Bakkali-Hassani2026} for investigating two-body bound states 
(further details and an alternative protocol can be found in App.~\ref{app:extended-expansion-dynamics}).

The protocol starts from the (adiabatically prepared) ground state of three particles confined to $3$ neighboring sites with zero on-site interaction ($U=0$), then the confinement is suddenly removed to let the system freely expand in a lattice of size $L\gg 3$, see Fig.~\ref{fig:5-6-density-vs-time-ground-state-expansion}(a). 
As this preparation scheme minimizes the energy on a subspace with the particles being close together, it is expected to have a large overlap with the energetically lowest three-body (quasi-) bound state if such a state exists. 
As a result of the fact that density-dependent Peierls phases dominate within this subspace, the prepared state is, moreover, expected to be centered around a finite COM quasimomentum $Q$. 
The large overlap of the so-prepared states with the three-body bound states is confirmed in~\cref{fig:5-6-density-vs-time-ground-state-expansion}(b).

From Fig.~\ref{fig:three-body-spectrum-full-AHM}(b,c), we know that the energies of finite-$Q$ three-body bound states possess a finite slope with respect to $Q$, indicating a finite group velocity. 
For the positive values of $\theta$ depicted in these plots, the velocity of bound states below the scattering continua is always negative, making them chiral in the sense that they will always be traveling to the left.
In turn, for negative $\theta$, the velocities are positive, and the bound triples move rightward. 
Thus, after releasing the particles from the three-site confinement suddenly, we expect a large part of the wave function to be projected onto a chiral bound state traveling in a direction dictated by $\theta$.

\begin{figure}[bt]
    \centering
    \hspace*{-.3cm}
    \includegraphics{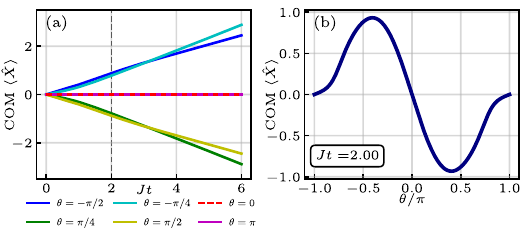}
    \caption{
    Asymmetric transport of three-body bound states.
    (a) The COM position $\ep{\hat{X}}$ as a function of time $t$ for various statistical angles $\theta$ (colors). 
    (b) The COM position $\ep{\hat{X}}$ as a function of the statistical angle $\theta$ at a fixed time of $Jt=2$ (cf.~gray vertical dashed line in (a)). 
    As in~\cref{fig:5-6-density-vs-time-ground-state-expansion} the initial state $\ket{\Psi_0}$ is the three particle ($N_b=3$) ground state prepared on $L_{\mathrm{small}}=3$ sites, after which the system is quenched into a larger AHM lattice with $L=31$ sites for $U=0$.
    }
    \label{fig:ground-state-expansion}
\end{figure}

We perform simulations of the aforementioned protocol via ED and present the results in~\cref{fig:5-6-density-vs-time-ground-state-expansion} and~\cref{fig:ground-state-expansion}~(the corresponding results for the experimentally constrained AHM are shown in~\cref{fig:5-6-density-vs-time-ground-state-expansion-experimental-AHM} and~\cref{fig:ground-state-expansion-experimental-AHM} in App.~\ref{app:experimentally-constrained-3body-AHM}). 
In particular, we compute the spatial density distribution $n_j=\ep{\hat{b}_{j}^{\dagger}\hat{b}_{j}}$ as well as the two- and three-particle correlations $\Gamma_{ij}=\ep{\hat{b}^{\dagger}_{j}\hat{b}_{i}^{\dagger}\hat{b}_{i}\hat{b}_{j}}$ and $\Gamma_{ijk}=\ep{\hat{b}^{\dagger}_{k}\hat{b}^{\dagger}_{j}\hat{b}_{i}^{\dagger}\hat{b}_{i}\hat{b}_{j}\hat{b}_{k}}$ as a function of time after the three-site confinement has been switched off suddenly.  
We note that these observables are readily accessible in quantum-gas microscopes via fluorescence imaging~\cite{Kwan2024,Bakkali-Hassani2026}. 
They are depicted in~\cref{fig:5-6-density-vs-time-ground-state-expansion}(c) for various values of $\theta$. 

The results of the upper panels show the density distribution $n_j(t)$ for various phases $\theta$. 
Already here, one can observe clear signatures of chiral bound states. 
Namely, for $\theta = 0$ and $\pi$, where no three-body bound states are predicted, we can see that the density distribution disperses with time and remains symmetric. 
In turn, for the values $\theta=\pi/2$ and $\theta = \pi/4$, for which three-body bound states exist, we find that part of the density travels at constant speed to the left without dispersing, while only a small fraction of the density disperses in the background. 
These results are consistent with the expectation that a large fraction of the initial state is projected onto a chiral three-body bound state. 

In the second row, we show the two-particle correlations $\Gamma_{ij}$ at a fixed time $Jt=5$. 
While bosonic ($\theta=0$) particles delocalize across the entire lattice individually, due to the large overlap with type-I scattering states, for $\theta\neq 0$, the weight of the correlation remains concentrated along the diagonal $\Gamma_{ii}$, indicating that two particles are likely to be found on the same site. 
For $\theta=\pi$ this is consistent with a large overlap of type-II scattering states, where two particles always form a bound pair. 
For $\theta=\pi/2$ and $\theta = \pi/4$, the correlations are even more pronounced, as all three particles are expected to remain close together, and we can again notice an asymmetry, as the peaks along the diagonal $\Gamma_{ii}$ are shifted to the lower left. 
We show in the third row the three-particle correlations $\Gamma_{ijk}$ at fixed $Jt=5$ and $\theta=\pi/2$, depicting 
the results as a function of $i$ and $j$ for fixed values of $k$. 
We find that $\Gamma_{ijk}$ exhibits a narrow peak~(at $i=j=k=10$ in the figure) showing that all three particles are highly likely to be found very close to each other, clearly indicating a three-body bound state. 

The evolution of the COM position $\ep{\hat{X}}=\sum_j j\ep{\hat{n}_{j}}/N_b$ of the three-particle wave packet is depicted in~\cref{fig:ground-state-expansion}(a) for different $\theta$. 
In both cases, we clearly see asymmetric transport ($\ep{\hat{X}}\neq 0$) for $\theta\neq 0,\pi$, with the transport direction changing with the sign of $\theta$~[\cref{fig:ground-state-expansion}(b)]. 
This asymmetric transport is directly related to the anyonic nature of the particles, as the transport is symmetric for bosons ($\theta=0$) and pseudo-fermions ($\theta=\pi$). 
Remarkably, the three-body bound state moves with a velocity that equals that of a single particle, whose dynamics is captured by the single-particle tunneling matrix element $J$. 
This can also be seen in the first row of~\cref{fig:5-6-density-vs-time-ground-state-expansion}(c), where the asymmetric chiral three-body bound states reach the edge of the light-cone corresponding to the ballistic wavefront of free particles. 
This is in contrast to clusters bound by on-site interactions alone~\cite{Winkler2006,Folling2007,Petrosyan2007,Piil2007,Wang2008,Valiente2008,Valiente2008a,Valiente2009,Valiente2010,Ronzheimer2013,Boschi2014,Weckesser2025}, for which binding and mobility are both tied to the same ratio $J/\abs{U}$.
In the strongly interacting regime $\abs{U}\gg J$, where $N$ particles are tightly bound with a binding energy of order $\sim\abs{U}$, the cluster can move only via $N$-th-order virtual tunneling, with a strongly suppressed effective matrix element $\sim J^{N}/\abs{U}^{N-1}$, so for $N=3$ such on-site trimers are practically immobile~\cite{Valiente2010a}.
Reducing $\abs{U}$ to a few $J$ can restore mobility, but the cluster is then only weakly bound~\cite{Valiente2008,Valiente2010a}. 
For the chiral anyon clusters, by contrast, mobility is decoupled from binding since the binding is kinematic and generated by the tunneling term itself, allowing the cluster to disperse with matrix elements on the order of $J$, independently of the size of its binding gap.

\section{Summary and Outlook}

In conclusion, we have shown that the one-dimensional anyon-Hubbard model does not only describe statistically-induced pairing, but also the formation of bound few-body clusters. 
This remains true also for the constrained model that can be realized in current experiments~\cite{Cardarelli2016,Kwan2024,Bakkali-Hassani2026}~{\footnote{The three-particle coincidences and the processes connecting to these configurations need not be excluded. Using the scheme proposed and utilized in Refs.~\cite{Cardarelli2016,Kwan2024,Bakkali-Hassani2026}, one would need to add two additional frequencies $\omega_{1/2}=\Delta\pm 2(U_0-U)$ with a required phase of $2\theta$ and $0$ for the process $\ket{120}\to \ket{030}$ and $\ket{030}\to \ket{021}$, respectively~[see Fig.~\ref{fig:three-body-spectrum-effective-AHM-U0-Binding-mechanism}(a)], where $\Delta$ is the energy tilt between adjacent sites, $U_0$ the bare on-site interaction strength, and $U$ the desired effective on-site interaction strength. Note, however, that the Peierls phases for the processes connecting configurations $\ket{012} \leftrightarrow \ket{021}$ still need to be excluded, as their transition frequencies are degenerate with the transition frequency $\Delta$ of a single particle hopping onto an unoccupied site.}}.
Although true three- and four-body bound states are only found outside the scattering continuum, quasi-bound states~(very long-lived or even stable in finite-size systems) may also lie within it. 
Finally, we have discussed an experimentally feasible quantum-walk protocol to probe three-body bound states. 
Remarkably, in contrast to slow strongly interaction-bound clusters, the chiral three-body bound states here are highly mobile, moving at similar speeds as free particles, while being bound at vanishing on-site interaction. 

Our results open several interesting questions. 
One of them concerns whether bound clusters of more than a few anyons can also be stabilized, while still moving at speeds comparable to those of single particles. 
A closely related question is whether it is possible to smoothly interpolate between the chiral few-body bound states of the AHM (investigated experimentally for two particles in Ref.~\cite{Bakkali-Hassani2026} and theoretically for three and four particles here) and the bright chiral solitons observed experimentally in~\Ccite{Frolian2022} in a superfluid realization of a chiral BF theory~\cite{Kundu1999,Aglietti1996,Chisholm2022}. 
One might envision two routes towards such highly mobile many-anyon clusters, to be investigated in future work: One possibility is the addition of weak attractive on-site interactions, $U<0$, which are just strong enough to stabilize the kinetic binding mechanism that we described here for few-body clusters also for clusters of more particles. 
At the same time, the interactions would need to remain small enough such that the so-stabilized clusters will not become heavy enough to lead to an effective tunneling matrix element that becomes exponentially small with the number of particles. 
However, whether such an intermediate regime exists is an open question. 
A second possible route towards chiral clusters of many anyons could rely on long-lived quasi-bound states in the continuum rather than on energetic stabilization. 
With increasing particle number $N$, the clusters will lie at a zero-dimensional corner of the $(N-1)$-dimensional relative coordinate space. 
This geometry might cause lifetimes that increase with $N$, asymptotically stabilizing the clusters in the limit of large $N$. 
This latter scenario also opens another question, namely whether the many-anyon model might feature weak ergodicity breaking or prethermalization, possibly related to the physics of correlated hopping models~\cite{Turner2018a,Chandran2023,Moudgalya2022,Hudomal2020,Serbyn2021,Banerjee2021,Aditya2024} or kinetically constrained models~\cite{Surace2021,Singh2021,Karle2021} with many-body (Aharonov-Bohm-like) Fock space cages~\cite{Jonay2026,Tan2025,Nicolau2026,Mohapatra2026,Dupont2026,Ben-Ami2025,Ben-Ami2026}, where the latter are already present in the simplest two-particle configuration space lattice~\cite{Bakkali-Hassani2026}. 
In turn, it will be interesting to employ the approaches developed in this work to explore the existence of few-body bound states in other correlated hopping models~\cite{Hudomal2020,Faugno2025,Faugno2022}, kinetically constrained models~\cite{Nicolau2026,Jonay2026,Brighi2023,Scherg2021}, or, in particular, in other exotic 1D anyonic models, such as the recently proposed traid-anyon-Hubbard model~\cite{Nagies2024} describing a discrete lattice version of traid anyons, which are intrinsic to 1D~\cite{Harshman2022,Harshman2020}.
\\

\begin{acknowledgments}
    We thank Nathan Harshman for insightful discussions and comments on an earlier version of the manuscript. 
    We are grateful for useful discussions with Alessio Celi, Leticia Tarruell, and Martin Bonkhoff. 
    Part of the work has been conducted using the QuSpin project~\cite{Weinberg2017,Weinberg2019}.
    L.S.\ acknowledges the support of the Deutsche Forschungsgemeinschaft (DFG, German Research Foundation) under Germany's Excellence Strategy -- EXC-2123 Quantum-Frontiers -- 390837967. 
    A.E.\ and IT acknowledge support by the Deutsche Forschungsgemeinschaft (DFG, German Research Foundation) via the Research Unit FOR 5688 (Project No. 521530974).  
    I.T.\ is grateful for support by the Studienstiftung des deutschen Volkes. 
    J.K., P.S., Y.L., M.G. and B.B.-H. acknowledge support from CUA.
    This publication is funded in part by the Gordon and Betty Moore Foundation through Grant GMBF 11521 to Harvard University to support the work of M.G.
\end{acknowledgments}

\section*{Competing interests}
    M.G. is a co-founder, shareholder, and consultant of QuEra Computing. All other authors declare no competing interests.

\section*{Data Availability}
The code used to generate the data and figures in this work is available from the corresponding author upon reasonable request. 

\clearpage
\makeatletter
\let\addcontentsline\origaddcontentsline
\makeatother
\appendix

\onecolumngrid
\begin{center}
   \Large\textbf{Appendices}
\end{center}
\twocolumngrid

\tableofcontents
\twocolumngrid

\section{Anyonic representation of the AHM}
\label{app:anyonic-representation-AHM}
By using a fractional Jordan-Wigner (JW) transformation relating the bosonic operators $\hat{b}_j$ to anyonic operators $\hat{a}_j = \exp(i \theta \sum_{k\leq j} \hat{n}_{k})\hat{b}_j$~\cite{Keilmann2011},
we can relate the bosonic representation of the AHM~\eqref{eq:AHM} to its anyonic representation
\begin{align}
    \hat{H}=-J\sum_{j}\left(\hat{a}_{j+1}^{\dagger}\hat{a}_j + \text{h.c.}\right)
    + \frac{U}{2} \sum_j \hat{n}_j (\hat{n}_j-1).
    \label{eq:AHM-anyon}
\end{align}
The anyonic operators $\hat{a}_j$ satisfy the generalized commutation relations~\cite{Keilmann2011}
\begin{align}
        \hat{a}_j \hat{a}_k^{\dagger} - e^{-i \theta \text{sgn}(j-k)} \hat{a}_k^{\dagger} \hat{a}_j &= \delta_{jk},  \nonumber \\
        \hat{a}_j \hat{a}_k - e^{i \theta \text{sgn}(j-k)} \hat{a}_k \hat{a}_j &= 0,
        \label{eq:CCR-anyon}
\end{align}
with the sign function $\mathrm{sgn}(\ell)$ taking values $-1$, $0$, $+1$ for $\ell<0$, $\ell=0$, $\ell>0$, respectively. Note that the particle density $n_j$ or any higher-order density correlators, such as $\Gamma_{ij}$, $\Gamma_{ijk}$ studied in the expansion dynamics in~\cref{sec:expansion-dynamics} of the main text, are invariant under the fractional JW transformation~\cite{Liu2018b}.

\section{Two-particle problem}
\label{app:2body-AHM-repr}
For completeness, we briefly review the 2-particle problem of the AHM following~\cite{Bakkali-Hassani2026}, 
and discuss the consequences of the chiral symmetry for $U=0$ on the two-particle spectrum, while also 
providing more details on the 2-body bound state in the continuum (BIC) solutions of the main text. 

In the two-particle problem of the AHM the configuration space representation of the two-particle problem is two-dimensional, as any Fock basis state $\ket{\{n_j}\}$ in the bosonic Fock space with two particles, i.e.,$\sum_j n_j=2$, can be represented by the basis state $\ket{n,m}$ in the configuration space, where $n$ is the position of the left particle and $m$ is the distance between the two particles, such that $n+m$ is the position of the right particle. In the case of an infinite 1D chain, which is the main focus of this work, we have $n\in \mathbb{Z}$ and for the relative distances $m\in \mathbb{N}$ to account for indistinguishability of the two particles~\footnote{In a finite lattice of $L$ sites with two particles, the allowed values are $n\in\{1,\ldots,L\}$ and $m\in\{0,\ldots,L-n\}$.}.

With this mapping, we can transform the two-particle AHM problem into a single-particle problem in the two-dimensional configuration space spanned by the basis states $\{\ket{n,m}\}$, which form a 2D-simplex (triangle) due to the indistinguishability of the two particles~\cite{Bakkali-Hassani2026}.
The AHM Hamiltonian in this representation reads
    \begin{align}
        \hat{H} = -&J \sum_{n} \Big[ \sqrt{2}\left\{
             e^{-i\theta}\ketbra{n,0}{n\!-\!1,1} +\ketbra{n,1}{n,0} 
            + \text{h.c.}  \right\} \nonumber\\
        + & \sum_{m\geq2}(\ketbra{n,m\!-\!1}{n,m} + \ketbra{n\!+\!1,m\!-\!1}{n,m}+ \text{h.c.})
        \Big]\nonumber\\
         + &U\sum_{n} \ketbra{n,0}{n,0}.
        \label{eq:app:2body-AHM-repr}
    \end{align}
    
Now, to study bound and scattering state solutions in the two-particle case, we assume an infinite 1D chain for the underlying AHM, such that $n\in\mathbb{Z}$ and $m\in\mathbb{N}$.
This introduces a translational invariance along the direction of the 1D chain, which in the configuration space representation becomes a translational invariance along the center-of-mass (COM) direction, where the COM position of the two particles is given by $n+\frac{m}{2}$.
Thus, due to this translational symmetry the COM quasimomentum $q$ is conserved and becomes a good quantum number, such that we can block-diagonalize the Hamiltonian~\eqref{eq:app:2body-AHM-repr} into blocks $\hat{H}_q$ where states with different COM quasimomenta $q$ decouple.
This is easiest done by first introducing hybrid basis states $\ket{q,m}$ which are characterized by a fixed COM quasimomentum $q$ and relative distance $m$, and reads $\ket{q,m} = \sum_{n} e^{i q (n+\frac{m}{2})}\ket{n,m}$ for $q\in[-\pi,\pi)$.
We can then transform the Hamiltonian~\eqref{eq:app:2body-AHM-repr} into the hybrid basis $\{\ket{q,m}\}$, which yields the block-diagonalized Hamiltonian $\hat{H} = \sum_q \hat{H}_q$, where each block $\hat{H}_q$ reads~\cite{Bakkali-Hassani2026}
    \begin{align}
        \hat{H}_q = &-J\Big\{\sqrt{2}\big[
        2\cos{\left(\frac{q+\theta}{2}\right)}e^{-i\theta/2}\ketbra{q,0}{q,1} +\text{h.c.}\big] \nonumber \\
        &+ \sum_{m\geq2}2\cos(q/2)\big(\ketbra{q,m}{q,m\!-\!1}  +\text{h.c.}\big)\Big\}\nonumber \\
        &+U\ketbra{q,0}{q,0}.
        \label{eq:app:2body-AHM-repr-hybrid}
    \end{align}
This Hamiltonian is the starting point to obtain the exact two-particle scattering and bound state solutions, which are found by solving the eigenvalue problem $\hat{H}_q\ket{\psi^{q,\kappa}} = E_{q,\kappa}\ket{\psi^{q,\kappa}}$ for each COM quasimomentum $q$, where $\kappa$ is a quantum number associated to the relative coordinate degree of freedom. 

\subsection{Scattering and bound-state spectra in the 2-body AHM}
\label{app:2body-bound-state}
For completeness, we briefly revisit the two-particle bound state solutions in the AHM.
The two-particle bound-state spectrum can be analytically obtained by solving the recurrence relations associated to the block-diagonalized two-particle AHM Hamiltonian~\eqref{eq:app:2body-AHM-repr-hybrid} resolved in the hybrid COM quasimomentum basis $\ket{q,m}$~\cite{Bakkali-Hassani2026}.

By using the eigenvalue equation $\hat{H}_q\ket{\psi^{q,\kappa}} = E_{q,\kappa}\ket{\psi^{q,\kappa}}$~\eqref{eq:app:2body-AHM-repr-hybrid}, and comparing the coefficients of the hybrid coordinate basis states $\ket{q,m}$, we can arrive at a linear set of equations 
for the relative coordinate coefficients $\psi^{q,\kappa}_m=\braket{q,m}{\psi^{q,\kappa}}$ \cite{Kwan2024,Greschner2018a,Cardarelli2016,Zhang2017b} (where $\kappa$ as quantum number labels the eigenstate in the $q$-block),
    \begin{align}
        (\epsilon_{q,\kappa}-u_q) \, \psi^{q,\kappa}_0 &=-\sqrt{2}\rho_q \, \psi^{q,\kappa}_1, \label{eq:AHM-rel-system-coeff1}\\
        \epsilon_{q,\kappa} \,\psi^{q,\kappa}_1 &= -\sqrt{2}\rho_q^* \psi^{q,\kappa}_0- c^{q,\kappa}_2, \label{eq:AHM-rel-system-coeff2}\\
        \epsilon_{q,\kappa} \, \psi^{q,\kappa}_m &= -(\psi^{q,\kappa}_{m-1}+\psi^{q,\kappa}_{m+1}), \quad m \geq 2, 
        \label{eq:AHM-rel-system-coeff3}
    \end{align}
where $\epsilon_{q,\kappa} =E_{q,\kappa}/\gamma_q$,
$u_q = U/\gamma_q$, 
$\rho_q= 2e^{-i\theta/2}\cos((q+\theta)/2)/\gamma_q$, and$\gamma_q=2J\cos(q/2)$.

For the exact solutions to the two-particle problem of the AHM,
one may directly start from the set 
of equations \eqref{eq:AHM-rel-system-coeff1}-\eqref{eq:AHM-rel-system-coeff3} 
and solve for the coefficients $\psi_m^{q,\kappa}$.
Generally, the 2nd order recurrence relation of 
\cref{eq:AHM-rel-system-coeff3} is solved by an ansatz of the form 
$\psi_m^{q,\kappa} = \lambda x^m_{+} + \mu x_{-}^m,\, \text{for}\, m\geq1$, where $x_{\pm}$ are the
two solutions of the associated characteristic polynomial equation,
$x^2 + \epsilon_{q,\kappa} x +1 =0$ \cite{Kwan2024}. 
Depending on the sign of the discriminant $\chi= \epsilon_{q,\kappa}^2 -4$, this solution will
yield complex plane wave solutions or real-valued exponential solutions. 

\emph{Scattering states.} The scattering states correspond to the complex-valued solutions of the form $\psi_m^{q,\kappa} = \lambda x^m_{+} + \mu x_{-}^m$, for $m\geq1$, which, in sufficiently large systems where the probability for two delocalized particles to meet is negligible, can be approximated by a product state of two free particles (even for finite $U$). 
The total energy of the system $E_{q,\kappa}$, which becomes exact in the thermodynamic limit, can be written as the sum of the kinetic energies,
$\epsilon(k)=-2J\cos(k)$, of the individual particles, i.e.,
$E_{q,\kappa} \equiv E_q(\kappa) = \epsilon(k_1) + \epsilon(k_2)$, with $k_1 = \frac{q}{2}+\kappa$ and $k_2 = \frac{q}{2}-\kappa$
being the momenta of the first and second particle respectively.
Here, $\kappa=(k_1-k_2)/2$ denotes the relative quasimomentum and $q=k_1+k_2$ 
the total COM quasimomentum \cite{Krutitsky2016,Valiente2008,Piil2007}.
Thus, we obtain the following energies for the scattering states
    \begin{align}
            E_q(\kappa) &= -2J\cos\left(\frac{q}{2}+\kappa\right)-2J\cos\left(\frac{q}{2}-\kappa\right) \nonumber \\
            &=-4J\cos\left(\frac{q}{2}\right)\cos\left(\kappa\right)=-2J\cos(\kappa)\gamma_q,
        \label{eq:AHM-Scattering-dispersion}
    \end{align}
with $\gamma_q=2J\cos(q/2)$ as above.
This continuum of solutions is bounded from below and above,
$\abs{E_q(\kappa)} \leq 4J \cos\left(\frac{q}{2}\right)$, where the lower (upper) bound 
corresponds to the relative momentum $\kappa=0$ ($\kappa=\pm\pi$).
The result of \cref{eq:AHM-Scattering-dispersion} may also be derived by assuming 
free plane wave solutions for the coefficients of the form $\psi^{q,\kappa}_m \sim e^{\pm i m \kappa}$, 
such that the recurrence relation \cref{eq:AHM-rel-system-coeff3} gives rise to
$\epsilon_{q,\kappa} = -2J\cos(\kappa)=E_q(\kappa)/\gamma_q$ yielding
the same energy dispersion relation \eqref{eq:AHM-Scattering-dispersion} \cite{Piil2007,Valiente2008}.
Thus, the scattering state solutions correspond to delocalized plane wave solutions in the bulk of the two-dimensional configuration space lattice given in~\eqref{eq:app:2body-AHM-repr}, which are delocalized along both the COM and relative coordinate directions.

\emph{Bound states.} The bound states correspond to (exponentially decaying) real-valued solutions of the form $\psi_m^{q,\kappa} = \lambda x_{\pm}^m, \, \text{for}\, m\geq1$, where $x$ is the solution of the characteristic polynomial equation. 
The corresponding solution of the bound state energy spectrum for arbitrary statistical angles $\theta$ and on-site interaction $U$ is given by~\cite{Kwan2024}
\begin{align}
            E_{\pm}(q)=&-\Bigg(U\left[\cos^2{\frac{q}{2}}-\cos^2\left(\frac{q+\theta}{2}\right) \right]
            \mp\cos^2\left(\frac{q+\theta}{2}\right) \nonumber\\ 
            &\times\sqrt{U^2+16J^2\left[2\cos^2\left(\frac{q+\theta}{2}\right)-\cos^2{\frac{q}{2}}\right]}\ \Bigg)
            \nonumber\\
            &\times\left[2\cos^2{\left(\frac{q+\theta}{2}\right)}-\cos^2{\frac{q}{2}}\right]^{-1},
    \label{eq:AHM-bound-states-dispersion}
\end{align}
which for $U=0$ simplifies to
\begin{align}
    E_{\pm}(q)=\pm\frac{
    4J\cos^2{\left(\frac{q+\theta}{2}\right)}}
    {\sqrt{2\cos^2{\left(\frac{q+\theta}{2}\right)}-\cos^2{\frac{q}{2}}}}.
    \label{eq:AHM-bound-states-dispersion-Uzero}
\end{align}

\subsection{Chiral symmetry of the two-particle AHM}\label{app:2body-chiral-symmetry}
Note that for $U=0$ the AHM Hamiltonian~\eqref{eq:AHM} admits a chiral symmetry for any particle number $N_b$ due to the bipartite nature of the configuration space lattice~\cite{Theel2025}, where the chiral symmetry operator $\hat{S}$ is given by $\hat{S} = \exp(i \pi \sum_{j} j\hat{n}_j) =\exp(i \pi N_b \hat{X})$, where $\hat{X}=\sum_j j \hat{n}_j/N_b$ is the COM position operator, which is always diagonal in the configuration space basis.
For $N_b=2$ particles the Hamiltonian $\hat{H}_q$~\eqref{eq:app:2body-AHM-repr-hybrid} acts on the one-dimensional bipartite relative coordinate space lattice spanned by $\ket{q,m}$, such that the chiral symmetry operator $\hat{S}$ is also diagonal in the hybrid COM quasimomentum basis $\ket{q,m}$. 
To see this, note that, by using $\hat{S}\ket{n,m}=e^{i\pi (2n+m)}\ket{n,m}=e^{i\pi m}\ket{n,m}$, it follows that 
\begin{align}
    \hat{S}\ket{q,m}=e^{i \pi m}\ket{q,m}.
\end{align}
The consequence of this is that the two-body bound-state spectrum satisfies the symmetry relation $E_{q,-}=-E_{q,+}$~\eqref{eq:AHM-bound-states-dispersion-Uzero} and, from the substitution $e^{im\kappa}\to e^{im(\kappa +\pi)}$ for the chiral partner, it also follows that the scattering state continuum energies satisfy $E_q(\kappa+\pi)=-E_q(\kappa)$~\eqref{eq:AHM-Scattering-dispersion}.

\subsection{Details on two-body bound states in the continuum of the AHM}
\label{app:2body-BIC}

Extending on the discussion of the main text (in~\cref{sec:2BodyBIC}), we provide more details on the two-body bound states in the continuum (BIC) solutions and the highly localized quasi-BICs in the proximity of the BIC solutions. 
To this end, note that we can partition our system, which is fully described by the Hamiltonian $\hat{H}_q$~\eqref{eq:app:2body-AHM-repr-hybrid} in the relative coordinate space, into two sectors: a deeply bound region, corresponding to the site $\ket{q,0}$, and the bulk of the system, corresponding to the sites $\ket{q,m \geq 1}$, which are only coupled by the single coupling term $\hat{V}_q = -2\sqrt{2}J\cos((q+\theta)/2)e^{-i\theta/2}\ketbra{q,0}{q,1}$ (and its Hermitian conjugate).
Denoting the projector onto deeply bound region by $\hat{P}=\ketbra{q,0}{q,0}$ and the projector onto the bulk of the system by $\hat{Q}=1-\hat{P} =\sum_m \ketbra{q,m}{q,m}$, we can write the Hamiltonian $\hat{H}_q$ in a block form with respect to the partitioning defined by $P$ and $Q$, which reads as follows
    \begin{align}
        \hat{H} = \begin{pmatrix}
            \hat{H}_{PP} & \hat{H}_{PQ} \\
            \hat{H}_{QP} & \hat{H}_{QQ}
        \end{pmatrix} = \begin{pmatrix}
            U\hat{P} & \hat{V}_q \\
            \hat{V}_q^{\dagger} & \hat{H}_{QQ}
        \end{pmatrix},
    \end{align}
with $\hat{H}_{PP} = \hat{P} \hat{H}_q \hat{P} = U \ketbra{q,0}{q,0}$, $\hat{H}_{PQ} = \hat{P} \hat{H}_q \hat{Q} = \hat{V}_q = -2\sqrt{2}J\cos((q+\theta)/2)e^{-i\theta/2}\ketbra{q,0}{q,1} \equiv v(q)\ketbra{q,0}{q,1} $, $\hat{H}_{QP} = \hat{Q} \hat{H}_q \hat{P} = \hat{V}_q^{\dagger}=v^*(q)\ketbra{q,1}{q,0}$, and $\hat{H}_{QQ} = \hat{Q} \hat{H}_q \hat{Q} = t(q) \sum_{m\geq 1} (\ketbra{q,m}{q,m-1} + \text{h.c.})$ which describes the bulk continuum of the system with $t(q)\equiv -2J\cos(q/2)=-\gamma_q$ and $v(q)=-2\sqrt{2}J\cos((q+\theta)/2)e^{-i\theta/2}$.

Generally, following the Fano-Feshbach-Schur projection method~\cite{Cohen-Tannoudji1998,Fano1961,Feshbach1958,Feshbach1962},  
it can be shown that the effect of the $Q$-space on the $P$-space region is captured by the projected resolvent into the $P$-space and given by
    \begin{align}
        \hat{G}_P(z) = \hat{P}\frac{1}{z-\hat{H}}\hat{P}= \frac{1}{z\hat{\mathbbm{1}}_P-\hat{P}\hat{H}\hat{P}-\hat{\Sigma}(z)},
        \label{eq:P-projected-resolvent-general}
    \end{align}
where $\hat{\mathbbm{1}}_P$ is the identity in the $P$-space and the self-energy $\hat{\Sigma}(z)$ is given by
\begin{align}
    \hat{\Sigma}(z) = \hat{P} \hat{H} \hat{Q} (z\hat{\mathbbm{1}}_Q-\hat{Q}\hat{H}\hat{Q})^{-1} \hat{Q} \hat{H} \hat{P},
\end{align} 
encoding, in an exact non-perturbative manner, all the virtual processes arising due to the coupling to the $Q$-space. It describes an effective potential for projected $P$-space Hamiltonian, which is generally non-local in $z$ (energy). 

Applied to our 2-body scenario, we can first compute the self-energy $\hat{\Sigma}(z)$ for the $P$-space region, which reads as follows
    \begin{align}
        \hat{\Sigma}(z) &=\hat{V}_q (z\hat{\mathbbm{1}}_Q-\hat{H}_{QQ})^{-1} \hat{V}_q^{\dagger} 
        \nonumber \\
        &= |v(q)|^2 \bra{q,1} (z\hat{\mathbbm{1}}_Q-\hat{H}_{QQ})^{-1} \ket{q,1}\projector{q,0} \nonumber \\
        &= |v(q)|^2 g(z) \projector{q,0}=|v(q)|^2 g(z)\hat{P}
        \label{eq:self-energy-2body-AHM}
    \end{align}
where we have defined the surface Green's function (resolvent) as $g(z) = \bra{q,1} (z\hat{\mathbbm{1}}_Q-\hat{H}_{QQ})^{-1} \ket{q,1}$. 
Thus, the projected resolvent into the $P$-space reads as follows
    \begin{align}
        \hat{G}^{-1}_P(z) = (z-U-|v(q)|^2 g(z))\hat{P},
        \label{eq:P-projected-resolvent-2body-AHM}
    \end{align}
where $|v(q)|^2=8J^2\cos^2([q+\theta]/2)$.
As $\hat{H}_{QQ}$ describes a uniform semi-infinite tight-binding chain, we can apply the same projection method within the block $\hat{H}_{QQ}$ iteratively. 
The first iteration starts with a new $\hat{P}^{(1)}=\projector{q,1}$, resulting in a new projected resolvent $\hat{G}^{-1}_{\hat{P}^{(1)}}(z) = z\hat{\mathbbm{1}}_{\hat{P}^{(1)}}-\hat{\Sigma}^{(1)}(z)=(z-t^2g^{(1)}(z))\hat{P}^{(1)}$ with $\hat{\Sigma}^{(1)}(z) = t^2 g^{(1)}(z)\hat{P}^{(1)}$ where $t\equiv t(q)$ is the hopping amplitude of the tight-binding chain and $g^{(1)}(z) = \bra{q,2} (z-\hat{H}^{(1)}_{QQ})^{-1} \ket{q,2}$ and $\hat{H}^{(1)}_{QQ}$ is $\hat{H}_{QQ}$ starting from $m=2$.
Following this iteration, we have for $n\geq 1$ that $\hat{G}^{-1}_{P^{(n)}}(z) = z\hat{\mathbbm{1}}_{P^{(n)}}-\hat{\Sigma}^{(n)}(z)$ with $\hat{\Sigma}^{(n)}(z) = t^2 g^{(n)}(z)\hat{P}^{(n)}$ where $g^{(n)}(z) = \bra{q,n+1} (z-\hat{H}^{(n)}_{QQ})^{-1} \ket{q,n+1}$ and $\hat{H}^{(n)}_{QQ}$ is $\hat{H}_{QQ}$ starting from $m=n+1$. 
Crucially, realizing that $g^{(n)}(z)=g(z)$ for all $n\geq 1$ due to the uniformity of the tight-binding chain and using~\eqref{eq:P-projected-resolvent-general} for the $\hat{P}^{(1)}$-space, i.e., $\hat{G}_{\hat{P}^{(1)}}(z) = \hat{P}^{(1)}(z\hat{\mathbbm{1}}_Q-\hat{H}^{(1)}_{QQ})^{-1}\hat{P}^{(1)}$, we arrive at the following recursion relation for the so-called surface Green's function $g(z)$
    \begin{align}
        g(z)&=\expval{q,1}{(z\hat{\mathbbm{1}}_Q-\hat{H}_{QQ})^{-1}} = \expval{q,1}{\hat{G}_{\hat{P}^{(1)}}(z)} \nonumber \\
        &= \frac{1}{z-t^2 g(z)}.
        \label{eq:two-body-surface-Green-function-equation}
    \end{align}
Solving this equation for $g(z)$, we obtain
    \begin{align}
        g(z) = \frac{z-\mathrm{sgn}(\mathrm{Re}\, z)\sqrt{z^2-4t^2}}{2t^2},
        \label{eq:two-body-surface-Green-function-solution}
    \end{align}
where we have chosen the principal square root with the branch cut on the negative real axis, such that $\mathrm{Im} g(z) < 0$ for $\mathrm{Im} z > 0$.
Here, for $z=E+i0^+$, inside the continuum $|E|<2|t(q)|=4J|\cos(q/2)|$, we have 
    \begin{align}
        g(E+i0^+) = \frac{E-i\sqrt{4t^2-E^2}}{2t^2}.
    \end{align}
Thus, inside the continuum, related to the imaginary part, we have a finite local density of states $\rho(E)$ given by
\begin{align}
    \rho(E) = -\frac{1}{\pi}\mathrm{Im} g(E+i0^+) = \frac{\sqrt{4t^2-E^2}}{2\pi t^2}, \quad |E|<2|t|,
\end{align}
and a finite real part $\mathrm{Re} g(E+i0^+) = E/2t^2$, which describes the energy shift of the $P$-space region due to the coupling to the bulk continuum.
Note that outside the continuum $|E|>2|t(q)|=4J|\cos(q/2)|$, $g(E) = [E-\mathrm{sgn}(E)\sqrt{E^2-4t^2}]/2t^2$ becomes purely real, which is consistent with the fact that outside the continuum there is no local density of states of the scattering states and thus no decay channels for a state localized in the $P$-space region. In this case, further note that the two-body bound-state spectrum outside the continuum~\eqref{eq:AHM-bound-states-dispersion} can be found by solving the pole equation $E-U-|v(q)|^2 g(E)=0$, i.e., vanishing of the denominator of $G_P(E)\equiv \expval{q,0}{\hat{G}_P(E)}$~\eqref{eq:P-projected-resolvent-2body-AHM}, for $|E|>2|t(q)|=4J|\cos(q/2)|$.

Equipped with this solution, we can compute the self-energy $\Sigma(E+i0^+,q)\equiv \expval{q,0}{\hat{\Sigma}(E+i0^+)}$~\eqref{eq:self-energy-2body-AHM} for the deeply bound $P$-space region
\begin{align}
    \Sigma(E+i0^+,q) &= |v(q)|^2 g(E+i0^+) \nonumber \\
    &= \Lambda(E,q)-\frac{i}{2}\Gamma(E,q),
    \label{eq:self-energy-2body-AHM-real-imaginary}
\end{align}
with the energy shift $\Lambda(E,q) = |v(q)|^2 \mathrm{Re} \,g(E+i0^+) =|v(q)|^2 E/(2t^2)$ and the decay rate 
    \begin{align}
        \Gamma(E,q) &= -2|v(q)|^2 \mathrm{Im} g(E+i0^+) = 2\pi |v(q)|^2 \rho(E) \nonumber \\
        &=16 \pi J^2 \cos^2\left(\frac{q+\theta}{2}\right) \rho(E)
        \label{eq:decay-width-2body-AHM}
    \end{align}
Eventually, using the projected resolvent in the deeply bound region $P$~\eqref{eq:P-projected-resolvent-2body-AHM}, we can compute the retarded Green's function $G_P(E+i0^+,q) \equiv \expval{q,0}{\hat{G}_P(E+i0^+)}$, from which we obtain the local spectral function 
in the deeply bound region $P$ as follows
    \begin{align}
        A_P(E,q) &= -2\mathrm{Im} G_P(E+i0^+,q) \nonumber \\
        &= \frac{\Gamma(E,q)}{(E-U-\Lambda(E,q))^2 + (\Gamma(E,q)/2)^2},
    \end{align}
describing a generalized Breit-Wigner-type spectral function~\cite{Breit1936}, which is correct whenever $|E|<2|t|=4J\cos(q/2)$ (inside the continuum).
The peak position occurs at the energy $\tilde{E}$ which makes denominator minimal, which is given by the self-consistent solution of $\tilde{E}=U+\Lambda(\tilde{E},q)$, which in our case is given by 
$\tilde{E}=U+|v(q)|^2 \tilde{E}/(2t^2)$ such that the solution is given by $\tilde{E}(q) = U/(1-|v(q)|^2/(2t^2))$.

Now, the exact two-body bound state in the continuum (BIC) solutions mentioned in the main text, occurring at $q=q^{\mathrm{BIC}}=n\pi-\theta, \forall \theta \neq 0$ (odd integer $n$), where $v(q^{\mathrm{BIC}})=0$, 
corresponds to the case where the self-energy~\eqref{eq:self-energy-2body-AHM-real-imaginary} vanishes exactly
\begin{align}
    \hat{\Sigma}(E,q^{\mathrm{BIC}}) = 0, \quad \forall E,
    \label{eq:self-energy-2body-AHM-BIC}
\end{align}
such that the projected resolvent into the deeply bound region $P$ becomes purely real and given by $G_P(E,q^{\mathrm{BIC}}) = 1/(E-U)$, with a simple pole at $E=U$ on the real axis such that the local spectral function is a delta peak 
    \begin{align}
        A_P(E,q^{\mathrm{BIC}}) = -2\mathrm{Im} G_P(E+i 0^+,q^{\mathrm{BIC}}) = 2\pi \delta(E-U),
        \label{eq:2-body-BIC-spectral-function}
    \end{align}
indicating the existence of a true bound state in the continuum with infinite lifetime, whenever its onsite energy $E_{q^{\mathrm{BIC}},\kappa}=U$, lies within the scattering band $|U| < 2|t(q^{\mathrm{BIC}})|=4J|\sin(\theta/2)|$.

We can also explain the highly localized two-body bound state solutions in the proximity of the BICs, 
the so-called, quasi-BICs, by the following argument.
Assuming $q=q^{\mathrm{BIC}}+\delta q$ with $|\delta q| \ll 1$, we have $\cos((q+\theta/2))=\mp \sin(\delta q/2) \approx \mp\delta q/2$ such that the coupling $|v(q)|^2$ between the deeply bound region $P$ and the bulk continuum becomes 
    \begin{align}
        |v(q)|^2 = 8J^2\cos^2\left(\frac{q+\theta}{2}\right) \approx 2J^2 \delta q^2,
    \end{align}
leading to the fact that the decay width~\eqref{eq:decay-width-2body-AHM} becomes
    \begin{align}
        \Gamma(\tilde{E},q)\approx 4\pi J^2 \delta q^2 \rho(\tilde{E}),
        \label{eq:2-body-quasi-BIC-decay-width}
    \end{align}
indicating that the lifetime $\tau(q)\sim 1/\Gamma \sim 1/\delta q^2$ diverges quadratically when approaching the BIC point ($\delta q \rightarrow 0$). 
Thus, the spectral function of the quasi-BICs becomes a narrow Lorentzian
    \begin{align}
        A_P(E,q) \approx \frac{4\pi J^2 \delta q^2 \rho(\tilde{E})}{(E-\tilde{E}(q))^2 + (2\pi J^2 \delta q^2 \rho(\tilde{E}))^2}.
        \label{eq:2-body-quasi-BIC-spectral-function}
    \end{align}
Here, we have used that, the self-energy varies slowly near the peak position $\tilde{E}$, i.e., $\Gamma \ll t \Leftrightarrow |v(q)|^2 \ll t^2$, which is satisfied here in the quasi-BIC regime. Thus, the spectral function is indeed a Lorentzian peaked at the renormalized energy $\tilde{E}(q)\approx U(1+|v(q)|^2/(2t^2))$ with a finite width $\Gamma(\tilde{E},q)$~\eqref{eq:2-body-quasi-BIC-decay-width}. 
The width $\Gamma(\tilde{E},q)$ shrinks quadratically as $J^2(\delta q)^2$ when approaching the BIC point, which explains the highly localized nature of the quasi-BICs in the proximity of the BICs. 
We see that the (simple) pole of the Green's function $G_P(z)$, which for the BIC solution was located on the real axis at $z=U$~\eqref{eq:2-body-BIC-spectral-function}, has now moved to $z=\tilde{E}(q)-i\Gamma(\tilde{E},q)/2$ into the lower half of the complex plane for the quasi-BIC with a non-zero imaginary part, which is a signature of the finite lifetime of the quasi-BICs due to the hybridization with the continuum.

In general, these quasi-BICs are resonances with a finite lifetime $\tau \sim (J\delta q)^{-2}$, for any $\delta q \neq 0$. It should be noted that, in the thermodynamic limit ($L\to \infty$), where continuum level spacing, $\delta E \sim J/L$ vanishes, the width $\Gamma(q)$ (for any fixed $\delta q \neq 0$) will at some point become larger than the level spacing, i.e., $\Gamma(q)/ \delta E\sim J \delta q^2L \gg 1$. In this case, the quasi-BICs will hybridize with the continuum and be indistinguishable from the scattering states. However, for any real finite system, the discrete level spacing can make the quasi-BICs appear well-resolved when the width $\Gamma(q)$ is smaller than the level spacing, i.e., $\Gamma(q)/\delta E \ll 1$.

Lastly, we can also make the hybridization of the quasi-BICs with the continuum explicit by computing the exact state correction $\psi_{m\geq 1}$ due to the coupling to the continuum. 
From the Fano-Feshbach $Q$-space equation, $\ket{\psi_Q} = \hat{G}_Q \hat{H}_{QP}\ket{\psi_P}$, one can obtain the exact and completely general expression
    \begin{align}
        \psi_m = \psi_0\, \frac{v^*(q)}{t(q)}\, x_-(z)^m, \quad m \geq 1, 
        \label{eq:two-body-general-wavefunction}
    \end{align}
valid for all $q$, where $x_-(z) = [z - \mathrm{sgn}(\mathrm{Re}\,z)\sqrt{z^2-4t^2}]/(2t)$ is the decaying root of the characteristic polynomial $tx^2 - zx + t = 0$ (derived in detail below~\eqref{eq:two-body-quasi-BIC-state-correction-GQ}), and $\psi_0$ is the $P$-space amplitude. This single formula unifies all three physically distinct regimes of the spectrum, depending only on the value of $x_-(E)$.

\emph{Outside the continuum:} $|E|>2|t(q)|$ (ordinary bound states $E_{\pm}(q)$): $x_-(E_\text{BS})\in(0,1)$ is real, so $\psi_m \propto x_-^m$ decays exponentially with decay length $\xi = -1/\ln x_-$, consistent with the bound-state ansatz of \cref{app:2body-bound-state}.

\emph{At the BIC point:} $q=q^\text{BIC}$ (perfect BIC): $v(q^\text{BIC})=0$, so the prefactor in~\eqref{eq:two-body-general-wavefunction} vanishes identically and $\psi_m = 0$ for all $m\geq 1$. The wave function collapses entirely onto $|q,0\rangle$ and decouples from the bulk, consistent with $\hat{\Sigma}=0$~\eqref{eq:self-energy-2body-AHM-BIC}.

\emph{Inside the continuum:} $|E|<2|t(q)|$ (quasi-BICs), $x_-(\tilde{E}+i0^+) = e^{-i\kappa}$ with $|x_-|=1$ and $\kappa = \arccos(\tilde{E}/2t)$, so
    \begin{align}
        \psi^{q,\kappa}_m
        = \psi_0\, \frac{v^*(q)}{t(q)}\, e^{-i\kappa m}
        = \psi_0\, \frac{\sqrt{2}\cos\!\left(\tfrac{q+\theta}{2}\right)}{\cos\!\left(\tfrac{q}{2}\right)}\, e^{i\left(\frac{\theta}{2} -\kappa m\right)},
        \label{eq:two-body-quasi-BIC-state-correction}
    \end{align}
for $m\geq 1$. 
The quasi-BIC leaks into the bulk as a traveling plane wave with momentum $\kappa$ with the amplitude $|v(q)/t(q)|$ controlling how strongly the state hybridizes with the continuum.
Close to the BIC point, $q=q^{\mathrm{BIC}}+\delta q$ with $|\delta q|\ll 1$, the plane-wave amplitude vanishes linearly: $v^*(q)/t(q)\approx -\delta q/(\sqrt{2}\sin(\theta/2))$, giving $\psi_m \approx -\psi_0\, \delta q\, e^{i(\theta/2-\kappa^{\mathrm{BIC}}m)}/(\sqrt{2}\sin(\theta/2))$ for $m\geq 1$, with $\kappa^\mathrm{BIC} = \arccos[U/(4J\sin(\theta/2))]$, where we have used $\tilde{E}(q)\approx U$ and $t(q^\mathrm{BIC})=\mp 2J\sin(\theta/2)$.

Let us now derive the above formula~\eqref{eq:two-body-general-wavefunction} for the state correction.
To this end, note that it follows directly from the partitioning of the Hilbert space that the $Q$-space component of the eigenstate is given by $\ket{\psi_Q} = \hat{G}_Q \hat{H}_{QP}\ket{\psi_P}$, which in our two-body case reads as follows
\begin{align}
        \ket{\psi_Q} &= v^*(q)\psi_0 (z\hat{\mathbbm{1}}_Q-\hat{H}_{QQ})^{-1}\ket{q,1}=\sum_{m\geq 1} \psi_m \ket{q,m},\nonumber \\
        \psi_m &= \braket{q,m}{\psi_Q} = v^*(q)\psi_0 \bra{q,m}(z\hat{\mathbbm{1}}_Q-\hat{H}_{QQ})^{-1}\ket{q,1} 
        \nonumber \\
        &\equiv \psi_0\, v^*(q) G_Q(m,1;z)
        \label{eq:two-body-quasi-BIC-state-correction-GQ}
    \end{align}
In the last line we have defined the off-diagonal Green's function $G_Q(m,1;z) \equiv \bra{q,m}(z\hat{\mathbbm{1}}_Q-\hat{H}_{QQ})^{-1}\ket{q,1}$ for $m\geq 1$, which we now compute to obtain the state correction. 
Projecting the identity $(z\hat{\mathbbm{1}}_Q-\hat{H}_{QQ})(z\hat{\mathbbm{1}}_Q-\hat{H}_{QQ})^{-1} = \hat{\mathbbm{1}}_Q$, onto the state $\ket{q,m}$ and $\ket{q,1}$ gives the following recursion relation $G_Q(m,1;z)$
    \begin{align}
        &z G_Q(m,1;z) - t( G_Q(m-1,1;z) + G_Q(m+1,1;z)) \nonumber \\ 
        &= \delta_{m,1},\, \forall m\geq 1, 
        \label{eq:GQ-recursion}
    \end{align}
with the boundary condition $G_Q(0,1;z)=0$.
The solution for the surface Green's function above $g(z) = G_Q(1,1;z)$~\eqref{eq:two-body-surface-Green-function-equation}, the $m=1$ case of the above equation gives $G_Q(2,1;z) = (z g(z)-1)/t$, and the $m\geq 2$ case gives $G_Q(m+1,1;z) -\tfrac{z}{t}G_Q(m,1;z) + G_Q(m-1,1;z)=0$, which is a second order homogeneous recurrence relation with the characteristic polynomial $x^2 -\tfrac{z}{t}x +1=0$, whose roots are given by $x_{\pm} = [z\pm \sqrt{z^2-4t^2}]/2t$ satisfying $x_+ x_- =1$ and $x_+ + x_- = z/t$, such that the general solution for $G_Q(m,1;z)$ for $m\geq 1$ is given by $G_Q(m,1;z) = Ax_+^{m} + B x_-^{m}$, where $A$ and $B$ are constants to be determined by the boundary conditions. 
For any $z$ with $\mathrm{Im}(z)>0$, one can verify from the branch structure of $\sqrt{z^2-4t^2}$ that $\abs{x_-(z)}<1<\abs{x_+(z)}$, so the general solution $G_Q(m,1;z)$ must have $A=0$ for $G_Q$ to remain bounded as
$m\to\infty$, where the retarded limit $z=E+i0^+$ is then taken afterward. 
Using $G_Q(m,1;z)=Bx_-^m$ into the $m=1$ equation of~\eqref{eq:GQ-recursion}, and using the characteristic equation $tx_{-}^2-zx_{-} +t=0$ gives $B=1/t$. 
Thus, we arrive at 
    \begin{align}
        G_Q(m,1;z) = \frac{x_-^{m}}{t} = \frac{x_-^{m-1}}{tx_+},
    \end{align}
which, inserted into~\eqref{eq:two-body-quasi-BIC-state-correction-GQ}, yields the general wave function~\eqref{eq:two-body-general-wavefunction}.

\section{Three-particle problem}
\label{app:3body-AHM}
In this appendix we provide more details regarding the three-particle AHM. Here, we discuss the model's configuration space and relative coordinate space representation, the two types of scattering state continua, and the effective model for the three-particle bound states. Then, we briefly mention the consequences of the chiral symmetry on the three-particle spectrum, details regarding the three-body bound states in the continuum, and present the results for the experimentally constrained three-particle AHM. Finally, we also show the three-particle spectrum for $U\neq 0$, and end with an extended analysis of the three-particle wave packet expansion dynamics. 

\subsection{3-particle AHM in configuration space and relative coordinate space}
\label{app:3body-AHM-repr}
The three-particle configuration space basis is given by $\{\ket{n,m,l}\}$ as introduced in the main text~[\cref{sec:3body-problem}]. For an infinite 1D lattice, relevant to this work, we have $n\in \mathbb{Z}$ and for the relative distances $m,l\in \mathbb{N}$ to account for indistinguishability of the three particles~\footnote{In a finite lattice of $L$ sites with three particles, the allowed values are $n\in\{1,\ldots,L\}$, $m\in\{0,\ldots,L-n\}$, and $l\in\{0,\ldots,L-n-m\}$.}.

The full expression for AHM for three particles in the configuration space basis~\eqref{eq:3body-AHM-repr} is given by 
    \begin{widetext}
    \begin{align}
        \hat{H}= -J& \sum_n \bigg[\sqrt{3}\ketbra{n,0,1}{n,0,0} + 2 e^{- i \theta}\ketbra{n,1,0}{n,0,1} +\sqrt{3} e^{- 2 i \theta}\ketbra{n\!+\!1,0,0}{n,1,0} \nonumber\\ 
        & + \sum_{m \geq 0, l \geq 1} \ketbra{n,m,l+1}{n,m,l} + \sum_{m \geq 1} \sqrt{2}\ketbra{n,m,1}{n,m,0} + \sum_{l \geq 2} \sqrt{2}\ketbra{n,1,l-1}{n,0,l} \nonumber \\ 
        &+ \sum_{m \geq 1} \sqrt{2} e^{- i \theta}\ketbra{n,m+1,0}{n,m,1} + \sum_{m \geq 1, l \geq 2} \ketbra{n,m+1,l-1}{n,m,l} \nonumber \\
        & + \sum_{l \geq 1} \sqrt{2} e^{- i \theta}\ketbra{n\!+\!1,0,l}{n,1,l}  + \sum_{m \geq 2, l \geq 0} \ketbra{n\!+\!1,m-1,l}{n,m,l}  + \text{h.c.}\bigg] \nonumber \\
        &+ U \sum_n \bigg(3\ketbra{n,0,0}{n,0,0} + \sum_{l \geq 1} \ketbra{n,0,l}{n,0,l} + \sum_{m \geq 1} \ketbra{n,m,0}{n,m,0}\bigg).
        \label{eq:3body-AHM-repr-full}
    \end{align}
    \end{widetext}
Assuming translational invariance along the direction of the COM position, we can transform this Hamiltonian~\eqref{eq:3body-AHM-repr-full} into the hybrid COM quasimomentum 
basis $\{\ket{Q,m,l}\}$~\eqref{eq:COM-MOM-hybrid-basis}, by which we obtain the AHM Hamiltonian in the relative coordinate space (quasimomentum space) representation
    \begin{widetext}
        \begin{align}
            \hat{H}=\sum_Q &\Big[\hat{H}_Q \Big] 
            =\sum_Q -J\bigg[
        \sqrt{3} e^{\frac{i Q}{3}}\ketbra{Q,0,0}{Q,0,1}  + \sqrt{3} e^{- \frac{i \left(Q + 6 \theta\right)}{3}}\ketbra{Q,0,0}{Q,1,0} + 2 e^{- i \left(\frac{Q}{3} + \theta\right)}\ketbra{Q,1,0}{Q,0,1} \nonumber \\
        &+ \sum_{m \geq 1, l \geq 0} e^{- \frac{i Q}{3}}\ketbra{Q,m,l}{Q,m+1,l} + \sum_{l \geq 1} \sqrt{2} e^{- i \left(\frac{Q}{3} + \theta\right)}\ketbra{Q,0,l}{Q,1,l} + \sum_{m \geq 0, l \geq 1} e^{\frac{i Q}{3}}\ketbra{Q,m,l}{Q,m,l+1} \nonumber  \\
        &+ \sum_{m \geq 1} \sqrt{2} e^{\frac{i Q}{3}}\ketbra{Q,m,0}{Q,m,1}  + \sum_{m \geq 2, l \geq 1} e^{- \frac{i Q}{3}}\ketbra{Q,m,l}{Q,m-1,l+1}  + \sum_{m \geq 2} \sqrt{2} e^{- i \left(\frac{Q}{3} + \theta\right)}\ketbra{Q,m,0}{Q,m-1,1} \nonumber  \\
        & + \sum_{l \geq 1} \sqrt{2} e^{- \frac{i Q}{3}}\ketbra{Q,1,l}{Q,0,l+1}  + \text{h.c.}\bigg] \nonumber  \\
        & + U\bigg(3\ketbra{Q,0,0}{Q,0,0} + \sum_{l \geq 1} \ketbra{Q,0,l}{Q,0,l} + \sum_{m \geq 1} \ketbra{Q,m,0}{Q,m,0}\bigg).
            \label{eq:3body-AHM-repr-hybrid-full}
        \end{align}
    \end{widetext}
    Note that the hybrid COM quasimomentum basis $\{\ket{Q,m,l}\}$~\eqref{eq:COM-MOM-hybrid-basis} allows us to express any eigenstate of the Hamiltonian~\eqref{eq:3body-AHM-repr-full} in the hybrid basis as
    \begin{align}
        \ket{\Psi^{Q,\kappa_1,\kappa_2}}&=\sum_{m,l} \Psi^{Q,\kappa_1,\kappa_2}_{m,l} \ket{Q,m,l}\nonumber \\
        &=\sum_{n,m,l}\Psi^{Q,\kappa_1,\kappa_2}_{m,l}e^{i Q \left(n+\frac{2m}{3}+\frac{l}{3}\right)} \ket{n,m,l}\nonumber \\ 
        &=\sum_{n,m,l} \Psi^{Q,\kappa_1,\kappa_2}_{n,m,l} \ket{n,m,l},
        \label{eq:eigenstate-in-hybrid-basis}
    \end{align}
    where $\kappa_1$ and $\kappa_2$ are yet to be determined quantum numbers associated with the state space spanned by relative coordinates $m$ and $l$, respectively, and where we have defined $\Psi^{Q,\kappa_1,\kappa_2}_{n,m,l}\equiv\Psi^{Q,\kappa_1,\kappa_2}_{m,l}e^{i Q \left(n+2m/3+l/3\right)}$. 
    The three-particle spectrum of the AHM~\eqref{eq:3body-AHM-repr-hybrid-full} can be obtained by solving the eigenvalue problem $\hat{H}_Q\ket{\Psi^{Q,\kappa_1,\kappa_2}} = E_{Q,\kappa_1,\kappa_2}\ket{\Psi^{Q,\kappa_1,\kappa_2}}$~\eqref{eq:3body-AHM-repr-hybrid} for each COM quasimomentum $Q$, a given statistical angle $\theta$ and interaction strength $U$, where $\kappa_{1,2}$ are the quantum numbers associated to the relative coordinates $m,l$ (labeling the eigenstates for a given $Q$). 
    Here, for numerically solving the eigen value problem via exact diagonalization (ED), the relative coordinates $m$ and $l$ have to be truncated to a finite maximum value $m_{\mathrm{max}}$ and $l_{\mathrm{max}}$, which has to be chosen large enough to capture the relevant bound and scattering states.
    In this work we always choose the relative coordinate axes to have the same maximum value which we label by $K$, i.e., $K=m_{\mathrm{max}}=l_{\mathrm{max}}$. 

    \subsection{Scattering state continua in the three-particle AHM}
    \label{app:3body-scattering-state-continuum}
    The eigenstates of the three-body AHM Hamiltonian $\hat{H}_Q$~\eqref{eq:3body-AHM-repr-hybrid-full} can reflect both scattering state and bound state solutions. 
    To extend the discussion of the main text, in the following, we will discuss what type of scattering states exist in the three-particle AHM.

    While bound state solutions are typically characterized by a plane-wave function along the COM direction~\eqref{eq:COM-MOM-hybrid-basis}, which is further modulated by an exponentially decaying wave function $\Psi^{Q}_{m,l}$ into the bulk of the relative coordinate space, the scattering state solutions are typically delocalized across the entire configuration space (omiting the indices $\kappa_i$ for now).
    Hence, they are given by a plane-wave function along the COM direction including a modulation function $\Psi^{Q}_{m,l}$, which is also spatially oscillatory in the relative coordinate space, i.e., in the directions of the relative coordinates $m$ and $l$.

    Since we have a three-particle problem, there are two types of scattering state continua possible.
    On the one hand, we have the single-particle continuum consisting of three free particles and the continuum consisting of two-particle bound states with one free particle, which we will refer to as scattering state continuum of type-I and II respectively in the following.
    We briefly review how these can be understood analytically.

    \subsubsection{Single-particle continuum via three free particles - type-I}
    \label{app:sub:3-body-single-part-continuum}
    The first continuum spectrum (scattering states of type-I in the main text), can be found by noting that in the case of an infinite 1D chain as considered here, the probability of finding the three particles to be close is negligible. 
    Hence, the scattering state solution can be approximated by a product state of three free particles.
    Consequently, for these scattering states the total energy $E(Q,\kappa_1,\kappa_2)$, which becomes exact in the thermodynamic limit, can be written as the sum of the kinetic energies~\cite{Valiente2010a}, $\epsilon(k_i)=-2J\cos(k_i)$, 
    of the three particles, i.e., 
    \begin{align}
        E(Q,\kappa_1,\kappa_2)=\epsilon(k_1)+\epsilon(k_2)+\epsilon(k_3),
        \label{eq:3body-scattering-state-total-energy}
    \end{align}
    where $k_i$ ($i\in\{1,2,3\}$) are the individual quasimomenta of the three particles.
    Note that the total energy of two free particles can be expressed by $E_q(\kappa)=\epsilon(k_1)+\epsilon(k_2)=-4J\cos(q/2)\cos(\kappa)$~\eqref{eq:AHM-Scattering-dispersion}, where $k_1=q/2+\kappa$ and $k_2=q/2-\kappa$ are the quasimomenta of the first and second particle respectively and where $q$ is the two-particle COM quasimomentum $q=k_1+k_2$ and $\kappa=(k_1-k_2)/2$ is the relative quasimomentum of the two particles~\cite{Krutitsky2016,Valiente2008,Piil2007}.
    With this, we can express the total energy~\eqref{eq:3body-scattering-state-total-energy} of the three-particle scattering state in terms of the three-particle COM quasimomentum $Q$ and the relative quasimomenta $\kappa_1$ and $\kappa_2$ as
    \begin{align}
        E(Q,\kappa_1,\kappa_2)=-4J\cos\left(\frac{q}{2}\right)\cos(\kappa)-2J\cos(Q-q)
        \label{eq:3body-scattering-state-total-energy-2}
    \end{align}
    Thus, the continuum spectrum of the free three-particle scattering states is bounded by
    \begin{align}
        \min_{k_1+k_2+k_3=Q}\left[\sum_{i=1}^3\epsilon(k_i)\right] 
        \leq E \leq 
        \max_{k_1+k_2+k_3=Q}\left[\sum_{i=1}^3\epsilon(k_i)\right].
        \label{eq:3body-free-scattering-state-energy-bounds}
    \end{align}
    Using a complementary approach discussed in the next section, it can be shown that the bounds for this continuum, i.e., the minimum and maximum energies of~\eqref{eq:3body-free-scattering-state-energy-bounds}, are given by $E_{\text{min}}(Q)=-6J\cos(Q/3)$ and $E_{\text{max}}(Q)=6J\cos(Q/3+\pi/3)$ for $Q\in[-\pi,0)$ and $E_{\text{max}}(Q)=6J\cos(Q/3-\pi/3)$ for $Q\in[0,\pi]$ respectively.
    
    \subsubsection{Single-particle continuum via plane wave in 2D triangular relative coordinate space lattice - type-I}
    \label{app:sub:3-body-single-part-continuum-plane-wave}
    Note that above, for the 2-particle case, we have shown that one can understand the scattering continuum energy of two free particles  $E_{q,\kappa}=-4J\cos(q/2)\cos\left(\kappa\right)$ by either taking the sum of two individual single-particle energy dispersions~\eqref{eq:AHM-Scattering-dispersion} or also by assuming a plane wave ansatz $\psi^{q,\kappa}_m\sim e^{\pm im\kappa}$ for a state deep inside the bulk of a translationally invariant chain in the 1D relative coordinate space, which essentially corresponds to the recurrence relation~\eqref{eq:AHM-rel-system-coeff3}.
    In the following, complementary to the discussion in Appendix~\ref{app:3body-scattering-state-continuum}, we also want to investigate the second approach for the three-particle case.

    The scattering state solutions of type-I correspond to delocalized 2D plane wave solutions deep in the bulk of the relative coordinate space associated with the coordinates $\ket{Q,m\geq 2,l \geq 2}$ in~\cref{fig:Relative-coordinate-space} of the main text. 
    The Hamiltonian governing these dynamics deep in the bulk is given by,
        \begin{align}
            \hat{H}_Q=&-J\sum_{m,l \geq 2}\big[e^{-i\frac{Q}{3}}\big(\ketbra{Q,m-1,l}{Q,m,l}\\
            &\quad +\ketbra{Q,m,l-1}{Q,m-1,l}\nonumber\\
            &\quad +\ketbra{Q,m,l}{Q,m,l-1}\big) + \text{h.c.} \big],\nonumber
            \label{eq:H-Q-3body-bulk}
        \end{align}
    corresponding to the translationally invariant 2D triangular lattice for $m,l \geq 2$ (see~\cref{fig:Relative-coordinate-space} of the main text).
    Due to the translational invariance along both $m$ and $l$, we can introduce the relative quasimomenta $\bm{\kappa}=(\kappa_1,\kappa_2)$ associated to the relative coordinates $m$ and $l$, such that we can diagonalize the Hamiltonian $\hat{H}_Q$~\eqref{eq:H-Q-3body-bulk} deep in the bulk to yield
        \begin{align}
            \hat{H}_Q=\sum_{\bm{\kappa}} E(Q,\bm{\kappa}) \ketbra{Q,\kappa_1,\kappa_2}{Q,\kappa_1,\kappa_2},
        \end{align}
    where $\Psi^{Q,\kappa_1,\kappa_2}_{m,l}=\braket{Q,m,l}{Q,\kappa_1,\kappa_2}\sim e^{\pm i(m\kappa_1+l\kappa_2)}$ are the plane wave eigenstates deep inside in the relative coordinate space and $\bm{\kappa}=(\kappa_1,\kappa_2)$ is the relative quasimomentum vector.
    Thus, the energy dispersion relation of the scattering states of type-I, $E(Q,\bm{\kappa})$, is thus the energy dispersion in a 2D triangular lattice
        \begin{align}
            E(Q,\bm{\kappa})=&-2J[\cos\left(\bm{\kappa}\cdot\bm{a}_1-Q/3\right)\nonumber\\
            &+\cos\left(\bm{\kappa}\cdot\bm{a}_2+Q/3\right)+\cos\left(\bm{\kappa}\cdot\bm{a}_3+Q/3\right)],
            \label{eq:3body-scattering-state-energy-dispersion-triangular-type-I}
        \end{align}
    where $\bm{a}_1$ is the lattice vector along the $m$-direction, connecting for instance $\ket{Q,2,2}$ to $\ket{Q,3,2}$, $\bm{a}_2$ is the lattice vector along the $l$-direction, connecting for instance $\ket{Q,2,2}$ to $\ket{Q,2,3}$, and $\bm{a}_3=\bm{a}_1-\bm{a}_2$. By rotating the relative coordinate space~\cref{fig:Relative-coordinate-space} we can identify $\bm{a}_1=(1,0)^T$, $\bm{a}_2=(0,1)^T$ and $\bm{a}_3=(1,-1)^T$, such that $E(Q,\bm{\kappa})= -2J\big[\cos\left(\kappa_1-Q/3\right)+\cos\left(\kappa_2+Q/3\right)+\cos\left(\kappa_1-\kappa_2+Q/3\right)\big]$. 
    Solutions of the scattering state continuum of type-I lie between the minimum and maximum of the energy dispersion relation $E(Q,\bm{\kappa})$. 
    The minimum of the energy dispersion relation is given by $E_{\text{min}}(Q,\bm{\kappa}=\bm{0})=-6J\cos(Q/3)$ when both relative quasimomenta are zero, $\kappa_1=\kappa_2=0$.
    The maximum is attained, for $Q\in[-\pi,0)$, at $\bm{\kappa}=(2\pi/3,-2\pi/3)$ with $E_{\text{max}}(Q,\bm{\kappa})=6J\cos((Q+\pi)/3)$, and, for $Q\in[0,\pi]$, at $\bm{\kappa}=(-2\pi/3,2\pi/3)$ with $E_{\text{max}}(Q,\bm{\kappa})=6J\cos((Q-\pi)/3)$, which can be compactly written as $E_{\text{max}}(Q,\bm{\kappa})=6J\cos\left(\frac{\pi-\abs{Q}}{3}\right)$ for $Q\in[-\pi,\pi)$.
    Finally, using $k_1 =-\kappa_1+Q/3$, $k_2 = \kappa_2+Q/3$ and $k_3 = \kappa_1-\kappa_2+Q/3$, where $k_i$ are the momenta of the three individual particles ($i=1,2,3$) such that the constraint $Q=k_1+k_2+k_3$ is satisfied, we can also write the energy dispersion relation as the sum of the individual single-particle energy dispersions, i.e., $E(Q,\bm{\kappa}) = \epsilon(k_1)+\epsilon(k_2)+\epsilon(k_3)$, with $\epsilon(k)=-2J\cos(k)$ being the energy dispersion of a single free particle, which matches the discussion above in Appendix~\ref{app:sub:3-body-single-part-continuum}.

    \subsubsection{Continuum of a two-particle bound state and one free particle - type-II}
    \label{app:sub:3-body-two-part-bound-one-free-continuum}
    Because the two-particle AHM spectrum hosts exact two-particle bound states in the continuum (BIC) for arbitrary statistical angles $\theta\neq0$ and on-site interaction $U$~\cite{Kwan2024}, we can also find a second type of continuum in the three-particle AHM.
    This second continuum spectrum consists of scattering state solutions, which can consist of a two-particle bound state which is accompanied by one free particle (scattering states of type-II in the main text)
    As the energy spectrum $E_{\pm}(q)$ of the two-particle bound state is analytically known for arbitrary statistical angles $\theta$ and on-site interaction $U$~\cite{Kwan2024}, we can express the total energy of the three-particle scattering state as the sum of the two-particle bound-state spectrum and the kinetic energy of a single free particle, i.e.,
    \begin{align}
        E^{\pm}(Q,q)=E_{\pm}(q)+\epsilon(Q-q)
        \label{eq:3body-scattering-state-total-energy-3}
    \end{align}
    Here, $Q=q+k_3$ is the total three-particle COM quasimomentum, $\epsilon(k_3)=-2J\cos(k_3)=-2J\cos(Q-q)$ is the kinetic energy of the third free particle and $E_{\pm}(q)$~\eqref{eq:AHM-bound-states-dispersion} is the exact energy of the two-particle bound state with the two-particle COM quasimomentum $q=k_1+k_2$ (cf.~\cref{app:2body-bound-state}).
    The upper and lower bound of the continuum spectrum with two-particle bound states are thus computed via
    \begin{align}
        \min_{q+k_3=Q}[\epsilon(k_3)+E_{\pm}(q)] \leq E \leq \max_{q+k_3=Q}[\epsilon(k_3)+E_{\pm}(q)],
        \label{eq:3body-bound-scattering-state-energy-bounds}
    \end{align}
    where, different to the free particle continuum~\eqref{eq:3body-free-scattering-state-energy-bounds}, this continuum explicitly depends on the statistical angle $\theta$ and the on-site interaction $U$.

    The free particle continuum \eqref{eq:3body-free-scattering-state-energy-bounds} and the continuum with two-particle bound states \eqref{eq:3body-bound-scattering-state-energy-bounds} are shown in 
    \cref{fig:three-body-spectrum-full-AHM} of the main text for different statistical angles $\theta$ and $U=0$.
    From an energetic perspective, any state, for a given COM quasimomentum $Q$, which lies outside both continua is considered to be a (three-body)-bound state.
    
    \subsection{Effective $4\times 4$ model for three-body bound states in the AHM}
    \label{app:3body-effective-model-4x4}
    In this appendix, we provide details on the derivation of the effective $4\times 4$ model for three-body bound states in the AHM discussed in~\cref{sec:approximate-spectrum-AHM} of the main text.
    The effective model is obtained by truncating the configuration space to tunneling processes between the four configurations $\ket{n,0,0}, \ket{n,1,0}, \ket{n,0,1}$ and $\ket{n,1,1}$ (as illustrated in~\cref{fig:three-body-spectrum-effective-AHM-U0-Binding-mechanism}(a) of the main text).
    The configuration space Hamiltonian~\eqref{eq:3body-AHM-repr-full} truncated to those processes reads,  
    \begin{widetext}
        \begin{align}
            \hat{H}^{\mathrm{eff}}_{4\times 4} = -J &\sum_n \bigg[
                \sqrt{3}\left( 
                    e^{-2i\theta} \ketbra{n,0,0}{n-1,1,0}+\ketbra{n,0,1}{n,0,0} +\text{h.c.} \right) 
                    \bigg] \nonumber\\
                    &+\left( e^{-i\theta}
                    \left( \sqrt{2}\ketbra{n,0,1}{n-1,1,1}+2\ketbra{n,1,0}{n,0,1}\right)+\sqrt{2}\ketbra{n,1,1}{n,1,0}+ \text{h.c.}\right)
                    \nonumber \\
                    &+U \sum_n \bigg(3\ketbra{n,0,0}{n,0,0} +  \ketbra{n,1,0}{n,1,0} + \ketbra{n,0,1}{n,0,1} \bigg).
                    \label{eq:3body-AHM-repr-effective-4x4}
                \end{align}
    Transforming into the hybrid COM quasimomentum basis $\{\ket{Q,m,l}\}$~\eqref{eq:3body-AHM-repr-hybrid-full}, we obtain the Hamiltonian corresponding to the deeply bound 3-body states region in the relative coordinate space~\cref{fig:Relative-coordinate-space}, where this effective Hamiltonian reads
    \begin{align}
        \hat{H}^{\mathrm{eff}}_{4\times 4} = \sum_Q  \Big[\hat{H}^{\mathrm{eff}}_{Q,4\times 4}\Big]
        =\sum_Q &\bigg[
            \sqrt{3}\left( 
                e^{-i\left(2\theta+\frac{Q}{3}\right)} \ketbra{Q,0,0}{Q,1,0}
                +e^{-i\frac{Q}{3}}\ketbra{Q,0,1}{Q,0,0} +\text{h.c.} \right) \nonumber\\
                &+\Big(e^{-i\left(\theta+\frac{Q}{3}\right)}
                \left( \sqrt{2}\ketbra{Q,0,1}{Q,1,1}+2\ketbra{Q,1,0}{Q,0,1}\right) \nonumber\\
                &+\sqrt{2}e^{-i\frac{Q}{3}}\ketbra{Q,1,1}{Q,1,0}+ \text{h.c.}\Big)
                \nonumber\\
                &+U \bigg(3\ketbra{Q,0,0}{Q,0,0} +  \ketbra{Q,1,0}{Q,1,0} + \ketbra{Q,0,1}{Q,0,1} \bigg)
                \bigg],
        \label{eq:3body-AHM-repr-effective-4x4-hybrid-full}
    \end{align}
    which can be expressed in matrix form as
    \begin{align}
         \hat{H}^{\mathrm{eff}}_{4\times 4} 
        &=\sum_Q
            \bm{u}_Q^{\dagger}
            \begin{pmatrix}
                3U & -\sqrt{3}J e^{-i\left(2\theta+\frac{Q}{3}\right)} & -\sqrt{3}J e^{i\frac{Q}{3}} & 0 \\
                -\sqrt{3}J e^{i\left(2\theta+\frac{Q}{3}\right)} & U & -2J e^{-i\left(\theta+\frac{Q}{3}\right)} & -\sqrt{2}J e^{i\frac{Q}{3}} \\
                -\sqrt{3}J e^{-i\frac{Q}{3}} & -2J e^{i\left(\theta+\frac{Q}{3}\right)} & U & -\sqrt{2}J e^{-i\left(\frac{Q}{3}+\theta\right)} \\
                0 & -\sqrt{2}J e^{-i\frac{Q}{3}} & -\sqrt{2}J e^{i\left(\frac{Q}{3}+\theta\right)} & 0
            \end{pmatrix}
            \bm{u}_Q,
            \label{eq:3body-AHM-repr-effective-4x4-hybrid-matrix}
        \end{align}
        where we have defined the vector
        $\bm{u}^{\dagger}_Q = \begin{pmatrix}
            \ket{Q,0,0} &
            \ket{Q,1,0} &
            \ket{Q,0,1} &
            \ket{Q,1,1}
        \end{pmatrix}$.
        
\end{widetext}

\subsection{Chiral symmetry of the three-particle AHM}\label{app:3body-chiral-symmetry}
    As per the discussion above in Appendix~\ref{app:2body-chiral-symmetry}, also the 3-particle AHM for $U=0$ admits a chiral symmetry that acts as $\hat{S}\ket{n,m,l} = e^{i\pi (3n+2m+l)}\ket{n,m,l} = e^{i\pi (n+l)}\ket{n,m,l}$ in the configuration space representation. 
    However, for the three-particle AHM, this chiral symmetry operator is \emph{not} diagonal in the hybrid COM quasimomentum space basis, i.e., 
    \begin{align}
        \hat{S}\ket{Q,m,l} = e^{i\frac{2}{3}\pi(l-m)}\ket{Q+\pi,m,l},
        \label{eq:3body-chiral-symmetry-hybrid-basis}
    \end{align}
    which is essentially because the Hamiltonian $\hat{H}_Q$~\eqref{eq:3body-AHM-repr-effective-3x3-hybrid-full} acts on a relative coordinate space lattice~[Fig.~\ref{fig:Relative-coordinate-space} in the main text] that is not bipartite. 
    The consequence of this symmetry on the scattering state continuum of type-I~\eqref{eq:3body-scattering-state-energy-dispersion-triangular-type-I} is $E(Q+\pi,\kappa_1-\frac{2\pi}{3},\kappa_2+\frac{2\pi}{3})=-E(Q,\kappa_1,\kappa_2)$. 
    For the scattering state continuum of type-II, using properties of the two-body case $E_{q,+}=-E_{q,-}$~\eqref{eq:AHM-bound-states-dispersion-Uzero} for $U=0$, we have that $E^{\mp}(Q+\pi,q)=-E^{\pm}(Q,q)$~\eqref{eq:3body-scattering-state-total-energy-3}, which is also consistent with the numerical results shown in~\cref{fig:three-body-spectrum-full-AHM} of the main text.
    Moreover, as $\hat{S}$ maps the $Q$-block to the $(Q+\pi)$-block it follows that each bound state band $n$ at COM quasimomentum $Q$ has a chiral partner band at COM quasimomentum $Q+\pi$ with energy $-E_n(Q)$, which again is confirmed by looking at~\cref{fig:three-body-spectrum-full-AHM} of the main text. 
    In particular, for the $4\times 4$ effective model~\eqref{eq:3body-AHM-repr-effective-4x4-hybrid-matrix}, with the four effective energy band solutions $E^{\mathrm{eff}}_{n}(Q)$, $n=1,2,3,4$, ordered as $E^{\mathrm{eff}}_{1}(Q)\leq \dots \leq E^{\mathrm{eff}}_{4}(Q)$, we have that $E^{\mathrm{eff}}_{4}(Q+\pi)=-E^{\mathrm{eff}}_{1}(Q)$ and $E^{\mathrm{eff}}_{3}(Q+\pi)=-E^{\mathrm{eff}}_{2}(Q)$, which further agrees with the results shown in~\cref{fig:three-body-spectrum-effective-AHM-U0-Binding-mechanism}(b) of the main text. 

\subsection{Details on the three-body (quasi-)bound states in the continuum of the AHM}{\label{app:3body-BIC-mechanism}}
\begin{figure}[b]
\includegraphics[width=\columnwidth]{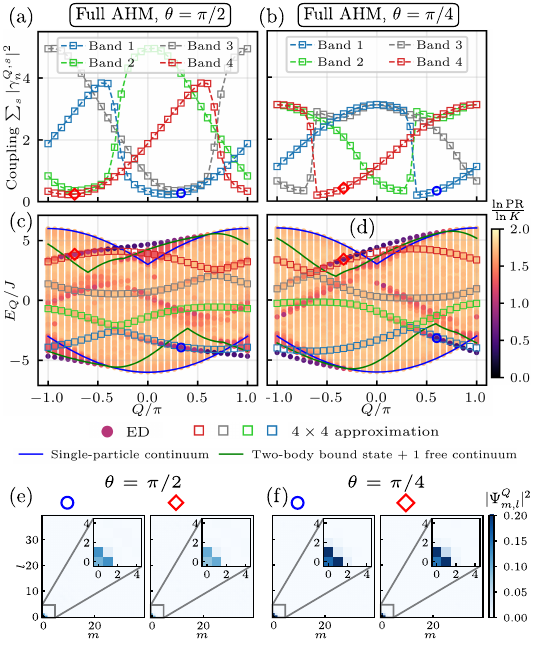}
    \caption{\textbf{Mechanism for three-particle bound states in the continuum for $U=0$.}
    (a)-(b) Total coupling $\sum_{m,l}|\gamma_{n}^{Q,m,l}|^2$~\eqref{eq:three-body-BIC-mechanism-couplings} for the four eigenstates $\ket{\psi^{Q,n}}$ of the deeply bound 3-body state region to the neighboring bulk sites as a function of the three-particle COM quasimomentum $Q$ for $\theta=\pi/2$ and $\theta=\pi/4$ respectively.
    (c)-(d) Full three-particle spectrum (ED) together with the $4\times 4$ approximate spectrum of~\eqref{eq:3body-AHM-repr-effective-4x4-hybrid-matrix} for $\theta=\pi/2$ and $\theta=\pi/4$ respectively, where the color of the dots again indicates the scaled participation ratio.
    (e)-(f) Density distributions of selected three-body BICs of the lower and upper bound state branch in the relative coordinate space.
    Other parameters are the same as in~\cref{fig:three-body-spectrum-full-AHM} of the main text.
    }
\label{fig:three-body-BIC-mechanism-couplings}
\end{figure}

We provide a more detailed discussion on the mechanism for the stabilization of three-body bound states inside the scattering state continua in the AHM, which was mentioned in~\cref{sec:3BodyBIC} of the main text. 
Similar to the two-body case in Appendix~\ref{app:2body-BIC}, here, the Hamiltonian $H_Q$ describing the three-particle AHM in the relative coordinate space~\eqref{eq:3body-AHM-repr-hybrid-full}, corresponding to the~\cref{fig:Relative-coordinate-space}, can be decomposed into two sectors, a deeply bound 3-body state region, defined by the projector $\hat{P}=\sum_{m,l\leq 1}\ketbra{Q,m,l}{Q,m,l}$, and a bulk region hosting the rest of the relative coordinate space, defined by the complement $\hat{\mathcal{Q}} = \mathbbm{1}-\hat{P}$. 
In this decomposition, the Hamiltonian $\hat{H}_Q$ takes the following block form,
\begin{align}
    \hat{H}_Q = \begin{pmatrix}
        \hat{H}_{\text{3-BS}} & \hat{V}_{Q} \\
        \hat{V}^\dagger_{Q} & \hat{H}_{\text{bulk}}
    \end{pmatrix},
\end{align}
where, here the subscript $Q$ (the three-particle COM quasimomentum) should not be confused with the notation of the complement projector $\hat{\mathcal{Q}}$.
The Hamiltonian $\hat{H}_{\text{3-BS}}=\hat{P}\hat{H}_Q\hat{P}$, which is given by~\eqref{eq:3body-AHM-repr-effective-4x4-hybrid-matrix} describes the deeply bound 3-body state region, while $\hat{H}_{\text{bulk}}=\hat{\mathcal{Q}}\hat{H}_Q\hat{\mathcal{Q}}$ the rest of the relative coordinate space, including the left and right triangular ladders, as well as the bulk of the 2D triangular lattice hosting the continuum of scattering states of type-I (see~\ref{app:sub:3-body-single-part-continuum-plane-wave}). 
The $\hat{V}_{Q}=\hat{P}\hat{H}_Q\hat{\mathcal{Q}}$ then describes the coupling between these two regions~[Fig.~\ref{fig:Relative-coordinate-space}] and is explicitly given by
\begin{widetext}
\begin{align}
    \hat{V}_{Q}= &-J[e^{iQ/3}(\ketbra{Q,0,1}{Q,0,2}+\ketbra{Q,1,1}{Q,1,2}) 
    +e^{-i(Q/3)}(\ketbra{Q,1,1}{Q,2,1}+\ketbra{Q,1,0}{Q,2,0})\nonumber \\
    &+ \sqrt{2}(e^{+i(Q/3+\theta)}\ketbra{Q,1,1}{Q,2,0}+e^{-iQ/3}\ketbra{Q,1,1}{Q,0,2})].
    \label{eq:3-body-BIC-mechanism-coupling-operator}
\end{align}
\end{widetext}
With this, our $P$-space projected Green's function (operator) reads
\begin{align}
    \hat{G}_P(E,Q) = \frac{1}{E\hat{P}-\hat{H}_{\text{3-BS}}-\hat{\Sigma}_P(E,Q)},
    \label{eq:3-body-BIC-mechanism-P-space-resolvent}
\end{align}
where the self-energy operator $\hat{\Sigma}_P(z,Q)$ given by
    \begin{align}
        \hat{\Sigma}_P(z,Q) &= \hat{V}_{Q}\frac{1}{z\hat{\mathcal{Q}}-\hat{H}_{\text{bulk}}}\hat{V}^\dagger_{Q}\nonumber \\
        &= \sum_k \frac{\hat{V}_Q|E_k^\text{bulk}\rangle\langle E_k^\text{bulk}|\hat{V}_Q^\dagger}{z - E_k^\text{bulk}}
        \label{eq:3-body-BIC-mechanism-self-energy-operator}
    \end{align}
with $|E_k^\text{bulk}\rangle$ and $E_k^\text{bulk}$ being the eigenstates and eigenenergies of the bulk Hamiltonian $\hat{H}_{\text{bulk}}$ respectively.
Next, we would like to represent the self-energy operator $\hat{\Sigma}_P(E,Q)$ in the eigenbasis of $\hat{H}_{\text{3-BS}}$, which are defined by the eigenvalue problem, $\hat{H}_{\text{3-BS}}\ket{\psi^{Q,n}} = E^{\text{eff}}_{Q,n}\ket{\psi^{Q,n}}$ with $n\in\{1,2,3,4\}$ labeling the four eigenstates corresponding to the four eigenvalues $E^{\text{eff}}_{Q,n}$. Now, we note that $V_Q^\dagger$~\eqref{eq:3-body-BIC-mechanism-coupling-operator} only couples the four boundary (bdy) bulk sites $(m,l)\in \{(2,0), (0,2), (2,1), (1,2)\}$ to the corner eigentstates of the deeply bound 3-body state region, i.e., $\matel{E^{\text{bulk}}_k}{V_Q^\dagger}{\psi^{Q,n}} = \sum_{m,l \in \mathrm{bdy}} \braket{E_k^\text{bulk}}{Q,m,l}\gamma_{n}^{Q,m,l}$, where we have defined the transition matrix elements $\gamma_{n}^{Q,m,l} \equiv \bra{Q,m,l}\hat{V}^\dagger_Q\ket{\psi^{Q,n}}$. 
Then, in the eigenbasis of $\hat{H}_{\text{3-BS}}$, the self-energy operator~\eqref{eq:3-body-BIC-mechanism-self-energy-operator} reads
\begin{widetext}
    \begin{align}
        [\Sigma_P(z,Q)]_{n,n'} = \bra{\psi^{Q,n}}\hat{\Sigma}_P(z,Q)\ket{\psi^{Q,n'}}
        &= \sum_{(m,l),(m',l')\in \mathrm{bdy}} (\gamma_{n}^{Q,m,l})^* 
        \matel{Q,m,l}{(z\hat{\mathcal{Q}}-\hat{H}_{\text{bulk}})^{-1}}{Q,m',l'}
        (\gamma_{n'}^{Q,m',l'}) \nonumber \\
        &\equiv \sum_{(m,l),(m',l')\in \mathrm{bdy}} (\gamma_{n}^{Q,m,l})^* 
        \big[G_{\mathcal{Q}}(z)\big]_{(m,l),(m',l')} (\gamma_{n'}^{Q,m',l'}),
        \label{eq:3-body-BIC-mechanism-self-energy-matrix-elements}
    \end{align}
\end{widetext}
With this, we have obtained a similar situation as for the two-body BIC case in Appendix~\ref{app:2body-BIC}, but with a more complicated self-energy operator $\hat{\Sigma}_P(E,Q)$ involving the projected Green's function of the bulk region, which is now a $4\times 4$ matrix in the boundary site basis, i.e., $\big[G_{\mathcal{Q}}(z)\big]_{(m,l),(m',l')} = \matel{Q,m,l}{(z\hat{\mathcal{Q}}-\hat{H}_{\text{bulk}})^{-1}}{Q,m',l'}$ with $(m,l),(m',l')\in \mathrm{bdy}$. So far everything is exact, and no approximation has been made yet. 

However, here in the 2D case, we do not have an exact analytical recursion relation like for the 1D two-body case, such that we need to employ an approximation or treat this impurity problem numerically. 
Let us start with a simplifying approximation where we assume the boundary sites to be locally equivalent and uncorrelated, such that we can approximate the projected Green's function as $\big[G_{\mathcal{Q}}(z)\big]_{(m,l),(m',l')} \approx \delta_{(m,l),(m',l')}g_{\mathrm{bulk}}(z)$, where $g_{\mathrm{bulk}}(z)$ is the local Green's function of the bulk region. At first this seems like a rather crude approximation as two of the four boundary sites lie on the type-II edge ladders while the other two are type-I bulk sites~[cf.~\cref{fig:Relative-coordinate-space}], but it turns out to be a very good approximation for our purposes as we will see below.
The accuracy of this local approximation is set by the off-diagonal part of the bulk resolvent and not by the coupling strength. Writing the exact self-energy~\eqref{eq:3-body-BIC-mechanism-self-energy-matrix-elements} as $\hat{\Sigma}_P(z) = \hat{\gamma}^{\dagger}\hat  G_{\mathcal{Q}}(z)\hat{\gamma}$, with $[\hat{\gamma}]_{(m,l),n}=\gamma_{n}^{Q,m,l}$ and $\hat G_{\mathcal{Q}}(z)$ the boundary-projected bulk resolvent, and splitting $\hat G_{\mathcal{Q}}(z) = g_{\mathrm{bulk}}(z)\mathbbm{1} + \delta G_{\mathcal{Q}}(z)$ into on-site and inter-site parts ($[\delta G_{\mathcal{Q}}(z)]_{(m,l),(m,l)}=0$) separates the local approximation from the neglected remainder,
\begin{align}
    \hat{\Sigma}_P(z) = \underbrace{g_{\mathrm{bulk}}(z)\,\hat{\gamma}^{\dagger}\hat{\gamma}}_{\hat{\Sigma}_P^{\mathrm{loc}}(z)} + \underbrace{\hat{\gamma}^{\dagger}\,\delta G_{\mathcal{Q}}(z)\,\hat{\gamma}}_{\delta\hat{\Sigma}_P(z)}.
    \label{eq:3body-BIC-self-energy-local-split}
\end{align}
Both terms are quadratic in $\hat{\gamma}$ and hence of the same order $\abs{\gamma}^2$, so weak coupling does not make the local part dominant; the neglected inter-site term $\delta\hat{\Sigma}_P$ is not parametrically small, as some boundary sites are indeed nearest neighbors~[\cref{fig:Relative-coordinate-space}]. 
We therefore use the local approximation only as a qualitative estimate of the energy-resolved spectral function. 
Our identification of three-body qBICs later does not depend on it. 
Since $\hat{\Sigma}_P^{\mathrm{loc}}$ and $\delta\hat{\Sigma}_P$ are both quadratic in $\hat{\gamma}$, they vanish together wherever a band's boundary couplings do, so the qBICs are located at the dips of the total coupling $\sum_{(m,l)\in\mathrm{bdy}}\abs{\gamma_{n}^{Q,m,l}}^2$ irrespective of the approximation. A perfect three-body BIC is fixed by the exact condition $\hat{V}_Q^{\dagger}\ket{\psi^{Q,n}}=0\Leftrightarrow\gamma_{n}^{Q,m,l}=0\ \forall(m,l)$~\eqref{eq:three-body-BIC-mechanism-couplings-individual}, and we confirm below that the dips of this indicator coincide with the three-body qBICs of the exact spectrum~[Fig.~\ref{fig:three-body-BIC-mechanism-couplings}]. 
The approximation enters only the energy-resolved spectral function, i.e., its widths and level shifts, not the COM momenta at which the qBICs occur. 

Now equipped with this local approximation, the self-energy operator factorizes in the eigenbasis of $\hat{H}_{\text{3-BS}}$ and reads $[\Sigma_P(z,Q)]_{n,n'} \approx g_{\mathrm{bulk}}(z)\sum_{(m,l)\in \mathrm{bdy}} (\gamma_{n}^{Q,m,l})^* (\gamma_{n'}^{Q,m,l})$. 
We see that this approximation neglects inter-site correlations between $(m,l)\leftrightarrow (m',l')$ in the bulk region, but keeps correlations between the different eigenstates $(n,n')$ encoded via the off-diagonal elements of the self-energy, $[\Sigma_P(z,Q)]_{n,n'}$.
On the real axis $z=E+i0^+$, we can then decompose $g_{\mathrm{bulk}}(E+i0^+) = \mathrm{Re}\{g_{\mathrm{bulk}}(E)\} - i\pi \rho_{\mathrm{bulk}}(E)$, such that we have 
$[\Sigma_P(E,Q)]_{n,n'} = \Lambda_{n,n'}(E,Q)-\tfrac{i}{2}\Gamma_{n,n'}(E,Q)$ with the energy shift matrix $\Lambda_{n,n'}(E,Q) = \mathrm{Re}\{g_{\mathrm{bulk}}(E)\}\sum_{(m,l)\in \mathrm{bdy}} (\gamma_{n}^{Q,m,l})^* (\gamma_{n'}^{Q,m,l})$ and the decoherence matrix $\Gamma_{n,n'}(E,Q) = 2\pi \rho_{\mathrm{bulk}}(E)\sum_{(m,l)\in \mathrm{bdy}} (\gamma_{n}^{Q,m,l})^* (\gamma_{n'}^{Q,m,l})$, where the diagonal elements give the individual decay rates 
    \begin{align}
        \Gamma_{nn}(E,Q) = 2\pi \rho_{\mathrm{bulk}}(E)\sum_{(m,l)\in \mathrm{bdy}} |\gamma_{n}^{Q,m,l}|^2.
        \label{eq:three-body-BIC-mechanism-self-energy-diagonal-decay-rates}
    \end{align}

Expressing the full projected resolvent~\eqref{eq:3-body-BIC-mechanism-P-space-resolvent} in the eigenbasis of $\hat{H}_{\text{3-BS}}$, while being far away from avoided crossings of the energy bands $E^{\text{eff}}_{Q,n}$ where the off-diagonal elements of the self-energy $[\Sigma_P(E,Q)]_{n,n'}$ can be neglected, and the matrix inversion becomes diagonal, we can obtain a spectral function for each band $n$ independently, which reads
    \begin{align}
        &A_n(E,Q) = -2\mathrm{Im} [G_P(E+i0^+,Q)]_{n,n} \nonumber\\
        &= \frac{\Gamma_{nn}(E,Q)}{(E - E^{\text{eff}}_{Q,n} - \Lambda_{nn}(E,Q))^2 + (\Gamma_{nn}(E,Q)/2)^2}.
    \end{align}
Near avoided crossings one must invert the full $4\times 4$ problem such that the spectral function becomes a superposition of coupled resonances.

Let us now test whether a perfect 3-body BIC is possible, which is characterized by a total vanishing of the self-energy matrix $[\Sigma_P(E,Q)]_{n,n'}=0~\eqref{eq:3-body-BIC-mechanism-self-energy-matrix-elements}$, as for the 2-body case above~\eqref{eq:self-energy-2body-AHM-BIC}. The vanishing of the self-energy matrix is equivalent to $V_Q^\dagger\ket{\psi^{Q,n}}=0$ such that the eigenstate $\ket{\psi^{Q,n}}$ corresponding to the energy $E^{\text{eff}}_{Q,n}$ of the deeply bound 3-body state region would be completely decoupled from the bulk and become an eigenstate of the full Hamiltonian $\hat{H}_Q$, and thus a perfect three-body BIC. 
By $V_Q^\dagger\ket{\psi^{Q,n}}=\sum_{(m,l)\in \mathrm{bdy}}\gamma_{n}^{Q,m,l} \ket{Q,m,l}=0$, this criterion becomes that each individual transition matrix element must vanish, i.e., $\gamma_{n}^{Q,m,l} =0$ for all boundary sites $(m,l)\in \{(2,0), (0,2), (2,1), (1,2)\}$~\eqref{eq:3-body-BIC-mechanism-coupling-operator}, which are explicitly given by 
    \begin{align}
        \gamma_{n}^{Q,2,0}
        &=-J e^{i{Q/3}}\left[\psi^{Q,n}_{1,0} + \sqrt{2}e^{-i{(2Q/3+\theta)}}\psi^{Q,n}_{1,1} \right], \nonumber \\
        \gamma_{n}^{Q,0,2}
        &=-J e^{-i{Q/3}}\left[\psi^{Q,n}_{0,1} + \sqrt{2}e^{i{(2Q/3)}}\psi^{Q,n}_{1,1} \right], \nonumber \\
        \gamma_{n}^{Q,2,1}&=-J e^{+i{Q/3}}\psi^{Q,n}_{1,1}, \nonumber \\
        \gamma_{n}^{Q,1,2}&=-J e^{-i{Q/3}}\psi^{Q,n}_{1,1}.
        \label{eq:three-body-BIC-mechanism-couplings-individual}
    \end{align}
Here, we have expanded the eigenstate $\ket{\psi^{Q,n}} = \sum_{m,l\leq 1} \psi^{Q,n}_{m,l}\ket{Q,m,l}$ in the original basis of the relative coordinate space. As we have also discussed in the main text pictorially via~\cref{fig:Relative-coordinate-space}, it follows from~\eqref{eq:three-body-BIC-mechanism-couplings-individual} that for the perfect 3-body BIC to exist, $\psi^{Q,n}_{1,1}=\psi^{Q,n}_{0,1}=\psi^{Q,n}_{1,0}=0$, such that only an eigenstate that is fully localized at $\psi^{Q,n}_{0,0}=1$ could be a perfect 3-body BIC. 
Note that these are $d=5$ real constraints (with normalization) with only one free parameter $Q$ such that we have an over-constrained problem suggesting that, the existence of a perfect 3-body BIC would require fine-tuning of the parameters. 
Moreover, since each $\abs{\gamma_{n}^{Q,m,l}}^2 \geq 0$, the vanishing of the total coupling $\sum_{m,l\in \mathrm{bdy}}|\gamma_{n}^{Q,m,l}|^2=0$ is equivalent to $\gamma_{n}^{Q,m,l} =0, \, \forall (m,l)\in \mathrm{bdy}$. 
Thus, in the following, we use the total coupling, given by
\begin{widetext}
    \begin{align}
        &\sum_{m,l\in \mathrm{bdy}}|\gamma_{n}^{Q,m,l}|^2
        =J^2\big[6\abs{\psi^{Q,n}_{1,1}}^2 + \abs{\psi^{Q,n}_{1,0}}^2 + \abs{\psi^{Q,n}_{0,1}}^2 
        +2\sqrt{2}\mathrm{Re}\Big\{(\psi^{Q,n}_{1,1})^*\Big[\psi^{Q,n}_{1,0}e^{i{(2Q/3+\theta)}} + \psi^{Q,n}_{0,1}e^{-i{(2Q/3)}}\Big]\Big\} \big],
        \label{eq:three-body-BIC-mechanism-couplings}
    \end{align}
\end{widetext}
as an indicator for the presence of three-body qBICs.

We test this hypothesis numerically and depict the results in Fig.~\ref{fig:three-body-BIC-mechanism-couplings}, where the panels (a),(c),(e) and (b),(d),(f) correspond to the cases of $\theta=\pi/2$ and $\theta=\pi/4$ respectively.
In Fig.~\ref{fig:three-body-BIC-mechanism-couplings}(a)-(b), we show the total coupling  $\sum_{m,l}|\gamma_{n}^{Q,m,l}|^2$~\eqref{eq:three-body-BIC-mechanism-couplings} for each of the four eigenstates $\ket{\psi^{Q,n}}$ of the deeply bound 3-body state region as a function of the three-particle COM quasimomentum $Q$.
Below, in the respective panels, in (c)-(d), the full three-particle spectrum (ED) is shown together with the $4\times 4$ approximate spectrum~\eqref{eq:3body-AHM-repr-effective-4x4-hybrid-matrix} [same as in Fig.~\ref{fig:three-body-spectrum-effective-AHM-U0-Binding-mechanism} of the main text].
Moreover, in (e)-(f), we plot the density distributions of selected three-body qBICs of the lower and upper bound state branch in the relative coordinate space.

We indeed find, that the total coupling for the effective eigenstate of the lower (band 1) and upper (band 4) bound state branch (blue and red-colored curves) indeed tends to zero at certain values of the COM quasimomentum $Q$, which coincides with the existence of three-body qBICs in the full spectrum. 
However, we also find that the total coupling never completely vanishes, thus no perfect three-body BICs exist in the AHM for these values of $\theta$ and $U=0$, which is consistent with the fact that the eigenstates of the deeply bound 3-body state region have non-zero weight outside $\ket{Q,0,0}$, as seen in the density distributions in Fig.~\ref{fig:three-body-BIC-mechanism-couplings}(e)-(f). 
In the proximity of the middle two effective branches (bands 2 and 3) there are no deeply bound three-particle bound states, even though the total coupling of the effective eigenstates of these branches to the bulk also tends to zero at certain values of $Q$~[Fig.~\ref{fig:three-body-BIC-mechanism-couplings}(a)-(d)].
The density distributions of the exact three-body qBICs in the lower and upper branches confirm that these states are predominantly localized in the deeply bound 3-body state region~[Fig.~\ref{fig:three-body-BIC-mechanism-couplings}(e)-(f)], with negligible weight on the bulk sites. 
Comparing the results for $\theta=\pi/2$ and $\theta=\pi/4$, we observe that three-body qBICs can exist for a slightly wider range of COM quasimomenta $Q$ for $\theta=\pi/2$ than for $\theta=\pi/4$.
Generally, we find that the total coupling~\eqref{eq:three-body-BIC-mechanism-couplings} is a good indicator for the existence of three-body qBICs, which becomes better in the regions of COM quasimomenta $Q$, where a sufficient energy gap between the individual effective energy bands is present. This latter condition is better satisfied for $\theta=\pi/2$ than for $\theta=\pi/4$, explaining the better agreement between the total coupling and the existence of three-body qBICs for $\theta=\pi/2$ than for $\theta=\pi/4$.

Far from these avoided crossings, each effective band behaves almost independently, so checking its own total coupling to bulk~\eqref{eq:three-body-BIC-mechanism-couplings} is meaningful, related to the diagonal elements of the self-energy~$[\Sigma_P(E,Q)]_{n,n'}$~\eqref{eq:three-body-BIC-mechanism-self-energy-diagonal-decay-rates}. However, close to avoided crossings, the bands strongly mix through continuum-mediated coupling processes encoded in the off-diagonal self-energy $[\Sigma_P(E,Q)]_{n,n'}$, where the leakage to the bulk is then governed by the mixture of the bare bands, not by either bare bands alone. 
To summarize, in the three-body case we only have approximate or quasi-BICs, which are still very robust and long-lived due to the small total coupling to the bulk. 
The finite lifetime of these qBICs is roughly set by the decay rate $\Gamma_{nn}(E,Q)$~\eqref{eq:three-body-BIC-mechanism-self-energy-diagonal-decay-rates}, fixed by the small total coupling and the bulk density of states $\rho_{\mathrm{bulk}}(E)\sim 1/W_{\mathrm{bulk}}$, with $W_{\mathrm{bulk}}$ the bandwidth of the type-I continuum. 
As in the two-body case~[App.~\ref{app:2body-BIC}], a qBIC will stay sharply resolved as long as $\Gamma_{nn}$ remains below the finite-size level spacing $\delta E\sim W_{\mathrm{bulk}}/K^2$, and turns into a genuine slowly-decaying resonance only in the thermodynamic limit $K\to\infty$.
This is actually remarkable, as from the over-constrained set of the equations, given by the individual couplings~\eqref{eq:three-body-BIC-mechanism-couplings-individual} having to vanish with one free parameter $Q$ for three-body BIC to exist, one would expect that no BICs at all can exist, but we find that the system is close enough to this idealized situation such that robust three-body qBICs can still exist in the AHM for $U=0$.

\subsection{Experimentally constrained three-particle AHM}
\label{app:experimentally-constrained-3body-AHM}
\begin{figure*}[t]
    \centering
    \includegraphics{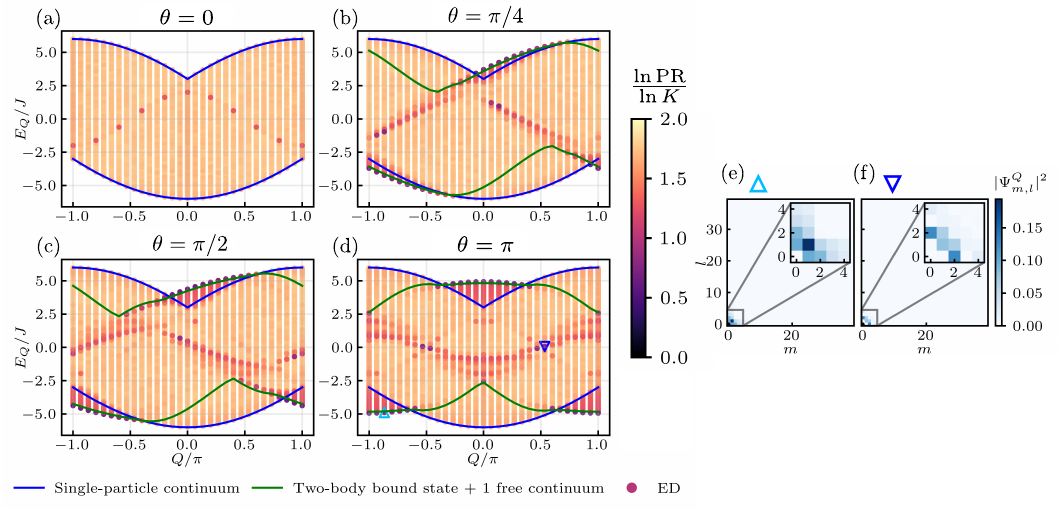}
    \caption{
        Three-particle energy spectrum of the experimentally constrained AHM for $U=0$.
        Three-particle energy spectrum $E_Q$ as a function of the COM quasimomentum $Q$ for $\theta=0,\pi/4,\pi/2,\pi$ is shown in (a), (b), (c), and (d) respectively.
        The blue solid lines indicate the edges of the continuum of three independent particles~\eqref{eq:3body-free-scattering-state-energy-bounds}, whereas the green solid lines provide the boundaries of the continuum spanned by two-particle bound states and one free particle~\eqref{eq:3body-bound-scattering-state-energy-bounds}.
        The exact diagonalization (ED) results for each $Q$ value are shown via the dots whose color indicates the scaled participation ratio.
        Two characteristic states are indicated by the cyan and blue triangles in (d), whose density distributions $|\Psi^{Q}_{m,l}|^2$ are shown in (e) and (f) respectively.
        They correspond to a ``regular'' bound state outside the continuum (e) and a bound state inside the continuum (f).
        Other parameters are the same as in~\cref{fig:three-body-spectrum-full-AHM} of the main text.
        }
    \label{fig:three-body-spectrum-experimental-AHM}
    \end{figure*}
\begin{figure}[hbt]
        \centering
        \hspace*{-.2cm}
        \includegraphics[width=\columnwidth]{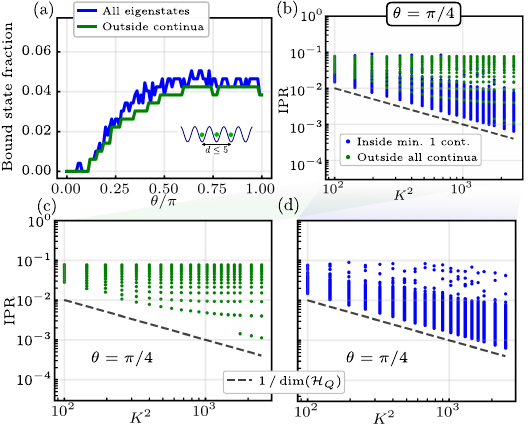}
        \caption{
            Three-body bound state fraction and finite-size scaling of the three-particle spectrum of the experimentally constrained AHM for $U=0$. 
            (a) Bound state fraction as a function of the statistical angle $\theta$ for the entire three-particle spectrum~\eqref{eq:3body-AHM-repr-hybrid-full} defined as the fraction of states with $\mathrm{PR}$~\eqref{eq:participation-ratio} below that of states where the left and right particle are at most separated by five sites for all eigenstates in blue and for eigenstates outside the continua in green. 
            (b) Inverse participation ratio $\mathrm{IPR}=1/\mathrm{PR}$~\eqref{eq:participation-ratio} for a fixed statistical angle $\theta=\pi/4$ as a function of the size of the relative coordinate space $K^2$ for the entire three-particle spectrum~\eqref{eq:3body-AHM-repr-hybrid-full} where $K^2$ ranges from $10^2$ to $50^2$ sites.
            The green (blue) markers correspond to eigenstates lying outside both continua (inside at least one of the two continua).
            (c) Same as (b) but only for states outside both continua. 
            (d) Same as (b) but only for states inside at least one of the two continua. 
            All other parameters are the same as in Fig.~\ref{fig:three-body-spectrum-experimental-AHM}, where in (a)~$K=40$ sites along each of the relative coordinate directions $m$ and $l$ are used.
            }
            \label{fig:Part-Ratio-Bound-state-fraction-U0-experimental-AHM}
        \end{figure}

Complementary to the results of the main text covering the full AHM, in this section we will present the results for the currently experimentally accessible, experimentally constrained version of the three-particle AHM~\cite{Cardarelli2016,Greschner2018a,Kwan2024,Bakkali-Hassani2026}. 
As described in the main text (see end of Section~\ref{sec:3body-bound-states-AHM}), it is obtained by excluding three-particle coincidences and the Peierls phases for the tunneling processes $\ket{\dots21\dots} \leftrightarrow \ket{\dots12\dots}$~(see~\cref{fig:three-body-spectrum-effective-AHM-U0-Binding-mechanism}(a) as well). 
We present the figures for the three-particle spectrum and the three-body bound state fraction for $U=0$ in~\cref{fig:three-body-spectrum-experimental-AHM} and~\cref{fig:Part-Ratio-Bound-state-fraction-U0-experimental-AHM} respectively, which are complementary to the results for the full AHM presented in~\cref{fig:three-body-spectrum-full-AHM} and~\cref{fig:Part-Ratio-Bound-state-fraction-U0} of the main text, respectively.
Then we will present the figures for expansion dynamics of three-particle bound states in the experimentally constrained AHM for $U=0$ in~\cref{fig:5-6-density-vs-time-ground-state-expansion-experimental-AHM} and~\cref{fig:ground-state-expansion-experimental-AHM}, which are complementary to the results for the full AHM presented in~\cref{fig:5-6-density-vs-time-ground-state-expansion} and~\cref{fig:ground-state-expansion} of the main text, respectively. 

\subsubsection{Three-particle spectrum for $U=0$}
\label{app:3body-spectrum-experimental-AHM}    
    The three-particle spectrum of the experimentally constrained AHM for various statistical angles $\theta$ and zero on-site interaction, $U=0$ is shown in~\cref{fig:three-body-spectrum-experimental-AHM}. 
    The other parameters are exactly the same as in~\cref{fig:three-body-spectrum-full-AHM} of the main text.
    We find that the three-particle spectrum of the experimentally constrained AHM~\cref{fig:three-body-spectrum-experimental-AHM} qualitatively agrees with the three-particle spectrum of the full AHM~\cref{fig:three-body-spectrum-full-AHM}. 
    In particular, we can also identify bound state solutions both outside the scattering state continua for the statistical angles $\theta=\pi/4,\pi/2$, which however appear are less energetically separated from the continua as compared to the full AHM. 
    Moreover, we also find bound states outside the scattering state continua for $\theta=\pi$, which is different from the full AHM case.
    Similar to the full AHM case, we can also identify various quasi-bound states inside the continua, which are, however, less localized as compared to the full AHM case. 
    
    \subsubsection{Three-body bound state fraction for $U=0$}
        The three-body bound state fraction and the finite-size scaling of the three-particle spectrum for the experimentally constrained AHM for $U=0$ is shown in~\cref{fig:Part-Ratio-Bound-state-fraction-U0-experimental-AHM}, which is complementary to the full AHM results presented in~\cref{fig:Part-Ratio-Bound-state-fraction-U0} of the main text. 
        We find that the bound state fraction is shifted to larger values of the statistical angle $\theta$, where the most localized states are found for $\theta > \pi/2$~[\cref{fig:Part-Ratio-Bound-state-fraction-U0-experimental-AHM}(a)]. 
        Generally, we find that the bound state are less localized as compared to the full AHM case (see slightly lower $\mathrm{IPR}$ in~[\cref{fig:Part-Ratio-Bound-state-fraction-U0-experimental-AHM}(b)-(d)] as compared to~\cref{fig:Part-Ratio-Bound-state-fraction-U0}(b)-(d) of the main text), which is consistent with the results for the three-particle spectrum in~\cref{fig:three-body-spectrum-experimental-AHM} showing that the bound states are less energetically separated from the continua as compared to the full AHM case. 
        The qBICs for the exp.~constr.~AHM remain less localized when increasing the system sizes although still robust on experimentally relevant system sizes~[\cref{fig:Part-Ratio-Bound-state-fraction-U0-experimental-AHM}(d)].
        
        \subsubsection{Expansion dynamics of three-particle wave packets}
        \label{app:expansion-dynamics-constrained-AHM}
        \begin{figure*}[bt]
            \centering
            \includegraphics[width=2.05\columnwidth]{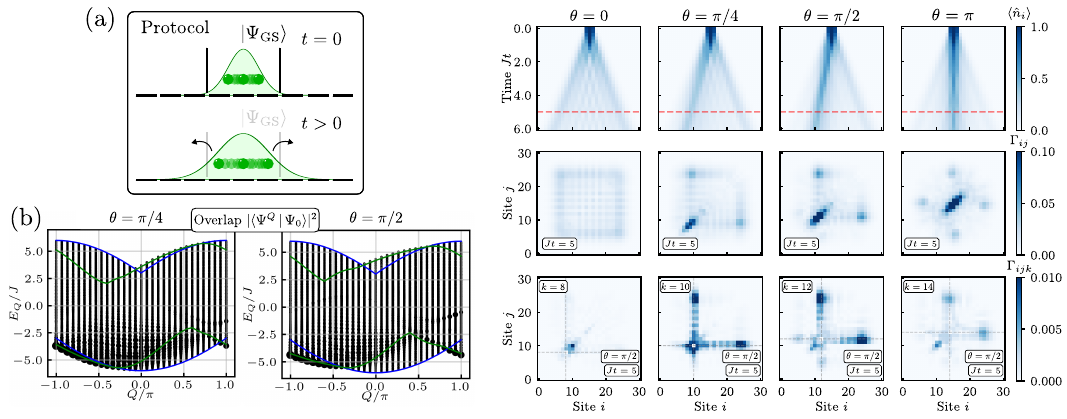} 
            \caption{
            Asymmetric expansion dynamics and density correlators of three-particle wave packets for the experimentally constrained AHM for $U=0$.
            (a) Illustration of the expansion protocol starting from the ground state of three particles on three sites. 
            (b) The three-particle spectrum $E_Q$(cf.~\cref{fig:three-body-spectrum-experimental-AHM}(b)-(c)) with the overlap $\mathcal{O}=\abs{\braket{\Psi^{Q}}{\Psi_0}}^2$ of the initial state $\ket{\Psi_0}$ with the entire three-particle spectrum $\{\ket{\Psi^{Q}}\}$~\eqref{eq:3body-AHM-repr-hybrid-full} for $\theta=\pi/4$ (left) and $\theta=\pi/2$ (right). 
            The overlap is illustrated by the size of the area of the filled circles, which scales quadratically with the magnitude of the overlap.
            The first row in both (c) shows the spatio-temporal density distribution 
            $\ep{\hat{n}_{j}}(t)=\expval{\Psi(t)}{\hat{n}_{j}}$ for the statistical angles $\theta=0$ (bosons), $\theta=\pi/4$, $\theta=\pi/2$, and $\theta=\pi$.
            The second row shows the second-order density correlator $\Gamma_{ij}=\ep{\hat{b}^{\dagger}_{j}\hat{b}_{i}^{\dagger}\hat{b}_{i}\hat{b}_{j}}$ at the fixed time $Jt=5$ (red dashed line in first row) for the same statistical angles $\theta$.
            The third and fourth row show the third-order density correlator $\Gamma_{ijk}=\ep{\hat{b}^{\dagger}_{k}\hat{b}^{\dagger}_{j}\hat{b}_{i}^{\dagger}\hat{b}_{i}\hat{b}_{j}\hat{b}_{k}}$ at the fixed time $Jt=5$ (red dashed line in the first row) for the fixed statistical angle $\theta=\pi/2$, where the index $k=8,10,12,14$ increases from left to right. 
            The gray dashed lines are a guide to the eye, with the vertical and horizontal lines corresponding to $i=k$ and $j=k$, respectively, and their crossing point indicating the three-particle coincidence point $i=j=k$.
            The initial state $\ket{\Psi_0}$ is the ground state of the AHM of $N_b=3$ bosons on $L_{\mathrm{small}}=3$ sites for $U=0$. 
            For the quench dynamics, the system is then evolved under the experimentally constrained AHM with $L=31$ sites and $U=0$. 
            }
            \label{fig:5-6-density-vs-time-ground-state-expansion-experimental-AHM}
        \end{figure*} 
        \begin{figure}[t]
            \centering
            \hspace*{-.3cm}
            \includegraphics{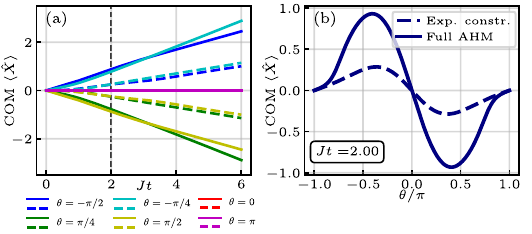}
            \caption{
            Asymmetric transport of three-body bound states for both the full and the experimentally constrained AHM. 
            (a) The COM position $\ep{\hat{X}}$ as a function of time $t$ for various statistical angles $\theta$ (colors). 
            The solid (dashed) lines indicate the results for the full (experimentally constrained) AHM.
            (b) The COM position $\ep{\hat{X}}$ as a function of the statistical angle $\theta$ at a fixed time of $Jt=2$ (cf.~gray vertical dashed line in (a)) for the full (solid) and experimentally constrained (dashed) AHM. As in~\cref{fig:5-6-density-vs-time-ground-state-expansion} (and~\cref{fig:5-6-density-vs-time-ground-state-expansion-experimental-AHM}) the initial state $\ket{\Psi_0}$ is the three particle ($N_b=3$) ground state prepared on $L_{\mathrm{small}}=3$ sites, after which the system is quenched into a larger AHM lattice with $L=31$ sites for $U=0$.
            }
            \label{fig:ground-state-expansion-experimental-AHM}
        \end{figure}
        Here, we present the results for the expansion dynamics of three-particle wave packets for the experimentally constrained AHM for $U=0$, complementary to the results for the full AHM presented in~\cref{sec:expansion-dynamics} of the main text, with~\cref{fig:5-6-density-vs-time-ground-state-expansion-experimental-AHM} showing the results for the expansion dynamics and density correlators, and~\cref{fig:ground-state-expansion-experimental-AHM} showing the results for the COM position $\ep{\hat{X}}$ as a function of time $t$ and statistical angle $\theta$. 

        \cref{fig:5-6-density-vs-time-ground-state-expansion-experimental-AHM} shows the results for the asymmetric expansion dynamics complementary to the results for the full AHM presented in~\cref{fig:5-6-density-vs-time-ground-state-expansion} of the main text. 
        Here, we find that the initial state (three-particle ground state on three sites) also has a significant overlap with the energetically lower three-body bound states branches exhibiting a finite slope and thus prepares a moving bound state. 
        However, we also find a non-negligible overlap with the scattering states of both type-I and II~[\cref{fig:5-6-density-vs-time-ground-state-expansion-experimental-AHM}(b)]. 
        This is expected as the bound state branches are less energetically separated from the continua as compared to the full AHM case~[cf.~\cref{fig:three-body-spectrum-experimental-AHM} compared to~\cref{fig:three-body-spectrum-full-AHM} of the main text]. 
        In~\cref{fig:5-6-density-vs-time-ground-state-expansion-experimental-AHM}(c), we show the spatio-temporal density distribution $\ep{\hat{n}_j}(t)$ for various statistical angles $\theta$, the second order density correlator $\Gamma_{ij}$ for the same statistical angles $\theta$ at a fixed time $Jt=5$, and the third-order density correlator $\Gamma_{ijk}$ for the fixed statistical angle $\theta=\pi/2$ at the same fixed time $Jt=5$. 
        We find qualitatively similar results as for the full AHM case. 
        In particular, we still observe an asymmetric density distribution $n_j(t)$ for $\theta \neq 0,\pi$, as expected from populating the moving bound states~[(b)].
        However, for the third-order density correlator $\Gamma_{ijk}$, the peak~(at $i=j=k=10$ in the figure) showing that all three particles are highly likely to be found very close to each other, is less pronounced as compared to the full AHM case~[\cref{fig:5-6-density-vs-time-ground-state-expansion}(c) of the main text]. 
        In addition to the three-particle coincidence for $i=j=k=10$, additional peaks 
        (at $i=k\approx 10$ and $j\approx 25$, and $j=k\approx 10$ and $i\approx 25$) indicate a small but non-negligible overlap with type-II scattering states~\footnote{Note that one further improve upon this by using optimal control techniques, as those in explored in~Refs.~\cite{Theel2025,Blatz2024}, to prepare an initial state with a larger overlap with the desired three-body bound states.}, consistent with the results in (b). 

        Lastly, we also show the COM position $\ep{\hat{X}}$ as a function of time $t$ for various statistical angles $\theta$ for both the full and the experimentally constrained AHM in~\cref{fig:ground-state-expansion-experimental-AHM}(a), and the COM position $\ep{\hat{X}}$ as a function of the statistical angle $\theta$ at a fixed time of $Jt=2$ for both models in~\cref{fig:ground-state-expansion-experimental-AHM}(b). 
        The experimentally constrained AHM shows the same asymmetry of $X$, with the transport changing its direction as $\theta$ switches its sign. 
        Compared to the full AHM, the speed of the prepared moving bound state is slightly reduced, but still faster than bound states localized by non-statistical interactions. 
    
    \subsection{Three-particle spectrum for finite on-site interaction}
    \label{app:3body-spectrum-finite-U}
\begin{figure*}[t]
    \centering
    \includegraphics{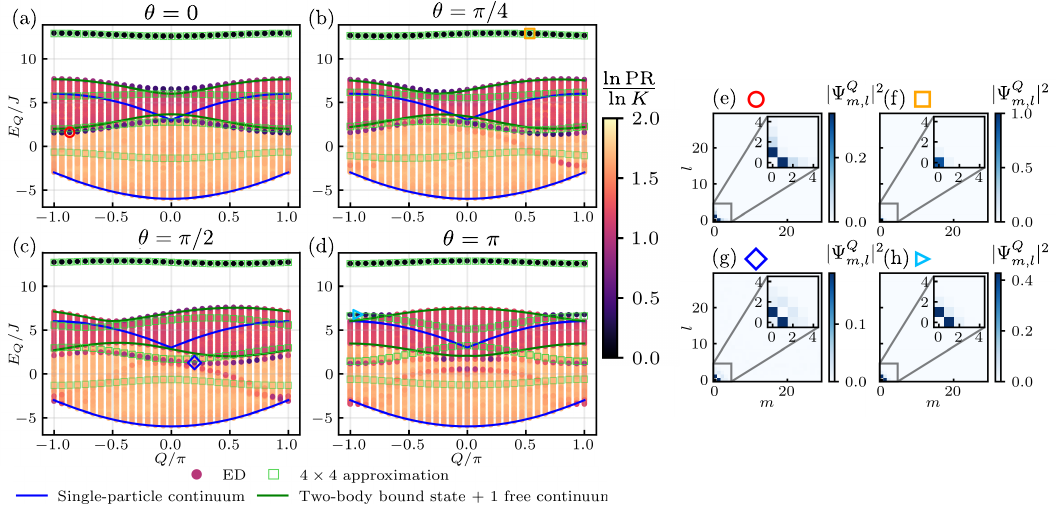}
        \caption{\textbf{Three-particle energy spectrum of AHM for $U=4J$.}
        Three-particle energy spectrum $E_Q$ as a function of the COM quasimomentum $Q$ for $\theta=0,\pi/4,\pi/2,\pi$ is shown in (a), (b), (c), and (d) respectively.
        The blue solid lines indicate the edges of the continuum of three independent particles~\eqref{eq:3body-free-scattering-state-energy-bounds}, whereas the green solid lines provide the boundaries of the continuum spanned by two-particle bound states and one free particle~\eqref{eq:3body-bound-scattering-state-energy-bounds}.
        The exact diagonalization (ED) results for each $Q$ value are shown via the dots whose color indicates the scaled participation ratio.
        The light green squares indicate the effective three-body bound state energy spectrum of the approximate $4\times 4$ model~\eqref{eq:3body-AHM-repr-effective-4x4-hybrid-matrix}.
        Four characteristic states are indicated by the red circle in (a), orange square in (b), by the blue diamond in (c), and cyan triangle in (d), whose density distributions $|\Psi^{Q}_{m,l}|^2$ are shown in (e), (f), (g), and (h) respectively.
        Other parameters are $K=30$, and the rest of the parameters are the same as in~\cref{fig:three-body-spectrum-full-AHM} of the main text.
        }
    \label{fig:three-body-spectrum-U4-AHM}
    \end{figure*}

\begin{figure*}[!tb]
        \centering
        \includegraphics{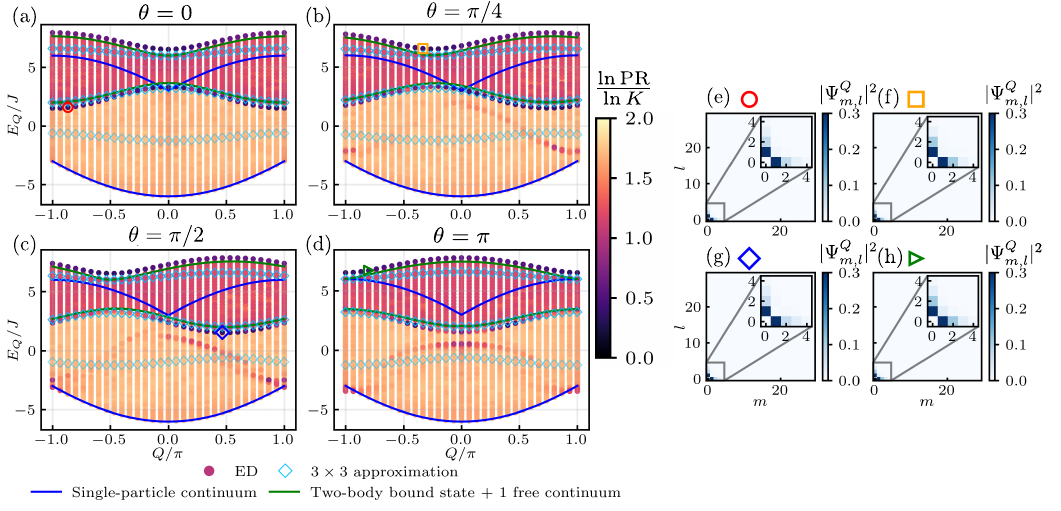}
            \caption{\textbf{Three-particle energy spectrum of the experimentally constrained AHM for $U=4J$.}
            Three-particle energy spectrum $E_Q$ as a function of the COM quasimomentum $Q$ for $\theta=0,\pi/4,\pi/2,\pi$ is shown in (a), (b), (c), and (d) respectively.
            The blue solid lines indicate the edges of the continuum of three independent particles~\eqref{eq:3body-free-scattering-state-energy-bounds}, whereas the green solid lines provide the boundaries of the continuum spanned by two-particle bound states and one free particle~\eqref{eq:3body-bound-scattering-state-energy-bounds}.
            The exact diagonalization (ED) results for each $Q$ value are shown via the dots whose color indicates the scaled participation ratio.
            The light blue diamonds indicate the effective three-body bound state energy spectrum corresponding to the approximate $3\times 3$ model~\eqref{eq:3body-AHM-repr-effective-3x3-hybrid-full}.
            Four characteristic states are indicated by the red circle in (a), orange square in (b), by the blue diamond in (c), and green triangle in (d), whose density distributions $|\Psi^{Q}_{m,l}|^2$ are shown in (e), (f), (g), and (h) respectively.
            Other parameters are the same as in~\cref{fig:three-body-spectrum-U4-AHM} above.
            }
        \label{fig:three-body-spectrum-U4-experimental-AHM}
        \end{figure*}

    Both in the main text and in~\cref{app:3body-spectrum-experimental-AHM}, we have presented the three-particle spectrum of the full AHM and the experimentally constrained AHM for zero on-site interaction $U=0$ (see~\cref{fig:three-body-spectrum-full-AHM} and ~\cref{fig:three-body-spectrum-experimental-AHM} respectively).
    Here, we complement these results by presenting the three-particle spectrum of both models for finite repulsive on-site interaction $U=4J$ in the following~\footnote{We note that the attractive case $U=-4J$ of the spectrum can be obtained by a mirror reflection against the $E=0$ axis with simultaneous shift of each COM quasimomentum $Q \to (Q+3\pi) \bmod 2\pi$~\cite{Valiente2010a}.}.
    We find a plethora of bound state solutions both outside and inside the scattering state continua for various statistical angles $\theta$, where, interestingly, we also find (quasi-) three-body bound states in the continuum (qBIC) for the Bose-Hubbard case $\theta=0$, whose lifetime for larger system sizes has to be investigated in future work. 
    For an exemplary value of $U=4J$ we present the three-particle spectrum of the full AHM in~\cref{fig:three-body-spectrum-U4-AHM} and the three-particle spectrum of the experimentally constrained AHM in~\cref{fig:three-body-spectrum-U4-experimental-AHM} for various statistical angles $\theta$ and exemplary bound state density distributions. 
    The three-particle spectrum of the full AHM for $U=4J$ in~\cref{fig:three-body-spectrum-U4-AHM} features triplon bound states above both scattering state continua for all statistical angles $\theta$ considered, where a strong occupation of the three-particle coincidence configuration $\ket{n,0,0}$ as visible in the density distribution~\cref{fig:three-body-spectrum-U4-AHM}(f).
    Moreover, we also find bound states directly outside the scattering state continuum of type-II (two-particle bound state + one free particle) for all the statistical angles $\theta$ shown in ~\cref{fig:three-body-spectrum-U4-AHM}(a)-(d). An example density distribution of such a bound state is shown in~\cref{fig:three-body-spectrum-U4-AHM}(e), showing that it consists of a dimer-monomer configuration, as already found in the Bose-Hubbard model~\cite{Valiente2010a}.
    Interestingly, for the value of $U=4J$ shown, we also find bound states inside scattering state continua of type-I (single-particle continuum) for all the statistical angles $\theta$ shown, with two exemplary density distributions shown in~\cref{fig:three-body-spectrum-U4-AHM}(e) and (g) for the qBICs of $\theta=0$ and $\theta=\pi/2$ respectively.
    These qBICs are also localized in the relative coordinate space, although slightly less strongly localized as compared to the regular bound states outside the continua. 
    Moreover, in~\cref{fig:three-body-spectrum-U4-AHM} we also show the effective $4\times 4$ model discussed in~\cref{app:3body-effective-model-4x4} of the full AHM on top of the exact diagonalization results.
    This effective model is perfectly able to capture the triplon bound states above both scattering state continua for all statistical angles $\theta$ shown, the bound states outside the continua for $\theta=\pi$, and also the qBICs for $\theta=0,\pi/4,\pi/2$ qualitatively well.
    
    The three-particle spectrum of the experimentally constrained AHM for $U=4J$ in~\cref{fig:three-body-spectrum-U4-experimental-AHM} features similar characteristics as the full AHM case, although with some differences.
    Most strikingly, there are no triplon bound states outside the scattering state continua, as expected since the three-particle coincidences $\ket{n,0,0}$ are explicitly excluded in this model.
    We still find the monomer-dimer bound states outside the scattering state continuum of type-II for all statistical angles $\theta$ shown, similar to the full AHM case.
    Two exemplary density distributions of such bound states are shown in~\cref{fig:three-body-spectrum-U4-experimental-AHM}(f) and (h) for $\theta=\pi/4$ and $\theta=\pi$ respectively.
    Moreover, we also find qBICs inside the scattering state continuum of type-I for all statistical angles $\theta$ shown, with two exemplary density distributions shown in~\cref{fig:three-body-spectrum-U4-experimental-AHM}(e) and (g) for the qBICs of $\theta=0$ and $\theta=\pi/2$ respectively.
    Here, we find that these qBICs are as localized in the relative coordinate space as their counterparts outside the continuum.
    Finally, in~\cref{fig:three-body-spectrum-U4-experimental-AHM}, we also show the effective three-body bound state model discussed in~\cref{app:3body-effective-model-4x4}, but adapted to the experimentally constrained AHM, on top of the exact diagonalization results.
    Due to the exclusion of the three-particle coincidence configuration $\ket{n,0,0}$ in the experimentally constrained AHM (and the Peierls phase connecting $\ket{n,0,1}$ to $\ket{n,1,0}$), the effective model becomes a $3\times 3$ model that is essentially described by the lower $3\times 3$ block of the Hamiltonian~\eqref{eq:3body-AHM-repr-effective-4x4-hybrid-matrix} with the Peierls phase between $\ket{Q,1,0}$ and $\ket{Q,0,1}$ set to zero, leading to the following COM-quasimomentum-diagonal effective Hamiltonian
    \begin{widetext}
        \begin{align}
            \hat{H}^{\mathrm{eff}}_{Q,3\times 3}
        &=\sum_Q
            \begin{pmatrix}
                \ket{Q,1,0} &
                \ket{Q,0,1} &
                \ket{Q,1,1}
            \end{pmatrix}
            \begin{pmatrix}
                U & -2J e^{-i\frac{Q}{3}} & -\sqrt{2}J e^{i\frac{Q}{3}} \\
                -2J e^{i\frac{Q}{3}} & U & -\sqrt{2}J e^{-i\left(\frac{Q}{3}+\theta\right)} \\
                -\sqrt{2}J e^{-i\frac{Q}{3}} & -\sqrt{2}J e^{i\left(\frac{Q}{3}+\theta\right)} & 0
            \end{pmatrix}
            \begin{pmatrix}
                \bra{Q,1,0} \\
                \bra{Q,0,1} \\
                \bra{Q,1,1}
            \end{pmatrix}.
            \label{eq:3body-AHM-repr-effective-3x3-hybrid-full}
        \end{align}
    \end{widetext}
    We find that this simple model is still able to qualitatively capture the monomer-dimer bound states inside the single-particle continuum for all statistical angles $\theta$ shown quite well~[\cref{fig:three-body-spectrum-U4-experimental-AHM}]. However, it fails to capture the bound states just outside the scattering state continua, which would require the inclusion of more configurations in the effective model.
    \linebreak
    \newpage
    \twocolumngrid
    \subsection{Extended analysis of three-particle wave packet expansion dynamics}
    \label{app:extended-expansion-dynamics}
    \begin{figure*}[b]
        \centering
        \includegraphics{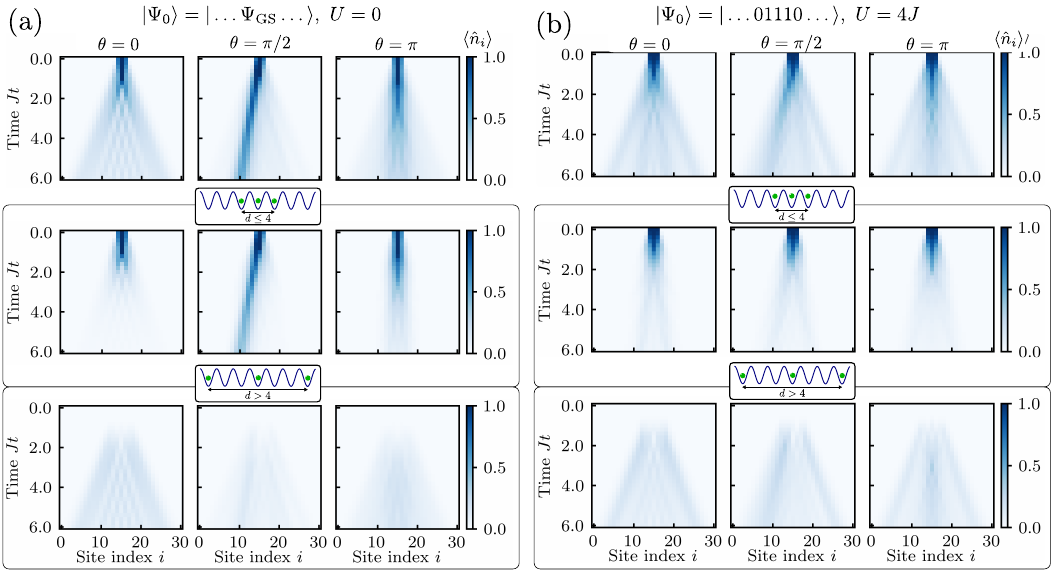}
        \caption{\textbf{Asymmetric expansion dynamics of the three-particle wave packets conditioned on bound state components.}
        For the initial state $\ket{\Psi_0}$ being the ground state of the AHM of $N_b=3$ bosons on $L_{\mathrm{small}}=3$ sites for $U=0$ (a) and an initial $N_b=3$-particle Fock state $\ket{\Psi_0}=\ket{\dots 0 111 0 \dots}$ (b), the spatio-temporal density distribution $\ep{\hat{n}_{j}}(t)=\expval{\Psi(t)}{\hat{n}_{j}}$ is shown in the first row for the statistical angles $\theta=0$ (bosons), $\theta=\pi/2$, and $\theta=\pi$, where for the time evolution system is evolved under the AHM~\eqref{eq:AHM} with $L=31$ sites for $U=0$ (a) and $U=4J$ (b).
        For the same statistical angles $\theta$, the second (third) rows show the spatio-temporal density distribution $\ep{\hat{n}_{j}}(t)=\expval{\Psi(t)}{\hat{n}_{j}}$ conditioned on the relative distance 
        $d\equiv m+l\leq 4$ ($d>4$) between the leftmost and rightmost particle, approximately separating
        the bound state (scattering state) components.
        }
        \label{fig:projected-asymmetric-expansion-U0GS-and-FockU4}
    \end{figure*}
        Here, we complement the discussion of the expansion dynamics of three-particle wave packets in the main text (\cref{sec:expansion-dynamics}).
    In particular, we present an extended analysis of the asymmetric expansion dynamics by conditioning the wave function of the three-particle wave packets on the relative distance between the leftmost and rightmost particle, leading to density distributions for near and distant particle configurations, respectively.
    We perform this analysis for the initial state of the quench protocol being the three-particle ground state of the AHM on three lattice sites for $U=0$ (discussed in~\cref{sec:expansion-dynamics} in the main text) and an initial three-particle Fock state (whose two-particle counterpart has been recently studied experimentally in~\Ccite{Kwan2024}), where for the latter consider both zero and finite on-site interaction $U=4J$.
    For the time evolution of starting from the initial three-particle Fock state with $U=0$, we further present an extended analysis of the width or root mean square (RMS) of the expanding wave packet for various statistical angles $\theta$.
    Lastly, we explain the asymmetric expansion dynamics present in the initial three-particle Fock state case with finite on-site interaction $U=4J$.
    
    \subsubsection{Conditioned expansion dynamics}
    \label{app:conditioned-expansion-dynamics}
    \begin{figure}[!t]
        \centering
        \includegraphics[width=\columnwidth]{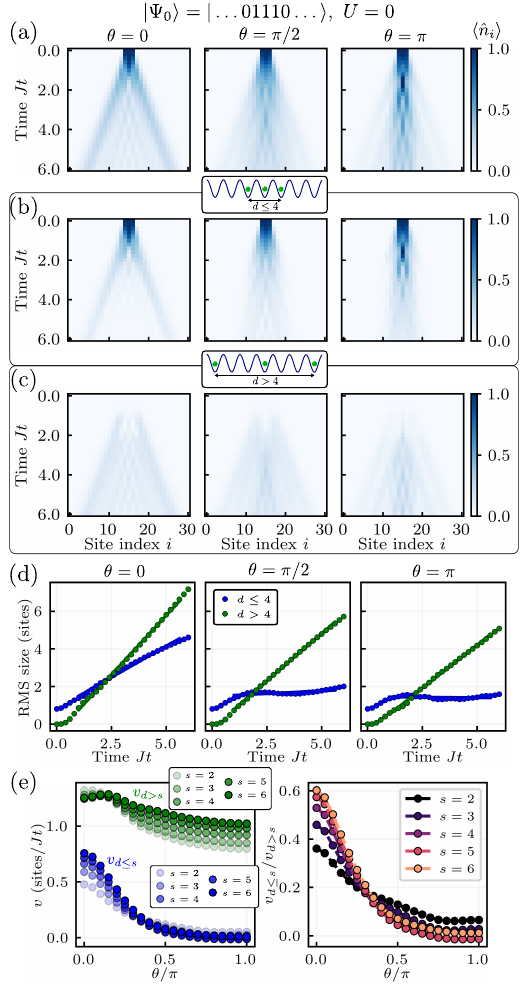}
            \caption{\textbf{Asymmetric expansion dynamics of the three-particle wave packets conditioned on bound state components.}
            (a) Spatio-temporal density distribution $\ep{\hat{n}_{j}}(t)=\expval{\Psi(t)}{\hat{n}_{j}}$ for the time evolution system under the AHM~\eqref{eq:AHM} with $L=31$ sites for $U=0$ starting from the initial $N_b=3$-particle Fock state $\ket{\Psi_0}=\ket{\dots 0 111 0 \dots}$ for the statistical angles $\theta=0$ (bosons), $\theta=\pi/2$, and $\theta=\pi$.
            (b) and (c) Spatio-temporal density distribution $\ep{\hat{n}_{j}}(t)=\expval{\Psi(t)}{\hat{n}_{j}}$ conditioned on the relative distance $d\equiv m+l\leq 4$  and $d>4$ between the leftmost and rightmost particle for the same statistical angles $\theta$ as in (a).
            (d) Root-mean-square size~\eqref{eq:RMS-size-definition} of the conditioned density distributions shown in (b) and (c) as function of evolution time shown in blue ($d\leq 4$) and green ($d>4$) respectively. Solid lines are linear fits to the data points to extract the spreading velocities.
            (e) Left and right panel: Spreading velocities $v_{d\leq s}$ (blue) and $v_{d>s}$ (green) and their ratio extracted from slopes of the linear fits to (d) for as function of the statistical angle $\theta$ for various cutoff distances $s=2,3,4,5,6$.
            }
            \label{fig:projected-expansion-FockU0-width-analysis}
        \end{figure}
            In~\cref{fig:projected-asymmetric-expansion-U0GS-and-FockU4}, we show the spatio-temporal density distributions after conditioning the three-particle wave function on the relative distance $m+l$ between the leftmost and rightmost particle starting from two different initial states for the quench protocol: (a) the three-particle ground state of the AHM on three lattice sites for $U=0$ and (b) an initial three-particle Fock state $\ket{\Psi_0}=\ket{\dots 0 111 0 \dots}$ for finite on-site interaction $U=4J$.
            The time evolution in both cases is performed under the AHM~\eqref{eq:AHM} on $L=31$ sites with $U=0$ (a) and $U=4J$ (b) respectively.
            The first rows of~\cref{fig:projected-asymmetric-expansion-U0GS-and-FockU4} show the spatio-temporal density distributions $\ep{\hat{n}_{j}}(t)=\expval{\Psi(t)}{\hat{n}_{j}}$, where the results for (a) reproduce the results of~\cref{fig:5-6-density-vs-time-ground-state-expansion}(a) in the main text.
            We observe that for both initial states (a) and (b) the asymmetric expansion dynamics are observed in the unconditioned density distributions $\ep{\hat{n}_{j}}(t)$, although the asymmetry for the initial Fock state case (b) is less pronounced.
            By projecting onto the small relative distance components $m+l\leq 4$, we argue that 
            the asymmetry mainly stems from the wave function components where the three particles are close to each other (bound state components), as clearly visible for (a), while for (b) the asymmetry is still present but less pronounced.
            The complementary projection onto components with $m+l>4$ shows a nearly symmetric expansion dynamic for both initial states.
            
            \subsubsection{Width expansion analysis for initial three-particle Fock state at $U=0$}
            \label{app:width-expansion-analysis-Fock-U0}
            In~\cref{fig:projected-expansion-FockU0-width-analysis}, we present a detailed analysis of the expansion dynamics starting from the initial three-particle Fock state $\ket{\Psi_0}=\ket{\dots 0 111 0 \dots}$ for zero on-site interaction $U=0$.
            Similar to~\cref{fig:projected-asymmetric-expansion-U0GS-and-FockU4}, we show in the first row of~\cref{fig:projected-expansion-FockU0-width-analysis}(a) the spatio-temporal density distribution $\ep{\hat{n}_{j}}(t)=\expval{\Psi(t)}{\hat{n}_{j}}$ of the unconditioned wave function for various statistical angles $\theta$.
            We find that, unlike for the two cases considered in~\cref{fig:projected-asymmetric-expansion-U0GS-and-FockU4}, the density profile is completely symmetric, but narrows as the statistical angle $\theta$ is increased from $0$ to $\pi$ showing a dominant internal cone in the density distribution.
            The second and third rows of~\cref{fig:projected-expansion-FockU0-width-analysis}(b) and (c) show the spatio-temporal density distributions for the state conditioned on the relative distance $d\equiv m+l\leq 4$ and $d>4$ between the leftmost and rightmost particles, respectively, which approximately separate the bound state and scattering state components of the wave function. 
            We observe that the emergence of the internal cone originates mainly from the bound state components ($d\leq 4$), while the scattering state components ($d>4$) mainly contribute to the outer parts of the density distribution.
            In~\cref{fig:projected-expansion-FockU0-width-analysis}(d), we then depict the width (RMS size) of the conditioned density distributions shown in (b) and (c) as function of evolution time $t$ for various statistical angles $\theta$, where we define the RMS size as the width of the density distribution via
            \begin{align}
                \Delta X = \sqrt{\sum_j p_j (j-j_0)^2+\bigg[\sum_j p_j (j-j_0)\bigg]^2},
                \label{eq:RMS-size-definition}
            \end{align}
            with $p_j = \ep{\hat{n}_{j}}(t)/N_b$ being the normalized density distribution at site $j$ at time $t$, $j_0$ the initial center position of the wave packet and $N_b=3$ is the total number of particles.
            We then determine the spreading velocities $v_{d\leq s}$ and $v_{d>s}$ of the conditioned density distributions by performing linear fits to the data points shown in~\cref{fig:projected-expansion-FockU0-width-analysis}(d), where the results are shown in the left panel of~\cref{fig:projected-expansion-FockU0-width-analysis}(e) as function of the statistical angle $\theta$ for various cutoff distances $s$ between the leftmost and rightmost particle.
            We find that the spreading velocity $v_{d\leq s}$ of the near particle configurations ($d\leq s$) decreases significantly as the statistical angle $\theta$ is increased from $0$ to $\pi$, 
            almost vanishing for $\theta=\pi$ for all cutoff distances $s$ considered.
            On the other hand, the spreading velocity $v_{d>s}$ of the distant particle configurations ($d>s$) slightly decreases also with increasing $\theta$ from $0$ to $\pi$, but remains finite even for $\theta=\pi$. 
            This is to be contrasted with the two-particle case reported in~\cite{Kwan2024}, where the spreading velocity of the distant particle configurations is independent of the statistical angle $\theta$, reflecting the two-particle scattering states in that case.
            Here, the slight decrease of $v_{d>s}$ as $\theta$ reaches $\pi$ is most likely due to the fact that the distant particle configurations ($d>s$) contain contributions from both types of three-particle scattering states discussed in~\cref{app:3body-scattering-state-continuum}, i.e., the single-particle continuum (type-I) and the continuum of two-particle bound states with one free particle (type-II). As the states of the latter continuum are most localized for $\theta=\pi$, since the contributing two-particle bound states~\eqref{eq:AHM-bound-states-dispersion} are most strongly localized for that angle, the spreading velocity $v_{d>s}$ slightly decreases as $\theta$ approaches $\pi$.
            Lastly, in the right panel of~\cref{fig:projected-expansion-FockU0-width-analysis}(e), we depict the ratio $v_{d\leq s}/v_{d>s}$ of the spreading velocities as a function of $\theta$ for various cutoff distances $s$, showing a significant decrease of this ratio as $\theta$ increases from $0$ to $\pi$, which reflects that the narrowing of the density configuration is mainly driven by the near particle configurations ($d\leq s$).
            
            \subsubsection{Explanation of asymmetric expansion dynamics for initial three-particle Fock state at $U=4J$}
            \label{app:explanation-asymmetric-expansion-Fock-U4}
\begin{figure}[tbh]
                \centering
                \includegraphics[width=\columnwidth]{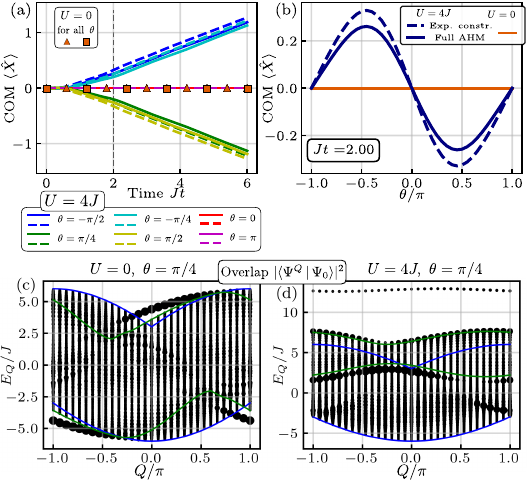}
                \caption{\textbf{Expansion dynamics starting from an initial three-particle Fock state.}
                (a) The center-of-mass (COM) position $\ep{\hat{X}}$ as a function of time $t$ for various statistical angles $\theta$ (colors), where the solid (dashed) lines indicate the results for the full (experimentally constrained) AHM for $U=4J$.
                For $U=0$ the results for the full (orange squares) and experimentally constrained (orange triangles) AHM coincide and the $\ep{\hat{X}}=0$ for all times and statistical angles $\theta$. 
                (b) The COM position $\ep{\hat{X}}$ as a function of the statistical angle $\theta$ at a fixed time of $Jt=2$ (cf.~gray vertical dashed line in (a)) for the full (blue solid) and experimentally constrained (blue dashed) AHM for $U=4J$. 
                The orange solid line indicates the COM position $\ep{\hat{X}}$ for $U=0$ for both the full and experimentally constrained AHM, whose results coincide.
                (c)-(d) The three-particle spectrum $E_Q$~\eqref{eq:3body-AHM-repr-hybrid} for $\theta=\pi/4$ (cf.~\cref{fig:three-body-spectrum-full-AHM}(b)) with the overlap $\mathcal{O}=\abs{\braket{\Psi^{Q}}{\Psi_0}}^2$ of the initial state $\ket{\Psi_0}$ with the entire three-particle spectrum $\{\ket{\Psi^{Q}}\}$~\eqref{eq:3body-AHM-repr-hybrid} for the full AHM.
                The magnitude of the overlap is illustrated by the size of the area of the filled circles (scales quadratically).
                For both $U=0$ and $U=4J$ the initial state $\ket{\Psi_0}$ for the time evolution under the AHM~\eqref{eq:AHM} with $L=31$ sites is given by the $N_b=3$-particle Fock state $\ket{\Psi_0}=\ket{\dots 0 111 0 \dots}$~(cf.~\cref{fig:projected-expansion-FockU0-width-analysis} and \cref{fig:projected-asymmetric-expansion-U0GS-and-FockU4}(b)).
                }
                \label{fig:fock-state-expansion-analysis}
                \end{figure}

            In~\cref{fig:fock-state-expansion-analysis}, we present an analysis of the asymmetric expansion dynamics starting from the initial three-particle Fock state $\ket{\Psi_0}=\ket{\dots 0 111 0 \dots}$ for zero ($U=0$) and finite on-site interaction ($U=4J$).
            In~\cref{fig:fock-state-expansion-analysis}(a) the center-of-mass (COM) position $\ep{\hat{X}}(t)=\sum_j j \ep{\hat{n}_j}(t)/N_b$ of the expanding wave packet is shown as a function of time $t$ for various statistical angles $\theta$ for both $U=0$ and $U=4J$. 
            In~\cref{fig:fock-state-expansion-analysis}(b) we show $\ep{\hat{X}}$ as a function of $\theta$ at an evolution time of $Jt=2$ for both the full AHM~\eqref{eq:AHM} and the experimentally constrained AHM, also for both $U=0$ and $U=4J$.
    We find completely symmetric expansion dynamics for $U=0$ in both models, i.e., $\ep{\hat{X}}(t)=0$ for all statistical angles $\theta$ (also see~\cref{fig:projected-expansion-FockU0-width-analysis}(a)), while for finite on-site interaction $U=4J$ we observe asymmetric expansion dynamics with a finite COM position $\ep{\hat{X}}(t)\neq0$ for both models when $\theta$ takes on fractional values, where, interestingly, the asymmetry is slightly more pronounced in the experimentally constrained AHM as compared to the full AHM.
    To explain the emergence of this symmetric (asymmetric) expansion of the three-particle Fock states $\ket{\Psi_0}$ for $U=0$ ($U=4J$), we analyze the overlap of the initial Fock state with the three-particle eigenstates $\{\ket{\Psi^{Q}}\}$~\eqref{eq:3body-AHM-repr-hybrid} for the full AHM, which for $\theta=\pi/2$ is shown in (c) for $U=0$ and (d) for $U=4J$ respectively.
    We observe that for $U=0$ the Fock state populates both the lower and upper three-body bound state branches, with equal weight, which possess opposite group velocities leading to a cancelling left-moving and right-moving contribution in the expansion dynamics, resulting in a symmetric expansion dynamics overall. This is to be contrasted with the case of the initial three-particle ground state on three sites for $U=0$ discussed in the main text (see~\cref{fig:ground-state-expansion}(d)), where only the lower three-body bound state branch is populated, which has led to the asymmetric expansion dynamics observed there.
    On the other hand, for finite on-site interaction $U=4J$, we find that the initial Fock state mostly populates the bound states inside the continuum (qBICs) in an asymmetric fashion, which have a finite but less pronounced slope in their energy dispersion as compared to the lower three-body bound state branch at $U=0$. 
    This results in less pronounced asymmetric expansion dynamics as compared to the initial three-particle ground state on three sites for $U=0$ discussed in the main text.

\section{Four-particle problem}
\label{app:4body-AHM}
In this section, we provide details regarding the four-particle problem of the AHM.
For the four-particle problem, we proceed as before in the two-and-three particle case. We represent the four-particle problem of the AHM in the configuration space representation, where we label basis states therein by $\ket{n,m,l,k}$, such that, when counted from the left, $n$, $n+m$, $n+m+l$, and $n+m+l+k$ are the positions of the first, second, third, and fourth particles, respectively, and hence $m,l,k \in \mathbb{N}$ the separation between the first and second, second and third, and third and fourth particles, respectively.
Again, in the case of an underlying infinite 1D lattice, we have $n\in \mathbb{Z}$ while distances $m,l,k\in \mathbb{N}$ to account for indistinguishability of the particles~\footnote{In a finite lattice of $L$ sites with four particles, the allowed values are $n\in\{1,\ldots,L\}$, $m\in\{0,\ldots,L-n\}$, $l\in\{0,\ldots,L-n-m\}$, and $k\in\{0,\ldots,L-n-m-l\}$}.

With this mapping, we can transform the four-particle AHM problem into a single-particle problem in the four-dimensional configuration space spanned by the basis states $\{\ket{n,m,l,k}\}$ forming a 4D-simplex due to the indistinguishability of the particles. 

\subsection{4-particle AHM in configuration space and relative coordinate space}
\label{app:4body-AHM-repr}
The full expression for AHM for four particles in the configuration space basis representation
reads as follows
    \begin{widetext}
    \begin{align}
    \hat{H} = &-J \sum_n \bigg[
    2\ketbra{n,0,0,1}{n,0,0,0} + \sqrt{6} e^{- i \theta}\ketbra{n,0,1,0}{n,0,0,1} + \sqrt{6} e^{- 2 i \theta}\ketbra{n,1,0,0}{n,0,1,0} \nonumber \\
    &+ 2 e^{- 3 i \theta}\ketbra{n\!+\!1,0,0,0}{n,1,0,0} + \sum_{m,l \geq 0, k \geq 1} \ketbra{n,m,l,k+1}{n,m,l,k} + \sum_{m \geq 0, l \geq 1} \sqrt{2}\ketbra{n,m,l,1}{n,m,l,0} \nonumber \\
    & + \sum_{m \geq 1} \sqrt{3}\ketbra{n,m,0,1}{n,m,0,0}  + \sum_{k \geq 2} \sqrt{3}\ketbra{n,0,1,k-1}{n,0,0,k}  + \sum_{m \geq 0, l \geq 1} \sqrt{2} e^{- i \theta}\ketbra{n,m,l+1,0}{n,m,l,1} \nonumber \\
    & + \sum_{m \geq 0, l \geq 1, k \geq 2} \ketbra{n,m,l+1,k-1}{n,m,l,k}  + \sum_{m \geq 1} 2 e^{- i \theta}\ketbra{n,m,1,0}{n,m,0,1}  \nonumber \\
    &+ \sum_{m \geq 1, k \geq 2} \sqrt{2}\ketbra{n,m,1,k-1}{n,m,0,k} + \sum_{k \geq 1} 2 e^{- i \theta}\ketbra{n,1,0,k}{n,0,1,k} \nonumber \\
    &+ \sum_{l \geq 2, k \geq 0} \sqrt{2}\ketbra{n,1,l-1,k}{n,0,l,k} + \sum_{m \geq 1} \sqrt{3} e^{- 2 i \theta}\ketbra{n,m+1,0,0}{n,m,1,0} \nonumber \\
    & + \sum_{m \geq 1, k \geq 1} \sqrt{2} e^{- i \theta}\ketbra{n,m+1,0,k}{n,m,1,k} + \sum_{m \geq 1, l \geq 2, k \geq 0} \ketbra{n,m+1,l-1,k}{n,m,l,k} \nonumber \\
    &+ \sum_{k \geq 1} \sqrt{3} e^{- 2 i \theta}\ketbra{n\!+\!1,0,0,k}{n,1,0,k} + \sum_{l \geq 1, k \geq 0} \sqrt{2} e^{- i \theta}\ketbra{n\!+\!1,0,l,k}{n,1,l,k} \nonumber \\
    & + \sum_{m \geq 2, l \geq 0, k \geq 0} \ketbra{n\!+\!1,m-1,l,k}{n,m,l,k}+ \text{h.c.}\bigg] \nonumber \\
    & + U \sum_n \bigg(6\ketbra{n,0,0,0}{n,0,0,0} + \sum_{k \geq 1} 3\ketbra{n,0,0,k}{n,0,0,k} + \sum_{m \geq 1} 3\ketbra{n,m,0,0}{n,m,0,0} \nonumber \\
    &\qquad+ \sum_{l \geq 1} 2\ketbra{n,0,l,0}{n,0,l,0} + \sum_{l \geq 1, k \geq 1} \ketbra{n,0,l,k}{n,0,l,k} + \sum_{m \geq 1, k \geq 1} \ketbra{n,m,0,k}{n,m,0,k} \nonumber \\
    &\qquad + \sum_{m \geq 1, l \geq 1} \ketbra{n,m,l,0}{n,m,l,0}\bigg)
    \label{eq:4body-AHM-repr-full}
    \end{align}
    \end{widetext}
Assuming translational invariance along the direction of the COM position 
$[n+(n+m)+(n+m+l)+(n+m+l+k)]/4=n+3m/4+l/2+k/4$, we can transform this Hamiltonian~\eqref{eq:4body-AHM-repr-full} into the hybrid COM quasimomentum basis, which is defined via
    \begin{align}
        \ket{Q,m,l,k} = \sum_n e^{i Q\left( n+\frac{3m}{4}+\frac{l}{2}+\frac{k}{4}\right)} \ket{n,m,l,k},
        \label{eq:COM-MOM-hybrid-basis-4body}
    \end{align}
such that we obtain the AHM Hamiltonian in the relative coordinate space (quasimomentum space) representation
    \begin{widetext}
        \begin{align}
        \hat{H}_{Q} = &-J\bigg[
        2 e^{\frac{i Q}{4}}\ketbra{Q,0,0,0}{Q,0,0,1} + 2 e^{- \frac{i \left(Q + 12 \theta\right)}{4}}\ketbra{Q,0,0,0}{Q,1,0,0} + \sqrt{6} e^{- i \left(\frac{Q}{4} + \theta\right)}\ketbra{Q,0,1,0}{Q,0,0,1} \nonumber \\
        &+ \sqrt{6} e^{- \frac{i \left(Q + 8 \theta\right)}{4}}\ketbra{Q,1,0,0}{Q,0,1,0} + \sum_{m \geq 1, l \geq 0, k \geq 0} e^{- \frac{i Q}{4}}\ketbra{Q,m,l,k}{Q,m+1,l,k} \nonumber \\
        &+ \sum_{l \geq 1, k \geq 0} \sqrt{2} e^{- i \left(\frac{Q}{4} + \theta\right)}\ketbra{Q,0,l,k}{Q,1,l,k} + \sum_{k \geq 1} \sqrt{3} e^{- \frac{i \left(Q + 8 \theta\right)}{4}}\ketbra{Q,0,0,k}{Q,1,0,k} \nonumber \\
        &+ \sum_{m \geq 0, l \geq 0, k \geq 1} e^{\frac{i Q}{4}}\ketbra{Q,m,l,k}{Q,m,l,k+1} + \sum_{m \geq 0, l \geq 1} \sqrt{2} e^{\frac{i Q}{4}}\ketbra{Q,m,l,0}{Q,m,l,1} \nonumber \\
        &+ \sum_{m \geq 1} \sqrt{3} e^{\frac{i Q}{4}}\ketbra{Q,m,0,0}{Q,m,0,1} + \sum_{m \geq 1} 2 e^{- i \left(\frac{Q}{4} + \theta\right)}\ketbra{Q,m,1,0}{Q,m,0,1} \nonumber \\
        &+ \sum_{m \geq 0, l \geq 2, k \geq 1} e^{- \frac{i Q}{4}}\ketbra{Q,m,l,k}{Q,m,l-1,k+1} + \sum_{m \geq 0, l \geq 2} \sqrt{2} e^{- i \left(\frac{Q}{4} + \theta\right)}\ketbra{Q,m,l,0}{Q,m,l-1,1} \nonumber \\
        &+ \sum_{m \geq 1, k \geq 1} \sqrt{2} e^{- \frac{i Q}{4}}\ketbra{Q,m,1,k}{Q,m,0,k+1} + \sum_{k \geq 1} \sqrt{3} e^{- \frac{i Q}{4}}\ketbra{Q,0,1,k}{Q,0,0,k+1} \nonumber \\
        &+ \sum_{k \geq 1} 2 e^{- i \left(\frac{Q}{4} + \theta\right)}\ketbra{Q,1,0,k}{Q,0,1,k} + \sum_{m \geq 2, l \geq 1, k \geq 0} e^{- \frac{i Q}{4}}\ketbra{Q,m,l,k}{Q,m-1,l+1,k} \nonumber \\
        &+ \sum_{m \geq 2, k \geq 1} \sqrt{2} e^{- i \left(\frac{Q}{4} + \theta\right)}\ketbra{Q,m,0,k}{Q,m-1,1,k} + \sum_{l \geq 1, k \geq 0} \sqrt{2} e^{- \frac{i Q}{4}}\ketbra{Q,1,l,k}{Q,0,l+1,k} \nonumber \\
        &+ \sum_{m \geq 2} \sqrt{3} e^{- \frac{i \left(Q + 8 \theta\right)}{4}}\ketbra{Q,m,0,0}{Q,m-1,1,0} + \text{h.c.}\bigg] \nonumber \\
        & + U\bigg(6\ketbra{Q,0,0,0}{Q,0,0,0} + \sum_{k \geq 1} 3\ketbra{Q,0,0,k}{Q,0,0,k} + \sum_{m \geq 1} 3\ketbra{Q,m,0,0}{Q,m,0,0} \nonumber \\
        &\qquad + \sum_{l \geq 1} 2\ketbra{Q,0,l,0}{Q,0,l,0} + \sum_{l \geq 1, k \geq 1} \ketbra{Q,0,l,k}{Q,0,l,k} + \sum_{m \geq 1, k \geq 1} \ketbra{Q,m,0,k}{Q,m,0,k} \nonumber \\
        &\qquad  + \sum_{m \geq 1, l \geq 1} \ketbra{Q,m,l,0}{Q,m,l,0}\bigg)
            \label{eq:4body-AHM-repr-hybrid-full}
        \end{align}
    \end{widetext}
    We can obtain the four-particle spectrum of the AHM~\eqref{eq:4body-AHM-repr-hybrid-full} by solving the eigenvalue problem $\hat{H}_Q\ket{\Psi^{Q,\bm{\kappa}}} = E_{Q,\bm{\kappa}}\ket{\Psi^{Q,\bm{\kappa}}}$~\eqref{eq:3body-AHM-repr-hybrid} for each COM quasimomentum $Q$, a given statistical angle $\theta$ and interaction strength $U$, where $\bm{\kappa}=(\kappa_1,\kappa_2,\kappa_3)$ are the quantum numbers associated to the relative coordinates $m,l,k$ (labeling the eigenstates for a given $Q$). The eigenvalue problem is solved via exact diagonalization (ED), where the relative coordinates $m,l,k$ are truncated to a finite maximum value $m_{\mathrm{max}}$, $l_{\mathrm{max}}$, $k_{\mathrm{max}}$, where, here in this work, we always choose each relative coordinate axis to have the same maximum value, which we label by $K\equiv m_{\mathrm{max}}=l_{\mathrm{max}}=k_{\mathrm{max}}$.

    \subsection{Scattering state continua in the four-particle AHM}
    \label{app:4body-scattering-state-continuum}
    We briefly discuss the different types of scattering state continua that can be expected to be present in the four-particle spectrum of the AHM. 
    
    \subsubsection{Single-particle continuum of four free particles}
    This continuum is formed by the scattering states of four free particles. 
    This continuum can be constructed by the sum of four single-particle dispersions
        \begin{align}
            E^{\mathrm{4free}}_{q_1,q_2,\kappa_1,\kappa_2} = \sum_{i=1}^4 -2J\cos(k_i) \nonumber \\
            =-4J\cos\left(\frac{q_1}{2}\right)\cos\left(\kappa_1\right) -4J\cos\left(\frac{q_2}{2}\right)\cos\left(\kappa_2\right),
             \label{eq:4body-continuum-four-free-particles}
        \end{align}
    where we have used that we can write the sum of two single particle dispersions as~\eqref{eq:AHM-Scattering-dispersion} with $k_{1,2,3,4}$ being the individual quasimomenta of the four free particles, and $q_{1}=k_1+k_2$ ($q_2=k_3+k_4$) and $\kappa_{1}=(k_1-k_2)/2$ ($\kappa_2=(k_3-k_4)/2$) being the total and relative quasimomentum of the first (second) pair of free particles, respectively. 
    The bounds of this continuum spectrum are thus given by minimizing and maximizing~\eqref{eq:4body-continuum-four-free-particles} over the relative quasimomenta $\kappa_{1,2}$ and under the constraint of the conservation of the total COM quasimomentum $Q=q_1+q_2=\sum_{i=1}^4 k_i$,
        \begin{align}
        \min_{Q=q_1+q_2, \kappa_{1,2}} [E^{\mathrm{4free}}_{q_1,q_2,\kappa_1,\kappa_2}] \leq E  
         \leq \max_{Q=q_1+q_2, \kappa_{1,2}} [E^{\mathrm{4free}}_{q_1,q_2,\kappa_1,\kappa_2}].
        \label{eq:4body-four-free-particles-scattering-state-energy-bounds}
        \end{align}
    These bounds of the four-body single-particle continuum can be solved analytically and are given by $E^{\mathrm{4free}}_{\mathrm{min/max}}=\mp 8J\cos(Q/4)$ for $Q\in[-\pi,\pi)$.
    
    \subsubsection{Continuum of a two-particle bound state and two free particles}
    This continuum is formed by one two-particle bound state with two free particles. This continuum can be constructed by the sum of one two-particle bound state dispersion $E^{\mathrm{2b}}_{q,\pm}$~\eqref{eq:AHM-bound-states-dispersion} and two single-particle dispersions $E=-2J\cos(k_1)-2J\cos(k_2)=-4J\cos\left(q/2\right)\cos\left(\kappa\right)$~\eqref{eq:AHM-Scattering-dispersion}, with $q=k_1+k_2$ and $\kappa=(k_1-k_2)/2$ being the total and relative quasimomentum of the two free particles, respectively, such that this total continuum's energy dispersion reads as 
        \begin{align}
            E^{\mathrm{2b+2free}}_{q_1,q_2,\kappa'\pm} = E^{\mathrm{2b}}_{q_1,\pm} -4J\cos\left(\frac{q_2}{2}\right)\cos\left(\kappa'\right).
             \label{eq:4body-continuum-two-body-bound-state-with-two-free-particles}
        \end{align}
    Here, $q_{1/2}$ is the total COM quasimomentum of the two-particle bound state and the two free particles, respectively, and $\kappa'$ is the relative quasimomentum of the two free particles.
    The upper and lower bounds of this continuum spectrum are thus given by minimizing and maximizing~\eqref{eq:4body-continuum-two-body-bound-state-with-two-free-particles} over the relative quasimomentum $\kappa'$ and under the constraint of the conservation of the total COM quasimomentum $Q=q_1+q_2$,
        \begin{align}
        &\min_{Q=q_1+q_2, \kappa'} [E^{\mathrm{2b}}_{q_1,\pm} -4J\cos\left(\frac{q_2}{2}\right)\cos\left(\kappa'\right)] \leq \nonumber \\  E  
        & \leq \max_{Q=q_1+q_2, \kappa'} [E^{\mathrm{2b}}_{q_1,\pm} -4J\cos\left(\frac{q_2}{2}\right)\cos\left(\kappa'\right)].
        \label{eq:4body-two-body-bound-with-two-free-particles-scattering-state-energy-bounds}
        \end{align}

    \subsubsection{Continuum of two independent two-particle bound states}
    This continuum is formed by two independent two-particle bound states, where each bound pair can freely roam through the lattice. This continuum can be constructed by the sum of two two-particle bound state dispersions $E^{\mathrm{2b}}_{q,\pm}$~\eqref{eq:AHM-bound-states-dispersion},
        \begin{align}
            E^{\mathrm{2b+2b}}_{q_1,q_2\pm,\pm} = E^{\mathrm{2b}}_{q_1,\pm} + E^{\mathrm{2b}}_{q_2,\pm},
             \label{eq:4body-continuum-two-two-body-bound-states}
        \end{align}
    where $q_1$ is the total COM quasimomentum of the first bound pair and $q_2$ the total COM quasimomentum of the second bound pair.
    The upper and lower bounds of this continuum spectrum with are thus given by minimizing and maximizing~\eqref{eq:4body-continuum-two-two-body-bound-states} under the constraint of the conservation of the total COM quasimomentum $Q=q_1+q_2$,
        \begin{align}
        \min_{Q=q_1+q_2} [E^{\mathrm{2b}}_{q_1,\pm} + E^{\mathrm{2b}}_{q_2,\pm}] \leq E  \leq \max_{Q=q_1+q_2} [E^{\mathrm{2b}}_{q_1,\pm} + E^{\mathrm{2b}}_{q_2,\pm}].
        \label{eq:4body-two-two-body-bound-scattering-state-energy-bounds}
        \end{align}
    Note as the two-particle bound states of the AHM only exist for certain values of the total COM quasimomentum $q$ and statistical angle $\theta$, the two independent two-body-bound continua will only be present for those values of $Q$ and $\theta$ where there exists a valid partitioning of $Q=q_1+q_2$ such that both $q_1$ and $q_2$ lie in the bound-state-supporting region for a given $\theta$. 
    For some values of $Q$ and $\theta$, there will be no valid partitioning of $Q=q_1+q_2$ where both momenta lie in the bound-state-supporting region for a given $\theta$, such that at those values the continuum of two independent two-particle bound states will be absent, since there at least one of the two pairs carries a momentum where the pair dissolves into the scattering continuum and hence does not form a bound state [see~\cref{fig:four-body-spectrum-U0-AHM} in the main text].
    
    \subsubsection{Continuum of a three-particle bound state with one free particle}
    This continuum is formed by the scattering states of a three-particle bound state with one free particle, which can be constructed by the sum of a three-particle bound state energy $E^{\mathrm{3b}}_{q}$ outside the continuum and a single-particle dispersion $\epsilon=-2J\cos(k)$, 
        \begin{align}
            E^{\mathrm{3b+1free}}_{q,k} = E^{\mathrm{3b}}_{q} - 2J\cos(k).
             \label{eq:4body-continuum-three-body-bound-state-with-one-free-particle}
        \end{align}
    with $q$ being the total COM quasimomentum of the three-particle bound state and $k$ the quasimomentum of the free particle. The upper and lower bounds of this continuum spectrum are thus given by minimizing and maximizing~\eqref{eq:4body-continuum-three-body-bound-state-with-one-free-particle} over the three-particle quasimomentum $q$ under the constraint of the conservation of the total COM quasimomentum
    $Q=q+k$, i.e.,
        \begin{align}
        &\min_{Q=q+k} [E^{\mathrm{3b}}_{q} - 2J\cos(Q-q)] \leq \nonumber \\  E  
        & \leq \max_{Q=q+k} [E^{\mathrm{3b}}_{q} - 2J\cos(Q-q)].
        \label{eq:4body-three-body-bound-with-one-free-particle-scattering-state-energy-bounds}
        \end{align}
    Note as there are no analytical expressions for the three-particle bound states of the AHM, we utilize the numerical three-particle bound-state solutions outside the continua, obtained by using ED to solve the eigenvalue problem of~\eqref{eq:3body-AHM-repr-hybrid-full} as discussed in~\cref{sec:3body-bound-states-AHM} of the main text.

    \subsection{Chiral symmetry of the four-particle AHM}\label{app:4body-chiral-symmetry}
    As per the discussion in Appendices~\ref{app:2body-chiral-symmetry} and~\ref{app:3body-chiral-symmetry}, also the four-particle AHM for $U=0$ admits a chiral symmetry~\cite{Theel2025}.
    The chiral symmetry operator $\hat{S} = \exp(i\pi\sum_j j\hat{n}_j)$ acts on the four-particle configuration space basis as
    $\hat{S}\ket{n,m,l,k} = e^{i\pi(4n+3m+2l+k)}\ket{n,m,l,k}= (-1)^{m+k}\ket{n,m,l,k}$. 
    As a consequence, the action of $\hat{S}$ on the hybrid COM quasimomentum basis states~\eqref{eq:COM-MOM-hybrid-basis-4body} reads
    \begin{align}
        \hat{S}\ket{Q,m,l,k} = (-1)^{m+k}\ket{Q,m,l,k},
        \label{eq:4body-chiral-symmetry-hybrid-basis}
    \end{align}
    i.e., $\hat{S}$ is \emph{diagonal} in the hybrid basis at \emph{fixed} $Q$. 
    This is the same situation as for $N_b=2$, and contrasts with the odd-particle-number case $N_b=3$, where $\hat{S}$ maps the $Q$-block to the $(Q+\pi)$-block~\eqref{eq:3body-chiral-symmetry-hybrid-basis}.

    The reason for this diagonal action is that the three-dimensional relative coordinate space lattice of the four-particle problem, spanned by $(m,l,k)\in\mathbb{N}^3$, is \emph{bipartite} under the coloring (grading) $(-1)^{m+k}$: every hopping process in~\eqref{eq:4body-AHM-repr-hybrid-full} changes $m+k$ by exactly $\pm 1$ (either $m$ or $k$ changes by $\pm 1$, while the other coordinates may change simultaneously to conserve particle number, but always such that the net change in $m+k$ is odd). 

    Since $\hat{S}$ is diagonal and squares to unity, the anticommutation relation $\hat{S}\hat{H}_Q\hat{S}^\dagger = -\hat{H}_Q$ holds block-by-block for $U=0$.
    Consequently, the four-particle spectrum satisfies the \emph{intra-block} pairing, i.e., for each eigenstate $\ket{\Psi^{Q,n}}$ with energy $E_n(Q)$, $\hat{S}\ket{\Psi^{Q,n}}$ is an eigenstate at the \emph{same} $Q$ with energy $-E_n(Q)$.
    Hence, the four-body bound state bands as well as all scattering state continua come in $\pm E$ pairs at each fixed $Q$, as can be nicely seen in~\cref{fig:four-body-spectrum-U0-AHM} of the main text.

    More generally, for even (odd) particle numbers $N_b$ we have that $(-1)^{N_b n}=1$ ($(-1)^{N_b n}=(-1)^n$) in the action of $\hat{S}$ on the hybrid basis states, and the relative coordinate lattice is bipartite (non-bipartite), such that the chiral symmetry operator $\hat{S}$ acts diagonally within each $Q$-block (or shifts the quasimomentum $Q$).
    \section{$N$-body problem}
    \label{app:N-body-AHM}
    Let us now briefly discuss the $N$-body generalization of the bound-state problem in the AHM. Note first of all that the $N$-body problem on the 1D AHM lattice can be mapped to a single-particle problem in the $N$-dimensional configuration space, given by the $N$-dimensional simplex, which is spanned by the following basis states $\ket{n,\bm{m}}$, with $\bm{m}=(m_1,\ldots m_{N-1})$. Labeling the particles positions from left to right by $x_1 < x_2 < \dots < x_N$, $n$ denotes the position of the first and leftmost particle, and $m_i=x_{i+1}-x_i\geq 0$ the separation between the $i$-th and $(i+1)$-th particle, such that the total COM position is given by $X = n + \sum_{i=1}^{N-1} (N-i)m_i/N$. 
    Assuming then an underlying infinite 1D AHM chain, the problem in the configuration space becomes translationally invariant along the COM direction, so that we can transform the single-particle problem in the $N$-dimensional configuration space into a single-particle problem in the $(N-1)$-dimensional relative coordinate space spanned by the basis states $\ket{Q,\bm{m}}$ with $\bm{m}=(m_1,m_2,\ldots,m_{N-1})$, where $Q$ is the COM quasimomentum, where the two bases are as before related via $\ket{Q,\bm{m}} = \sum_n e^{i Q X}\ket{n,\bm{m}}$. 
    The relative coordinate space is given by the region 
    $\mathcal{W}_{N-1}^+ = \{ (m_1,m_2,\ldots,m_{N-1}) \in \mathbb{Z}^{N-1} : m_i \geq 0 \}$,
    i.e., the positive orthant of the $(N-1)$-dimensional integer lattice $\mathbb{Z}^{N-1}$.

    Now, let us first note that the structure of the relative coordinate space lattice and its single-particle continuum admit a natural $N$-body generalization. 
    Each right-ward tunneling process of particle $j$ in the original 1D lattice corresponds to a single-particle hopping process in the relative coordinate space, where the hopping structure is determined by the $N$ displacement vectors $\bm{\delta}_j = \mathbf{e}_{j-1}-\mathbf{e}_j$ ($j=1,\ldots,N$, $\mathbf{e}_0=\mathbf{e}_N=\mathbf{0}$), where $\bm{e}_j$ are the standard basis vectors of $\mathbb{Z}^{N-1}$. Including Hermitian conjugate processes, the set of displacement vectors reachable by a single hopping process is given by $\mathcal{S}=\{\pm \bm{\delta}_j\}_{j=1}^N$, which contains at most $2N$ distinct vectors. The coordination number $z=\abs{\mathcal{S}}$ of the relative coordinate space (i.e., the number of nearest neighbors of each site in the relative coordinate space) is thus upper bounded by $2N$. 
    For $N=2$, we only have $\mathcal{S} = \{-\bm{e}_1, \bm{e}_1\}$, where $\bm{\delta}_2 = -\bm{\delta}_1$, which corresponds to the 1D chain of~\eqref{eq:app:2body-AHM-repr-hybrid} with $z=2$. 
    For $N=3$, we have the three distinct displacement vectors $\bm{\delta}_1 = -\bm{e}_1$, $\bm{\delta}_2 = \bm{e}_1 - \bm{e}_2$, and $\bm{\delta}_3 = \bm{e}_2$, such that $\mathcal{S} = \{\pm \bm{\delta}_1, \pm \bm{\delta}_2, \pm \bm{\delta}_3\}$ contains $6$ distinct vectors, which corresponds to the triangular lattice structure of the relative coordinate space for three particles as shown in~\cref{fig:Relative-coordinate-space} of the main text (see also App.~\ref{app:sub:3-body-single-part-continuum-plane-wave}). 
    In general for $N \geq 3$, the set $\mathcal{S}$ only contains distinct vectors, and the coordination number is always $z=2N$.

    \begin{widetext}
        \begin{center}
            \begin{table}[th!]
                \begin{tabular}{c|c|c|c|c|c}
                    $N$ & Rel.\ coord.\ dim.\ & Bulk lattice & $z$ & $E_{\mathrm{min}}(Q)$ & $E_{\mathrm{max}}(Q)$ \\
                    \hline
                    2 & 1 & Chain & 2 & $-4J\cos(Q/2)$ & $+4J\cos(Q/2)$ \\
                    3 & 2 & Triangular & 6& $-6J\cos(Q/3)$ & $+6J\cos\left( \frac{\pi-\abs{Q}}{3} \right)$
                    \\
                    4 & 3 & $\mathcal{W}^+_{3}$ & 8 & $-8J\cos(Q/4)$ & $+8J\cos(Q/4)$ \\
                    $\vdots$ & $\vdots$ & $\vdots$ & $\vdots$ & $\vdots$ & $\vdots$ \\
                    $N$ & $N-1$ & $\mathcal{W}^+_{N-1}$ & $2N$ & $-2NJ\cos(Q/N)$ & 
                    \makecell{$+2NJ\cos(Q/N)$ if $N$ even, \\ $+2NJ\cos\left( \frac{\pi-\abs{Q}}{N} \right)$ if $N$ is odd}\\
                \end{tabular}
                \caption{Summary of the $N$-body relative coordinate space lattice structure and the single-particle continuum edges $E_{\mathrm{min}/\mathrm{max}}(Q)$ for $N$ free particles at fixed COM quasimomentum $Q$. The coordination number $z=2N$ for $N\geq 3$ counts the $N$ distinct hopping directions in relative coordinate space lattice.}
                \label{tab:Nbody-lattice-and-singlepart-continuum}
            \end{table}
        \end{center}
    \end{widetext}
    \subsection{Single-particle continuum of $N$ free particles}
    \label{subsec:Nbody-single-particle-continuum}
    Deep in the bulk ($m_i \gg 1$, all edge effects negligible), single particle can hop independently, and the resulting hopping structure in the relative coordinate space is determined by the $N$ displacement vectors $\bm{\delta}_j$. Assuming a plane-wave ansatz $\Psi_{\mathbf{m}} \sim e^{i\bm{\kappa}\cdot\mathbf{m}}$ deep in the bulk, the energy dispersion relation of the $N$-body single-particle continuum (i.e., the continuum of $N$ free particles) at fixed COM quasimomentum $Q$ reads
    \begin{align}
        E(Q,\bm{\kappa}) = -2J\sum_{j=1}^{N}\cos\!\left(\bm{\kappa}\cdot\bm{\delta}_j + \frac{Q}{N}\right) = \sum_{j=1}^{N}\epsilon(k_j),
        \label{eq:Nbody-scattering-dispersion-general}
    \end{align}
    where $\epsilon(k)=-2J\cos(k)$ is the single-particle dispersion, $\bm{\kappa}=(\kappa_1,\ldots,\kappa_{N-1})$ is the $(N-1)$-dimensional relative quasimomentum vector, and the individual quasimomenta $k_j = \bm{\kappa}\cdot\bm{\delta}_j + Q/N$ satisfy the COM constraint $\sum_{j=1}^N k_j = Q$.
    The extrema of~\eqref{eq:Nbody-scattering-dispersion-general} under the constraint $\sum_j k_j = Q$ can be found by a Lagrange multiplier argument. The minimum is attained when all particles carry equal quasimomenta $k_j=Q/N$, 
    \begin{align}
        E_{\mathrm{min}}(Q) = -2NJ\cos\!\left(\frac{Q}{N}\right).
        \label{eq:Nbody-continuum-minimum}
    \end{align}
    while the maximum depends on the parity of $N$. For even $N$, we have
     \begin{align}
        E_{\mathrm{max}}(Q)\rvert_{N,\mathrm{even}} = +2NJ\cos(Q/N) = -E_{\mathrm{min}}(Q), 
     \end{align}
    obtained from flipping all momenta $k_j \to k_j + \pi$ for all $j$ with the constraint $\sum_j (k_j +\pi) = Q\bmod 2\pi$ still being satisfied.
    For odd $N$, the maximum is attained at a mixed configuration, i.e., it depends on the sign of $Q$, and is given by 
     \begin{align}
        E_{\mathrm{max}}(Q)\rvert_{N,\mathrm{odd}} = +2NJ\cos\!\left(\frac{\pi-\abs{Q}}{N}\right).
     \end{align}
    These results are summarized in~\cref{tab:Nbody-lattice-and-singlepart-continuum}.

    \subsection{Bound states outside the continua in the $N$-body problem of the AHM}
    \label{subsec:Nbody-bound-states-outside-continua}
    Let us now argue why there is a critical particle number $N^*$ beyond which no bound states outside the continua in the AHM for $U=0$ are to be expected. 
    The organizing principle is given by the competition, between the possible $N$-body bound state and all relevant continua in the $N$-body problem, to minimize their respective kinetic energy at finite COM quasimomentum $Q$. 
    The different types of continua are affected in different ways by the statistical angle $\theta$ and the bosonic enhancement factors and can delocalize over regions in the relative coordinate space lattice of varying dimensionality. 
    In contrast, the $N$-body bound state, although more strongly affected by $\theta$ and bosonic enhancement factors, is always localized around the origin of the relative coordinate space lattice. 
    This is similar to the discussion of the binding mechanism of the three-body bound states in~\cref{sec:Binding-mechanism-outside-AHM} of the main text. 
    
    For $N$ particles, the distinct types of scattering continua correspond to integer partitions of $N$ with $k\geq 2$ parts, $\lambda_1 \geq \lambda_2 \geq \dots \geq \lambda_k \geq 1$ and $\sum_{i=1}^k \lambda_i = N$. Here, each partition $(\lambda_1,\ldots,\lambda_k)$ with $k\geq 2$ parts defines a scattering channel where $k$ sub-clusters propagate with individual COM momenta $q_1,\ldots,q_k$ subject to the total COM constraint $Q=q_1+\dots+q_k$, so that there is a freedom to vary the $k-1$ relative momenta subject to the total COM constraint, giving rise to a continuum band of scattering states at each $Q$. 
    For instance, for $N=2$, we have only one partition, $(1^2)$ with $\lambda_1=\lambda_2=1$, which corresponds to the continuum of two single particles, while $N=3$, we have two partitions, $(1^3)$ with $\lambda_1=\lambda_2=\lambda_3=1$ and $(2,1)$ with $\lambda_1=2$ and $\lambda_2=1$, which correspond to the continuum of three single particles and the continuum of a two-particle bound state with one free particle, respectively, and so on for higher $N$.
    Our notation chosen for the partition $(\lambda_1,\ldots,\lambda_k)$ lists the parts in non-increasing order, writing repeated values explicitly as powers, such that the exponential notation $\lambda^r$ is a shorthand for the value $\lambda$ appears $r$ times as a part.
    Thus, each cluster can be either a free particle (if $\lambda_i=1$) or a bound state of $\lambda_i$ particles (if $\lambda_i > 1$); see the summarizing table for the different types of continua in~\cref{tab:Nbody-continua}.

    \begin{center}
        \begin{table}[b]
            \begin{tabular}{c|c|c}
                $N$ & Number of partitions & Partitions \\
                \hline 
                2 & 1 & $(1^2)$ \\
                3 & 2 & $(1^3)$, $(2,1)$ \\
                4 & 4 & $(1^4)$, $(2,1^2)$, $(2^2)$, $(3,1)$ \\
                5 & 6 & \makecell{ $(1^5)$, $(2,1^3)$, $(2^2,1)$,\\  $(3,1^2)$, $(3,2)$, $(4,1)$ }\\
                $\vdots$  & $\vdots$ & $\vdots$  \\
                $N$ & $p(N)-1$ & \makecell{All partitions of $N$ \\ with at least two parts.}
            \end{tabular}
            \caption{Summary of the different types of scattering continua in the $N$-body problem of the AHM, which correspond to integer partitions of $N$ with at least two parts. The partition $(\lambda_1,\ldots,\lambda_k)$ corresponds to a continuum formed by $k$ clusters of $\lambda_1,\ldots,\lambda_k$ particles, where each cluster can be either a free particle (if $\lambda_i=1$) or a $\lambda_i$-body bound state (if $\lambda_i > 1$). There are $p(N)-1$ integer partitions of $N$, 
            which grows exponentially with $N$ as $p(N)$ for large $N$~\cite{Hardy1918}.}
            \label{tab:Nbody-continua}
        \end{table}
     \end{center}
     The partition corresponding to the single-particle continuum of $N$ free particles, $(1^N)$, is the most delocalized one, as it is given by plane waves spread over all $N$ dimensions of the relative coordinate space lattice (see App.~\ref{subsec:Nbody-single-particle-continuum}). 
     However, it is also the only one which is unaffected by both the statistical angle $\theta$ and the bosonic enhancement factors. 
     The comparison relevant for the existence of $N$-body bound states must be made against every partition at each $Q$. 
     We expect the hardest constraint to be set by the $(N-1,1)$ partition, since it has the largest sub-cluster among all two-cluster partitions, that is most affected by $\theta$ and bosonic enhancement factors (this is confirmed in the three- and four-body cases in Fig.~\ref{fig:three-body-spectrum-full-AHM} and Fig.~\ref{fig:four-body-spectrum-U0-AHM} of the main text, respectively). 
     Thus, we can expect the energy minimum of the $(N-1,1)$ partition continuum to be closest to that of a potential $N$-body bound state, and hence the binding gap condition to be set by the energy difference between the $N$-body bound state and the lower edge of the $(N-1,1)$ partition continuum. 
     The $N$-body bound state is localized around the origin of the relative coordinate space lattice, and the $(N-1,1)$ partition continuum is delocalized over the individual directions corresponding to a single free particle in the partition. 
     The number of directions along which the $(N-1,1)$ partition can delocalize grows linearly with $N-1$ as $N$ increases, since the AHM does not dictate a priori which of the $N$ particles should be in the single-particle cluster, e.g. for $N=3$ particles the $(2,1)$ cluster can delocalize along both the $m$ and $l$ direction in the relative coordinate space lattice~(see Fig.~\ref{fig:Relative-coordinate-space} and Fig.~\ref{fig:three-body-spectrum-full-AHM}(h) in the main text). 
     For a relative coordinate space of size $K^N$ (with $K$ sites along each relative coordinate axis), the $(N-1,1)$ partition continuum can delocalize over $\sim K(N-1)$ sites, while the $N$-body bound state can only delocalize over $\mathcal{O}(1)$ sites around the origin (although having slightly more bosonic enhancement factors). 
     Hence, one can expect there to be a finite critical particle number $N^*$ beyond which no $N$-body bound states outside the continua are found. 

\bibliographystyle{apsrev4-2-arxiv-notitlelink} 
\bibliography{main}

\begin{thebibliography}{151}%
\makeatletter
\providecommand \@ifxundefined [1]{%
 \@ifx{#1\undefined}
}%
\providecommand \@ifnum [1]{%
 \ifnum #1\expandafter \@firstoftwo
 \else \expandafter \@secondoftwo
 \fi
}%
\providecommand \@ifx [1]{%
 \ifx #1\expandafter \@firstoftwo
 \else \expandafter \@secondoftwo
 \fi
}%
\providecommand \natexlab [1]{#1}%
\providecommand \enquote  [1]{``#1''}%
\providecommand \bibnamefont  [1]{#1}%
\providecommand \bibfnamefont [1]{#1}%
\providecommand \citenamefont [1]{#1}%
\providecommand \href@noop [0]{\@secondoftwo}%
\providecommand \href [0]{\begingroup \@sanitize@url \@href}%
\providecommand \@href[1]{\@@startlink{#1}\@@href}%
\providecommand \@@href[1]{\endgroup#1\@@endlink}%
\providecommand \@sanitize@url [0]{\catcode `\\12\catcode `\$12\catcode `\&12\catcode `\#12\catcode `\^12\catcode `\_12\catcode `\%12\relax}%
\providecommand \@@startlink[1]{}%
\providecommand \@@endlink[0]{}%
\providecommand \url  [0]{\begingroup\@sanitize@url \@url }%
\providecommand \@url [1]{\endgroup\@href {#1}{\urlprefix }}%
\providecommand \urlprefix  [0]{URL }%
\providecommand \Eprint [0]{\href }%
\providecommand \doibase [0]{https://doi.org/}%
\providecommand \selectlanguage [0]{\@gobble}%
\providecommand \bibinfo  [0]{\@secondoftwo}%
\providecommand \bibfield  [0]{\@secondoftwo}%
\providecommand \translation [1]{[#1]}%
\providecommand \BibitemOpen [0]{}%
\providecommand \bibitemStop [0]{}%
\providecommand \bibitemNoStop [0]{.\EOS\space}%
\providecommand \EOS [0]{\spacefactor3000\relax}%
\providecommand \BibitemShut  [1]{\csname bibitem#1\endcsname}%
\let\auto@bib@innerbib\@empty
\bibitem [{\citenamefont {Kwan}\ \emph {et~al.}(2024)\citenamefont {Kwan}, \citenamefont {Segura}, \citenamefont {Li}, \citenamefont {Kim}, \citenamefont {Gorshkov}, \citenamefont {Eckardt}, \citenamefont {{Bakkali-Hassani}},\ and\ \citenamefont {Greiner}}]{Kwan2024}%
  \BibitemOpen
  \bibfield  {author} {\bibinfo {author} {\bibfnamefont {J.}~\bibnamefont {Kwan}}, \bibinfo {author} {\bibfnamefont {P.}~\bibnamefont {Segura}}, \bibinfo {author} {\bibfnamefont {Y.}~\bibnamefont {Li}}, \bibinfo {author} {\bibfnamefont {S.}~\bibnamefont {Kim}}, \bibinfo {author} {\bibfnamefont {A.~V.}\ \bibnamefont {Gorshkov}}, \bibinfo {author} {\bibfnamefont {A.}~\bibnamefont {Eckardt}}, \bibinfo {author} {\bibfnamefont {B.}~\bibnamefont {{Bakkali-Hassani}}},\ and\ \bibinfo {author} {\bibfnamefont {M.}~\bibnamefont {Greiner}},\ }\bibfield  {title} {\bibinfo {title} {Realization of one-dimensional anyons with arbitrary statistical phase},\ }\href {https://doi.org/10.1126/science.adi3252} {\bibfield  {journal} {\bibinfo  {journal} {Science}\ }\textbf {\bibinfo {volume} {386}},\ \bibinfo {pages} {1055} (\bibinfo {year} {2024})}\BibitemShut {NoStop}%
\bibitem [{\citenamefont {Dhar}\ \emph {et~al.}(2025)\citenamefont {Dhar}, \citenamefont {Wang}, \citenamefont {Horvath}, \citenamefont {Vashisht}, \citenamefont {Zeng}, \citenamefont {Zvonarev}, \citenamefont {Goldman}, \citenamefont {Guo}, \citenamefont {Landini},\ and\ \citenamefont {N{\"a}gerl}}]{Dhar2025}%
  \BibitemOpen
  \bibfield  {author} {\bibinfo {author} {\bibfnamefont {S.}~\bibnamefont {Dhar}}, \bibinfo {author} {\bibfnamefont {B.}~\bibnamefont {Wang}}, \bibinfo {author} {\bibfnamefont {M.}~\bibnamefont {Horvath}}, \bibinfo {author} {\bibfnamefont {A.}~\bibnamefont {Vashisht}}, \bibinfo {author} {\bibfnamefont {Y.}~\bibnamefont {Zeng}}, \bibinfo {author} {\bibfnamefont {M.~B.}\ \bibnamefont {Zvonarev}}, \bibinfo {author} {\bibfnamefont {N.}~\bibnamefont {Goldman}}, \bibinfo {author} {\bibfnamefont {Y.}~\bibnamefont {Guo}}, \bibinfo {author} {\bibfnamefont {M.}~\bibnamefont {Landini}},\ and\ \bibinfo {author} {\bibfnamefont {H.-C.}\ \bibnamefont {N{\"a}gerl}},\ }\bibfield  {title} {\bibinfo {title} {Observing anyonization of bosons in a quantum gas},\ }\href {https://doi.org/10/g9m86b} {\bibfield  {journal} {\bibinfo  {journal} {Nature}\ }\textbf {\bibinfo {volume} {642}},\ \bibinfo {pages} {53} (\bibinfo {year} {2025})}\BibitemShut {NoStop}%
\bibitem [{\citenamefont {{Bakkali-Hassani}}\ \emph {et~al.}(2026)\citenamefont {{Bakkali-Hassani}}, \citenamefont {Kwan}, \citenamefont {Segura}, \citenamefont {Li}, \citenamefont {Tesfaye}, \citenamefont {{Valent{\'i}-Rojas}}, \citenamefont {Eckardt},\ and\ \citenamefont {Greiner}}]{Bakkali-Hassani2026}%
  \BibitemOpen
  \bibfield  {author} {\bibinfo {author} {\bibfnamefont {B.}~\bibnamefont {{Bakkali-Hassani}}}, \bibinfo {author} {\bibfnamefont {J.}~\bibnamefont {Kwan}}, \bibinfo {author} {\bibfnamefont {P.}~\bibnamefont {Segura}}, \bibinfo {author} {\bibfnamefont {Y.}~\bibnamefont {Li}}, \bibinfo {author} {\bibfnamefont {I.}~\bibnamefont {Tesfaye}}, \bibinfo {author} {\bibfnamefont {G.}~\bibnamefont {{Valent{\'i}-Rojas}}}, \bibinfo {author} {\bibfnamefont {A.}~\bibnamefont {Eckardt}},\ and\ \bibinfo {author} {\bibfnamefont {M.}~\bibnamefont {Greiner}},\ }\bibinfo {title} {Revealing {{Pseudo-Fermionization}} and {{Chiral Binding}} of {{One-Dimensional Anyons}} using {{Adiabatic State Preparation}}} (\bibinfo {year} {2026}),\ \Eprint {https://arxiv.org/abs/2602.20421} {arXiv:2602.20421} \BibitemShut {NoStop}%
\bibitem [{\citenamefont {Wilczek}(1982)}]{Wilczek1982}%
  \BibitemOpen
  \bibfield  {author} {\bibinfo {author} {\bibfnamefont {F.}~\bibnamefont {Wilczek}},\ }\bibfield  {title} {\bibinfo {title} {Magnetic flux, angular momentum, and statistics},\ }\href {https://doi.org/10.1103/PhysRevLett.48.1144} {\bibfield  {journal} {\bibinfo  {journal} {Phys. Rev. Lett.}\ }\textbf {\bibinfo {volume} {48}},\ \bibinfo {pages} {1144} (\bibinfo {year} {1982})}\BibitemShut {NoStop}%
\bibitem [{\citenamefont {Leinaas}\ and\ \citenamefont {Myrheim}(1977)}]{Leinaas1977}%
  \BibitemOpen
  \bibfield  {author} {\bibinfo {author} {\bibfnamefont {J.~M.}\ \bibnamefont {Leinaas}}\ and\ \bibinfo {author} {\bibfnamefont {J.}~\bibnamefont {Myrheim}},\ }\bibfield  {title} {\bibinfo {title} {On the theory of identical particles},\ }\href {https://doi.org/10.1007/BF02727953} {\bibfield  {journal} {\bibinfo  {journal} {Il Nuovo Cimento B 1971-1996}\ }\textbf {\bibinfo {volume} {37}},\ \bibinfo {pages} {1} (\bibinfo {year} {1977})}\BibitemShut {NoStop}%
\bibitem [{\citenamefont {Wu}(1984{\natexlab{a}})}]{Wu1984}%
  \BibitemOpen
  \bibfield  {author} {\bibinfo {author} {\bibfnamefont {Y.-S.}\ \bibnamefont {Wu}},\ }\bibfield  {title} {\bibinfo {title} {Multiparticle quantum mechanics obeying fractional statistics},\ }\href {https://doi.org/10.1103/PhysRevLett.53.111} {\bibfield  {journal} {\bibinfo  {journal} {Phys. Rev. Lett.}\ }\textbf {\bibinfo {volume} {53}},\ \bibinfo {pages} {111} (\bibinfo {year} {1984}{\natexlab{a}})}\BibitemShut {NoStop}%
\bibitem [{\citenamefont {Wu}(1984{\natexlab{b}})}]{Wu1984a}%
  \BibitemOpen
  \bibfield  {author} {\bibinfo {author} {\bibfnamefont {Y.-S.}\ \bibnamefont {Wu}},\ }\bibfield  {title} {\bibinfo {title} {General theory for quantum statistics in two dimensions},\ }\href {https://doi.org/10.1103/PhysRevLett.52.2103} {\bibfield  {journal} {\bibinfo  {journal} {Phys. Rev. Lett.}\ }\textbf {\bibinfo {volume} {52}},\ \bibinfo {pages} {2103} (\bibinfo {year} {1984}{\natexlab{b}})}\BibitemShut {NoStop}%
\bibitem [{\citenamefont {Greiter}\ and\ \citenamefont {Wilczek}(2024)}]{Greiter2024}%
  \BibitemOpen
  \bibfield  {author} {\bibinfo {author} {\bibfnamefont {M.}~\bibnamefont {Greiter}}\ and\ \bibinfo {author} {\bibfnamefont {F.}~\bibnamefont {Wilczek}},\ }\bibfield  {title} {\bibinfo {title} {Fractional {{Statistics}}},\ }\href {https://doi.org/10/gts959} {\bibfield  {journal} {\bibinfo  {journal} {Annu. Rev. Condens. Matter Phys.}\ }\textbf {\bibinfo {volume} {15}},\ \bibinfo {pages} {131} (\bibinfo {year} {2024})}\BibitemShut {NoStop}%
\bibitem [{\citenamefont {Fr{\"o}hlich}\ and\ \citenamefont {Marchetti}(1988)}]{Frohlich1988}%
  \BibitemOpen
  \bibfield  {author} {\bibinfo {author} {\bibfnamefont {J.}~\bibnamefont {Fr{\"o}hlich}}\ and\ \bibinfo {author} {\bibfnamefont {P.-A.}\ \bibnamefont {Marchetti}},\ }\bibfield  {title} {\bibinfo {title} {Quantum field theory of anyons},\ }\href {https://doi.org/10.1007/BF00402043} {\bibfield  {journal} {\bibinfo  {journal} {Lett. Math. Phys.}\ }\textbf {\bibinfo {volume} {16}},\ \bibinfo {pages} {347} (\bibinfo {year} {1988})}\BibitemShut {NoStop}%
\bibitem [{\citenamefont {Fr{\"o}hlich}(1976)}]{Frohlich1976}%
  \BibitemOpen
  \bibfield  {author} {\bibinfo {author} {\bibfnamefont {J.}~\bibnamefont {Fr{\"o}hlich}},\ }\bibfield  {title} {\bibinfo {title} {New super-selection sectors (``soliton-states'') in two dimensional {{Bose}} quantum field models},\ }\href {https://doi.org/10/bqq69t} {\bibfield  {journal} {\bibinfo  {journal} {Commun. Math. Phys.}\ }\textbf {\bibinfo {volume} {47}},\ \bibinfo {pages} {269} (\bibinfo {year} {1976})}\BibitemShut {NoStop}%
\bibitem [{\citenamefont {Tsui}\ \emph {et~al.}(1982)\citenamefont {Tsui}, \citenamefont {Stormer},\ and\ \citenamefont {Gossard}}]{Tsui1982}%
  \BibitemOpen
  \bibfield  {author} {\bibinfo {author} {\bibfnamefont {D.~C.}\ \bibnamefont {Tsui}}, \bibinfo {author} {\bibfnamefont {H.~L.}\ \bibnamefont {Stormer}},\ and\ \bibinfo {author} {\bibfnamefont {A.~C.}\ \bibnamefont {Gossard}},\ }\bibfield  {title} {\bibinfo {title} {Two-dimensional magnetotransport in the extreme quantum limit},\ }\href {https://doi.org/10.1103/PhysRevLett.48.1559} {\bibfield  {journal} {\bibinfo  {journal} {Phys. Rev. Lett.}\ }\textbf {\bibinfo {volume} {48}},\ \bibinfo {pages} {1559} (\bibinfo {year} {1982})}\BibitemShut {NoStop}%
\bibitem [{\citenamefont {Laughlin}(1983)}]{Laughlin1983}%
  \BibitemOpen
  \bibfield  {author} {\bibinfo {author} {\bibfnamefont {R.~B.}\ \bibnamefont {Laughlin}},\ }\bibfield  {title} {\bibinfo {title} {Anomalous quantum hall effect: {{An}} incompressible quantum fluid with fractionally charged excitations},\ }\href {https://doi.org/10.1103/PhysRevLett.50.1395} {\bibfield  {journal} {\bibinfo  {journal} {Phys. Rev. Lett.}\ }\textbf {\bibinfo {volume} {50}},\ \bibinfo {pages} {1395} (\bibinfo {year} {1983})}\BibitemShut {NoStop}%
\bibitem [{\citenamefont {Arovas}\ \emph {et~al.}(1984)\citenamefont {Arovas}, \citenamefont {Schrieffer},\ and\ \citenamefont {Wilczek}}]{Arovas1984}%
  \BibitemOpen
  \bibfield  {author} {\bibinfo {author} {\bibfnamefont {D.}~\bibnamefont {Arovas}}, \bibinfo {author} {\bibfnamefont {J.~R.}\ \bibnamefont {Schrieffer}},\ and\ \bibinfo {author} {\bibfnamefont {F.}~\bibnamefont {Wilczek}},\ }\bibfield  {title} {\bibinfo {title} {Fractional statistics and the quantum hall effect},\ }\href {https://doi.org/10.1103/PhysRevLett.53.722} {\bibfield  {journal} {\bibinfo  {journal} {Phys. Rev. Lett.}\ }\textbf {\bibinfo {volume} {53}},\ \bibinfo {pages} {722} (\bibinfo {year} {1984})}\BibitemShut {NoStop}%
\bibitem [{\citenamefont {Bartolomei}\ \emph {et~al.}(2020)\citenamefont {Bartolomei}, \citenamefont {Kumar}, \citenamefont {Bisognin}, \citenamefont {Marguerite}, \citenamefont {Berroir}, \citenamefont {Bocquillon}, \citenamefont {Pla{\c c}ais}, \citenamefont {Cavanna}, \citenamefont {Dong}, \citenamefont {Gennser}, \citenamefont {Jin},\ and\ \citenamefont {F{\`e}ve}}]{Bartolomei2020}%
  \BibitemOpen
  \bibfield  {author} {\bibinfo {author} {\bibfnamefont {H.}~\bibnamefont {Bartolomei}}, \bibinfo {author} {\bibfnamefont {M.}~\bibnamefont {Kumar}}, \bibinfo {author} {\bibfnamefont {R.}~\bibnamefont {Bisognin}}, \bibinfo {author} {\bibfnamefont {A.}~\bibnamefont {Marguerite}}, \bibinfo {author} {\bibfnamefont {J.-M.}\ \bibnamefont {Berroir}}, \bibinfo {author} {\bibfnamefont {E.}~\bibnamefont {Bocquillon}}, \bibinfo {author} {\bibfnamefont {B.}~\bibnamefont {Pla{\c c}ais}}, \bibinfo {author} {\bibfnamefont {A.}~\bibnamefont {Cavanna}}, \bibinfo {author} {\bibfnamefont {Q.}~\bibnamefont {Dong}}, \bibinfo {author} {\bibfnamefont {U.}~\bibnamefont {Gennser}}, \bibinfo {author} {\bibfnamefont {Y.}~\bibnamefont {Jin}},\ and\ \bibinfo {author} {\bibfnamefont {G.}~\bibnamefont {F{\`e}ve}},\ }\bibfield  {title} {\bibinfo {title} {Fractional statistics in anyon collisions},\ }\href {https://doi.org/10.1126/science.aaz5601} {\bibfield  {journal} {\bibinfo  {journal} {Science}\ }\textbf {\bibinfo {volume} {368}},\ \bibinfo {pages} {173} (\bibinfo {year} {2020})}\BibitemShut {NoStop}%
\bibitem [{\citenamefont {Nakamura}\ \emph {et~al.}(2020)\citenamefont {Nakamura}, \citenamefont {Liang}, \citenamefont {Gardner},\ and\ \citenamefont {Manfra}}]{Nakamura2020}%
  \BibitemOpen
  \bibfield  {author} {\bibinfo {author} {\bibfnamefont {J.}~\bibnamefont {Nakamura}}, \bibinfo {author} {\bibfnamefont {S.}~\bibnamefont {Liang}}, \bibinfo {author} {\bibfnamefont {G.~C.}\ \bibnamefont {Gardner}},\ and\ \bibinfo {author} {\bibfnamefont {M.~J.}\ \bibnamefont {Manfra}},\ }\bibfield  {title} {\bibinfo {title} {Direct observation of anyonic braiding statistics},\ }\href {https://doi.org/10.1038/s41567-020-1019-1} {\bibfield  {journal} {\bibinfo  {journal} {Nat. Phys.}\ }\textbf {\bibinfo {volume} {16}},\ \bibinfo {pages} {931} (\bibinfo {year} {2020})}\BibitemShut {NoStop}%
\bibitem [{\citenamefont {Halperin}(1984)}]{Halperin1984}%
  \BibitemOpen
  \bibfield  {author} {\bibinfo {author} {\bibfnamefont {B.~I.}\ \bibnamefont {Halperin}},\ }\bibfield  {title} {\bibinfo {title} {Statistics of quasiparticles and the hierarchy of fractional quantized hall states},\ }\href {https://doi.org/10.1103/PhysRevLett.52.1583} {\bibfield  {journal} {\bibinfo  {journal} {Phys. Rev. Lett.}\ }\textbf {\bibinfo {volume} {52}},\ \bibinfo {pages} {1583} (\bibinfo {year} {1984})}\BibitemShut {NoStop}%
\bibitem [{\citenamefont {Coldea}\ \emph {et~al.}(2001)\citenamefont {Coldea}, \citenamefont {Tennant}, \citenamefont {Tsvelik},\ and\ \citenamefont {Tylczynski}}]{Coldea2001}%
  \BibitemOpen
  \bibfield  {author} {\bibinfo {author} {\bibfnamefont {R.}~\bibnamefont {Coldea}}, \bibinfo {author} {\bibfnamefont {D.~A.}\ \bibnamefont {Tennant}}, \bibinfo {author} {\bibfnamefont {A.~M.}\ \bibnamefont {Tsvelik}},\ and\ \bibinfo {author} {\bibfnamefont {Z.}~\bibnamefont {Tylczynski}},\ }\bibfield  {title} {\bibinfo {title} {Experimental {{Realization}} of a {{2D Fractional Quantum Spin Liquid}}},\ }\href {https://doi.org/10/bnd7w7} {\bibfield  {journal} {\bibinfo  {journal} {Phys. Rev. Lett.}\ }\textbf {\bibinfo {volume} {86}},\ \bibinfo {pages} {1335} (\bibinfo {year} {2001})}\BibitemShut {NoStop}%
\bibitem [{\citenamefont {Kitaev}(2006)}]{Kitaev2006a}%
  \BibitemOpen
  \bibfield  {author} {\bibinfo {author} {\bibfnamefont {A.}~\bibnamefont {Kitaev}},\ }\bibfield  {title} {\bibinfo {title} {Anyons in an exactly solved model and beyond},\ }\href {https://doi.org/10.1016/j.aop.2005.10.005} {\bibfield  {journal} {\bibinfo  {journal} {Ann. Phys.}\ }\textbf {\bibinfo {volume} {321}},\ \bibinfo {pages} {2} (\bibinfo {year} {2006})}\BibitemShut {NoStop}%
\bibitem [{\citenamefont {Semeghini}\ \emph {et~al.}(2021)\citenamefont {Semeghini}, \citenamefont {Levine}, \citenamefont {Keesling}, \citenamefont {Ebadi}, \citenamefont {Wang}, \citenamefont {Bluvstein}, \citenamefont {Verresen}, \citenamefont {Pichler}, \citenamefont {Kalinowski}, \citenamefont {Samajdar}, \citenamefont {Omran}, \citenamefont {Sachdev}, \citenamefont {Vishwanath}, \citenamefont {Greiner}, \citenamefont {Vuleti{\'c}},\ and\ \citenamefont {Lukin}}]{Semeghini2021}%
  \BibitemOpen
  \bibfield  {author} {\bibinfo {author} {\bibfnamefont {G.}~\bibnamefont {Semeghini}}, \bibinfo {author} {\bibfnamefont {H.}~\bibnamefont {Levine}}, \bibinfo {author} {\bibfnamefont {A.}~\bibnamefont {Keesling}}, \bibinfo {author} {\bibfnamefont {S.}~\bibnamefont {Ebadi}}, \bibinfo {author} {\bibfnamefont {T.~T.}\ \bibnamefont {Wang}}, \bibinfo {author} {\bibfnamefont {D.}~\bibnamefont {Bluvstein}}, \bibinfo {author} {\bibfnamefont {R.}~\bibnamefont {Verresen}}, \bibinfo {author} {\bibfnamefont {H.}~\bibnamefont {Pichler}}, \bibinfo {author} {\bibfnamefont {M.}~\bibnamefont {Kalinowski}}, \bibinfo {author} {\bibfnamefont {R.}~\bibnamefont {Samajdar}}, \bibinfo {author} {\bibfnamefont {A.}~\bibnamefont {Omran}}, \bibinfo {author} {\bibfnamefont {S.}~\bibnamefont {Sachdev}}, \bibinfo {author} {\bibfnamefont {A.}~\bibnamefont {Vishwanath}}, \bibinfo {author} {\bibfnamefont {M.}~\bibnamefont {Greiner}}, \bibinfo {author} {\bibfnamefont {V.}~\bibnamefont {Vuleti{\'c}}},\ and\ \bibinfo {author} {\bibfnamefont {M.~D.}\ \bibnamefont {Lukin}},\ }\bibfield  {title} {\bibinfo {title} {Probing topological spin liquids on a programmable quantum simulator},\ }\href {https://doi.org/10.1126/science.abi8794} {\bibfield  {journal} {\bibinfo  {journal} {Science}\ }\textbf {\bibinfo {volume} {374}},\ \bibinfo {pages} {1242} (\bibinfo {year} {2021})}\BibitemShut {NoStop}%
\bibitem [{\citenamefont {Fradkin}(2013)}]{Fradkin2013}%
  \BibitemOpen
  \bibfield  {author} {\bibinfo {author} {\bibfnamefont {E.}~\bibnamefont {Fradkin}},\ }\href {https://doi.org/10.1017/CBO9781139015509} {\emph {\bibinfo {title} {Field {{Theories}} of {{Condensed Matter Physics}}}}},\ \bibinfo {edition} {2nd}\ ed.\ (\bibinfo  {publisher} {Cambridge University Press},\ \bibinfo {address} {Cambridge},\ \bibinfo {year} {2013})\BibitemShut {NoStop}%
\bibitem [{\citenamefont {Nayak}\ \emph {et~al.}(2008)\citenamefont {Nayak}, \citenamefont {Simon}, \citenamefont {Stern}, \citenamefont {Freedman},\ and\ \citenamefont {Das~Sarma}}]{Nayak2008}%
  \BibitemOpen
  \bibfield  {author} {\bibinfo {author} {\bibfnamefont {C.}~\bibnamefont {Nayak}}, \bibinfo {author} {\bibfnamefont {S.~H.}\ \bibnamefont {Simon}}, \bibinfo {author} {\bibfnamefont {A.}~\bibnamefont {Stern}}, \bibinfo {author} {\bibfnamefont {M.}~\bibnamefont {Freedman}},\ and\ \bibinfo {author} {\bibfnamefont {S.}~\bibnamefont {Das~Sarma}},\ }\bibfield  {title} {\bibinfo {title} {Non-{{Abelian}} anyons and topological quantum computation},\ }\href {https://doi.org/10.1103/RevModPhys.80.1083} {\bibfield  {journal} {\bibinfo  {journal} {Rev. Mod. Phys.}\ }\textbf {\bibinfo {volume} {80}},\ \bibinfo {pages} {1083} (\bibinfo {year} {2008})}\BibitemShut {NoStop}%
\bibitem [{\citenamefont {Kitaev}(2003)}]{Kitaev2003}%
  \BibitemOpen
  \bibfield  {author} {\bibinfo {author} {\bibfnamefont {{\relax A.Yu}.}~\bibnamefont {Kitaev}},\ }\bibfield  {title} {\bibinfo {title} {Fault-tolerant quantum computation by anyons},\ }\href {https://doi.org/10.1016/s0003-4916(02)00018-0} {\bibfield  {journal} {\bibinfo  {journal} {Ann. Phys.}\ }\textbf {\bibinfo {volume} {303}},\ \bibinfo {pages} {2} (\bibinfo {year} {2003})}\BibitemShut {NoStop}%
\bibitem [{\citenamefont {Bravyi}(2006)}]{Bravyi2006}%
  \BibitemOpen
  \bibfield  {author} {\bibinfo {author} {\bibfnamefont {S.}~\bibnamefont {Bravyi}},\ }\bibfield  {title} {\bibinfo {title} {Universal quantum computation with the \$\textbackslash ensuremath\textbraceleft\textbackslash nu\textbraceright =5/2\$ fractional quantum {{Hall}} state},\ }\href {https://doi.org/10/crp7zk} {\bibfield  {journal} {\bibinfo  {journal} {Phys. Rev. A}\ }\textbf {\bibinfo {volume} {73}},\ \bibinfo {pages} {042313} (\bibinfo {year} {2006})}\BibitemShut {NoStop}%
\bibitem [{\citenamefont {Clarke}\ \emph {et~al.}(2013)\citenamefont {Clarke}, \citenamefont {Alicea},\ and\ \citenamefont {Shtengel}}]{Clarke2013}%
  \BibitemOpen
  \bibfield  {author} {\bibinfo {author} {\bibfnamefont {D.~J.}\ \bibnamefont {Clarke}}, \bibinfo {author} {\bibfnamefont {J.}~\bibnamefont {Alicea}},\ and\ \bibinfo {author} {\bibfnamefont {K.}~\bibnamefont {Shtengel}},\ }\bibfield  {title} {\bibinfo {title} {Exotic non-{{Abelian}} anyons from conventional fractional quantum {{Hall}} states},\ }\href {https://doi.org/10/f4m8g7} {\bibfield  {journal} {\bibinfo  {journal} {Nat. Commun.}\ }\textbf {\bibinfo {volume} {4}},\ \bibinfo {pages} {1348} (\bibinfo {year} {2013})}\BibitemShut {NoStop}%
\bibitem [{\citenamefont {Andersen}\ \emph {et~al.}(2023)\citenamefont {Andersen}, \citenamefont {Lensky}, \citenamefont {Kechedzhi}, \citenamefont {Drozdov}, \citenamefont {Bengtsson}, \citenamefont {Hong}, \citenamefont {Morvan}, \citenamefont {Mi}, \citenamefont {Opremcak}, \citenamefont {Acharya}, \citenamefont {Allen}, \citenamefont {Ansmann}, \citenamefont {Arute}, \citenamefont {Arya}, \citenamefont {Asfaw}, \citenamefont {Atalaya}, \citenamefont {Babbush}, \citenamefont {Bacon}, \citenamefont {Bardin}, \citenamefont {Bortoli}, \citenamefont {Bourassa}, \citenamefont {Bovaird}, \citenamefont {Brill}, \citenamefont {Broughton}, \citenamefont {Buckley}, \citenamefont {Buell}, \citenamefont {Burger}, \citenamefont {Burkett}, \citenamefont {Bushnell}, \citenamefont {Chen}, \citenamefont {Chiaro}, \citenamefont {Chik}, \citenamefont {Chou}, \citenamefont {Cogan}, \citenamefont {Collins}, \citenamefont {Conner}, \citenamefont {Courtney}, \citenamefont {Crook}, \citenamefont {Curtin}, \citenamefont {Debroy}, \citenamefont {Del Toro~Barba}, \citenamefont {Demura}, \citenamefont {Dunsworth}, \citenamefont {Eppens}, \citenamefont {Erickson}, \citenamefont {Faoro}, \citenamefont {Farhi}, \citenamefont {Fatemi}, \citenamefont {Ferreira}, \citenamefont {Burgos}, \citenamefont {Forati}, \citenamefont {Fowler}, \citenamefont {Foxen}, \citenamefont {Giang}, \citenamefont {Gidney}, \citenamefont {Gilboa}, \citenamefont {Giustina}, \citenamefont {Gosula}, \citenamefont {Dau}, \citenamefont {Gross}, \citenamefont {Habegger}, \citenamefont {Hamilton}, \citenamefont {Hansen}, \citenamefont {Harrigan}, \citenamefont {Harrington}, \citenamefont {Heu}, \citenamefont {Hilton}, \citenamefont {Hoffmann}, \citenamefont {Huang}, \citenamefont {Huff}, \citenamefont {Huggins}, \citenamefont {Ioffe}, \citenamefont {Isakov}, \citenamefont {Iveland}, \citenamefont {Jeffrey}, \citenamefont {Jiang}, \citenamefont {Jones}, \citenamefont {Juhas}, \citenamefont {Kafri}, \citenamefont {Khattar}, \citenamefont {Khezri}, \citenamefont {Kieferov{\'a}}, \citenamefont {Kim}, \citenamefont {Kitaev}, \citenamefont {Klimov}, \citenamefont {Klots}, \citenamefont {Korotkov}, \citenamefont {Kostritsa}, \citenamefont {Kreikebaum}, \citenamefont {Landhuis}, \citenamefont {Laptev}, \citenamefont {Lau}, \citenamefont {Laws}, \citenamefont {Lee}, \citenamefont {Lee}, \citenamefont {Lester}, \citenamefont {Lill}, \citenamefont {Liu}, \citenamefont {Locharla}, \citenamefont {Lucero}, \citenamefont {Malone}, \citenamefont {Martin}, \citenamefont {McClean}, \citenamefont {McCourt}, \citenamefont {McEwen}, \citenamefont {Miao}, \citenamefont {Mieszala}, \citenamefont {Mohseni}, \citenamefont {Montazeri}, \citenamefont {Mount}, \citenamefont {Movassagh}, \citenamefont {Mruczkiewicz}, \citenamefont {Naaman}, \citenamefont {Neeley}, \citenamefont {Neill}, \citenamefont {Nersisyan}, \citenamefont {Newman}, \citenamefont {Ng}, \citenamefont {Nguyen}, \citenamefont {Nguyen}, \citenamefont {Niu}, \citenamefont {O'Brien}, \citenamefont {Omonije}, \citenamefont {Petukhov}, \citenamefont {Potter}, \citenamefont {Pryadko}, \citenamefont {Quintana}, \citenamefont {Rocque}, \citenamefont {Rubin}, \citenamefont {Saei}, \citenamefont {Sank}, \citenamefont {Sankaragomathi}, \citenamefont {Satzinger}, \citenamefont {Schurkus}, \citenamefont {Schuster}, \citenamefont {Shearn}, \citenamefont {Shorter}, \citenamefont {Shutty}, \citenamefont {Shvarts}, \citenamefont {Skruzny}, \citenamefont {Smith}, \citenamefont {Somma}, \citenamefont {Sterling}, \citenamefont {Strain}, \citenamefont {Szalay}, \citenamefont {Torres}, \citenamefont {Vidal}, \citenamefont {Villalonga}, \citenamefont {Heidweiller}, \citenamefont {White}, \citenamefont {Woo}, \citenamefont {Xing}, \citenamefont {Yao}, \citenamefont {Yeh}, \citenamefont {Yoo}, \citenamefont {Young}, \citenamefont {Zalcman}, \citenamefont {Zhang}, \citenamefont {Zhu}, \citenamefont {Zobrist}, \citenamefont {Neven}, \citenamefont {Boixo}, \citenamefont {Megrant}, \citenamefont {Kelly}, \citenamefont {Chen}, \citenamefont {Smelyanskiy}, \citenamefont {Kim}, \citenamefont {Aleiner}, \citenamefont {Roushan},\ and\ \citenamefont {{Google Quantum AI and Collaborators}}}]{Andersen2023}%
  \BibitemOpen
  \bibfield  {author} {\bibinfo {author} {\bibfnamefont {T.~I.}\ \bibnamefont {Andersen}}, \bibinfo {author} {\bibfnamefont {Y.~D.}\ \bibnamefont {Lensky}}, \bibinfo {author} {\bibfnamefont {K.}~\bibnamefont {Kechedzhi}}, \bibinfo {author} {\bibfnamefont {I.~K.}\ \bibnamefont {Drozdov}}, \bibinfo {author} {\bibfnamefont {A.}~\bibnamefont {Bengtsson}}, \bibinfo {author} {\bibfnamefont {S.}~\bibnamefont {Hong}}, \bibinfo {author} {\bibfnamefont {A.}~\bibnamefont {Morvan}}, \bibinfo {author} {\bibfnamefont {X.}~\bibnamefont {Mi}}, \bibinfo {author} {\bibfnamefont {A.}~\bibnamefont {Opremcak}}, \bibinfo {author} {\bibfnamefont {R.}~\bibnamefont {Acharya}}, \bibinfo {author} {\bibfnamefont {R.}~\bibnamefont {Allen}}, \bibinfo {author} {\bibfnamefont {M.}~\bibnamefont {Ansmann}}, \bibinfo {author} {\bibfnamefont {F.}~\bibnamefont {Arute}}, \bibinfo {author} {\bibfnamefont {K.}~\bibnamefont {Arya}}, \bibinfo {author} {\bibfnamefont {A.}~\bibnamefont {Asfaw}}, \bibinfo {author} {\bibfnamefont {J.}~\bibnamefont {Atalaya}}, \bibinfo {author} {\bibfnamefont {R.}~\bibnamefont {Babbush}}, \bibinfo {author} {\bibfnamefont {D.}~\bibnamefont {Bacon}}, \bibinfo {author} {\bibfnamefont {J.~C.}\ \bibnamefont {Bardin}}, \bibinfo {author} {\bibfnamefont {G.}~\bibnamefont {Bortoli}}, \bibinfo {author} {\bibfnamefont {A.}~\bibnamefont {Bourassa}}, \bibinfo {author} {\bibfnamefont {J.}~\bibnamefont {Bovaird}}, \bibinfo {author} {\bibfnamefont {L.}~\bibnamefont {Brill}}, \bibinfo {author} {\bibfnamefont {M.}~\bibnamefont {Broughton}}, \bibinfo {author} {\bibfnamefont {B.~B.}\ \bibnamefont {Buckley}}, \bibinfo {author} {\bibfnamefont {D.~A.}\ \bibnamefont {Buell}}, \bibinfo {author} {\bibfnamefont {T.}~\bibnamefont {Burger}}, \bibinfo {author} {\bibfnamefont {B.}~\bibnamefont {Burkett}}, \bibinfo {author} {\bibfnamefont {N.}~\bibnamefont {Bushnell}}, \bibinfo {author} {\bibfnamefont {Z.}~\bibnamefont {Chen}}, \bibinfo {author} {\bibfnamefont {B.}~\bibnamefont {Chiaro}}, \bibinfo {author} {\bibfnamefont {D.}~\bibnamefont {Chik}}, \bibinfo {author} {\bibfnamefont {C.}~\bibnamefont {Chou}}, \bibinfo {author} {\bibfnamefont {J.}~\bibnamefont {Cogan}}, \bibinfo {author} {\bibfnamefont {R.}~\bibnamefont {Collins}}, \bibinfo {author} {\bibfnamefont {P.}~\bibnamefont {Conner}}, \bibinfo {author} {\bibfnamefont {W.}~\bibnamefont {Courtney}}, \bibinfo {author} {\bibfnamefont {A.~L.}\ \bibnamefont {Crook}}, \bibinfo {author} {\bibfnamefont {B.}~\bibnamefont {Curtin}}, \bibinfo {author} {\bibfnamefont {D.~M.}\ \bibnamefont {Debroy}}, \bibinfo {author} {\bibfnamefont {A.}~\bibnamefont {Del Toro~Barba}}, \bibinfo {author} {\bibfnamefont {S.}~\bibnamefont {Demura}}, \bibinfo {author} {\bibfnamefont {A.}~\bibnamefont {Dunsworth}}, \bibinfo {author} {\bibfnamefont {D.}~\bibnamefont {Eppens}}, \bibinfo {author} {\bibfnamefont {C.}~\bibnamefont {Erickson}}, \bibinfo {author} {\bibfnamefont {L.}~\bibnamefont {Faoro}}, \bibinfo {author} {\bibfnamefont {E.}~\bibnamefont {Farhi}}, \bibinfo {author} {\bibfnamefont {R.}~\bibnamefont {Fatemi}}, \bibinfo {author} {\bibfnamefont {V.~S.}\ \bibnamefont {Ferreira}}, \bibinfo {author} {\bibfnamefont {L.~F.}\ \bibnamefont {Burgos}}, \bibinfo {author} {\bibfnamefont {E.}~\bibnamefont {Forati}}, \bibinfo {author} {\bibfnamefont {A.~G.}\ \bibnamefont {Fowler}}, \bibinfo {author} {\bibfnamefont {B.}~\bibnamefont {Foxen}}, \bibinfo {author} {\bibfnamefont {W.}~\bibnamefont {Giang}}, \bibinfo {author} {\bibfnamefont {C.}~\bibnamefont {Gidney}}, \bibinfo {author} {\bibfnamefont {D.}~\bibnamefont {Gilboa}}, \bibinfo {author} {\bibfnamefont {M.}~\bibnamefont {Giustina}}, \bibinfo {author} {\bibfnamefont {R.}~\bibnamefont {Gosula}}, \bibinfo {author} {\bibfnamefont {A.~G.}\ \bibnamefont {Dau}}, \bibinfo {author} {\bibfnamefont {J.~A.}\ \bibnamefont {Gross}}, \bibinfo {author} {\bibfnamefont {S.}~\bibnamefont {Habegger}}, \bibinfo {author} {\bibfnamefont {M.~C.}\ \bibnamefont {Hamilton}}, \bibinfo {author} {\bibfnamefont {M.}~\bibnamefont {Hansen}}, \bibinfo {author} {\bibfnamefont {M.~P.}\ \bibnamefont {Harrigan}}, \bibinfo {author} {\bibfnamefont {S.~D.}\ \bibnamefont {Harrington}}, \bibinfo {author} {\bibfnamefont {P.}~\bibnamefont {Heu}}, \bibinfo {author} {\bibfnamefont {J.}~\bibnamefont {Hilton}}, \bibinfo {author} {\bibfnamefont {M.~R.}\ \bibnamefont {Hoffmann}}, \bibinfo {author} {\bibfnamefont {T.}~\bibnamefont {Huang}}, \bibinfo {author} {\bibfnamefont {A.}~\bibnamefont {Huff}}, \bibinfo {author} {\bibfnamefont {W.~J.}\ \bibnamefont {Huggins}}, \bibinfo {author} {\bibfnamefont {L.~B.}\ \bibnamefont {Ioffe}}, \bibinfo {author} {\bibfnamefont {S.~V.}\ \bibnamefont {Isakov}}, \bibinfo {author} {\bibfnamefont {J.}~\bibnamefont {Iveland}}, \bibinfo {author} {\bibfnamefont {E.}~\bibnamefont {Jeffrey}}, \bibinfo {author} {\bibfnamefont {Z.}~\bibnamefont {Jiang}}, \bibinfo {author} {\bibfnamefont {C.}~\bibnamefont {Jones}}, \bibinfo {author} {\bibfnamefont {P.}~\bibnamefont {Juhas}}, \bibinfo {author} {\bibfnamefont {D.}~\bibnamefont {Kafri}}, \bibinfo {author} {\bibfnamefont {T.}~\bibnamefont {Khattar}}, \bibinfo {author} {\bibfnamefont {M.}~\bibnamefont {Khezri}}, \bibinfo {author} {\bibfnamefont {M.}~\bibnamefont {Kieferov{\'a}}}, \bibinfo {author} {\bibfnamefont {S.}~\bibnamefont {Kim}}, \bibinfo {author} {\bibfnamefont {A.}~\bibnamefont {Kitaev}}, \bibinfo {author} {\bibfnamefont {P.~V.}\ \bibnamefont {Klimov}}, \bibinfo {author} {\bibfnamefont {A.~R.}\ \bibnamefont {Klots}}, \bibinfo {author} {\bibfnamefont {A.~N.}\ \bibnamefont {Korotkov}}, \bibinfo {author} {\bibfnamefont {F.}~\bibnamefont {Kostritsa}}, \bibinfo {author} {\bibfnamefont {J.~M.}\ \bibnamefont {Kreikebaum}}, \bibinfo {author} {\bibfnamefont {D.}~\bibnamefont {Landhuis}}, \bibinfo {author} {\bibfnamefont {P.}~\bibnamefont {Laptev}}, \bibinfo {author} {\bibfnamefont {K.-M.}\ \bibnamefont {Lau}}, \bibinfo {author} {\bibfnamefont {L.}~\bibnamefont {Laws}}, \bibinfo {author} {\bibfnamefont {J.}~\bibnamefont {Lee}}, \bibinfo {author} {\bibfnamefont {K.~W.}\ \bibnamefont {Lee}}, \bibinfo {author} {\bibfnamefont {B.~J.}\ \bibnamefont {Lester}}, \bibinfo {author} {\bibfnamefont {A.~T.}\ \bibnamefont {Lill}}, \bibinfo {author} {\bibfnamefont {W.}~\bibnamefont {Liu}}, \bibinfo {author} {\bibfnamefont {A.}~\bibnamefont {Locharla}}, \bibinfo {author} {\bibfnamefont {E.}~\bibnamefont {Lucero}}, \bibinfo {author} {\bibfnamefont {F.~D.}\ \bibnamefont {Malone}}, \bibinfo {author} {\bibfnamefont {O.}~\bibnamefont {Martin}}, \bibinfo {author} {\bibfnamefont {J.~R.}\ \bibnamefont {McClean}}, \bibinfo {author} {\bibfnamefont {T.}~\bibnamefont {McCourt}}, \bibinfo {author} {\bibfnamefont {M.}~\bibnamefont {McEwen}}, \bibinfo {author} {\bibfnamefont {K.~C.}\ \bibnamefont {Miao}}, \bibinfo {author} {\bibfnamefont {A.}~\bibnamefont {Mieszala}}, \bibinfo {author} {\bibfnamefont {M.}~\bibnamefont {Mohseni}}, \bibinfo {author} {\bibfnamefont {S.}~\bibnamefont {Montazeri}}, \bibinfo {author} {\bibfnamefont {E.}~\bibnamefont {Mount}}, \bibinfo {author} {\bibfnamefont {R.}~\bibnamefont {Movassagh}}, \bibinfo {author} {\bibfnamefont {W.}~\bibnamefont {Mruczkiewicz}}, \bibinfo {author} {\bibfnamefont {O.}~\bibnamefont {Naaman}}, \bibinfo {author} {\bibfnamefont {M.}~\bibnamefont {Neeley}}, \bibinfo {author} {\bibfnamefont {C.}~\bibnamefont {Neill}}, \bibinfo {author} {\bibfnamefont {A.}~\bibnamefont {Nersisyan}}, \bibinfo {author} {\bibfnamefont {M.}~\bibnamefont {Newman}}, \bibinfo {author} {\bibfnamefont {J.~H.}\ \bibnamefont {Ng}}, \bibinfo {author} {\bibfnamefont {A.}~\bibnamefont {Nguyen}}, \bibinfo {author} {\bibfnamefont {M.}~\bibnamefont {Nguyen}}, \bibinfo {author} {\bibfnamefont {M.~Y.}\ \bibnamefont {Niu}}, \bibinfo {author} {\bibfnamefont {T.~E.}\ \bibnamefont {O'Brien}}, \bibinfo {author} {\bibfnamefont {S.}~\bibnamefont {Omonije}}, \bibinfo {author} {\bibfnamefont {A.}~\bibnamefont {Petukhov}}, \bibinfo {author} {\bibfnamefont {R.}~\bibnamefont {Potter}}, \bibinfo {author} {\bibfnamefont {L.~P.}\ \bibnamefont {Pryadko}}, \bibinfo {author} {\bibfnamefont {C.}~\bibnamefont {Quintana}}, \bibinfo {author} {\bibfnamefont {C.}~\bibnamefont {Rocque}}, \bibinfo {author} {\bibfnamefont {N.~C.}\ \bibnamefont {Rubin}}, \bibinfo {author} {\bibfnamefont {N.}~\bibnamefont {Saei}}, \bibinfo {author} {\bibfnamefont {D.}~\bibnamefont {Sank}}, \bibinfo {author} {\bibfnamefont {K.}~\bibnamefont {Sankaragomathi}}, \bibinfo {author} {\bibfnamefont {K.~J.}\ \bibnamefont {Satzinger}}, \bibinfo {author} {\bibfnamefont {H.~F.}\ \bibnamefont {Schurkus}}, \bibinfo {author} {\bibfnamefont {C.}~\bibnamefont {Schuster}}, \bibinfo {author} {\bibfnamefont {M.~J.}\ \bibnamefont {Shearn}}, \bibinfo {author} {\bibfnamefont {A.}~\bibnamefont {Shorter}}, \bibinfo {author} {\bibfnamefont {N.}~\bibnamefont {Shutty}}, \bibinfo {author} {\bibfnamefont {V.}~\bibnamefont {Shvarts}}, \bibinfo {author} {\bibfnamefont {J.}~\bibnamefont {Skruzny}}, \bibinfo {author} {\bibfnamefont {W.~C.}\ \bibnamefont {Smith}}, \bibinfo {author} {\bibfnamefont {R.}~\bibnamefont {Somma}}, \bibinfo {author} {\bibfnamefont {G.}~\bibnamefont {Sterling}}, \bibinfo {author} {\bibfnamefont {D.}~\bibnamefont {Strain}}, \bibinfo {author} {\bibfnamefont {M.}~\bibnamefont {Szalay}}, \bibinfo {author} {\bibfnamefont {A.}~\bibnamefont {Torres}}, \bibinfo {author} {\bibfnamefont {G.}~\bibnamefont {Vidal}}, \bibinfo {author} {\bibfnamefont {B.}~\bibnamefont {Villalonga}}, \bibinfo {author} {\bibfnamefont {C.~V.}\ \bibnamefont {Heidweiller}}, \bibinfo {author} {\bibfnamefont {T.}~\bibnamefont {White}}, \bibinfo {author} {\bibfnamefont {B.~W.~K.}\ \bibnamefont {Woo}}, \bibinfo {author} {\bibfnamefont {C.}~\bibnamefont {Xing}}, \bibinfo {author} {\bibfnamefont {Z.~J.}\ \bibnamefont {Yao}}, \bibinfo {author} {\bibfnamefont {P.}~\bibnamefont {Yeh}}, \bibinfo {author} {\bibfnamefont {J.}~\bibnamefont {Yoo}}, \bibinfo {author} {\bibfnamefont {G.}~\bibnamefont {Young}}, \bibinfo {author} {\bibfnamefont {A.}~\bibnamefont {Zalcman}}, \bibinfo {author}
  {\bibfnamefont {Y.}~\bibnamefont {Zhang}}, \bibinfo {author} {\bibfnamefont {N.}~\bibnamefont {Zhu}}, \bibinfo {author} {\bibfnamefont {N.}~\bibnamefont {Zobrist}}, \bibinfo {author} {\bibfnamefont {H.}~\bibnamefont {Neven}}, \bibinfo {author} {\bibfnamefont {S.}~\bibnamefont {Boixo}}, \bibinfo {author} {\bibfnamefont {A.}~\bibnamefont {Megrant}}, \bibinfo {author} {\bibfnamefont {J.}~\bibnamefont {Kelly}}, \bibinfo {author} {\bibfnamefont {Y.}~\bibnamefont {Chen}}, \bibinfo {author} {\bibfnamefont {V.}~\bibnamefont {Smelyanskiy}}, \bibinfo {author} {\bibfnamefont {E.-A.}\ \bibnamefont {Kim}}, \bibinfo {author} {\bibfnamefont {I.}~\bibnamefont {Aleiner}}, \bibinfo {author} {\bibfnamefont {P.}~\bibnamefont {Roushan}},\ and\ \bibinfo {author} {\bibnamefont {{Google Quantum AI and Collaborators}}},\ }\bibfield  {title} {\bibinfo {title} {Non-{{Abelian}} braiding of graph vertices in a superconducting processor},\ }\href {https://doi.org/10/gr749s} {\bibfield  {journal} {\bibinfo  {journal} {Nature}\ }\textbf {\bibinfo {volume} {618}},\ \bibinfo {pages} {264} (\bibinfo {year} {2023})}\BibitemShut {NoStop}%
\bibitem [{\citenamefont {Harshman}\ and\ \citenamefont {Knapp}(2020)}]{Harshman2020}%
  \BibitemOpen
  \bibfield  {author} {\bibinfo {author} {\bibfnamefont {N.}~\bibnamefont {Harshman}}\ and\ \bibinfo {author} {\bibfnamefont {A.}~\bibnamefont {Knapp}},\ }\bibfield  {title} {\bibinfo {title} {Anyons from three-body hard-core interactions in one dimension},\ }\href {https://doi.org/10.1016/j.aop.2019.168003} {\bibfield  {journal} {\bibinfo  {journal} {Ann. Phys.}\ }\textbf {\bibinfo {volume} {412}},\ \bibinfo {pages} {168003} (\bibinfo {year} {2020})}\BibitemShut {NoStop}%
\bibitem [{\citenamefont {Bonkhoff}\ \emph {et~al.}(2021)\citenamefont {Bonkhoff}, \citenamefont {J{\"a}gering}, \citenamefont {Eggert}, \citenamefont {Pelster}, \citenamefont {Thorwart},\ and\ \citenamefont {Posske}}]{Bonkhoff2021}%
  \BibitemOpen
  \bibfield  {author} {\bibinfo {author} {\bibfnamefont {M.}~\bibnamefont {Bonkhoff}}, \bibinfo {author} {\bibfnamefont {K.}~\bibnamefont {J{\"a}gering}}, \bibinfo {author} {\bibfnamefont {S.}~\bibnamefont {Eggert}}, \bibinfo {author} {\bibfnamefont {A.}~\bibnamefont {Pelster}}, \bibinfo {author} {\bibfnamefont {M.}~\bibnamefont {Thorwart}},\ and\ \bibinfo {author} {\bibfnamefont {T.}~\bibnamefont {Posske}},\ }\bibfield  {title} {\bibinfo {title} {Bosonic continuum theory of one-dimensional lattice anyons},\ }\href {https://doi.org/10.1103/PhysRevLett.126.163201} {\bibfield  {journal} {\bibinfo  {journal} {Phys. Rev. Lett.}\ }\textbf {\bibinfo {volume} {126}},\ \bibinfo {pages} {163201} (\bibinfo {year} {2021})}\BibitemShut {NoStop}%
\bibitem [{\citenamefont {Kundu}(1999)}]{Kundu1999}%
  \BibitemOpen
  \bibfield  {author} {\bibinfo {author} {\bibfnamefont {A.}~\bibnamefont {Kundu}},\ }\bibfield  {title} {\bibinfo {title} {Exact solution of double {{$\delta$}} function bose gas through an interacting anyon gas},\ }\href {https://doi.org/10.1103/PhysRevLett.83.1275} {\bibfield  {journal} {\bibinfo  {journal} {Phys. Rev. Lett.}\ }\textbf {\bibinfo {volume} {83}},\ \bibinfo {pages} {1275} (\bibinfo {year} {1999})}\BibitemShut {NoStop}%
\bibitem [{\citenamefont {Rabello}(1995)}]{Rabello1995}%
  \BibitemOpen
  \bibfield  {author} {\bibinfo {author} {\bibfnamefont {S.~J.}\ \bibnamefont {Rabello}},\ }\bibfield  {title} {\bibinfo {title} {A gauge theory of one-dimensional anyons},\ }\href {https://doi.org/10/dvtt6b} {\bibfield  {journal} {\bibinfo  {journal} {Phys. Lett. B}\ }\textbf {\bibinfo {volume} {363}},\ \bibinfo {pages} {180} (\bibinfo {year} {1995})}\BibitemShut {NoStop}%
\bibitem [{\citenamefont {Zhu}\ and\ \citenamefont {Wang}(1996)}]{Zhu1996}%
  \BibitemOpen
  \bibfield  {author} {\bibinfo {author} {\bibfnamefont {J.-X.}\ \bibnamefont {Zhu}}\ and\ \bibinfo {author} {\bibfnamefont {Z.~D.}\ \bibnamefont {Wang}},\ }\bibfield  {title} {\bibinfo {title} {Topological effects associated with fractional statistics in one-dimensional mesoscopic rings},\ }\href {https://doi.org/10.1103/PhysRevA.53.600} {\bibfield  {journal} {\bibinfo  {journal} {Phys. Rev. At. Mol. Opt. Phys.}\ }\textbf {\bibinfo {volume} {53}},\ \bibinfo {pages} {600} (\bibinfo {year} {1996})}\BibitemShut {NoStop}%
\bibitem [{\citenamefont {Keilmann}\ \emph {et~al.}(2011)\citenamefont {Keilmann}, \citenamefont {Lanzmich}, \citenamefont {McCulloch},\ and\ \citenamefont {Roncaglia}}]{Keilmann2011}%
  \BibitemOpen
  \bibfield  {author} {\bibinfo {author} {\bibfnamefont {T.}~\bibnamefont {Keilmann}}, \bibinfo {author} {\bibfnamefont {S.}~\bibnamefont {Lanzmich}}, \bibinfo {author} {\bibfnamefont {I.}~\bibnamefont {McCulloch}},\ and\ \bibinfo {author} {\bibfnamefont {M.}~\bibnamefont {Roncaglia}},\ }\bibfield  {title} {\bibinfo {title} {Statistically induced phase transitions and anyons in {{1D}} optical lattices},\ }\href {https://doi.org/10.1038/ncomms1353} {\bibfield  {journal} {\bibinfo  {journal} {Nat. Commun.}\ }\textbf {\bibinfo {volume} {2}},\ \bibinfo {pages} {361} (\bibinfo {year} {2011})}\BibitemShut {NoStop}%
\bibitem [{\citenamefont {Greschner}\ and\ \citenamefont {Santos}(2015)}]{Greschner2015a}%
  \BibitemOpen
  \bibfield  {author} {\bibinfo {author} {\bibfnamefont {S.}~\bibnamefont {Greschner}}\ and\ \bibinfo {author} {\bibfnamefont {L.}~\bibnamefont {Santos}},\ }\bibfield  {title} {\bibinfo {title} {Anyon hubbard model in one-dimensional optical lattices},\ }\href {https://doi.org/10.1103/PhysRevLett.115.053002} {\bibfield  {journal} {\bibinfo  {journal} {Phys. Rev. Lett.}\ }\textbf {\bibinfo {volume} {115}},\ \bibinfo {pages} {053002} (\bibinfo {year} {2015})}\BibitemShut {NoStop}%
\bibitem [{\citenamefont {Nagies}\ \emph {et~al.}(2024)\citenamefont {Nagies}, \citenamefont {Wang}, \citenamefont {Knapp}, \citenamefont {Eckardt},\ and\ \citenamefont {Harshman}}]{Nagies2024}%
  \BibitemOpen
  \bibfield  {author} {\bibinfo {author} {\bibfnamefont {S.}~\bibnamefont {Nagies}}, \bibinfo {author} {\bibfnamefont {B.}~\bibnamefont {Wang}}, \bibinfo {author} {\bibfnamefont {A.~C.}\ \bibnamefont {Knapp}}, \bibinfo {author} {\bibfnamefont {A.}~\bibnamefont {Eckardt}},\ and\ \bibinfo {author} {\bibfnamefont {N.~L.}\ \bibnamefont {Harshman}},\ }\bibfield  {title} {\bibinfo {title} {Beyond braid statistics: {{Constructing}} a lattice model for anyons with exchange statistics intrinsic to one dimension},\ }\href {https://doi.org/10.21468/SciPostPhys.16.3.086} {\bibfield  {journal} {\bibinfo  {journal} {SciPost Phys.}\ }\textbf {\bibinfo {volume} {16}},\ \bibinfo {pages} {086} (\bibinfo {year} {2024})}\BibitemShut {NoStop}%
\bibitem [{\citenamefont {Haldane}(1991)}]{Haldane1991}%
  \BibitemOpen
  \bibfield  {author} {\bibinfo {author} {\bibfnamefont {F.~D.~M.}\ \bibnamefont {Haldane}},\ }\bibfield  {title} {\bibinfo {title} {``{{Fractional}} statistics'' in arbitrary dimensions: {{A}} generalization of the {{Pauli}} principle},\ }\href {https://doi.org/10.1103/PhysRevLett.67.937} {\bibfield  {journal} {\bibinfo  {journal} {Phys. Rev. Lett.}\ }\textbf {\bibinfo {volume} {67}},\ \bibinfo {pages} {937} (\bibinfo {year} {1991})}\BibitemShut {NoStop}%
\bibitem [{\citenamefont {Shastry}(1988)}]{Shastry1988}%
  \BibitemOpen
  \bibfield  {author} {\bibinfo {author} {\bibfnamefont {B.~S.}\ \bibnamefont {Shastry}},\ }\bibfield  {title} {\bibinfo {title} {Exact solution of an {{S}}=1/2 {{Heisenberg}} antiferromagnetic chain with long-ranged interactions},\ }\href {https://doi.org/10/fwxtw7} {\bibfield  {journal} {\bibinfo  {journal} {Phys. Rev. Lett.}\ }\textbf {\bibinfo {volume} {60}},\ \bibinfo {pages} {639} (\bibinfo {year} {1988})}\BibitemShut {NoStop}%
\bibitem [{\citenamefont {Haldane}\ and\ \citenamefont {Rezayi}(1988)}]{Haldane1988}%
  \BibitemOpen
  \bibfield  {author} {\bibinfo {author} {\bibfnamefont {F.~D.~M.}\ \bibnamefont {Haldane}}\ and\ \bibinfo {author} {\bibfnamefont {E.~H.}\ \bibnamefont {Rezayi}},\ }\bibfield  {title} {\bibinfo {title} {Spin-singlet wave function for the half-integral quantum {{Hall}} effect},\ }\href {https://doi.org/10/dd2pzb} {\bibfield  {journal} {\bibinfo  {journal} {Phys. Rev. Lett.}\ }\textbf {\bibinfo {volume} {60}},\ \bibinfo {pages} {956} (\bibinfo {year} {1988})}\BibitemShut {NoStop}%
\bibitem [{\citenamefont {Arikawa}\ \emph {et~al.}(2001)\citenamefont {Arikawa}, \citenamefont {Saiga},\ and\ \citenamefont {Kuramoto}}]{Arikawa2001}%
  \BibitemOpen
  \bibfield  {author} {\bibinfo {author} {\bibfnamefont {M.}~\bibnamefont {Arikawa}}, \bibinfo {author} {\bibfnamefont {Y.}~\bibnamefont {Saiga}},\ and\ \bibinfo {author} {\bibfnamefont {Y.}~\bibnamefont {Kuramoto}},\ }\bibfield  {title} {\bibinfo {title} {Electron {{Addition Spectrum}} in the {{Supersymmetric}} \$\textbackslash mathit\textbraceleft t\textbraceright\textbackslash ensuremath\textbraceleft -\textbraceright\textbackslash mathit\textbraceleft{{J}}\textbraceright\$ {{Model}} with {{Inverse-Square Interaction}}},\ }\href {https://doi.org/10.1103/PhysRevLett.86.3096} {\bibfield  {journal} {\bibinfo  {journal} {Phys. Rev. Lett.}\ }\textbf {\bibinfo {volume} {86}},\ \bibinfo {pages} {3096} (\bibinfo {year} {2001})}\BibitemShut {NoStop}%
\bibitem [{\citenamefont {Kato}(1998)}]{Kato1998}%
  \BibitemOpen
  \bibfield  {author} {\bibinfo {author} {\bibfnamefont {Y.}~\bibnamefont {Kato}},\ }\bibfield  {title} {\bibinfo {title} {Hole {{Dynamics}} of the {{One-Dimensional Supersymmetric}} \$\textbackslash mathit\textbraceleft t\textbraceright\textbackslash ensuremath\textbraceleft -\textbraceright\textbackslash mathit\textbraceleft{{J}}\textbraceright\$ {{Model}} with a {{Long-Range Interaction}}: {{An Exact Result}}},\ }\href {https://doi.org/10/btgd2k} {\bibfield  {journal} {\bibinfo  {journal} {Phys. Rev. Lett.}\ }\textbf {\bibinfo {volume} {81}},\ \bibinfo {pages} {5402} (\bibinfo {year} {1998})}\BibitemShut {NoStop}%
\bibitem [{\citenamefont {Kuramoto}\ and\ \citenamefont {Kato}(1995)}]{Kuramoto1995}%
  \BibitemOpen
  \bibfield  {author} {\bibinfo {author} {\bibfnamefont {Y.}~\bibnamefont {Kuramoto}}\ and\ \bibinfo {author} {\bibfnamefont {Y.}~\bibnamefont {Kato}},\ }\bibfield  {title} {\bibinfo {title} {Spin-{{Charge Separation}} at {{Finite Temperature}} in the {{Supersymmetric}} t-{{J Model}} with {{Long-Range Interactions}}},\ }\href {https://doi.org/10/cdrcgw} {\bibfield  {journal} {\bibinfo  {journal} {J. Phys. Soc. Jpn.}\ }\textbf {\bibinfo {volume} {64}},\ \bibinfo {pages} {4518} (\bibinfo {year} {1995})}\BibitemShut {NoStop}%
\bibitem [{\citenamefont {Kuramoto}\ and\ \citenamefont {Yokoyama}(1991)}]{Kuramoto1991}%
  \BibitemOpen
  \bibfield  {author} {\bibinfo {author} {\bibfnamefont {Y.}~\bibnamefont {Kuramoto}}\ and\ \bibinfo {author} {\bibfnamefont {H.}~\bibnamefont {Yokoyama}},\ }\bibfield  {title} {\bibinfo {title} {Exactly soluble supersymmetric t-{{J-type}} model with long-range exchange and transfer},\ }\href {https://doi.org/10/dhmbw7} {\bibfield  {journal} {\bibinfo  {journal} {Phys. Rev. Lett.}\ }\textbf {\bibinfo {volume} {67}},\ \bibinfo {pages} {1338} (\bibinfo {year} {1991})}\BibitemShut {NoStop}%
\bibitem [{\citenamefont {Ha}(1994)}]{Ha1994}%
  \BibitemOpen
  \bibfield  {author} {\bibinfo {author} {\bibfnamefont {Z.~N.~C.}\ \bibnamefont {Ha}},\ }\bibfield  {title} {\bibinfo {title} {Exact {{Dynamical Correlation Functions}} of {{Calogero-Sutherland Model}} and {{One-Dimensional Fractional Statistics}}},\ }\href {https://doi.org/10/c2p8r9} {\bibfield  {journal} {\bibinfo  {journal} {Phys. Rev. Lett.}\ }\textbf {\bibinfo {volume} {73}},\ \bibinfo {pages} {1574} (\bibinfo {year} {1994})}\BibitemShut {NoStop}%
\bibitem [{\citenamefont {Ha}(1995)}]{Ha1995}%
  \BibitemOpen
  \bibfield  {author} {\bibinfo {author} {\bibfnamefont {Z.}~\bibnamefont {Ha}},\ }\bibfield  {title} {\bibinfo {title} {Fractional statistics in one dimension: View from an exactly solvable model},\ }\href {https://doi.org/10.1016/0550-3213(94)00537-o} {\bibfield  {journal} {\bibinfo  {journal} {Nucl. Phys. B}\ }\textbf {\bibinfo {volume} {435}},\ \bibinfo {pages} {604} (\bibinfo {year} {1995})}\BibitemShut {NoStop}%
\bibitem [{\citenamefont {Murthy}\ and\ \citenamefont {Shankar}(1994)}]{Murthy1994}%
  \BibitemOpen
  \bibfield  {author} {\bibinfo {author} {\bibfnamefont {M.~V.~N.}\ \bibnamefont {Murthy}}\ and\ \bibinfo {author} {\bibfnamefont {R.}~\bibnamefont {Shankar}},\ }\bibfield  {title} {\bibinfo {title} {Thermodynamics of a {{One-Dimensional Ideal Gas}} with {{Fractional Exclusion Statistics}}},\ }\href {https://doi.org/10/d9spms} {\bibfield  {journal} {\bibinfo  {journal} {Phys. Rev. Lett.}\ }\textbf {\bibinfo {volume} {73}},\ \bibinfo {pages} {3331} (\bibinfo {year} {1994})}\BibitemShut {NoStop}%
\bibitem [{\citenamefont {Batchelor}\ \emph {et~al.}(2006)\citenamefont {Batchelor}, \citenamefont {Guan},\ and\ \citenamefont {Oelkers}}]{Batchelor2006}%
  \BibitemOpen
  \bibfield  {author} {\bibinfo {author} {\bibfnamefont {M.~T.}\ \bibnamefont {Batchelor}}, \bibinfo {author} {\bibfnamefont {X.-W.}\ \bibnamefont {Guan}},\ and\ \bibinfo {author} {\bibfnamefont {N.}~\bibnamefont {Oelkers}},\ }\bibfield  {title} {\bibinfo {title} {One-dimensional interacting anyon gas: {{Low-energy}} properties and haldane exclusion statistics},\ }\href {https://doi.org/10.1103/PhysRevLett.96.210402} {\bibfield  {journal} {\bibinfo  {journal} {Phys. Rev. Lett.}\ }\textbf {\bibinfo {volume} {96}},\ \bibinfo {pages} {210402} (\bibinfo {year} {2006})}\BibitemShut {NoStop}%
\bibitem [{\citenamefont {P{\^a}{\c t}u}\ \emph {et~al.}(2007)\citenamefont {P{\^a}{\c t}u}, \citenamefont {Korepin},\ and\ \citenamefont {Averin}}]{Patu2007}%
  \BibitemOpen
  \bibfield  {author} {\bibinfo {author} {\bibfnamefont {O.~I.}\ \bibnamefont {P{\^a}{\c t}u}}, \bibinfo {author} {\bibfnamefont {V.~E.}\ \bibnamefont {Korepin}},\ and\ \bibinfo {author} {\bibfnamefont {D.~V.}\ \bibnamefont {Averin}},\ }\bibfield  {title} {\bibinfo {title} {Correlation functions of one-dimensional {{Lieb}}--{{Liniger}} anyons},\ }\href {https://doi.org/10.1088/1751-8113/40/50/004} {\bibfield  {journal} {\bibinfo  {journal} {J. Phys. Math. Theor.}\ }\textbf {\bibinfo {volume} {40}},\ \bibinfo {pages} {14963} (\bibinfo {year} {2007})}\BibitemShut {NoStop}%
\bibitem [{\citenamefont {Posske}\ \emph {et~al.}(2017)\citenamefont {Posske}, \citenamefont {Trauzettel},\ and\ \citenamefont {Thorwart}}]{Posske2017}%
  \BibitemOpen
  \bibfield  {author} {\bibinfo {author} {\bibfnamefont {T.}~\bibnamefont {Posske}}, \bibinfo {author} {\bibfnamefont {B.}~\bibnamefont {Trauzettel}},\ and\ \bibinfo {author} {\bibfnamefont {M.}~\bibnamefont {Thorwart}},\ }\bibfield  {title} {\bibinfo {title} {Second quantization of {{Leinaas-Myrheim}} anyons in one dimension and their relation to the {{Lieb-Liniger}} model},\ }\href {https://doi.org/10.1103/PhysRevB.96.195422} {\bibfield  {journal} {\bibinfo  {journal} {Phys. Rev. B}\ }\textbf {\bibinfo {volume} {96}},\ \bibinfo {pages} {195422} (\bibinfo {year} {2017})}\BibitemShut {NoStop}%
\bibitem [{\citenamefont {P{\^a}{\c t}{\c t}u}(2019)}]{Pattu2019}%
  \BibitemOpen
  \bibfield  {author} {\bibinfo {author} {\bibfnamefont {O.~I.}\ \bibnamefont {P{\^a}{\c t}{\c t}u}},\ }\bibfield  {title} {\bibinfo {title} {Correlation functions of one-dimensional strongly interacting two-component gases},\ }\href {https://doi.org/10.1103/PhysRevA.100.063635} {\bibfield  {journal} {\bibinfo  {journal} {Phys. Rev. At. Mol. Opt. Phys.}\ }\textbf {\bibinfo {volume} {100}},\ \bibinfo {pages} {063635} (\bibinfo {year} {2019})}\BibitemShut {NoStop}%
\bibitem [{\citenamefont {Hao}\ and\ \citenamefont {Chen}(2012)}]{Hao2012}%
  \BibitemOpen
  \bibfield  {author} {\bibinfo {author} {\bibfnamefont {Y.}~\bibnamefont {Hao}}\ and\ \bibinfo {author} {\bibfnamefont {S.}~\bibnamefont {Chen}},\ }\bibfield  {title} {\bibinfo {title} {Dynamical properties of hard-core anyons in one-dimensional optical lattices},\ }\href {https://doi.org/10.1103/PhysRevA.86.043631} {\bibfield  {journal} {\bibinfo  {journal} {Phys. Rev. At. Mol. Opt. Phys.}\ }\textbf {\bibinfo {volume} {86}},\ \bibinfo {pages} {043631} (\bibinfo {year} {2012})}\BibitemShut {NoStop}%
\bibitem [{\citenamefont {Zinner}(2015)}]{Zinner2015}%
  \BibitemOpen
  \bibfield  {author} {\bibinfo {author} {\bibfnamefont {N.~T.}\ \bibnamefont {Zinner}},\ }\bibfield  {title} {\bibinfo {title} {Strongly interacting mesoscopic systems of anyons in one dimension},\ }\href {https://doi.org/10.1103/PhysRevA.92.063634} {\bibfield  {journal} {\bibinfo  {journal} {Phys. Rev. At. Mol. Opt. Phys.}\ }\textbf {\bibinfo {volume} {92}},\ \bibinfo {pages} {063634} (\bibinfo {year} {2015})}\BibitemShut {NoStop}%
\bibitem [{\citenamefont {Valiente}(2021)}]{Valiente2021}%
  \BibitemOpen
  \bibfield  {author} {\bibinfo {author} {\bibfnamefont {M.}~\bibnamefont {Valiente}},\ }\bibfield  {title} {\bibinfo {title} {Universal duality transformations in interacting one-dimensional quantum systems},\ }\href {https://doi.org/10.1103/PhysRevA.103.L021302} {\bibfield  {journal} {\bibinfo  {journal} {Phys. Rev. At. Mol. Opt. Phys.}\ }\textbf {\bibinfo {volume} {103}},\ \bibinfo {pages} {L021302} (\bibinfo {year} {2021})}\BibitemShut {NoStop}%
\bibitem [{\citenamefont {Greschner}\ \emph {et~al.}(2014)\citenamefont {Greschner}, \citenamefont {Sun}, \citenamefont {Poletti},\ and\ \citenamefont {Santos}}]{Greschner2014}%
  \BibitemOpen
  \bibfield  {author} {\bibinfo {author} {\bibfnamefont {S.}~\bibnamefont {Greschner}}, \bibinfo {author} {\bibfnamefont {G.}~\bibnamefont {Sun}}, \bibinfo {author} {\bibfnamefont {D.}~\bibnamefont {Poletti}},\ and\ \bibinfo {author} {\bibfnamefont {L.}~\bibnamefont {Santos}},\ }\bibfield  {title} {\bibinfo {title} {Density-dependent synthetic gauge fields using periodically modulated interactions},\ }\href {https://doi.org/10.1103/PhysRevLett.113.215303} {\bibfield  {journal} {\bibinfo  {journal} {Phys. Rev. Lett.}\ }\textbf {\bibinfo {volume} {113}},\ \bibinfo {pages} {215303} (\bibinfo {year} {2014})}\BibitemShut {NoStop}%
\bibitem [{\citenamefont {Cardarelli}\ \emph {et~al.}(2016)\citenamefont {Cardarelli}, \citenamefont {Greschner},\ and\ \citenamefont {Santos}}]{Cardarelli2016}%
  \BibitemOpen
  \bibfield  {author} {\bibinfo {author} {\bibfnamefont {L.}~\bibnamefont {Cardarelli}}, \bibinfo {author} {\bibfnamefont {S.}~\bibnamefont {Greschner}},\ and\ \bibinfo {author} {\bibfnamefont {L.}~\bibnamefont {Santos}},\ }\bibfield  {title} {\bibinfo {title} {Engineering interactions and anyon statistics by multicolor lattice-depth modulations},\ }\href {https://doi.org/10.1103/PhysRevA.94.023615} {\bibfield  {journal} {\bibinfo  {journal} {Phys. Rev. At. Mol. Opt. Phys.}\ }\textbf {\bibinfo {volume} {94}},\ \bibinfo {pages} {023615} (\bibinfo {year} {2016})}\BibitemShut {NoStop}%
\bibitem [{\citenamefont {Str{\"a}ter}\ \emph {et~al.}(2016)\citenamefont {Str{\"a}ter}, \citenamefont {Srivastava},\ and\ \citenamefont {Eckardt}}]{Strater2016}%
  \BibitemOpen
  \bibfield  {author} {\bibinfo {author} {\bibfnamefont {C.}~\bibnamefont {Str{\"a}ter}}, \bibinfo {author} {\bibfnamefont {S.~C.~L.}\ \bibnamefont {Srivastava}},\ and\ \bibinfo {author} {\bibfnamefont {A.}~\bibnamefont {Eckardt}},\ }\bibfield  {title} {\bibinfo {title} {Floquet {{Realization}} and {{Signatures}} of {{One-Dimensional Anyons}} in an {{Optical Lattice}}},\ }\href {https://doi.org/10.1103/PhysRevLett.117.205303} {\bibfield  {journal} {\bibinfo  {journal} {Phys. Rev. Lett.}\ }\textbf {\bibinfo {volume} {117}},\ \bibinfo {pages} {205303} (\bibinfo {year} {2016})}\BibitemShut {NoStop}%
\bibitem [{\citenamefont {Greschner}\ \emph {et~al.}(2018)\citenamefont {Greschner}, \citenamefont {Cardarelli},\ and\ \citenamefont {Santos}}]{Greschner2018a}%
  \BibitemOpen
  \bibfield  {author} {\bibinfo {author} {\bibfnamefont {S.}~\bibnamefont {Greschner}}, \bibinfo {author} {\bibfnamefont {L.}~\bibnamefont {Cardarelli}},\ and\ \bibinfo {author} {\bibfnamefont {L.}~\bibnamefont {Santos}},\ }\bibfield  {title} {\bibinfo {title} {Probing the exchange statistics of one-dimensional anyon models},\ }\href {https://doi.org/10.1103/PhysRevA.97.053605} {\bibfield  {journal} {\bibinfo  {journal} {Phys. Rev. At. Mol. Opt. Phys.}\ }\textbf {\bibinfo {volume} {97}},\ \bibinfo {pages} {053605} (\bibinfo {year} {2018})}\BibitemShut {NoStop}%
\bibitem [{\citenamefont {Yannouleas}\ and\ \citenamefont {Landman}(2019)}]{Yannouleas2019}%
  \BibitemOpen
  \bibfield  {author} {\bibinfo {author} {\bibfnamefont {C.}~\bibnamefont {Yannouleas}}\ and\ \bibinfo {author} {\bibfnamefont {U.}~\bibnamefont {Landman}},\ }\bibfield  {title} {\bibinfo {title} {Anyon optics with time-of-flight two-particle interference of double-well-trapped interacting ultracold atoms},\ }\href {https://doi.org/10/g9p5qk} {\bibfield  {journal} {\bibinfo  {journal} {Phys. Rev. A}\ }\textbf {\bibinfo {volume} {100}},\ \bibinfo {pages} {013605} (\bibinfo {year} {2019})}\BibitemShut {NoStop}%
\bibitem [{\citenamefont {G{\"o}rg}\ \emph {et~al.}(2019)\citenamefont {G{\"o}rg}, \citenamefont {Sandholzer}, \citenamefont {Minguzzi}, \citenamefont {Desbuquois}, \citenamefont {Messer},\ and\ \citenamefont {Esslinger}}]{Gorg2019}%
  \BibitemOpen
  \bibfield  {author} {\bibinfo {author} {\bibfnamefont {F.}~\bibnamefont {G{\"o}rg}}, \bibinfo {author} {\bibfnamefont {K.}~\bibnamefont {Sandholzer}}, \bibinfo {author} {\bibfnamefont {J.}~\bibnamefont {Minguzzi}}, \bibinfo {author} {\bibfnamefont {R.}~\bibnamefont {Desbuquois}}, \bibinfo {author} {\bibfnamefont {M.}~\bibnamefont {Messer}},\ and\ \bibinfo {author} {\bibfnamefont {T.}~\bibnamefont {Esslinger}},\ }\bibfield  {title} {\bibinfo {title} {Realization of density-dependent {{Peierls}} phases to engineer quantized gauge fields coupled to ultracold matter},\ }\href {https://doi.org/10/ggbrxv} {\bibfield  {journal} {\bibinfo  {journal} {Nat. Phys.}\ }\textbf {\bibinfo {volume} {15}},\ \bibinfo {pages} {1161} (\bibinfo {year} {2019})}\BibitemShut {NoStop}%
\bibitem [{\citenamefont {Schweizer}\ \emph {et~al.}(2019)\citenamefont {Schweizer}, \citenamefont {Grusdt}, \citenamefont {Berngruber}, \citenamefont {Barbiero}, \citenamefont {Demler}, \citenamefont {Goldman}, \citenamefont {Bloch},\ and\ \citenamefont {Aidelsburger}}]{Schweizer2019}%
  \BibitemOpen
  \bibfield  {author} {\bibinfo {author} {\bibfnamefont {C.}~\bibnamefont {Schweizer}}, \bibinfo {author} {\bibfnamefont {F.}~\bibnamefont {Grusdt}}, \bibinfo {author} {\bibfnamefont {M.}~\bibnamefont {Berngruber}}, \bibinfo {author} {\bibfnamefont {L.}~\bibnamefont {Barbiero}}, \bibinfo {author} {\bibfnamefont {E.}~\bibnamefont {Demler}}, \bibinfo {author} {\bibfnamefont {N.}~\bibnamefont {Goldman}}, \bibinfo {author} {\bibfnamefont {I.}~\bibnamefont {Bloch}},\ and\ \bibinfo {author} {\bibfnamefont {M.}~\bibnamefont {Aidelsburger}},\ }\bibfield  {title} {\bibinfo {title} {Floquet approach to {{$\mathbb{Z}$2}} lattice gauge theories with ultracold atoms in optical lattices},\ }\href {https://doi.org/10.1038/s41567-019-0649-7} {\bibfield  {journal} {\bibinfo  {journal} {Nat. Phys.}\ }\textbf {\bibinfo {volume} {15}},\ \bibinfo {pages} {1168} (\bibinfo {year} {2019})}\BibitemShut {NoStop}%
\bibitem [{\citenamefont {Clark}\ \emph {et~al.}(2018)\citenamefont {Clark}, \citenamefont {Anderson}, \citenamefont {Feng}, \citenamefont {Gaj}, \citenamefont {Levin},\ and\ \citenamefont {Chin}}]{Clark2018}%
  \BibitemOpen
  \bibfield  {author} {\bibinfo {author} {\bibfnamefont {L.~W.}\ \bibnamefont {Clark}}, \bibinfo {author} {\bibfnamefont {B.~M.}\ \bibnamefont {Anderson}}, \bibinfo {author} {\bibfnamefont {L.}~\bibnamefont {Feng}}, \bibinfo {author} {\bibfnamefont {A.}~\bibnamefont {Gaj}}, \bibinfo {author} {\bibfnamefont {K.}~\bibnamefont {Levin}},\ and\ \bibinfo {author} {\bibfnamefont {C.}~\bibnamefont {Chin}},\ }\bibfield  {title} {\bibinfo {title} {Observation of {{Density-Dependent Gauge Fields}} in a {{Bose-Einstein Condensate Based}} on {{Micromotion Control}} in a {{Shaken Two-Dimensional Lattice}}},\ }\href {https://doi.org/10/gdtmgw} {\bibfield  {journal} {\bibinfo  {journal} {Phys. Rev. Lett.}\ }\textbf {\bibinfo {volume} {121}},\ \bibinfo {pages} {030402} (\bibinfo {year} {2018})}\BibitemShut {NoStop}%
\bibitem [{\citenamefont {Lienhard}\ \emph {et~al.}(2020)\citenamefont {Lienhard}, \citenamefont {Scholl}, \citenamefont {Weber}, \citenamefont {Barredo}, \citenamefont {De~L{\'e}s{\'e}leuc}, \citenamefont {Bai}, \citenamefont {Lang}, \citenamefont {Fleischhauer}, \citenamefont {B{\"u}chler}, \citenamefont {Lahaye},\ and\ \citenamefont {Browaeys}}]{Lienhard2020}%
  \BibitemOpen
  \bibfield  {author} {\bibinfo {author} {\bibfnamefont {V.}~\bibnamefont {Lienhard}}, \bibinfo {author} {\bibfnamefont {P.}~\bibnamefont {Scholl}}, \bibinfo {author} {\bibfnamefont {S.}~\bibnamefont {Weber}}, \bibinfo {author} {\bibfnamefont {D.}~\bibnamefont {Barredo}}, \bibinfo {author} {\bibfnamefont {S.}~\bibnamefont {De~L{\'e}s{\'e}leuc}}, \bibinfo {author} {\bibfnamefont {R.}~\bibnamefont {Bai}}, \bibinfo {author} {\bibfnamefont {N.}~\bibnamefont {Lang}}, \bibinfo {author} {\bibfnamefont {M.}~\bibnamefont {Fleischhauer}}, \bibinfo {author} {\bibfnamefont {H.~P.}\ \bibnamefont {B{\"u}chler}}, \bibinfo {author} {\bibfnamefont {T.}~\bibnamefont {Lahaye}},\ and\ \bibinfo {author} {\bibfnamefont {A.}~\bibnamefont {Browaeys}},\ }\bibfield  {title} {\bibinfo {title} {Realization of a {{Density-Dependent Peierls Phase}} in a {{Synthetic}}, {{Spin-Orbit Coupled Rydberg System}}},\ }\href {https://doi.org/10.1103/PhysRevX.10.021031} {\bibfield  {journal} {\bibinfo  {journal} {Phys. Rev. X}\ }\textbf {\bibinfo {volume} {10}},\ \bibinfo {pages} {021031} (\bibinfo {year} {2020})}\BibitemShut {NoStop}%
\bibitem [{\citenamefont {Yao}\ \emph {et~al.}(2022)\citenamefont {Yao}, \citenamefont {Zhang},\ and\ \citenamefont {Chin}}]{Yao2022}%
  \BibitemOpen
  \bibfield  {author} {\bibinfo {author} {\bibfnamefont {K.-X.}\ \bibnamefont {Yao}}, \bibinfo {author} {\bibfnamefont {Z.}~\bibnamefont {Zhang}},\ and\ \bibinfo {author} {\bibfnamefont {C.}~\bibnamefont {Chin}},\ }\bibfield  {title} {\bibinfo {title} {Domain-wall dynamics in {{Bose}}--{{Einstein}} condensates with synthetic gauge fields},\ }\href {https://doi.org/10.1038/s41586-021-04250-3} {\bibfield  {journal} {\bibinfo  {journal} {Nature}\ }\textbf {\bibinfo {volume} {602}},\ \bibinfo {pages} {68} (\bibinfo {year} {2022})}\BibitemShut {NoStop}%
\bibitem [{\citenamefont {Barbiero}\ \emph {et~al.}(2019)\citenamefont {Barbiero}, \citenamefont {Schweizer}, \citenamefont {Aidelsburger}, \citenamefont {Demler}, \citenamefont {Goldman},\ and\ \citenamefont {Grusdt}}]{Barbiero2019}%
  \BibitemOpen
  \bibfield  {author} {\bibinfo {author} {\bibfnamefont {L.}~\bibnamefont {Barbiero}}, \bibinfo {author} {\bibfnamefont {C.}~\bibnamefont {Schweizer}}, \bibinfo {author} {\bibfnamefont {M.}~\bibnamefont {Aidelsburger}}, \bibinfo {author} {\bibfnamefont {E.}~\bibnamefont {Demler}}, \bibinfo {author} {\bibfnamefont {N.}~\bibnamefont {Goldman}},\ and\ \bibinfo {author} {\bibfnamefont {F.}~\bibnamefont {Grusdt}},\ }\bibfield  {title} {\bibinfo {title} {Coupling ultracold matter to dynamical gauge fields in optical lattices: {{From}} flux attachment to {{$\mathbb{Z}$2}} lattice gauge theories},\ }\href {https://doi.org/10/ghrjrs} {\bibfield  {journal} {\bibinfo  {journal} {Sci. Adv.}\ }\textbf {\bibinfo {volume} {5}},\ \bibinfo {pages} {eaav7444} (\bibinfo {year} {2019})}\BibitemShut {NoStop}%
\bibitem [{\citenamefont {{Arcila-Forero}}\ \emph {et~al.}(2016)\citenamefont {{Arcila-Forero}}, \citenamefont {Franco},\ and\ \citenamefont {{Silva-Valencia}}}]{Arcila-Forero2016}%
  \BibitemOpen
  \bibfield  {author} {\bibinfo {author} {\bibfnamefont {J.}~\bibnamefont {{Arcila-Forero}}}, \bibinfo {author} {\bibfnamefont {R.}~\bibnamefont {Franco}},\ and\ \bibinfo {author} {\bibfnamefont {J.}~\bibnamefont {{Silva-Valencia}}},\ }\bibfield  {title} {\bibinfo {title} {Critical points of the anyon-{{Hubbard}} model},\ }\href {https://doi.org/10.1103/PhysRevA.94.013611} {\bibfield  {journal} {\bibinfo  {journal} {Phys. Rev. A}\ }\textbf {\bibinfo {volume} {94}},\ \bibinfo {pages} {013611} (\bibinfo {year} {2016})}\BibitemShut {NoStop}%
\bibitem [{\citenamefont {Lange}\ \emph {et~al.}(2017{\natexlab{a}})\citenamefont {Lange}, \citenamefont {Ejima},\ and\ \citenamefont {Fehske}}]{Lange2017}%
  \BibitemOpen
  \bibfield  {author} {\bibinfo {author} {\bibfnamefont {F.}~\bibnamefont {Lange}}, \bibinfo {author} {\bibfnamefont {S.}~\bibnamefont {Ejima}},\ and\ \bibinfo {author} {\bibfnamefont {H.}~\bibnamefont {Fehske}},\ }\bibfield  {title} {\bibinfo {title} {Strongly repulsive anyons in one dimension},\ }\href {https://doi.org/10/g9p5q7} {\bibfield  {journal} {\bibinfo  {journal} {Phys. Rev. A}\ }\textbf {\bibinfo {volume} {95}},\ \bibinfo {pages} {063621} (\bibinfo {year} {2017}{\natexlab{a}})}\BibitemShut {NoStop}%
\bibitem [{\citenamefont {Zhang}\ \emph {et~al.}(2017)\citenamefont {Zhang}, \citenamefont {Greschner}, \citenamefont {Fan}, \citenamefont {Scott},\ and\ \citenamefont {Zhang}}]{Zhang2017b}%
  \BibitemOpen
  \bibfield  {author} {\bibinfo {author} {\bibfnamefont {W.}~\bibnamefont {Zhang}}, \bibinfo {author} {\bibfnamefont {S.}~\bibnamefont {Greschner}}, \bibinfo {author} {\bibfnamefont {E.}~\bibnamefont {Fan}}, \bibinfo {author} {\bibfnamefont {T.~C.}\ \bibnamefont {Scott}},\ and\ \bibinfo {author} {\bibfnamefont {Y.}~\bibnamefont {Zhang}},\ }\bibfield  {title} {\bibinfo {title} {Ground-state properties of the one-dimensional unconstrained pseudo-anyon {{Hubbard}} model},\ }\href {https://doi.org/10.1103/PhysRevA.95.053614} {\bibfield  {journal} {\bibinfo  {journal} {Phys. Rev. A}\ }\textbf {\bibinfo {volume} {95}},\ \bibinfo {pages} {053614} (\bibinfo {year} {2017})}\BibitemShut {NoStop}%
\bibitem [{\citenamefont {Bonkhoff}\ \emph {et~al.}(2025)\citenamefont {Bonkhoff}, \citenamefont {J{\"a}gering}, \citenamefont {Hu}, \citenamefont {Pelster}, \citenamefont {Eggert},\ and\ \citenamefont {Schneider}}]{Bonkhoff2025}%
  \BibitemOpen
  \bibfield  {author} {\bibinfo {author} {\bibfnamefont {M.}~\bibnamefont {Bonkhoff}}, \bibinfo {author} {\bibfnamefont {K.}~\bibnamefont {J{\"a}gering}}, \bibinfo {author} {\bibfnamefont {S.}~\bibnamefont {Hu}}, \bibinfo {author} {\bibfnamefont {A.}~\bibnamefont {Pelster}}, \bibinfo {author} {\bibfnamefont {S.}~\bibnamefont {Eggert}},\ and\ \bibinfo {author} {\bibfnamefont {I.}~\bibnamefont {Schneider}},\ }\bibfield  {title} {\bibinfo {title} {Anyonic {{Phase Transitions}} in the {{1D Extended Hubbard Model}} with {{Fractional Statistics}}},\ }\href {https://doi.org/10.1103/7n1c-vq2p} {\bibfield  {journal} {\bibinfo  {journal} {Phys. Rev. Lett.}\ }\textbf {\bibinfo {volume} {135}},\ \bibinfo {pages} {036601} (\bibinfo {year} {2025})}\BibitemShut {NoStop}%
\bibitem [{\citenamefont {Theel}\ \emph {et~al.}(2025)\citenamefont {Theel}, \citenamefont {Bonkhoff}, \citenamefont {Schmelcher}, \citenamefont {Posske},\ and\ \citenamefont {Harshman}}]{Theel2025}%
  \BibitemOpen
  \bibfield  {author} {\bibinfo {author} {\bibfnamefont {F.}~\bibnamefont {Theel}}, \bibinfo {author} {\bibfnamefont {M.}~\bibnamefont {Bonkhoff}}, \bibinfo {author} {\bibfnamefont {P.}~\bibnamefont {Schmelcher}}, \bibinfo {author} {\bibfnamefont {T.}~\bibnamefont {Posske}},\ and\ \bibinfo {author} {\bibfnamefont {N.~L.}\ \bibnamefont {Harshman}},\ }\bibfield  {title} {\bibinfo {title} {Chirally {{Protected State Manipulation}} by {{Tuning One-Dimensional Statistics}}},\ }\href {https://doi.org/10.1103/kzf6-yx24} {\bibfield  {journal} {\bibinfo  {journal} {Phys. Rev. Lett.}\ }\textbf {\bibinfo {volume} {135}},\ \bibinfo {pages} {063401} (\bibinfo {year} {2025})}\BibitemShut {NoStop}%
\bibitem [{\citenamefont {Yang}\ \emph {et~al.}(2026)\citenamefont {Yang}, \citenamefont {Wang}, \citenamefont {Liu}, \citenamefont {Yang}, \citenamefont {Yuan},\ and\ \citenamefont {Li}}]{Yang2026}%
  \BibitemOpen
  \bibfield  {author} {\bibinfo {author} {\bibfnamefont {X.-C.}\ \bibnamefont {Yang}}, \bibinfo {author} {\bibfnamefont {B.}~\bibnamefont {Wang}}, \bibinfo {author} {\bibfnamefont {J.}~\bibnamefont {Liu}}, \bibinfo {author} {\bibfnamefont {B.}~\bibnamefont {Yang}}, \bibinfo {author} {\bibfnamefont {J.}~\bibnamefont {Yuan}},\ and\ \bibinfo {author} {\bibfnamefont {Y.}~\bibnamefont {Li}},\ }\bibinfo {title} {Statistics-governed dynamical scaling in interacting anyonic chains} (\bibinfo {year} {2026}),\ \Eprint {https://arxiv.org/abs/2603.15972} {arXiv:2603.15972} \BibitemShut {NoStop}%
\bibitem [{\citenamefont {Bonkhoff}\ \emph {et~al.}(2026)\citenamefont {Bonkhoff}, \citenamefont {Gurney}, \citenamefont {Theel}, \citenamefont {Schmelcher}, \citenamefont {Harshman},\ and\ \citenamefont {Posske}}]{Bonkhoff2026}%
  \BibitemOpen
  \bibfield  {author} {\bibinfo {author} {\bibfnamefont {M.}~\bibnamefont {Bonkhoff}}, \bibinfo {author} {\bibfnamefont {G.}~\bibnamefont {Gurney}}, \bibinfo {author} {\bibfnamefont {F.}~\bibnamefont {Theel}}, \bibinfo {author} {\bibfnamefont {P.}~\bibnamefont {Schmelcher}}, \bibinfo {author} {\bibfnamefont {N.~L.}\ \bibnamefont {Harshman}},\ and\ \bibinfo {author} {\bibfnamefont {T.}~\bibnamefont {Posske}},\ }\bibinfo {title} {Symmetry and integrability in the anyon-{{Hubbard}} model} (\bibinfo {year} {2026}),\ \Eprint {https://arxiv.org/abs/2605.28956} {arXiv:2605.28956} \BibitemShut {NoStop}%
\bibitem [{\citenamefont {Wei}\ \emph {et~al.}(2026)\citenamefont {Wei}, \citenamefont {Zhang}, \citenamefont {Ye}, \citenamefont {Wang}, \citenamefont {Wang}, \citenamefont {Zhou},\ and\ \citenamefont {Yao}}]{Wei2026}%
  \BibitemOpen
  \bibfield  {author} {\bibinfo {author} {\bibfnamefont {F.}~\bibnamefont {Wei}}, \bibinfo {author} {\bibfnamefont {C.}~\bibnamefont {Zhang}}, \bibinfo {author} {\bibfnamefont {Z.}~\bibnamefont {Ye}}, \bibinfo {author} {\bibfnamefont {D.}~\bibnamefont {Wang}}, \bibinfo {author} {\bibfnamefont {B.}~\bibnamefont {Wang}}, \bibinfo {author} {\bibfnamefont {X.}~\bibnamefont {Zhou}},\ and\ \bibinfo {author} {\bibfnamefont {H.}~\bibnamefont {Yao}},\ }\bibinfo {title} {Observation of anyonic thermodynamics and generalized {{Pauli}} principle} (\bibinfo {year} {2026}),\ \Eprint {https://arxiv.org/abs/2606.19009} {arXiv:2606.19009} \BibitemShut {NoStop}%
\bibitem [{\citenamefont {Wang}\ \emph {et~al.}(2026)\citenamefont {Wang}, \citenamefont {Hu}, \citenamefont {You},\ and\ \citenamefont {Sun}}]{Wang2026d}%
  \BibitemOpen
  \bibfield  {author} {\bibinfo {author} {\bibfnamefont {Y.-N.}\ \bibnamefont {Wang}}, \bibinfo {author} {\bibfnamefont {Q.-M.}\ \bibnamefont {Hu}}, \bibinfo {author} {\bibfnamefont {W.-L.}\ \bibnamefont {You}},\ and\ \bibinfo {author} {\bibfnamefont {G.}~\bibnamefont {Sun}},\ }\bibinfo {title} {Krylov complexity of anyons} (\bibinfo {year} {2026}),\ \Eprint {https://arxiv.org/abs/2608.16595} {arXiv:2608.16595} \BibitemShut {NoStop}%
\bibitem [{\citenamefont {Zhou}\ \emph {et~al.}(2026)\citenamefont {Zhou}, \citenamefont {Guo}, \citenamefont {Wu}, \citenamefont {Hu},\ and\ \citenamefont {Luo}}]{Zhou2026}%
  \BibitemOpen
  \bibfield  {author} {\bibinfo {author} {\bibfnamefont {Z.}~\bibnamefont {Zhou}}, \bibinfo {author} {\bibfnamefont {C.}~\bibnamefont {Guo}}, \bibinfo {author} {\bibfnamefont {H.}~\bibnamefont {Wu}}, \bibinfo {author} {\bibfnamefont {Q.}~\bibnamefont {Hu}},\ and\ \bibinfo {author} {\bibfnamefont {X.}~\bibnamefont {Luo}},\ }\bibinfo {title} {Tunable {{Statistics-Induced Caging}} in the {{Anyon-Hubbard Model}}} (\bibinfo {year} {2026}),\ \Eprint {https://arxiv.org/abs/2608.15739} {arXiv:2608.15739} \BibitemShut {NoStop}%
\bibitem [{\citenamefont {Lyngfelt}\ \emph {et~al.}(2026)\citenamefont {Lyngfelt}, \citenamefont {{Fern{\'a}ndez-Pend{\'a}s}}, \citenamefont {Johansson},\ and\ \citenamefont {{Garci{\'a}-{\'A}lvarez}}}]{Lyngfelt2026}%
  \BibitemOpen
  \bibfield  {author} {\bibinfo {author} {\bibfnamefont {I.}~\bibnamefont {Lyngfelt}}, \bibinfo {author} {\bibfnamefont {J.}~\bibnamefont {{Fern{\'a}ndez-Pend{\'a}s}}}, \bibinfo {author} {\bibfnamefont {G.}~\bibnamefont {Johansson}},\ and\ \bibinfo {author} {\bibfnamefont {L.}~\bibnamefont {{Garci{\'a}-{\'A}lvarez}}},\ }\bibinfo {title} {Analog quantum simulation of bosonic and anyonic models with flux-driven transmons} (\bibinfo {year} {2026}),\ \Eprint {https://arxiv.org/abs/2609.03737} {arXiv:2609.03737} \BibitemShut {NoStop}%
\bibitem [{\citenamefont {Lange}\ \emph {et~al.}(2017{\natexlab{b}})\citenamefont {Lange}, \citenamefont {Ejima},\ and\ \citenamefont {Fehske}}]{Lange2017a}%
  \BibitemOpen
  \bibfield  {author} {\bibinfo {author} {\bibfnamefont {F.}~\bibnamefont {Lange}}, \bibinfo {author} {\bibfnamefont {S.}~\bibnamefont {Ejima}},\ and\ \bibinfo {author} {\bibfnamefont {H.}~\bibnamefont {Fehske}},\ }\bibfield  {title} {\bibinfo {title} {Anyonic {{Haldane Insulator}} in {{One Dimension}}},\ }\href {https://doi.org/10.1103/PhysRevLett.118.120401} {\bibfield  {journal} {\bibinfo  {journal} {Phys. Rev. Lett.}\ }\textbf {\bibinfo {volume} {118}},\ \bibinfo {pages} {120401} (\bibinfo {year} {2017}{\natexlab{b}})}\BibitemShut {NoStop}%
\bibitem [{\citenamefont {Liu}\ \emph {et~al.}(2018)\citenamefont {Liu}, \citenamefont {Garrison}, \citenamefont {Deng}, \citenamefont {Gong},\ and\ \citenamefont {Gorshkov}}]{Liu2018b}%
  \BibitemOpen
  \bibfield  {author} {\bibinfo {author} {\bibfnamefont {F.}~\bibnamefont {Liu}}, \bibinfo {author} {\bibfnamefont {J.~R.}\ \bibnamefont {Garrison}}, \bibinfo {author} {\bibfnamefont {D.-L.}\ \bibnamefont {Deng}}, \bibinfo {author} {\bibfnamefont {Z.-X.}\ \bibnamefont {Gong}},\ and\ \bibinfo {author} {\bibfnamefont {A.~V.}\ \bibnamefont {Gorshkov}},\ }\bibfield  {title} {\bibinfo {title} {Asymmetric particle transport and light-cone dynamics induced by anyonic statistics},\ }\href {https://doi.org/10.1103/PhysRevLett.121.250404} {\bibfield  {journal} {\bibinfo  {journal} {Phys. Rev. Lett.}\ }\textbf {\bibinfo {volume} {121}},\ \bibinfo {pages} {250404} (\bibinfo {year} {2018})}\BibitemShut {NoStop}%
\bibitem [{\citenamefont {Tang}\ \emph {et~al.}(2015)\citenamefont {Tang}, \citenamefont {Eggert},\ and\ \citenamefont {Pelster}}]{Tang2015}%
  \BibitemOpen
  \bibfield  {author} {\bibinfo {author} {\bibfnamefont {G.}~\bibnamefont {Tang}}, \bibinfo {author} {\bibfnamefont {S.}~\bibnamefont {Eggert}},\ and\ \bibinfo {author} {\bibfnamefont {A.}~\bibnamefont {Pelster}},\ }\bibfield  {title} {\bibinfo {title} {Ground-state properties of anyons in a one-dimensional lattice},\ }\href {https://doi.org/10.1088/1367-2630/17/12/123016} {\bibfield  {journal} {\bibinfo  {journal} {New J. Phys.}\ }\textbf {\bibinfo {volume} {17}},\ \bibinfo {pages} {123016} (\bibinfo {year} {2015})}\BibitemShut {NoStop}%
\bibitem [{\citenamefont {Aglietti}\ \emph {et~al.}(1996)\citenamefont {Aglietti}, \citenamefont {Griguolo}, \citenamefont {Jackiw}, \citenamefont {Pi},\ and\ \citenamefont {Seminara}}]{Aglietti1996}%
  \BibitemOpen
  \bibfield  {author} {\bibinfo {author} {\bibfnamefont {U.}~\bibnamefont {Aglietti}}, \bibinfo {author} {\bibfnamefont {L.}~\bibnamefont {Griguolo}}, \bibinfo {author} {\bibfnamefont {R.}~\bibnamefont {Jackiw}}, \bibinfo {author} {\bibfnamefont {S.-Y.}\ \bibnamefont {Pi}},\ and\ \bibinfo {author} {\bibfnamefont {D.}~\bibnamefont {Seminara}},\ }\bibfield  {title} {\bibinfo {title} {Anyons and chiral solitons on a line},\ }\href {https://doi.org/10.1103/PhysRevLett.77.4406} {\bibfield  {journal} {\bibinfo  {journal} {Phys. Rev. Lett.}\ }\textbf {\bibinfo {volume} {77}},\ \bibinfo {pages} {4406} (\bibinfo {year} {1996})}\BibitemShut {NoStop}%
\bibitem [{\citenamefont {Chisholm}\ \emph {et~al.}(2022)\citenamefont {Chisholm}, \citenamefont {Fr{\"o}lian}, \citenamefont {Neri}, \citenamefont {Ramos}, \citenamefont {Tarruell},\ and\ \citenamefont {Celi}}]{Chisholm2022}%
  \BibitemOpen
  \bibfield  {author} {\bibinfo {author} {\bibfnamefont {C.~S.}\ \bibnamefont {Chisholm}}, \bibinfo {author} {\bibfnamefont {A.}~\bibnamefont {Fr{\"o}lian}}, \bibinfo {author} {\bibfnamefont {E.}~\bibnamefont {Neri}}, \bibinfo {author} {\bibfnamefont {R.}~\bibnamefont {Ramos}}, \bibinfo {author} {\bibfnamefont {L.}~\bibnamefont {Tarruell}},\ and\ \bibinfo {author} {\bibfnamefont {A.}~\bibnamefont {Celi}},\ }\bibfield  {title} {\bibinfo {title} {Encoding a one-dimensional topological gauge theory in a {{Raman-coupled Bose-Einstein}} condensate},\ }\href {https://doi.org/10.1103/PhysRevResearch.4.043088} {\bibfield  {journal} {\bibinfo  {journal} {Phys. Rev. Res.}\ }\textbf {\bibinfo {volume} {4}},\ \bibinfo {pages} {043088} (\bibinfo {year} {2022})}\BibitemShut {NoStop}%
\bibitem [{\citenamefont {Fr{\"o}lian}\ \emph {et~al.}(2022)\citenamefont {Fr{\"o}lian}, \citenamefont {Chisholm}, \citenamefont {Neri}, \citenamefont {Cabrera}, \citenamefont {Ramos}, \citenamefont {Celi},\ and\ \citenamefont {Tarruell}}]{Frolian2022}%
  \BibitemOpen
  \bibfield  {author} {\bibinfo {author} {\bibfnamefont {A.}~\bibnamefont {Fr{\"o}lian}}, \bibinfo {author} {\bibfnamefont {C.~S.}\ \bibnamefont {Chisholm}}, \bibinfo {author} {\bibfnamefont {E.}~\bibnamefont {Neri}}, \bibinfo {author} {\bibfnamefont {C.~R.}\ \bibnamefont {Cabrera}}, \bibinfo {author} {\bibfnamefont {R.}~\bibnamefont {Ramos}}, \bibinfo {author} {\bibfnamefont {A.}~\bibnamefont {Celi}},\ and\ \bibinfo {author} {\bibfnamefont {L.}~\bibnamefont {Tarruell}},\ }\bibfield  {title} {\bibinfo {title} {Realizing a {{1D}} topological gauge theory in an optically dressed {{BEC}}},\ }\href {https://doi.org/10.1038/s41586-022-04943-3} {\bibfield  {journal} {\bibinfo  {journal} {Nature}\ }\textbf {\bibinfo {volume} {608}},\ \bibinfo {pages} {293} (\bibinfo {year} {2022})}\BibitemShut {NoStop}%
\bibitem [{\citenamefont {Wang}\ \emph {et~al.}(2025)\citenamefont {Wang}, \citenamefont {Vashisht}, \citenamefont {Guo}, \citenamefont {Dhar}, \citenamefont {Landini}, \citenamefont {N{\"a}gerl},\ and\ \citenamefont {Goldman}}]{Wang2025}%
  \BibitemOpen
  \bibfield  {author} {\bibinfo {author} {\bibfnamefont {B.}~\bibnamefont {Wang}}, \bibinfo {author} {\bibfnamefont {A.}~\bibnamefont {Vashisht}}, \bibinfo {author} {\bibfnamefont {Y.}~\bibnamefont {Guo}}, \bibinfo {author} {\bibfnamefont {S.}~\bibnamefont {Dhar}}, \bibinfo {author} {\bibfnamefont {M.}~\bibnamefont {Landini}}, \bibinfo {author} {\bibfnamefont {H.-C.}\ \bibnamefont {N{\"a}gerl}},\ and\ \bibinfo {author} {\bibfnamefont {N.}~\bibnamefont {Goldman}},\ }\bibfield  {title} {\bibinfo {title} {Anyonization of {{Bosons}} in {{One Dimension}}: {{An Effective Swap Model}}},\ }\href {https://doi.org/10.1103/2np8-mp39} {\bibfield  {journal} {\bibinfo  {journal} {Phys. Rev. Lett.}\ }\textbf {\bibinfo {volume} {135}},\ \bibinfo {pages} {253403} (\bibinfo {year} {2025})}\BibitemShut {NoStop}%
\bibitem [{\citenamefont {Piil}\ and\ \citenamefont {M{\o}lmer}(2007)}]{Piil2007}%
  \BibitemOpen
  \bibfield  {author} {\bibinfo {author} {\bibfnamefont {R.}~\bibnamefont {Piil}}\ and\ \bibinfo {author} {\bibfnamefont {K.}~\bibnamefont {M{\o}lmer}},\ }\bibfield  {title} {\bibinfo {title} {Tunneling couplings in discrete lattices, single-particle band structure, and eigenstates of interacting atom pairs},\ }\href {https://doi.org/10.1103/PhysRevA.76.023607} {\bibfield  {journal} {\bibinfo  {journal} {Phys. Rev. A}\ }\textbf {\bibinfo {volume} {76}},\ \bibinfo {pages} {023607} (\bibinfo {year} {2007})}\BibitemShut {NoStop}%
\bibitem [{\citenamefont {Valiente}\ and\ \citenamefont {Petrosyan}(2008{\natexlab{a}})}]{Valiente2008}%
  \BibitemOpen
  \bibfield  {author} {\bibinfo {author} {\bibfnamefont {M.}~\bibnamefont {Valiente}}\ and\ \bibinfo {author} {\bibfnamefont {D.}~\bibnamefont {Petrosyan}},\ }\bibfield  {title} {\bibinfo {title} {Two-particle states in the {{Hubbard}} model},\ }\href {https://doi.org/10.1088/0953-4075/41/16/161002} {\bibfield  {journal} {\bibinfo  {journal} {J. Phys. B At. Mol. Opt. Phys.}\ }\textbf {\bibinfo {volume} {41}},\ \bibinfo {pages} {161002} (\bibinfo {year} {2008}{\natexlab{a}})}\BibitemShut {NoStop}%
\bibitem [{\citenamefont {Valiente}\ \emph {et~al.}(2010)\citenamefont {Valiente}, \citenamefont {Petrosyan},\ and\ \citenamefont {Saenz}}]{Valiente2010a}%
  \BibitemOpen
  \bibfield  {author} {\bibinfo {author} {\bibfnamefont {M.}~\bibnamefont {Valiente}}, \bibinfo {author} {\bibfnamefont {D.}~\bibnamefont {Petrosyan}},\ and\ \bibinfo {author} {\bibfnamefont {A.}~\bibnamefont {Saenz}},\ }\bibfield  {title} {\bibinfo {title} {Three-body bound states in a lattice},\ }\href {https://doi.org/10.1103/PhysRevA.81.011601} {\bibfield  {journal} {\bibinfo  {journal} {Phys. Rev. A}\ }\textbf {\bibinfo {volume} {81}},\ \bibinfo {pages} {011601} (\bibinfo {year} {2010})}\BibitemShut {NoStop}%
\bibitem [{\citenamefont {Winkler}\ \emph {et~al.}(2006)\citenamefont {Winkler}, \citenamefont {Thalhammer}, \citenamefont {Lang}, \citenamefont {Grimm}, \citenamefont {Hecker~Denschlag}, \citenamefont {Daley}, \citenamefont {Kantian}, \citenamefont {B{\"u}chler},\ and\ \citenamefont {Zoller}}]{Winkler2006}%
  \BibitemOpen
  \bibfield  {author} {\bibinfo {author} {\bibfnamefont {K.}~\bibnamefont {Winkler}}, \bibinfo {author} {\bibfnamefont {G.}~\bibnamefont {Thalhammer}}, \bibinfo {author} {\bibfnamefont {F.}~\bibnamefont {Lang}}, \bibinfo {author} {\bibfnamefont {R.}~\bibnamefont {Grimm}}, \bibinfo {author} {\bibfnamefont {J.}~\bibnamefont {Hecker~Denschlag}}, \bibinfo {author} {\bibfnamefont {A.~J.}\ \bibnamefont {Daley}}, \bibinfo {author} {\bibfnamefont {A.}~\bibnamefont {Kantian}}, \bibinfo {author} {\bibfnamefont {H.~P.}\ \bibnamefont {B{\"u}chler}},\ and\ \bibinfo {author} {\bibfnamefont {P.}~\bibnamefont {Zoller}},\ }\bibfield  {title} {\bibinfo {title} {Repulsively bound atom pairs in an optical lattice},\ }\href {https://doi.org/10.1038/nature04918} {\bibfield  {journal} {\bibinfo  {journal} {Nature}\ }\textbf {\bibinfo {volume} {441}},\ \bibinfo {pages} {853} (\bibinfo {year} {2006})}\BibitemShut {NoStop}%
\bibitem [{\citenamefont {Fukuhara}\ \emph {et~al.}(2013)\citenamefont {Fukuhara}, \citenamefont {Schau{\ss}}, \citenamefont {Endres}, \citenamefont {Hild}, \citenamefont {Cheneau}, \citenamefont {Bloch},\ and\ \citenamefont {Gross}}]{Fukuhara2013}%
  \BibitemOpen
  \bibfield  {author} {\bibinfo {author} {\bibfnamefont {T.}~\bibnamefont {Fukuhara}}, \bibinfo {author} {\bibfnamefont {P.}~\bibnamefont {Schau{\ss}}}, \bibinfo {author} {\bibfnamefont {M.}~\bibnamefont {Endres}}, \bibinfo {author} {\bibfnamefont {S.}~\bibnamefont {Hild}}, \bibinfo {author} {\bibfnamefont {M.}~\bibnamefont {Cheneau}}, \bibinfo {author} {\bibfnamefont {I.}~\bibnamefont {Bloch}},\ and\ \bibinfo {author} {\bibfnamefont {C.}~\bibnamefont {Gross}},\ }\bibfield  {title} {\bibinfo {title} {Microscopic observation of magnon bound states and their dynamics},\ }\href {https://doi.org/10.1038/nature12541} {\bibfield  {journal} {\bibinfo  {journal} {Nature}\ }\textbf {\bibinfo {volume} {502}},\ \bibinfo {pages} {76} (\bibinfo {year} {2013})}\BibitemShut {NoStop}%
\bibitem [{\citenamefont {Kranzl}\ \emph {et~al.}(2023)\citenamefont {Kranzl}, \citenamefont {Birnkammer}, \citenamefont {Joshi}, \citenamefont {Bastianello}, \citenamefont {Blatt}, \citenamefont {Knap},\ and\ \citenamefont {Roos}}]{Kranzl2023}%
  \BibitemOpen
  \bibfield  {author} {\bibinfo {author} {\bibfnamefont {F.}~\bibnamefont {Kranzl}}, \bibinfo {author} {\bibfnamefont {S.}~\bibnamefont {Birnkammer}}, \bibinfo {author} {\bibfnamefont {M.~K.}\ \bibnamefont {Joshi}}, \bibinfo {author} {\bibfnamefont {A.}~\bibnamefont {Bastianello}}, \bibinfo {author} {\bibfnamefont {R.}~\bibnamefont {Blatt}}, \bibinfo {author} {\bibfnamefont {M.}~\bibnamefont {Knap}},\ and\ \bibinfo {author} {\bibfnamefont {C.~F.}\ \bibnamefont {Roos}},\ }\bibfield  {title} {\bibinfo {title} {Observation of {{Magnon Bound States}} in the {{Long-Range}}, {{Anisotropic Heisenberg Model}}},\ }\href {https://doi.org/10.1103/PhysRevX.13.031017} {\bibfield  {journal} {\bibinfo  {journal} {Phys. Rev. X}\ }\textbf {\bibinfo {volume} {13}},\ \bibinfo {pages} {031017} (\bibinfo {year} {2023})}\BibitemShut {NoStop}%
\bibitem [{\citenamefont {Krutitsky}(2016)}]{Krutitsky2016}%
  \BibitemOpen
  \bibfield  {author} {\bibinfo {author} {\bibfnamefont {K.~V.}\ \bibnamefont {Krutitsky}},\ }\bibfield  {title} {\bibinfo {title} {Ultracold bosons with short-range interaction in regular optical lattices},\ }\href {https://doi.org/10.1016/j.physrep.2015.10.004} {\bibfield  {journal} {\bibinfo  {journal} {Phys. Rep.}\ }\textbf {\bibinfo {volume} {607}},\ \bibinfo {pages} {1} (\bibinfo {year} {2016})}\BibitemShut {NoStop}%
\bibitem [{Note1()}]{Note1}%
  \BibitemOpen
  \bibinfo {note} {While preparing this manuscript, a complementary study of these two-body bound states in the continuum appeared~\cite {Bonkhoff2026}.}\BibitemShut {Stop}%
\bibitem [{\citenamefont {Evers}\ and\ \citenamefont {Mirlin}(2008)}]{Evers2008}%
  \BibitemOpen
  \bibfield  {author} {\bibinfo {author} {\bibfnamefont {F.}~\bibnamefont {Evers}}\ and\ \bibinfo {author} {\bibfnamefont {A.~D.}\ \bibnamefont {Mirlin}},\ }\bibfield  {title} {\bibinfo {title} {Anderson transitions},\ }\href {https://doi.org/10.1103/RevModPhys.80.1355} {\bibfield  {journal} {\bibinfo  {journal} {Rev. Mod. Phys.}\ }\textbf {\bibinfo {volume} {80}},\ \bibinfo {pages} {1355} (\bibinfo {year} {2008})}\BibitemShut {NoStop}%
\bibitem [{\citenamefont {Hsu}\ \emph {et~al.}(2016)\citenamefont {Hsu}, \citenamefont {Zhen}, \citenamefont {Stone}, \citenamefont {Joannopoulos},\ and\ \citenamefont {Solja{\v c}i{\'c}}}]{Hsu2016}%
  \BibitemOpen
  \bibfield  {author} {\bibinfo {author} {\bibfnamefont {C.~W.}\ \bibnamefont {Hsu}}, \bibinfo {author} {\bibfnamefont {B.}~\bibnamefont {Zhen}}, \bibinfo {author} {\bibfnamefont {A.~D.}\ \bibnamefont {Stone}}, \bibinfo {author} {\bibfnamefont {J.~D.}\ \bibnamefont {Joannopoulos}},\ and\ \bibinfo {author} {\bibfnamefont {M.}~\bibnamefont {Solja{\v c}i{\'c}}},\ }\bibfield  {title} {\bibinfo {title} {Bound states in the continuum},\ }\href {https://doi.org/10/gfwccr} {\bibfield  {journal} {\bibinfo  {journal} {Nat. Rev. Mater.}\ }\textbf {\bibinfo {volume} {1}},\ \bibinfo {pages} {1} (\bibinfo {year} {2016})}\BibitemShut {NoStop}%
\bibitem [{\citenamefont {{von Neuman}}\ and\ \citenamefont {Wigner}(1929)}]{vonNeuman1929}%
  \BibitemOpen
  \bibfield  {author} {\bibinfo {author} {\bibfnamefont {J.}~\bibnamefont {{von Neuman}}}\ and\ \bibinfo {author} {\bibfnamefont {E.}~\bibnamefont {Wigner}},\ }\bibfield  {title} {\bibinfo {title} {Uber merkw\"urdige diskrete {{Eigenwerte}}. {{Uber}} das {{Verhalten}} von {{Eigenwerten}} bei adiabatischen {{Prozessen}}},\ }\href {https://ui.adsabs.harvard.edu/abs/1929PhyZ...30..467V} {\bibfield  {journal} {\bibinfo  {journal} {Phys. Z.}\ }\textbf {\bibinfo {volume} {30}},\ \bibinfo {pages} {467} (\bibinfo {year} {1929})}\BibitemShut {NoStop}%
\bibitem [{\citenamefont {Friedrich}\ and\ \citenamefont {Wintgen}(1985)}]{Friedrich1985}%
  \BibitemOpen
  \bibfield  {author} {\bibinfo {author} {\bibfnamefont {H.}~\bibnamefont {Friedrich}}\ and\ \bibinfo {author} {\bibfnamefont {D.}~\bibnamefont {Wintgen}},\ }\bibfield  {title} {\bibinfo {title} {Interfering resonances and bound states in the continuum},\ }\href {https://doi.org/10.1103/PhysRevA.32.3231} {\bibfield  {journal} {\bibinfo  {journal} {Phys. Rev. A}\ }\textbf {\bibinfo {volume} {32}},\ \bibinfo {pages} {3231} (\bibinfo {year} {1985})}\BibitemShut {NoStop}%
\bibitem [{\citenamefont {Plotnik}\ \emph {et~al.}(2011)\citenamefont {Plotnik}, \citenamefont {Peleg}, \citenamefont {Dreisow}, \citenamefont {Heinrich}, \citenamefont {Nolte}, \citenamefont {Szameit},\ and\ \citenamefont {Segev}}]{Plotnik2011}%
  \BibitemOpen
  \bibfield  {author} {\bibinfo {author} {\bibfnamefont {Y.}~\bibnamefont {Plotnik}}, \bibinfo {author} {\bibfnamefont {O.}~\bibnamefont {Peleg}}, \bibinfo {author} {\bibfnamefont {F.}~\bibnamefont {Dreisow}}, \bibinfo {author} {\bibfnamefont {M.}~\bibnamefont {Heinrich}}, \bibinfo {author} {\bibfnamefont {S.}~\bibnamefont {Nolte}}, \bibinfo {author} {\bibfnamefont {A.}~\bibnamefont {Szameit}},\ and\ \bibinfo {author} {\bibfnamefont {M.}~\bibnamefont {Segev}},\ }\bibfield  {title} {\bibinfo {title} {Experimental {{Observation}} of {{Optical Bound States}} in the {{Continuum}}},\ }\href {https://doi.org/10.1103/PhysRevLett.107.183901} {\bibfield  {journal} {\bibinfo  {journal} {Phys. Rev. Lett.}\ }\textbf {\bibinfo {volume} {107}},\ \bibinfo {pages} {183901} (\bibinfo {year} {2011})}\BibitemShut {NoStop}%
\bibitem [{\citenamefont {Zhang}\ \emph {et~al.}(2012)\citenamefont {Zhang}, \citenamefont {Braak},\ and\ \citenamefont {Kollar}}]{Zhang2012}%
  \BibitemOpen
  \bibfield  {author} {\bibinfo {author} {\bibfnamefont {J.~M.}\ \bibnamefont {Zhang}}, \bibinfo {author} {\bibfnamefont {D.}~\bibnamefont {Braak}},\ and\ \bibinfo {author} {\bibfnamefont {M.}~\bibnamefont {Kollar}},\ }\bibfield  {title} {\bibinfo {title} {Bound {{States}} in the {{Continuum Realized}} in the {{One-Dimensional Two-Particle Hubbard Model}} with an {{Impurity}}},\ }\href {https://doi.org/10/f38dx6} {\bibfield  {journal} {\bibinfo  {journal} {Phys. Rev. Lett.}\ }\textbf {\bibinfo {volume} {109}},\ \bibinfo {pages} {116405} (\bibinfo {year} {2012})}\BibitemShut {NoStop}%
\bibitem [{\citenamefont {Zhang}\ \emph {et~al.}(2013)\citenamefont {Zhang}, \citenamefont {Braak},\ and\ \citenamefont {Kollar}}]{Zhang2013}%
  \BibitemOpen
  \bibfield  {author} {\bibinfo {author} {\bibfnamefont {J.~M.}\ \bibnamefont {Zhang}}, \bibinfo {author} {\bibfnamefont {D.}~\bibnamefont {Braak}},\ and\ \bibinfo {author} {\bibfnamefont {M.}~\bibnamefont {Kollar}},\ }\bibfield  {title} {\bibinfo {title} {Bound states in the one-dimensional two-particle {{Hubbard}} model with an impurity},\ }\href {https://doi.org/10/gkjrmx} {\bibfield  {journal} {\bibinfo  {journal} {Phys. Rev. A}\ }\textbf {\bibinfo {volume} {87}},\ \bibinfo {pages} {023613} (\bibinfo {year} {2013})}\BibitemShut {NoStop}%
\bibitem [{\citenamefont {Longhi}(2007)}]{Longhi2007}%
  \BibitemOpen
  \bibfield  {author} {\bibinfo {author} {\bibfnamefont {S.}~\bibnamefont {Longhi}},\ }\bibfield  {title} {\bibinfo {title} {Bound states in the continuum in a single-level {{Fano-Anderson}} model},\ }\href {https://doi.org/10.1140/epjb/e2007-00143-2} {\bibfield  {journal} {\bibinfo  {journal} {Eur. Phys. J. B}\ }\textbf {\bibinfo {volume} {57}},\ \bibinfo {pages} {45} (\bibinfo {year} {2007})}\BibitemShut {NoStop}%
\bibitem [{\citenamefont {Cerjan}\ \emph {et~al.}(2019)\citenamefont {Cerjan}, \citenamefont {Hsu},\ and\ \citenamefont {Rechtsman}}]{Cerjan2019}%
  \BibitemOpen
  \bibfield  {author} {\bibinfo {author} {\bibfnamefont {A.}~\bibnamefont {Cerjan}}, \bibinfo {author} {\bibfnamefont {C.~W.}\ \bibnamefont {Hsu}},\ and\ \bibinfo {author} {\bibfnamefont {M.~C.}\ \bibnamefont {Rechtsman}},\ }\bibfield  {title} {\bibinfo {title} {Bound {{States}} in the {{Continuum}} through {{Environmental Design}}},\ }\href {https://doi.org/10/gkr878} {\bibfield  {journal} {\bibinfo  {journal} {Phys. Rev. Lett.}\ }\textbf {\bibinfo {volume} {123}},\ \bibinfo {pages} {023902} (\bibinfo {year} {2019})}\BibitemShut {NoStop}%
\bibitem [{\citenamefont {Cerjan}\ \emph {et~al.}(2020)\citenamefont {Cerjan}, \citenamefont {J{\"u}rgensen}, \citenamefont {Benalcazar}, \citenamefont {Mukherjee},\ and\ \citenamefont {Rechtsman}}]{Cerjan2020}%
  \BibitemOpen
  \bibfield  {author} {\bibinfo {author} {\bibfnamefont {A.}~\bibnamefont {Cerjan}}, \bibinfo {author} {\bibfnamefont {M.}~\bibnamefont {J{\"u}rgensen}}, \bibinfo {author} {\bibfnamefont {W.~A.}\ \bibnamefont {Benalcazar}}, \bibinfo {author} {\bibfnamefont {S.}~\bibnamefont {Mukherjee}},\ and\ \bibinfo {author} {\bibfnamefont {M.~C.}\ \bibnamefont {Rechtsman}},\ }\bibfield  {title} {\bibinfo {title} {Observation of a {{Higher-Order Topological Bound State}} in the {{Continuum}}},\ }\href {https://doi.org/10.1103/PhysRevLett.125.213901} {\bibfield  {journal} {\bibinfo  {journal} {Phys. Rev. Lett.}\ }\textbf {\bibinfo {volume} {125}},\ \bibinfo {pages} {213901} (\bibinfo {year} {2020})}\BibitemShut {NoStop}%
\bibitem [{\citenamefont {Huang}\ \emph {et~al.}(2024)\citenamefont {Huang}, \citenamefont {Ke}, \citenamefont {Zhong}, \citenamefont {Kivshar},\ and\ \citenamefont {Lee}}]{Huang2024}%
  \BibitemOpen
  \bibfield  {author} {\bibinfo {author} {\bibfnamefont {B.}~\bibnamefont {Huang}}, \bibinfo {author} {\bibfnamefont {Y.}~\bibnamefont {Ke}}, \bibinfo {author} {\bibfnamefont {H.}~\bibnamefont {Zhong}}, \bibinfo {author} {\bibfnamefont {Y.~S.}\ \bibnamefont {Kivshar}},\ and\ \bibinfo {author} {\bibfnamefont {C.}~\bibnamefont {Lee}},\ }\bibfield  {title} {\bibinfo {title} {Interaction-{{Induced Multiparticle Bound States}} in the {{Continuum}}},\ }\href {https://doi.org/10/g839jz} {\bibfield  {journal} {\bibinfo  {journal} {Phys. Rev. Lett.}\ }\textbf {\bibinfo {volume} {133}},\ \bibinfo {pages} {140202} (\bibinfo {year} {2024})}\BibitemShut {NoStop}%
\bibitem [{\citenamefont {Qin}\ \emph {et~al.}(2024)\citenamefont {Qin}, \citenamefont {Chen}, \citenamefont {Zhang}, \citenamefont {Zhang}, \citenamefont {Pan}, \citenamefont {Li}, \citenamefont {Shi}, \citenamefont {Zi},\ and\ \citenamefont {Zhang}}]{Qin2024}%
  \BibitemOpen
  \bibfield  {author} {\bibinfo {author} {\bibfnamefont {H.}~\bibnamefont {Qin}}, \bibinfo {author} {\bibfnamefont {S.}~\bibnamefont {Chen}}, \bibinfo {author} {\bibfnamefont {W.}~\bibnamefont {Zhang}}, \bibinfo {author} {\bibfnamefont {H.}~\bibnamefont {Zhang}}, \bibinfo {author} {\bibfnamefont {R.}~\bibnamefont {Pan}}, \bibinfo {author} {\bibfnamefont {J.}~\bibnamefont {Li}}, \bibinfo {author} {\bibfnamefont {L.}~\bibnamefont {Shi}}, \bibinfo {author} {\bibfnamefont {J.}~\bibnamefont {Zi}},\ and\ \bibinfo {author} {\bibfnamefont {X.}~\bibnamefont {Zhang}},\ }\bibfield  {title} {\bibinfo {title} {Optical moir\'e bound states in the continuum},\ }\href {https://doi.org/10/g8xkv4} {\bibfield  {journal} {\bibinfo  {journal} {Nat. Commun.}\ }\textbf {\bibinfo {volume} {15}},\ \bibinfo {pages} {9080} (\bibinfo {year} {2024})}\BibitemShut {NoStop}%
\bibitem [{\citenamefont {Qian}\ \emph {et~al.}(2024)\citenamefont {Qian}, \citenamefont {Zhang}, \citenamefont {Sun},\ and\ \citenamefont {Zhang}}]{Qian2024}%
  \BibitemOpen
  \bibfield  {author} {\bibinfo {author} {\bibfnamefont {L.}~\bibnamefont {Qian}}, \bibinfo {author} {\bibfnamefont {W.}~\bibnamefont {Zhang}}, \bibinfo {author} {\bibfnamefont {H.}~\bibnamefont {Sun}},\ and\ \bibinfo {author} {\bibfnamefont {X.}~\bibnamefont {Zhang}},\ }\bibfield  {title} {\bibinfo {title} {Non-{{Abelian Topological Bound States}} in the {{Continuum}}},\ }\href {https://doi.org/10/g82x78} {\bibfield  {journal} {\bibinfo  {journal} {Phys. Rev. Lett.}\ }\textbf {\bibinfo {volume} {132}},\ \bibinfo {pages} {046601} (\bibinfo {year} {2024})}\BibitemShut {NoStop}%
\bibitem [{Note2()}]{Note2}%
  \BibitemOpen
  \bibinfo {note} {Note the parity symmetry is broken in this case~\protect ~\cite {Liu2018b} and neither the anti-unitary time reversal symmetry~\protect ~\cite {Lange2017a} nor the chiral symmetry (for $U=0$)~\protect ~\cite {Theel2025} of the AHM, block-diagonalize the Hamiltonian to allow for symmetry-decoupled subspaces. The qBICs are also present when the chiral symmetry is broken by any finite on-site interaction $U$ (also see Fig.~\ref {fig:three-body-spectrum-U4-AHM}).}\BibitemShut {Stop}%
\bibitem [{\citenamefont {Greschner}\ and\ \citenamefont {{Heidrich-Meisner}}(2018)}]{Greschner2018}%
  \BibitemOpen
  \bibfield  {author} {\bibinfo {author} {\bibfnamefont {S.}~\bibnamefont {Greschner}}\ and\ \bibinfo {author} {\bibfnamefont {F.}~\bibnamefont {{Heidrich-Meisner}}},\ }\bibfield  {title} {\bibinfo {title} {Quantum phases of strongly interacting bosons on a two-leg {{Haldane}} ladder},\ }\href {https://doi.org/10.1103/PhysRevA.97.033619} {\bibfield  {journal} {\bibinfo  {journal} {Phys. Rev. A}\ }\textbf {\bibinfo {volume} {97}},\ \bibinfo {pages} {033619} (\bibinfo {year} {2018})}\BibitemShut {NoStop}%
\bibitem [{Note3()}]{Note3}%
  \BibitemOpen
  \bibinfo {note} {We note that two-particle anyonic BICs in a 1D lattice were previously discussed in~\cite {Zhang2023a}. However, as pointed out in~\cite {Zheng2024}, the analysis of Ref.~\cite {Zhang2023a} relies on an overcomplete, non-orthogonal basis for the two-anyon Hilbert space. This leads to non-physical redundant eigenstates whose wave function amplitudes violate the anyonic commutation relations~\cite {Zheng2024}, complicating the identification of genuine BIC there. Our analysis avoids this issue entirely by working directly in the \protect \emph {bosonic} relative coordinate space representation~\protect \eqref {eq:2body-AHM-repr-hybrid}, which is formulated in terms of the orthogonal and complete basis $\protect \ket {q,m}$ so that only physical eigenstates are obtained ab initio. This representation also makes the BIC condition analytically transparent: a perfect two-body BIC arises exactly when $\cos \protect \!\left ([q+\theta ]/2\right )=0$, as derived below.}\BibitemShut {Stop}%
\bibitem [{\citenamefont {F{\"o}lling}\ \emph {et~al.}(2007)\citenamefont {F{\"o}lling}, \citenamefont {Trotzky}, \citenamefont {Cheinet}, \citenamefont {Feld}, \citenamefont {Saers}, \citenamefont {Widera}, \citenamefont {M{\"u}ller},\ and\ \citenamefont {Bloch}}]{Folling2007}%
  \BibitemOpen
  \bibfield  {author} {\bibinfo {author} {\bibfnamefont {S.}~\bibnamefont {F{\"o}lling}}, \bibinfo {author} {\bibfnamefont {S.}~\bibnamefont {Trotzky}}, \bibinfo {author} {\bibfnamefont {P.}~\bibnamefont {Cheinet}}, \bibinfo {author} {\bibfnamefont {M.}~\bibnamefont {Feld}}, \bibinfo {author} {\bibfnamefont {R.}~\bibnamefont {Saers}}, \bibinfo {author} {\bibfnamefont {A.}~\bibnamefont {Widera}}, \bibinfo {author} {\bibfnamefont {T.}~\bibnamefont {M{\"u}ller}},\ and\ \bibinfo {author} {\bibfnamefont {I.}~\bibnamefont {Bloch}},\ }\bibfield  {title} {\bibinfo {title} {Direct observation of second-order atom tunnelling},\ }\href {https://doi.org/10/dr38s4} {\bibfield  {journal} {\bibinfo  {journal} {Nature}\ }\textbf {\bibinfo {volume} {448}},\ \bibinfo {pages} {1029} (\bibinfo {year} {2007})}\BibitemShut {NoStop}%
\bibitem [{\citenamefont {Petrosyan}\ \emph {et~al.}(2007)\citenamefont {Petrosyan}, \citenamefont {Schmidt}, \citenamefont {Anglin},\ and\ \citenamefont {Fleischhauer}}]{Petrosyan2007}%
  \BibitemOpen
  \bibfield  {author} {\bibinfo {author} {\bibfnamefont {D.}~\bibnamefont {Petrosyan}}, \bibinfo {author} {\bibfnamefont {B.}~\bibnamefont {Schmidt}}, \bibinfo {author} {\bibfnamefont {J.~R.}\ \bibnamefont {Anglin}},\ and\ \bibinfo {author} {\bibfnamefont {M.}~\bibnamefont {Fleischhauer}},\ }\bibfield  {title} {\bibinfo {title} {Quantum liquid of repulsively bound pairs of particles in a lattice},\ }\href {https://doi.org/10.1103/PhysRevA.76.033606} {\bibfield  {journal} {\bibinfo  {journal} {Phys. Rev. A}\ }\textbf {\bibinfo {volume} {76}},\ \bibinfo {pages} {033606} (\bibinfo {year} {2007})}\BibitemShut {NoStop}%
\bibitem [{\citenamefont {Wang}\ \emph {et~al.}(2008)\citenamefont {Wang}, \citenamefont {Hao},\ and\ \citenamefont {Chen}}]{Wang2008}%
  \BibitemOpen
  \bibfield  {author} {\bibinfo {author} {\bibfnamefont {L.}~\bibnamefont {Wang}}, \bibinfo {author} {\bibfnamefont {Y.}~\bibnamefont {Hao}},\ and\ \bibinfo {author} {\bibfnamefont {S.}~\bibnamefont {Chen}},\ }\bibfield  {title} {\bibinfo {title} {Quantum dynamics of repulsively bound atom pairs in the {{Bose-Hubbard}} model},\ }\href {https://doi.org/10.1140/epjd/e2008-00077-3} {\bibfield  {journal} {\bibinfo  {journal} {Eur. Phys. J. D}\ }\textbf {\bibinfo {volume} {48}},\ \bibinfo {pages} {229} (\bibinfo {year} {2008})}\BibitemShut {NoStop}%
\bibitem [{\citenamefont {Valiente}\ and\ \citenamefont {Petrosyan}(2008{\natexlab{b}})}]{Valiente2008a}%
  \BibitemOpen
  \bibfield  {author} {\bibinfo {author} {\bibfnamefont {M.}~\bibnamefont {Valiente}}\ and\ \bibinfo {author} {\bibfnamefont {D.}~\bibnamefont {Petrosyan}},\ }\bibfield  {title} {\bibinfo {title} {Quantum dynamics of one and two bosonic atoms in a combined tight-binding periodic and weak parabolic potential},\ }\href {https://doi.org/10.1209/0295-5075/83/30007} {\bibfield  {journal} {\bibinfo  {journal} {Europhys. Lett.}\ }\textbf {\bibinfo {volume} {83}},\ \bibinfo {pages} {30007} (\bibinfo {year} {2008}{\natexlab{b}})}\BibitemShut {NoStop}%
\bibitem [{\citenamefont {Valiente}\ and\ \citenamefont {Petrosyan}(2009)}]{Valiente2009}%
  \BibitemOpen
  \bibfield  {author} {\bibinfo {author} {\bibfnamefont {M.}~\bibnamefont {Valiente}}\ and\ \bibinfo {author} {\bibfnamefont {D.}~\bibnamefont {Petrosyan}},\ }\bibfield  {title} {\bibinfo {title} {Scattering resonances and two-particle bound states of the extended {{Hubbard}} model},\ }\href {https://doi.org/10/d8xbvg} {\bibfield  {journal} {\bibinfo  {journal} {J. Phys. B At. Mol. Opt. Phys.}\ }\textbf {\bibinfo {volume} {42}},\ \bibinfo {pages} {121001} (\bibinfo {year} {2009})}\BibitemShut {NoStop}%
\bibitem [{\citenamefont {Valiente}(2010)}]{Valiente2010}%
  \BibitemOpen
  \bibfield  {author} {\bibinfo {author} {\bibfnamefont {M.}~\bibnamefont {Valiente}},\ }\bibfield  {title} {\bibinfo {title} {Lattice two-body problem with arbitrary finite-range interactions},\ }\href {https://doi.org/10.1103/PhysRevA.81.042102} {\bibfield  {journal} {\bibinfo  {journal} {Phys. Rev. A}\ }\textbf {\bibinfo {volume} {81}},\ \bibinfo {pages} {042102} (\bibinfo {year} {2010})}\BibitemShut {NoStop}%
\bibitem [{\citenamefont {Ronzheimer}\ \emph {et~al.}(2013)\citenamefont {Ronzheimer}, \citenamefont {Schreiber}, \citenamefont {Braun}, \citenamefont {Hodgman}, \citenamefont {Langer}, \citenamefont {McCulloch}, \citenamefont {{Heidrich-Meisner}}, \citenamefont {Bloch},\ and\ \citenamefont {Schneider}}]{Ronzheimer2013}%
  \BibitemOpen
  \bibfield  {author} {\bibinfo {author} {\bibfnamefont {J.~P.}\ \bibnamefont {Ronzheimer}}, \bibinfo {author} {\bibfnamefont {M.}~\bibnamefont {Schreiber}}, \bibinfo {author} {\bibfnamefont {S.}~\bibnamefont {Braun}}, \bibinfo {author} {\bibfnamefont {S.~S.}\ \bibnamefont {Hodgman}}, \bibinfo {author} {\bibfnamefont {S.}~\bibnamefont {Langer}}, \bibinfo {author} {\bibfnamefont {I.~P.}\ \bibnamefont {McCulloch}}, \bibinfo {author} {\bibfnamefont {F.}~\bibnamefont {{Heidrich-Meisner}}}, \bibinfo {author} {\bibfnamefont {I.}~\bibnamefont {Bloch}},\ and\ \bibinfo {author} {\bibfnamefont {U.}~\bibnamefont {Schneider}},\ }\bibfield  {title} {\bibinfo {title} {Expansion {{Dynamics}} of {{Interacting Bosons}} in {{Homogeneous Lattices}} in {{One}} and {{Two Dimensions}}},\ }\href {https://doi.org/10.1103/PhysRevLett.110.205301} {\bibfield  {journal} {\bibinfo  {journal} {Phys. Rev. Lett.}\ }\textbf {\bibinfo {volume} {110}},\ \bibinfo {pages} {205301} (\bibinfo {year} {2013})}\BibitemShut {NoStop}%
\bibitem [{\citenamefont {Boschi}\ \emph {et~al.}(2014)\citenamefont {Boschi}, \citenamefont {Ercolessi}, \citenamefont {Ferrari}, \citenamefont {Naldesi}, \citenamefont {Ortolani},\ and\ \citenamefont {Taddia}}]{Boschi2014}%
  \BibitemOpen
  \bibfield  {author} {\bibinfo {author} {\bibfnamefont {C.~D.~E.}\ \bibnamefont {Boschi}}, \bibinfo {author} {\bibfnamefont {E.}~\bibnamefont {Ercolessi}}, \bibinfo {author} {\bibfnamefont {L.}~\bibnamefont {Ferrari}}, \bibinfo {author} {\bibfnamefont {P.}~\bibnamefont {Naldesi}}, \bibinfo {author} {\bibfnamefont {F.}~\bibnamefont {Ortolani}},\ and\ \bibinfo {author} {\bibfnamefont {L.}~\bibnamefont {Taddia}},\ }\bibfield  {title} {\bibinfo {title} {Bound states and expansion dynamics of interacting bosons on a one-dimensional lattice},\ }\href {https://doi.org/10.1103/PhysRevA.90.043606} {\bibfield  {journal} {\bibinfo  {journal} {Phys. Rev. A}\ }\textbf {\bibinfo {volume} {90}},\ \bibinfo {pages} {043606} (\bibinfo {year} {2014})}\BibitemShut {NoStop}%
\bibitem [{\citenamefont {Weckesser}\ \emph {et~al.}(2025)\citenamefont {Weckesser}, \citenamefont {Srakaew}, \citenamefont {Blatz}, \citenamefont {Wei}, \citenamefont {Adler}, \citenamefont {Agrawal}, \citenamefont {Bohrdt}, \citenamefont {Bloch},\ and\ \citenamefont {Zeiher}}]{Weckesser2025}%
  \BibitemOpen
  \bibfield  {author} {\bibinfo {author} {\bibfnamefont {P.}~\bibnamefont {Weckesser}}, \bibinfo {author} {\bibfnamefont {K.}~\bibnamefont {Srakaew}}, \bibinfo {author} {\bibfnamefont {T.}~\bibnamefont {Blatz}}, \bibinfo {author} {\bibfnamefont {D.}~\bibnamefont {Wei}}, \bibinfo {author} {\bibfnamefont {D.}~\bibnamefont {Adler}}, \bibinfo {author} {\bibfnamefont {S.}~\bibnamefont {Agrawal}}, \bibinfo {author} {\bibfnamefont {A.}~\bibnamefont {Bohrdt}}, \bibinfo {author} {\bibfnamefont {I.}~\bibnamefont {Bloch}},\ and\ \bibinfo {author} {\bibfnamefont {J.}~\bibnamefont {Zeiher}},\ }\bibfield  {title} {\bibinfo {title} {Realization of a {{Rydberg-dressed}} extended {{Bose-Hubbard}} model},\ }\href {https://doi.org/10.1126/science.adq7082} {\bibfield  {journal} {\bibinfo  {journal} {Science}\ }\textbf {\bibinfo {volume} {390}},\ \bibinfo {pages} {849} (\bibinfo {year} {2025})}\BibitemShut {NoStop}%
\bibitem [{Note4()}]{Note4}%
  \BibitemOpen
  \bibinfo {note} {The three-particle coincidences and the processes connecting to these configurations need not be excluded. Using the scheme proposed and utilized in Refs.~\cite {Cardarelli2016,Kwan2024,Bakkali-Hassani2026}, one would need to add two additional frequencies $\omega _{1/2}=\Delta \pm 2(U_0-U)$ with a required phase of $2\theta $ and $0$ for the process $\protect \ket {120}\to \protect \ket {030}$ and $\protect \ket {030}\to \protect \ket {021}$, respectively~[see Fig.~\ref {fig:three-body-spectrum-effective-AHM-U0-Binding-mechanism}(a)], where $\Delta $ is the energy tilt between adjacent sites, $U_0$ the bare on-site interaction strength, and $U$ the desired effective on-site interaction strength. Note, however, that the Peierls phases for the processes connecting configurations $\protect \ket {012} \leftrightarrow \protect \ket {021}$ still need to be excluded, as their transition frequencies are degenerate with the transition frequency $\Delta $ of a single particle hopping onto an unoccupied site.}\BibitemShut {Stop}%
\bibitem [{\citenamefont {Turner}\ \emph {et~al.}(2018)\citenamefont {Turner}, \citenamefont {Michailidis}, \citenamefont {Abanin}, \citenamefont {Serbyn},\ and\ \citenamefont {Papi{\'c}}}]{Turner2018a}%
  \BibitemOpen
  \bibfield  {author} {\bibinfo {author} {\bibfnamefont {C.~J.}\ \bibnamefont {Turner}}, \bibinfo {author} {\bibfnamefont {A.~A.}\ \bibnamefont {Michailidis}}, \bibinfo {author} {\bibfnamefont {D.~A.}\ \bibnamefont {Abanin}}, \bibinfo {author} {\bibfnamefont {M.}~\bibnamefont {Serbyn}},\ and\ \bibinfo {author} {\bibfnamefont {Z.}~\bibnamefont {Papi{\'c}}},\ }\bibfield  {title} {\bibinfo {title} {Weak ergodicity breaking from quantum many-body scars},\ }\href {https://doi.org/10/gdxjn6} {\bibfield  {journal} {\bibinfo  {journal} {Nat. Phys.}\ }\textbf {\bibinfo {volume} {14}},\ \bibinfo {pages} {745} (\bibinfo {year} {2018})}\BibitemShut {NoStop}%
\bibitem [{\citenamefont {Chandran}\ \emph {et~al.}(2023)\citenamefont {Chandran}, \citenamefont {Iadecola}, \citenamefont {Khemani},\ and\ \citenamefont {Moessner}}]{Chandran2023}%
  \BibitemOpen
  \bibfield  {author} {\bibinfo {author} {\bibfnamefont {A.}~\bibnamefont {Chandran}}, \bibinfo {author} {\bibfnamefont {T.}~\bibnamefont {Iadecola}}, \bibinfo {author} {\bibfnamefont {V.}~\bibnamefont {Khemani}},\ and\ \bibinfo {author} {\bibfnamefont {R.}~\bibnamefont {Moessner}},\ }\bibfield  {title} {\bibinfo {title} {Quantum {{Many-Body Scars}}: {{A Quasiparticle Perspective}}},\ }\href {https://doi.org/10/g82vrb} {\bibfield  {journal} {\bibinfo  {journal} {Annu. Rev. Condens. Matter Phys.}\ }\textbf {\bibinfo {volume} {14}},\ \bibinfo {pages} {443} (\bibinfo {year} {2023})}\BibitemShut {NoStop}%
\bibitem [{\citenamefont {Moudgalya}\ \emph {et~al.}(2022)\citenamefont {Moudgalya}, \citenamefont {Bernevig},\ and\ \citenamefont {Regnault}}]{Moudgalya2022}%
  \BibitemOpen
  \bibfield  {author} {\bibinfo {author} {\bibfnamefont {S.}~\bibnamefont {Moudgalya}}, \bibinfo {author} {\bibfnamefont {B.~A.}\ \bibnamefont {Bernevig}},\ and\ \bibinfo {author} {\bibfnamefont {N.}~\bibnamefont {Regnault}},\ }\bibfield  {title} {\bibinfo {title} {Quantum many-body scars and {{Hilbert}} space fragmentation: A review of exact results},\ }\href {https://doi.org/10.1088/1361-6633/ac73a0} {\bibfield  {journal} {\bibinfo  {journal} {Rep. Prog. Phys.}\ }\textbf {\bibinfo {volume} {85}},\ \bibinfo {pages} {086501} (\bibinfo {year} {2022})}\BibitemShut {NoStop}%
\bibitem [{\citenamefont {Hudomal}\ \emph {et~al.}(2020)\citenamefont {Hudomal}, \citenamefont {Vasi{\'c}}, \citenamefont {Regnault},\ and\ \citenamefont {Papi{\'c}}}]{Hudomal2020}%
  \BibitemOpen
  \bibfield  {author} {\bibinfo {author} {\bibfnamefont {A.}~\bibnamefont {Hudomal}}, \bibinfo {author} {\bibfnamefont {I.}~\bibnamefont {Vasi{\'c}}}, \bibinfo {author} {\bibfnamefont {N.}~\bibnamefont {Regnault}},\ and\ \bibinfo {author} {\bibfnamefont {Z.}~\bibnamefont {Papi{\'c}}},\ }\bibfield  {title} {\bibinfo {title} {Quantum scars of bosons with correlated hopping},\ }\href {https://doi.org/10/ghrjwt} {\bibfield  {journal} {\bibinfo  {journal} {Commun. Phys.}\ }\textbf {\bibinfo {volume} {3}},\ \bibinfo {pages} {1} (\bibinfo {year} {2020})}\BibitemShut {NoStop}%
\bibitem [{\citenamefont {Serbyn}\ \emph {et~al.}(2021)\citenamefont {Serbyn}, \citenamefont {Abanin},\ and\ \citenamefont {Papi{\'c}}}]{Serbyn2021}%
  \BibitemOpen
  \bibfield  {author} {\bibinfo {author} {\bibfnamefont {M.}~\bibnamefont {Serbyn}}, \bibinfo {author} {\bibfnamefont {D.~A.}\ \bibnamefont {Abanin}},\ and\ \bibinfo {author} {\bibfnamefont {Z.}~\bibnamefont {Papi{\'c}}},\ }\bibfield  {title} {\bibinfo {title} {Quantum many-body scars and weak breaking of ergodicity},\ }\href {https://doi.org/10/gj8f65} {\bibfield  {journal} {\bibinfo  {journal} {Nat. Phys.}\ }\textbf {\bibinfo {volume} {17}},\ \bibinfo {pages} {675} (\bibinfo {year} {2021})}\BibitemShut {NoStop}%
\bibitem [{\citenamefont {Banerjee}\ and\ \citenamefont {Sen}(2021)}]{Banerjee2021}%
  \BibitemOpen
  \bibfield  {author} {\bibinfo {author} {\bibfnamefont {D.}~\bibnamefont {Banerjee}}\ and\ \bibinfo {author} {\bibfnamefont {A.}~\bibnamefont {Sen}},\ }\bibfield  {title} {\bibinfo {title} {Quantum {{Scars}} from {{Zero Modes}} in an {{Abelian Lattice Gauge Theory}} on {{Ladders}}},\ }\href {https://doi.org/10.1103/PhysRevLett.126.220601} {\bibfield  {journal} {\bibinfo  {journal} {Phys. Rev. Lett.}\ }\textbf {\bibinfo {volume} {126}},\ \bibinfo {pages} {220601} (\bibinfo {year} {2021})}\BibitemShut {NoStop}%
\bibitem [{\citenamefont {Aditya}\ \emph {et~al.}(2024)\citenamefont {Aditya}, \citenamefont {Dhar},\ and\ \citenamefont {Sen}}]{Aditya2024}%
  \BibitemOpen
  \bibfield  {author} {\bibinfo {author} {\bibfnamefont {S.}~\bibnamefont {Aditya}}, \bibinfo {author} {\bibfnamefont {D.}~\bibnamefont {Dhar}},\ and\ \bibinfo {author} {\bibfnamefont {D.}~\bibnamefont {Sen}},\ }\bibfield  {title} {\bibinfo {title} {Subspace-restricted thermalization in a correlated-hopping model with strong {{Hilbert}} space fragmentation characterized by irreducible strings},\ }\href {https://doi.org/10.1103/PhysRevB.110.045418} {\bibfield  {journal} {\bibinfo  {journal} {Phys. Rev. B}\ }\textbf {\bibinfo {volume} {110}},\ \bibinfo {pages} {045418} (\bibinfo {year} {2024})}\BibitemShut {NoStop}%
\bibitem [{\citenamefont {Surace}\ \emph {et~al.}(2021)\citenamefont {Surace}, \citenamefont {Votto}, \citenamefont {Lazo}, \citenamefont {Silva}, \citenamefont {Dalmonte},\ and\ \citenamefont {Giudici}}]{Surace2021}%
  \BibitemOpen
  \bibfield  {author} {\bibinfo {author} {\bibfnamefont {F.~M.}\ \bibnamefont {Surace}}, \bibinfo {author} {\bibfnamefont {M.}~\bibnamefont {Votto}}, \bibinfo {author} {\bibfnamefont {E.~G.}\ \bibnamefont {Lazo}}, \bibinfo {author} {\bibfnamefont {A.}~\bibnamefont {Silva}}, \bibinfo {author} {\bibfnamefont {M.}~\bibnamefont {Dalmonte}},\ and\ \bibinfo {author} {\bibfnamefont {G.}~\bibnamefont {Giudici}},\ }\bibfield  {title} {\bibinfo {title} {Exact many-body scars and their stability in constrained quantum chains},\ }\href {https://doi.org/10/g6tbk9} {\bibfield  {journal} {\bibinfo  {journal} {Phys. Rev. B}\ }\textbf {\bibinfo {volume} {103}},\ \bibinfo {pages} {104302} (\bibinfo {year} {2021})}\BibitemShut {NoStop}%
\bibitem [{\citenamefont {Singh}\ \emph {et~al.}(2021)\citenamefont {Singh}, \citenamefont {Ware}, \citenamefont {Vasseur},\ and\ \citenamefont {Friedman}}]{Singh2021}%
  \BibitemOpen
  \bibfield  {author} {\bibinfo {author} {\bibfnamefont {H.}~\bibnamefont {Singh}}, \bibinfo {author} {\bibfnamefont {B.~A.}\ \bibnamefont {Ware}}, \bibinfo {author} {\bibfnamefont {R.}~\bibnamefont {Vasseur}},\ and\ \bibinfo {author} {\bibfnamefont {A.~J.}\ \bibnamefont {Friedman}},\ }\bibfield  {title} {\bibinfo {title} {Subdiffusion and {{Many-Body Quantum Chaos}} with {{Kinetic Constraints}}},\ }\href {https://doi.org/10.1103/PhysRevLett.127.230602} {\bibfield  {journal} {\bibinfo  {journal} {Phys. Rev. Lett.}\ }\textbf {\bibinfo {volume} {127}},\ \bibinfo {pages} {230602} (\bibinfo {year} {2021})}\BibitemShut {NoStop}%
\bibitem [{\citenamefont {Karle}\ \emph {et~al.}(2021)\citenamefont {Karle}, \citenamefont {Serbyn},\ and\ \citenamefont {Michailidis}}]{Karle2021}%
  \BibitemOpen
  \bibfield  {author} {\bibinfo {author} {\bibfnamefont {V.}~\bibnamefont {Karle}}, \bibinfo {author} {\bibfnamefont {M.}~\bibnamefont {Serbyn}},\ and\ \bibinfo {author} {\bibfnamefont {A.~A.}\ \bibnamefont {Michailidis}},\ }\bibfield  {title} {\bibinfo {title} {Area-{{Law Entangled Eigenstates}} from {{Nullspaces}} of {{Local Hamiltonians}}},\ }\href {https://doi.org/10.1103/PhysRevLett.127.060602} {\bibfield  {journal} {\bibinfo  {journal} {Phys. Rev. Lett.}\ }\textbf {\bibinfo {volume} {127}},\ \bibinfo {pages} {060602} (\bibinfo {year} {2021})}\BibitemShut {NoStop}%
\bibitem [{\citenamefont {Jonay}\ and\ \citenamefont {Pollmann}(2026)}]{Jonay2026}%
  \BibitemOpen
  \bibfield  {author} {\bibinfo {author} {\bibfnamefont {C.}~\bibnamefont {Jonay}}\ and\ \bibinfo {author} {\bibfnamefont {F.}~\bibnamefont {Pollmann}},\ }\bibfield  {title} {\bibinfo {title} {Localized {{Fock}} space cages in kinetically constrained models},\ }\href {https://doi.org/10.1103/wz33-vczt} {\bibfield  {journal} {\bibinfo  {journal} {Phys. Rev. B}\ }\textbf {\bibinfo {volume} {113}},\ \bibinfo {pages} {134313} (\bibinfo {year} {2026})}\BibitemShut {NoStop}%
\bibitem [{\citenamefont {Tan}\ and\ \citenamefont {Huang}(2025)}]{Tan2025}%
  \BibitemOpen
  \bibfield  {author} {\bibinfo {author} {\bibfnamefont {T.-L.}\ \bibnamefont {Tan}}\ and\ \bibinfo {author} {\bibfnamefont {Y.-P.}\ \bibnamefont {Huang}},\ }\bibinfo {title} {Interference-caged quantum many-body scars: The {{Fock}} space topological localization and interference zeros} (\bibinfo {year} {2025}),\ \Eprint {https://arxiv.org/abs/2504.07780} {arXiv:2504.07780} \BibitemShut {NoStop}%
\bibitem [{\citenamefont {Nicolau}\ \emph {et~al.}(2026)\citenamefont {Nicolau}, \citenamefont {Ljubotina},\ and\ \citenamefont {Serbyn}}]{Nicolau2026}%
  \BibitemOpen
  \bibfield  {author} {\bibinfo {author} {\bibfnamefont {E.}~\bibnamefont {Nicolau}}, \bibinfo {author} {\bibfnamefont {M.}~\bibnamefont {Ljubotina}},\ and\ \bibinfo {author} {\bibfnamefont {M.}~\bibnamefont {Serbyn}},\ }\bibfield  {title} {\bibinfo {title} {Fragmentation, {{Zero Modes}}, and {{Collective Bound States}} in {{Constrained Models}}},\ }\href {https://doi.org/10.1103/sl79-1xgb} {\bibfield  {journal} {\bibinfo  {journal} {PRX Quantum}\ }\textbf {\bibinfo {volume} {7}},\ \bibinfo {pages} {010352} (\bibinfo {year} {2026})}\BibitemShut {NoStop}%
\bibitem [{\citenamefont {Mohapatra}\ \emph {et~al.}(2026)\citenamefont {Mohapatra}, \citenamefont {Moudgalya},\ and\ \citenamefont {Balram}}]{Mohapatra2026}%
  \BibitemOpen
  \bibfield  {author} {\bibinfo {author} {\bibfnamefont {S.}~\bibnamefont {Mohapatra}}, \bibinfo {author} {\bibfnamefont {S.}~\bibnamefont {Moudgalya}},\ and\ \bibinfo {author} {\bibfnamefont {A.~C.}\ \bibnamefont {Balram}},\ }\bibfield  {title} {\bibinfo {title} {Additional quantum many-body scars of the spin-1 \${{XY}}\$ model with {{Fock-space}} cages and commutant algebras},\ }\href {https://doi.org/10.1103/4tv9-q7g7} {\bibfield  {journal} {\bibinfo  {journal} {Phys. Rev. B}\ }\textbf {\bibinfo {volume} {113}},\ \bibinfo {pages} {054310} (\bibinfo {year} {2026})}\BibitemShut {NoStop}%
\bibitem [{\citenamefont {Dupont}\ \emph {et~al.}(2026)\citenamefont {Dupont}, \citenamefont {Peaudecerf}, \citenamefont {{Gu{\'e}ry-Odelin}}, \citenamefont {Lemari{\'e}}, \citenamefont {Georgeot}, \citenamefont {Miniatura},\ and\ \citenamefont {Goldman}}]{Dupont2026}%
  \BibitemOpen
  \bibfield  {author} {\bibinfo {author} {\bibfnamefont {N.}~\bibnamefont {Dupont}}, \bibinfo {author} {\bibfnamefont {B.}~\bibnamefont {Peaudecerf}}, \bibinfo {author} {\bibfnamefont {D.}~\bibnamefont {{Gu{\'e}ry-Odelin}}}, \bibinfo {author} {\bibfnamefont {G.}~\bibnamefont {Lemari{\'e}}}, \bibinfo {author} {\bibfnamefont {B.}~\bibnamefont {Georgeot}}, \bibinfo {author} {\bibfnamefont {C.}~\bibnamefont {Miniatura}},\ and\ \bibinfo {author} {\bibfnamefont {N.}~\bibnamefont {Goldman}},\ }\bibinfo {title} {Many-body dynamical localization in {{Fock}} space} (\bibinfo {year} {2026}),\ \Eprint {https://arxiv.org/abs/2604.09224} {arXiv:2604.09224} \BibitemShut {NoStop}%
\bibitem [{\citenamefont {{Ben-Ami}}\ \emph {et~al.}(2025)\citenamefont {{Ben-Ami}}, \citenamefont {Heyl},\ and\ \citenamefont {Moessner}}]{Ben-Ami2025}%
  \BibitemOpen
  \bibfield  {author} {\bibinfo {author} {\bibfnamefont {T.}~\bibnamefont {{Ben-Ami}}}, \bibinfo {author} {\bibfnamefont {M.}~\bibnamefont {Heyl}},\ and\ \bibinfo {author} {\bibfnamefont {R.}~\bibnamefont {Moessner}},\ }\bibinfo {title} {Many-body cages: Disorder-free glassiness from flat bands in {{Fock}} space, and many-body {{Rabi}} oscillations} (\bibinfo {year} {2025}),\ \Eprint {https://arxiv.org/abs/2504.13086} {arXiv:2504.13086} \BibitemShut {NoStop}%
\bibitem [{\citenamefont {{Ben-Ami}}\ \emph {et~al.}(2026)\citenamefont {{Ben-Ami}}, \citenamefont {Moessner},\ and\ \citenamefont {Heyl}}]{Ben-Ami2026}%
  \BibitemOpen
  \bibfield  {author} {\bibinfo {author} {\bibfnamefont {T.}~\bibnamefont {{Ben-Ami}}}, \bibinfo {author} {\bibfnamefont {R.}~\bibnamefont {Moessner}},\ and\ \bibinfo {author} {\bibfnamefont {M.}~\bibnamefont {Heyl}},\ }\bibinfo {title} {Floquet {{Many-Body Cages}}} (\bibinfo {year} {2026}),\ \Eprint {https://arxiv.org/abs/2604.13027} {arXiv:2604.13027} \BibitemShut {NoStop}%
\bibitem [{\citenamefont {Faugno}\ \emph {et~al.}(2025)\citenamefont {Faugno}, \citenamefont {Katsura},\ and\ \citenamefont {Ozawa}}]{Faugno2025}%
  \BibitemOpen
  \bibfield  {author} {\bibinfo {author} {\bibfnamefont {W.~N.}\ \bibnamefont {Faugno}}, \bibinfo {author} {\bibfnamefont {H.}~\bibnamefont {Katsura}},\ and\ \bibinfo {author} {\bibfnamefont {T.}~\bibnamefont {Ozawa}},\ }\bibinfo {title} {Non-equilibirum physics of density-difference dependent {{Hamiltonian}}: {{Quantum Scarring}} from {{Emergent Chiral Symmetry}}} (\bibinfo {year} {2025}),\ \Eprint {https://arxiv.org/abs/2503.05252} {arXiv:2503.05252} \BibitemShut {NoStop}%
\bibitem [{\citenamefont {Faugno}\ and\ \citenamefont {Ozawa}(2022)}]{Faugno2022}%
  \BibitemOpen
  \bibfield  {author} {\bibinfo {author} {\bibfnamefont {W.~N.}\ \bibnamefont {Faugno}}\ and\ \bibinfo {author} {\bibfnamefont {T.}~\bibnamefont {Ozawa}},\ }\bibfield  {title} {\bibinfo {title} {Interaction-{{Induced Non-Hermitian Topological Phases}} from a {{Dynamical Gauge Field}}},\ }\href {https://doi.org/10.1103/PhysRevLett.129.180401} {\bibfield  {journal} {\bibinfo  {journal} {Phys. Rev. Lett.}\ }\textbf {\bibinfo {volume} {129}},\ \bibinfo {pages} {180401} (\bibinfo {year} {2022})}\BibitemShut {NoStop}%
\bibitem [{\citenamefont {Brighi}\ \emph {et~al.}(2023)\citenamefont {Brighi}, \citenamefont {Ljubotina},\ and\ \citenamefont {Serbyn}}]{Brighi2023}%
  \BibitemOpen
  \bibfield  {author} {\bibinfo {author} {\bibfnamefont {P.}~\bibnamefont {Brighi}}, \bibinfo {author} {\bibfnamefont {M.}~\bibnamefont {Ljubotina}},\ and\ \bibinfo {author} {\bibfnamefont {M.}~\bibnamefont {Serbyn}},\ }\bibfield  {title} {\bibinfo {title} {Hilbert space fragmentation and slow dynamics in particle-conserving quantum {{East}} models},\ }\href {https://doi.org/10/g85hjc} {\bibfield  {journal} {\bibinfo  {journal} {SciPost Phys.}\ }\textbf {\bibinfo {volume} {15}},\ \bibinfo {pages} {093} (\bibinfo {year} {2023})}\BibitemShut {NoStop}%
\bibitem [{\citenamefont {Scherg}\ \emph {et~al.}(2021)\citenamefont {Scherg}, \citenamefont {Kohlert}, \citenamefont {Sala}, \citenamefont {Pollmann}, \citenamefont {Hebbe~Madhusudhana}, \citenamefont {Bloch},\ and\ \citenamefont {Aidelsburger}}]{Scherg2021}%
  \BibitemOpen
  \bibfield  {author} {\bibinfo {author} {\bibfnamefont {S.}~\bibnamefont {Scherg}}, \bibinfo {author} {\bibfnamefont {T.}~\bibnamefont {Kohlert}}, \bibinfo {author} {\bibfnamefont {P.}~\bibnamefont {Sala}}, \bibinfo {author} {\bibfnamefont {F.}~\bibnamefont {Pollmann}}, \bibinfo {author} {\bibfnamefont {B.}~\bibnamefont {Hebbe~Madhusudhana}}, \bibinfo {author} {\bibfnamefont {I.}~\bibnamefont {Bloch}},\ and\ \bibinfo {author} {\bibfnamefont {M.}~\bibnamefont {Aidelsburger}},\ }\bibfield  {title} {\bibinfo {title} {Observing non-ergodicity due to kinetic constraints in tilted {{Fermi-Hubbard}} chains},\ }\href {https://doi.org/10/gmvc6m} {\bibfield  {journal} {\bibinfo  {journal} {Nat. Commun.}\ }\textbf {\bibinfo {volume} {12}},\ \bibinfo {pages} {4490} (\bibinfo {year} {2021})}\BibitemShut {NoStop}%
\bibitem [{\citenamefont {Harshman}\ and\ \citenamefont {Knapp}(2022)}]{Harshman2022}%
  \BibitemOpen
  \bibfield  {author} {\bibinfo {author} {\bibfnamefont {N.~L.}\ \bibnamefont {Harshman}}\ and\ \bibinfo {author} {\bibfnamefont {A.~C.}\ \bibnamefont {Knapp}},\ }\bibfield  {title} {\bibinfo {title} {Topological exchange statistics in one dimension},\ }\href {https://doi.org/10.1103/PhysRevA.105.052214} {\bibfield  {journal} {\bibinfo  {journal} {Phys. Rev. At. Mol. Opt. Phys.}\ }\textbf {\bibinfo {volume} {105}},\ \bibinfo {pages} {052214} (\bibinfo {year} {2022})}\BibitemShut {NoStop}%
\bibitem [{\citenamefont {Weinberg}\ and\ \citenamefont {Bukov}(2017)}]{Weinberg2017}%
  \BibitemOpen
  \bibfield  {author} {\bibinfo {author} {\bibfnamefont {P.}~\bibnamefont {Weinberg}}\ and\ \bibinfo {author} {\bibfnamefont {M.}~\bibnamefont {Bukov}},\ }\bibfield  {title} {\bibinfo {title} {{{QuSpin}}: A {{Python}} package for dynamics and exact diagonalisation of quantum many body systems part {{I}}: Spin chains},\ }\href {https://doi.org/10.21468/SciPostPhys.2.1.003} {\bibfield  {journal} {\bibinfo  {journal} {SciPost Phys.}\ }\textbf {\bibinfo {volume} {2}},\ \bibinfo {pages} {003} (\bibinfo {year} {2017})}\BibitemShut {NoStop}%
\bibitem [{\citenamefont {Weinberg}\ and\ \citenamefont {Bukov}(2019)}]{Weinberg2019}%
  \BibitemOpen
  \bibfield  {author} {\bibinfo {author} {\bibfnamefont {P.}~\bibnamefont {Weinberg}}\ and\ \bibinfo {author} {\bibfnamefont {M.}~\bibnamefont {Bukov}},\ }\bibfield  {title} {\bibinfo {title} {{{QuSpin}}: A {{Python}} package for dynamics and exact diagonalisation of quantum many body systems. {{Part II}}: Bosons, fermions and higher spins},\ }\href {https://doi.org/10.21468/SciPostPhys.7.2.020} {\bibfield  {journal} {\bibinfo  {journal} {SciPost Phys.}\ }\textbf {\bibinfo {volume} {7}},\ \bibinfo {pages} {020} (\bibinfo {year} {2019})}\BibitemShut {NoStop}%
\bibitem [{Note5()}]{Note5}%
  \BibitemOpen
  \bibinfo {note} {In a finite lattice of $L$ sites with two particles, the allowed values are $n\in \{1,\protect \ldots ,L\}$ and $m\in \{0,\protect \ldots ,L-n\}$.}\BibitemShut {Stop}%
\bibitem [{\citenamefont {{Cohen-Tannoudji}}\ \emph {et~al.}(1998)\citenamefont {{Cohen-Tannoudji}}, \citenamefont {{Dupont-Roc}},\ and\ \citenamefont {Grynberg}}]{Cohen-Tannoudji1998}%
  \BibitemOpen
  \bibfield  {author} {\bibinfo {author} {\bibfnamefont {C.}~\bibnamefont {{Cohen-Tannoudji}}}, \bibinfo {author} {\bibfnamefont {J.}~\bibnamefont {{Dupont-Roc}}},\ and\ \bibinfo {author} {\bibfnamefont {G.}~\bibnamefont {Grynberg}},\ }\href {https://doi.org/10.1002/9783527617197} {\emph {\bibinfo {title} {Atom---{{Photon Interactions}}}}},\ \bibinfo {edition} {1st}\ ed.\ (\bibinfo  {publisher} {John Wiley \& Sons, Ltd},\ \bibinfo {year} {1998})\BibitemShut {NoStop}%
\bibitem [{\citenamefont {Fano}(1961)}]{Fano1961}%
  \BibitemOpen
  \bibfield  {author} {\bibinfo {author} {\bibfnamefont {U.}~\bibnamefont {Fano}},\ }\bibfield  {title} {\bibinfo {title} {Effects of {{Configuration Interaction}} on {{Intensities}} and {{Phase Shifts}}},\ }\href {https://doi.org/10.1103/PhysRev.124.1866} {\bibfield  {journal} {\bibinfo  {journal} {Phys. Rev.}\ }\textbf {\bibinfo {volume} {124}},\ \bibinfo {pages} {1866} (\bibinfo {year} {1961})}\BibitemShut {NoStop}%
\bibitem [{\citenamefont {Feshbach}(1958)}]{Feshbach1958}%
  \BibitemOpen
  \bibfield  {author} {\bibinfo {author} {\bibfnamefont {H.}~\bibnamefont {Feshbach}},\ }\bibfield  {title} {\bibinfo {title} {Unified theory of nuclear reactions},\ }\href {https://doi.org/10.1016/0003-4916(58)90007-1} {\bibfield  {journal} {\bibinfo  {journal} {Ann. Phys.}\ }\textbf {\bibinfo {volume} {5}},\ \bibinfo {pages} {357} (\bibinfo {year} {1958})}\BibitemShut {NoStop}%
\bibitem [{\citenamefont {Feshbach}(1962)}]{Feshbach1962}%
  \BibitemOpen
  \bibfield  {author} {\bibinfo {author} {\bibfnamefont {H.}~\bibnamefont {Feshbach}},\ }\bibfield  {title} {\bibinfo {title} {A unified theory of nuclear reactions. {{II}}},\ }\href {https://doi.org/10.1016/0003-4916(62)90221-X} {\bibfield  {journal} {\bibinfo  {journal} {Ann. Phys.}\ }\textbf {\bibinfo {volume} {19}},\ \bibinfo {pages} {287} (\bibinfo {year} {1962})}\BibitemShut {NoStop}%
\bibitem [{\citenamefont {Breit}\ and\ \citenamefont {Wigner}(1936)}]{Breit1936}%
  \BibitemOpen
  \bibfield  {author} {\bibinfo {author} {\bibfnamefont {G.}~\bibnamefont {Breit}}\ and\ \bibinfo {author} {\bibfnamefont {E.}~\bibnamefont {Wigner}},\ }\bibfield  {title} {\bibinfo {title} {Capture of {{Slow Neutrons}}},\ }\href {https://doi.org/10.1103/PhysRev.49.519} {\bibfield  {journal} {\bibinfo  {journal} {Phys. Rev.}\ }\textbf {\bibinfo {volume} {49}},\ \bibinfo {pages} {519} (\bibinfo {year} {1936})}\BibitemShut {NoStop}%
\bibitem [{Note6()}]{Note6}%
  \BibitemOpen
  \bibinfo {note} {In a finite lattice of $L$ sites with three particles, the allowed values are $n\in \{1,\protect \ldots ,L\}$, $m\in \{0,\protect \ldots ,L-n\}$, and $l\in \{0,\protect \ldots ,L-n-m\}$.}\BibitemShut {Stop}%
\bibitem [{Note7()}]{Note7}%
  \BibitemOpen
  \bibinfo {note} {Note that one further improve upon this by using optimal control techniques, as those in explored in~Refs.~\cite {Theel2025,Blatz2024}, to prepare an initial state with a larger overlap with the desired three-body bound states.}\BibitemShut {Stop}%
\bibitem [{Note8()}]{Note8}%
  \BibitemOpen
  \bibinfo {note} {We note that the attractive case $U=-4J$ of the spectrum can be obtained by a mirror reflection against the $E=0$ axis with simultaneous shift of each COM quasimomentum $Q \to (Q+3\pi ) \protect \bmod 2\pi $~\cite {Valiente2010a}.}\BibitemShut {Stop}%
\bibitem [{Note9()}]{Note9}%
  \BibitemOpen
  \bibinfo {note} {In a finite lattice of $L$ sites with four particles, the allowed values are $n\in \{1,\protect \ldots ,L\}$, $m\in \{0,\protect \ldots ,L-n\}$, $l\in \{0,\protect \ldots ,L-n-m\}$, and $k\in \{0,\protect \ldots ,L-n-m-l\}$}\BibitemShut {NoStop}%
\bibitem [{\citenamefont {Hardy}\ and\ \citenamefont {Ramanujan}(1918)}]{Hardy1918}%
  \BibitemOpen
  \bibfield  {author} {\bibinfo {author} {\bibfnamefont {G.~H.}\ \bibnamefont {Hardy}}\ and\ \bibinfo {author} {\bibfnamefont {S.}~\bibnamefont {Ramanujan}},\ }\bibfield  {title} {\bibinfo {title} {Asymptotic {{Formula\ae}} in {{Combinatory Analysis}}},\ }\href {https://doi.org/10.1112/plms/s2-17.1.75} {\bibfield  {journal} {\bibinfo  {journal} {Proc. Lond. Math. Soc.}\ }\textbf {\bibinfo {volume} {s2-17}},\ \bibinfo {pages} {75} (\bibinfo {year} {1918})}\BibitemShut {NoStop}%
\bibitem [{\citenamefont {Zhang}\ \emph {et~al.}(2023)\citenamefont {Zhang}, \citenamefont {Qian}, \citenamefont {Sun},\ and\ \citenamefont {Zhang}}]{Zhang2023a}%
  \BibitemOpen
  \bibfield  {author} {\bibinfo {author} {\bibfnamefont {W.}~\bibnamefont {Zhang}}, \bibinfo {author} {\bibfnamefont {L.}~\bibnamefont {Qian}}, \bibinfo {author} {\bibfnamefont {H.}~\bibnamefont {Sun}},\ and\ \bibinfo {author} {\bibfnamefont {X.}~\bibnamefont {Zhang}},\ }\bibfield  {title} {\bibinfo {title} {Anyonic bound states in the continuum},\ }\href {https://doi.org/10/gtzx3k} {\bibfield  {journal} {\bibinfo  {journal} {Commun. Phys.}\ }\textbf {\bibinfo {volume} {6}},\ \bibinfo {pages} {1} (\bibinfo {year} {2023})}\BibitemShut {NoStop}%
\bibitem [{\citenamefont {Zheng}\ \emph {et~al.}(2024)\citenamefont {Zheng}, \citenamefont {Xie}, \citenamefont {Zhang}, \citenamefont {Chen},\ and\ \citenamefont {Zhang}}]{Zheng2024}%
  \BibitemOpen
  \bibfield  {author} {\bibinfo {author} {\bibfnamefont {C.}~\bibnamefont {Zheng}}, \bibinfo {author} {\bibfnamefont {J.}~\bibnamefont {Xie}}, \bibinfo {author} {\bibfnamefont {M.}~\bibnamefont {Zhang}}, \bibinfo {author} {\bibfnamefont {Y.}~\bibnamefont {Chen}},\ and\ \bibinfo {author} {\bibfnamefont {Y.}~\bibnamefont {Zhang}},\ }\bibfield  {title} {\bibinfo {title} {Necessity of orthogonal basis vectors for the two-anyon problem in a one-dimensional lattice},\ }\href {https://doi.org/10/g9m4h7} {\bibfield  {journal} {\bibinfo  {journal} {Commun. Theor. Phys.}\ }\textbf {\bibinfo {volume} {76}},\ \bibinfo {pages} {125103} (\bibinfo {year} {2024})}\BibitemShut {NoStop}%
\bibitem [{\citenamefont {Blatz}\ \emph {et~al.}(2024)\citenamefont {Blatz}, \citenamefont {Kwan}, \citenamefont {L{\'e}onard},\ and\ \citenamefont {Bohrdt}}]{Blatz2024}%
  \BibitemOpen
  \bibfield  {author} {\bibinfo {author} {\bibfnamefont {T.}~\bibnamefont {Blatz}}, \bibinfo {author} {\bibfnamefont {J.}~\bibnamefont {Kwan}}, \bibinfo {author} {\bibfnamefont {J.}~\bibnamefont {L{\'e}onard}},\ and\ \bibinfo {author} {\bibfnamefont {A.}~\bibnamefont {Bohrdt}},\ }\bibfield  {title} {\bibinfo {title} {Bayesian {{Optimization}} for {{Robust State Preparation}} in {{Quantum Many-Body Systems}}},\ }\href {https://doi.org/10.22331/q-2024-06-27-1388} {\bibfield  {journal} {\bibinfo  {journal} {Quantum}\ }\textbf {\bibinfo {volume} {8}},\ \bibinfo {pages} {1388} (\bibinfo {year} {2024})}\BibitemShut {NoStop}%
\end{thebibliography}%
\end{document}